\documentclass[binding=0.6cm,oneside,noexaminfo]{sapthesis}
\usepackage{microtype}
\usepackage{hyperref}
\hypersetup{pdftitle={Numerical aspects of black hole superradiance},pdfauthor={Giuseppe Lingetti}}
\title{Numerical aspects of black hole superradiance}
\author{Giuseppe Lingetti}
\IDnumber{1555443}
\course{Dottorato di Ricerca in Fisica}
\courseorganizer{Scuola Dottorale in Scienze Astronomiche, Chimiche, Fisiche, Matematiche e della Terra "Vito Volterra"}
\cycle{XXXV}
\AcademicYear{2022/2023}
\advisor{Prof. Paolo Pani}
\authoremail{giuseppe.lingetti@gmail.com}
\copyyear{2023}
\thesistype{PhD thesis}

\usepackage[utf8]{inputenc}
\usepackage{graphicx}
\usepackage{amssymb}
\usepackage{amsmath}
\usepackage{cprotect}
\usepackage{subcaption}
\usepackage{wasysym}
\usepackage{hyperref}
\usepackage{cite}
\usepackage{blindtext}
\usepackage{mathtools}

\DeclarePairedDelimiter\floor{\lfloor}{\rfloor}
\numberwithin{equation}{section}
\DeclareUnicodeCharacter{2009}{\,}
\DeclareMathAlphabet\mathbfcal{OMS}{cmsy}{b}{n}
\usepackage{mdframed}
\usepackage{float}
\newmdtheoremenv{theo}{Algorithm}

\usepackage{microtype}

\usepackage{hyperref}

\graphicspath{{images/}}

\usepackage{dutchcal} 

\renewtagform{default}{\normalsize(}{)}
\usetagform{default}
\begin{document}
\frontmatter
\maketitle

\begin{acknowledgments}
    Working for the PhD has been more painful than I could ever expect. No one could conceive the possibility that a global catastrophe such as the Covid-19 Pandemic would hit all of us just a few months after enrolling. The first year of PhD has been truly dystopian: when the Italian government was not locking us in our houses I was locking myself voluntarily, pushed by fear and the perceived lack of safety of my office. The interplay between Covid Crisis and the chronic issues related with the under-staffed bureaucracy of our university\footnote{We are all extremely grateful to the Italian government for cutting our resources!} caused a huge delay in getting the required instruments for carrying computational research. As a matter of fact, I got my PC delivered by Physics Department eight months after my colleagues' - i.e. twelve months after my enrolling - because they "forgot"\footnote{Obviously, I express my full solidarity to the lone worker who has to carry \textbf{all the tasks} of the PhD Office of the Physics Department.}, which meant that in the first year I had insufficient computational power for testing the cumbersome codes required for computing everything you will read about in this thesis. The psychological trauma of the Covid Crisis has haunted me during all the PhD years, causing me retiring the participation to conferences in many occasions. The only conference located outside Rome I attended in person after the pandemic hit was the one during which I got infected by covid for the first time, causing six months of post-covid issues. The Italian government granted only three months of paid covid extension for the PhD students of my Cycle\footnote{We are all extremely grateful to the Italian government!}, while colleagues who started earlier got extensions up to one year. Consequently, in order to conclude the PhD with an acceptable thesis, I also had to apply for an extra non-paid extension, thus had to fund myself through unemployment benefits\footnote{Luckly, the \href{https://dottorato.it/}{Italian Doctorate Association (ADI)} struggled for making also PhD students access to these benefits.}. I think I could have produced better quality results if we had not had to face such difficult times, but at the same time I feel like managing to conclude the PhD despite all of this is a small miracle. In addition, I must say that having had a stable income from my fellowship during the worst moments of the Covid Crisis has been a huge luck\footnote{Fun fact: in normal conditions the pay PhD students get in Italy is considered a shame, given that it is below the average Italian salary.}, given that many people risked facing or had to face poverty as a direct consequence of the pandemic.

    Though, I would not be honest if I asserted that what went wrong was just covid. Personally, when I enrolled I was fully aware of the shameful precarity characterizing the academic career and had a clear picture of the disastrous state of the Italian universitary system. Older people had informed me about all of these issues and I had seen the huge mass protests by students and researchers back in 2008 and 2010, against the destruction of our school and universitary systems carried out by the government. But that was not enough for being seriously ready: when you feel all of it personally the perspective changes completely. In academia it is well known that globally there are way more people who want to and could do research than available positions, and finding funding has become quite problematic. Moreover, people know that in any country of the world they are forced to sign many precarious one-year or two-years contracts for many years and publish a lot of papers\footnote{The most used metric for evaluating somebody's research is the \href{https://theconversation.com/why-the-h-index-is-a-bogus-measure-of-academic-impact-141684}{controversial h-index}, giving the "impact factor" of researchers through the evaluation of the citations they got. My personal comment is that evaluating research by using the unit "kilograms of papers" would be almost equivalent, i.e. the scientific community is blindly promoting quantity over quality.} in order to survive and get a stable position someday (maybe). In Italy we have our specific issues, whose interplay with the global context rises the level of academic dysfunctionality\footnote{In many cases in our country the condition of academia could be labelled as "late-feudal".}, but today's situation of academia is worrying worldwide. What I felt during my PhD is that there is not enough awareness about the consequences of how the academic system works today, especially about the impact on the scientific quality itself and how people live the overall situation.
    
    Firstly, the way funding is assigned has became corporate-like and thus Principal Investigators (PIs)\footnote{PIs are the senior researchers who lead scientific projects.} are forced to spend half of their time in the quest for getting money for their research projects by preparing countless applications, making them more akin to contemporary entrepreneurs than mid-20th century academics\footnote{Obviously academy has been changing with time in accordance to how society itself evolves, but change is not necessarily positive by itself.}. Consequently on average todays' PIs materially do much less research than in the past: most of the work is done by junior profiles hired through precarious contracts - i.e. mainly post-docs / temporary research fellows and in many cases also PhD students - which are therefore exploited in a sort of "neo-fordist" fashion. In Italy we have had cases of academics being able to publish one paper per week\footnote{I suggest checking the publications of Prof. Cingolani and Prof. Schillaci, just to give some famous "Italian-style" examples.}, which would not be possible without extreme levels of labour exploitation. Secondly, the gradual rise of peer review in the last 70 years in addition with the recent transformation of scientific journals into their online versions have made academic publishers accumulate huge power which distorts how academia works, heavily influencing how research is organized. The "imprimatur-like" power they get from peer review\footnote{Many people - both inside and outside academia - truly believe that peer review is what "certifies" if a paper is scientific or not. On the contrary, scientific truth has always been the product of the debate in the scientific community via accumulation of multiple and differing papers over time and discussion in conferences/meetings.}, the substitution of printed journals with online ones, the over-production of papers generated by "publish or perish" and extreme competition between researchers have been offering \href{https://www.theguardian.com/science/2017/jun/27/profitable-business-scientific-publishing-bad-for-science}{huge profit margins to academic publishers}, \href{https://tidsskriftet.no/en/2020/08/kronikk/money-behind-academic-publishing}{rivaling the ones of high-tech corporations such as Google or Microsoft}\footnote{Academic publishers can be considered the final and most powerful exploiters of academic labour, filling a role akin to the ones bankers have in "standard" capitalist accumulation. As a matter of fact, they make profits through the mechanisms of circulation of academic research, while in the financial system profit is made through the circulation of capitals.}. The most evident consequence on scientific production you get from the mechanisms described here is that research on average becomes more short-sighted, because the price of risk in an extremely competitive and marketized context can become very high. Moreover, from the specific point of view of theoretical research - fundamental research in particular - the proper intellectual activity one would expect in such academic fields is mostly sterilized: the activity of researchers is mostly reduced to technical studies or making extremely small gradual improvements, without enough questioning about where we are going and why\footnote{Recently some researchers claimed that \href{https://www.science.org/content/blog-post/decline-scientific-innovation}{scientific innovation has been declining} since the second half of the 20th century. I would not be surprised if what they claim is the global effect of what I am denouncing here.}. When I decided to become a theoretical physicist I was expecting to become an intellectual, but I found myself confined in doing extremely technical work. The same happened to many colleagues and can be easily checked by reading the pre-prints published every day on arxiv. When we start a PhD we do so because we would like to become academics. If we wanted to become either entrepreneurs or technicians we would have taken some other path, maybe less  riskful for our financial stability and our future perspectives. We obviously need to carry also technical work or be able to manage projects, but at the end it all should be finalized for carrying research in the best possible way while making researchers live a decent life: this is not happening at all. What is driving how the academic community organizes its work nowadays does not arise from scientific needs and is not scientifically motivated at all, it is just the product of violent capitalist subsumption. Moreover, given that making somebody reach the education level of a PhD has huge collective costs and that the striking majority of PhDs will be forced to leave research, academia reveals to be extremely inefficient in managing human resources even from a corporate-like point of view\footnote{One might assert that a PhD can help you get better jobs or can in general be useful for many other things other than research. Personally I think that in most cases it is useless outside academia, especially in a country like Italy which fails completely in investing in research and development.}. From the point of view of a young PhD student, full of enthusiasm and commitment towards the exploration of the unknown, the pressure coming from the perverse mechanisms I described here can be devastating and necessarily tend to produce disillusionment. Why would someone aspiring to be an intellectual accept to work in terrible conditions in academia as a sort of technician for little money, while outside academia you can get the same type of jobs but with better pays and conditions? Personally I am happy about having had the possibility of satisfying my scientific curiosity through a PhD despite all the problems I have exposed, but the price I paid is high and I see little life inside academia.

    I would like to thank all the people who supported and helped me in facing all the complex difficulties of these years of PhD. First of all my family, who always supported my choices and materially gave me resources which - unfortunately - are not always available to everybody in the unjust society we live in. I thank all the friends and colleagues in the Physics Department who have been pivotal in giving me moral support and ideas for the scientific activity, i.e. all the nice people in "$\pi$ room" and in the "$G_{\mu\nu}$" theoretical gravity group, but also all the nice people I have met since the bachelor. Finally, I would like to thank the comrades and collectives that are still struggling in Sapienza for a better society, who give me hope despite all the darkness we see around.
    
\end{acknowledgments}

\begin{abstract}
In this work we explore a numerical technique, based on the spherical harmonic decomposition and the discretization of the radial coordinate through Čebyšëv polynomial interpolation, for the computation of quasi-bound states of linear massive scalar and vector perturbations in spinning black hole spacetimes in General Relativity. The aim is studying black hole superradiant instabilities, an energy-extraction mechanism triggered by the presence of massive bosonic fields near black holes, which finds wide applications in constraining scenarios beyond Standard Model and General Relativity. This method does not rely on any separation ansätze, thus it can have wide applications. Consequently we extend the technique so that it can be applied also to the computation of massive tensor quasi-bound states in spinning black holes in General Relativity, whose separability ansatz is currently unknown. We also apply it to spinning black holes in scalar-tensor theory non-linearly interacting with plasma, wherein the massless scalar perturbations acquires an effective mass, finding a novel way for constraining scalar-tensor theories. 
\end{abstract}

\tableofcontents

\chapter*{Abbreviations}
ADAF = Advection-Dominated Accretion Flow\\
ALPs = Axion-Like Particles\\
BH(s) = black hole(s)\\
BD ghost = Boulware-Deser ghost\\
DECIGO = Deci-hertz Interferometer Gravitational wave Observatory\\
dRGT theory = de-Rham-Gabadadze-Tolley theory of massive gravity\\
EEP = Einstein Equivalence Principle\\
EFT(s) = Effective Field Theor(y/ies)\\
EH = Einstein-Hilbert\\
GR = General Relativity\\
GW(s) = gravitational wave(s)\\
ISCO = Innermost Stable Circular Orbit\\
LIGO = Laser Interferometer Gravitational-Wave Observatory\\
LISA = Laser Interferometer Space Antenna\\
NP = Newman-Penrose\\
QBS(s) = quasi-bound State(s)\\
QCD = Quantum Chromodynamics\\
QNM(s) = quasi-normal mode(s)\\
SEP = Strong Equivalence Principle\\
SM = Standard Model\\
SMBH(s) = super-massive black hole(s)\\
vDVZ discontinuity = van-Dam-Veltmann-Zakharov discontinuity\\
WEP = Weak Equivalence Principle\\
WKB = Wentzel–Kramers–Brillouin\\
ZAMO = Zero Angular Momentum Observer\\

\mainmatter

\chapter*{Introduction}
\addcontentsline{toc}{chapter}{Introduction}

Since their gradual theoretical discovery in the early XX century\footnote{Some key moments: the discovery of Schwarzschild's metric in 1915\cite{schwarzschild1999gravitational,Wald_General_relativity,carroll_spacetime_geometry,Ferrari_GR}, understanding in 1958 what the event horizon is and implies\cite{event-horizon_discovery}, Kerr's breakthrough about spinning black holes in 1963\cite{Kerr_1963}.}, black holes (BHs) have fascinated humanity due to their mysterious "exotic" nature and the theoretical paradoxes they bring. They are a direct theoretical prediction of General Relativity (GR) but at the same time they suggest the breakdown of our standard gravitational theory in their inside due to the presence of singularities\cite{Penrose_grav_collapse, Wald_General_relativity,carroll_spacetime_geometry,Ferrari_GR}, hence stimulating us in the continuous quest for expanding our knowledge of gravity. Because of their extreme gravitational and astrophysical nature, they are also macroscopic windows to the microscopic physical world, thus connecting astrophysics and particle physics\cite{Brito_evolution_SR_instabilities}.

In this work we explore and extend a numerical technique, first applied in \cite{Spectra_grav_atom} by \textit{Baumann et al.}, for the computation of quasi-bound states (QBSs) of massive bosonic perturbations propagating in spinning BH spacetimes, i.e. (classical) states describing integer-spin particles confined by gravity in the proximity of BHs. The aim is gathering valid mathematical and numerical instruments to be applied to the study of what can be considered one of the most interesting and investigated BH phenomena, superradiance, which is expected to play a key role in extending our knowledge about gravity and possible new interactions to be discovered.

In the context of BH theory, superradiance is an energy-extraction process which, under specific conditions, can be triggered by the presence of some ultralight-mass bosonic particles\cite{Brito_SR} in the proximity of BHs. In Chapter \ref{chapter:BH_superradiance}, after recalling some basic knowledge about BHs and BH perturbation theory in GR, we introduce spinning BH superradiance is, how it works, its key equations and what it implies. We show how this phenomenon is closely related to detecting hypothetical ultralight bosons through BH physics, thus in Chapter \ref{chapter:ultralight_bosons} we introduce what ultralight bosons are, where they come from in models beyond Standard Model (SM), and their role as dark matter candidates. In particular, we focus on their mathematical formulation from a classical field theory point of view, thus reviewing massive scalar, vector and tensor field theory, and sketch the main observational consequences they bring.

In Chapter \ref{chapter:spherical_harmonic} we start diving into the technique developed in \cite{Spectra_grav_atom}: we introduce the formalism applied by \textit{Baumann et al.} for the spherical harmonic decomposition of massive scalar and vector fields in GR spinning BHs and show how it can be extended also to the tensor case. In that chapter we show our first result, i.e. a complete and consistent fully general-relativistic framework for the multipolar expansion of massive fields up to the tensor case. Before \textit{Baumann et al.}'s work in \cite{Spectra_grav_atom} the only framework known for spherical harmonic decompositions was the one working in the weak-field limit developed by Thorne\cite{Thorne_multipole_expansions,Maggiore_GWs:vol_1}. In the general-relativistic case, instead, massless fields can be treated through Teukolsky's formalism\cite{Teukolsky_1972,Teukolsky_perturb_1,Teukolsky_perturb_2,Teukolsky_perturb_3}\footnote{Though this formalism works also for massive Klein-Gordon fields.}, while the massive vector fields can be managed through the formalism recently discovered by \textit{Frolov et al.}\cite{Spin-1_separability}. In Chapter \ref{chapter:spherical_harmonic} we also show a parametric technique we developed for the computation of the couplings among the spherical harmonics modes arising from the decomposition\cite{Lingetti_spherical_overlaps}, thus optimizing the computation time.

The spherical harmonic decomposition transforms the field equations describing the perturbative problem into an infinite cascade of radial equations: in Chapter \ref{chapter:numerical_SR_GR} we show how \textit{Baumann et al.} in \cite{Spectra_grav_atom} transformed them in a matrix problem through radial discretization via Čebyšëv polynomial interpolation. In that context we show the extension to the tensor case of the massive vector ansatz described in \cite{Spectra_grav_atom}, in order to apply it to the computation of massive tensor BH perturbations in GR. All the framework developed is therefore applied to the numerical computation of QBSs: we successfully reproduced already known results about massive scalar and vector perturbations, hence also confirming the validity of the parametrized approach for the computation of spherical harmonic couplings. The computation of massive tensor perturbations, instead, has not been completed yet, consequently we cannot show complete results, though we checked the consistency of the technique by reproducing a non-spinning specific mode. Although, we include the results got through an alternative technique developed by collaborators of ours in \cite{dias2023black}, which we expect to confirm and extend soon through our method. The full general-relativistic computation of massive tensor perturbations of GR spinning BHs has been an open problem for a long time, but now we can finally say it is basically closed.

In the last chapter, i.e. \ref{chapter:plasma_SR}, we apply the numerical technique by \textit{Baumann et al.} to BH superradiance triggered by non-linear interactions with plasma in a beyond-GR scenario, specifically in the context of scalar-tensor theory\cite{BH_scalar_tensor, scalar-tensor_theory_GWs,Brans-Dicke,Faraoni:2004pi}. There are many theoretical and observational problems which push us in searching for some models able to extend GR\cite{testing_GR,modified_gravity,Horndeski_modified_gravity}, among which the simplest proposals are the scalar-tensor theories. In Chapter \ref{chapter:plasma_SR} we compute scalar QBSs in scalar-tensor models, showing how superradiance can be used for constraining scalar-tensor extensions to GR\cite{Scalar_plasma_SR}. In this case the scalar field is massless and the mass is an effective term arising from the non-linear interactions between the extra scalar field and the matter surrounding the compact object.

The developments exposed in this work have been published in the following papers:
\begin{itemize}
    \item{G. Lingetti and P. Pani, “General spherical harmonic bra-ket overlap
    integrals of trigonometric functions,” \href{https://dx.doi.org/10.1088/1361-6382/acb880}{\textit{Classical and Quantum Gravity 40
    no. 5, (Feb, 2023) 057001.}}}
    \item{O. J. C. Dias, G. Lingetti, P. Pani, and J. E. Santos, “Black hole superradiant instability for massive spin-2 fields,” \href{https://doi.org/10.1103/PhysRevD.108.L041502}{\textit{Phys. Rev. D 108 (August 2023) L041502.}}}
    \item{G. Lingetti, E. Cannizzaro, and P. Pani, “Superradiant instabilities by
accretion disks in scalar-tensor theories,” \href{https://link.aps.org/doi/10.1103/PhysRevD.106.024007}{\textit{Phys. Rev. D 106 (Jul, 2022)
024007.}}}
\end{itemize}

\chapter{Superradiance in spinning black holes}\label{chapter:BH_superradiance}

Superradiance is a radiation amplification mechanism involving dissipative systems, which plays an important role in many fields of physics, like quantum mechanics, optics, astrophysics and GR\cite{Brito_SR}. This phenomenon consists in the extraction of energy from a system through the interaction with some radiation mediated by a dissipative mechanism, with the latter causing the energy transfer from the system by amplifying the interacting radiation.

In the context of GR, superradiance plays a prominent role in BH physics. What is peculiar about BH superradiance is its process of energy extraction from vacuum, while in other classical physics contexts the dissipative mechanism involves some material medium\cite{Brito_SR}. This is possible because of the presence of a region near spinning and charged BHs which allows for classical negative energy states, the ergoregion, while the BH event horizon acts as a viscous one-way membrane\cite{membrane_paradigm} dumping these negative states. In the context of GR, BH superradiance is possible when the compact object is spinning and/or electrically charged, thus it consists in the extraction of angular momentum/electric charge. This is closely related with the laws of BH mechanics, in particular with the BH horizon area law, which at the classical level allow the extraction of angular and charge energy only\cite{Wald_BH_thermo,Damour_BH_entropy,Bekenstein_BH_entropy,Page_BH_thermodynamics}. From a phenomenological point of view, though, astrophysical BHs are expected to have negligible electric charge \cite{Gibbons_neutral_BHs,Blandford_EM_extraction,Hanni_limits_charge,Eardley_BH_processes,Gong_BH_neutralization,Barausse_environmental_effects} due to effects such as charge neutralization by astrophysical plasma, quantum discharge, electron-positron pair production, thus in this text we will focus only on neutral spinning BHs. The radiation involved in BH superradiance is always bosonic, due to fermions not being able to trigger energy extraction\cite{Iyer_no_fermi_superradiance,Martellini_no_fermi_superradiance}. The reason is the following: fermions have positive definite current densities and bounded transmission amplitudes when scattered on BHs, while bosons can have negative current densities and transmission amplitudes. Negative energy states are pivotal in making BH superradiance possible, thus only bosonic fields are able to trigger the extraction of energy.

Through some confinement mechanism, e.g. produced by mass terms, non-linear interactions or anti-de Sitter boundaries, the radiation amplified by a BH can be trapped, thus getting continuously enhanced and consequently giving rise to instabilities \cite{Press_BH_bomb,Damour_BH_instability,Detweiler_scalar_instability,Dolan_spin-1_instability,Brito_magnetic_SR,Cardoso_AdS-Kerr_unstable,Cardoso_superradiant_instability,Brito_spin-2_SR_slow_rot,Brito_hydrogenic_spin-2_SR,Spectra_grav_atom, Brito_SR}. BH superradiance is triggered only when the radiation scattered off the BH satisfies a specific mathematical relationship, called superradiant condition, involving the frequency of the scattered wave, its azimuthal number and, in the case of spinning BHs, the angular momentum of the compact object. Thus, when superradiance triggers an instability, energy is extracted from the compact object while the superradiant condition is met, therefore causing the condensation of a bosonic cloud enveloping the BH \cite{East_non-linear_SR_1,East_non-linear_SR_2,Herdeiro_Kerr_Hair,Yoshino_axion_cloud_SR,Brito_evolution_SR_instabilities}. While the process is active the frequency window allowing superradiance gradually becomes smaller, because of the on-going reduction of energy that can be extracted. Consequently superradiance is suddenly shut off and the instability is arrested when the maximum possible frequency for energy extraction becomes smaller than the frequency of the radiation. After this first phase, the cloud evolves towards a stationary state: the bosonic condensate in fact features time-dependent multipolar moments causing the emission of gravitational radiation, therefore the cloud loses energy through continuous emissions until reaching equilibrium\cite{East_non-linear_SR_1,East_non-linear_SR_2,Herdeiro_Kerr_Hair,Yoshino_axion_cloud_SR}. The end state is expected to be a hairy black hole in most of the cases, i.e. a BH featuring a bosonic charge \cite{Herdeiro_Kerr_Hair,massive_gravity_hairy_BHs}.

BH superradiance offers a rich phenomenology potentially having astrophysical impact, ranging from gaps in the BH angular momentum\cite{Constraint_bosons_BH_spin,BH_spin_constraint_axion,Brito_hydrogenic_spin-2_SR} and signatures in BH binaries\cite{Binary_BHs_bosons_1,Binary_BHs_bosons_2,Binary_BHs_bosons_3} to continuous gravitational emissions and stochastic gravitational wave backgrounds\cite{GWs_signatures,Stochastic_GW_1,Stochastic_GW_2,GW_searches_ultralight_bosons}. The numerical study of such phenomenology can give the theoretical instruments for designing methods for testing the existence of exotic bosons through astrophysical observations\cite{Ultralight_vector_SR_signatures,Constraint_bosons_BH_spin,BH_spin_constraint_axion,Cardoso_constrsaint_dark_photon_SR,EHT_constraint_DM,GW_searches_ultralight_bosons,Brito_hydrogenic_spin-2_SR,discovering_axion_BHs}, thus constraining existing proposals of beyond BH particles or extensions of GR.

In this chapter BH superradiance is introduced, focusing on the instability phase in spinning BHs in GR, treated through perturbation theory.

\section{Kerr black holes}

GR is described by the Einstein-Hilbert action \cite{Hilbert_1915,carroll_spacetime_geometry,Wald_General_relativity,Ferrari_GR}, here expressed in geometrized $G=c=1$ units:
\begin{equation}\label{Einstein-Hilbert}
S_{EH}[g_{\mu\nu},\Psi]=\frac{1}{16\pi}\int d^4 x \sqrt{-g} R+S_m[g_{\mu\nu},\Psi]\,,
\end{equation}
where $g_{\mu\nu}$ is the metric tensor, $g$ is its determinant, $R$ is the Ricci scalar curvature, $\Psi$ is a generic field describing matter/energy coupled with gravity and $S_m[g_{\mu\nu},\Phi]$ is the action of the non-gravitational sector. By varying this action we get the Einstein Field Equations \footnote{No doubt these are the most beautiful equations in theoretical physics.}, describing the dynamics of gravity in GR\cite{Einstein_foundation_GR,carroll_spacetime_geometry,Wald_General_relativity,Ferrari_GR},
\begin{equation}
R_{\mu\nu}-\frac{1}{2}g_{\mu\nu} R=8\pi T_{\mu\nu}
\end{equation}
where $R_{\mu\nu}$ is the Ricci curvature tensor and $T_{\mu\nu}$ is the stress-energy tensor arising from $S_m[g_{\mu\nu},\Phi]$, which describes the energy/matter sourcing gravity.
Spinning BHs in GR are described by the Kerr solution\cite{Kerr_1963,Kerr_BH_monograph,Chandrasekhar1983} to the vacuum Einstein Field Equations (i.e. $R_{\mu\nu}=0$):
\begin{equation}\label{Kerr_metric}
\small ds^2=\frac{\Delta(r)}{\Sigma(r,\theta)}(dt-a \sin^2\theta d\phi)^2-\Sigma(r,\theta)\left[\frac{dr^2}{\Delta(r)}+d\theta^2\right]-\frac{\sin^2\theta}{\Sigma(r,\theta)}[a~ dt-(r^2+a^2) d\phi]^2\,,
\end{equation}
\[\Delta(r)=r^2-2  M r+a^2~,~ \Sigma(r,\theta)=r^2+a^2 \cos^2\theta\]
where $ds$ is the differential of spacetime distance, $x^\mu=(t,r,\theta,\phi)$ are the  Boyer-Lindquist coordinates, $a$ is the BH spin parameter, related with the angular momentum $J=a M$, and $M$ is the mass of the compact object\footnote{In all the text we will always use metric signature $(+,-,-,-)$.}. BHs can also feature an electric charge: the Kerr metric, in fact, belongs to a wider family of spacetimes, described by the Kerr-Newman solution to the equations of GR coupled with electromagnetism in curved spacetime (i.e. Einstein-Maxwell theory), describing electrically charged spinning BHs\cite{Kerr-Newman_BH,Chandrasekhar1983}. In the 1960s Hawking, Israel and Carter discovered that BHs in GR are uniquely described by the Kerr-Newman solution (Uniqueness Theorems)\cite{Bekenstein_no-hair,Carter_no-hair,Chrusciel_no-hair,Robinson_no-hair,Cardoso_no-hair}, i.e. mass, angular momentum and electric charge uniquely define a BH ("No-hair theorem")\footnote{Research on BHs beyond GR and SM try to challenge the "No-hair Theorem" by working on extensions which introduce new "hair" for BHs\cite{Cardoso_no-hair}.}. 

Despite not explicitly depending on $t$, i.e. stationarity, the Kerr metric is not static because it features non-null $t-\phi$ components that cannot be made vanishing through coordinate transformation. If we compare the Kerr metric with the non-spinning case ($a=0$, thus Schwarzschild metric), the rotation breaks the spherical simmetry characterizing non-rotating BHs, reducing it to axisymmetry in the $a \neq 0$ case (due to the metric not explicitly depending on $\phi$). The invariance under simultaneous $t$ and $\phi$ reversals demonstrates how a time reversal is associated with changing the direction of rotation. For the same reason the metric is also invariant under the trasformation ($a\rightarrow -a$ , $\phi\rightarrow -\phi$), thus we can consider only positive values for $a$ without any loss of generality.

The non-staticity of the spacetime becomes more clear if we focus on what happens on observers travelling near a spinning BH. If we consider a zero angular momentum observer (ZAMO) moving with time-like four-velocity $v^\mu$, we will have $L=v_\phi=v^\phi g_{\phi\phi}+v^t g_{t\phi}=0$ \footnote{In a geodesic motion in Kerr metric, $L=v_\phi$ is the conserved quantity associated to invariance under rotations in the $\phi$ angle. If $v^\mu$ is time-like, $L$ is the angular momentum per unit mass. When we consider a photon, instead, we cannot use the proper time $\tau$ or the spacetime differential displacement $ds$, thus we define some $u^\mu=\frac{dx^\mu}{d\lambda}$ for a given choice of the parameter $\lambda$, having $u_\mu u^\mu=0$. In this case the conserved quantity $u_\phi$ is proportional to the angular momentum of the photon and can be made equal to it if for a proper choice of the parameter $\lambda$.}, thus $v^\phi=-\frac{g_{t\phi}}{g_{\phi\phi}}v^t\neq 0$. At infinity $v^\phi\rightarrow 0$, therefore when the observer gets closer to the BH it acquires angular velocity, defined as $\Omega=\frac{d\phi}{dt}=\frac{v^\phi}{v^t}=-\frac{g_{t\phi}}{g_{\phi\phi}}$. This phenomenon is called frame-dragging, which means that the spacetime is "rotating" because of the BH spin.

For $\Sigma(r,\theta)=0$, thus for $r=0$ and $\theta=\pi/2$, the Kretschmann invariant $R_{\alpha\beta\mu\nu}R^{\alpha\beta\mu\nu}$ is singular, i.e. the metric shows a ring-shaped curvature singularity. The disk which is bordered by the ring singularity, instead, which is defined by $r=0$ and $0<\theta<\pi$, is a coordinate singularity \footnote{Boyer-Lindquist coordinates are ellipsoidal, thus near the center of the metric they behave very differently with respect to spherical coordinates. In fact surfaces of the type $r=const$ are ellipsoids, consequently the surface $r=0$ is a "degenerate ellipsoid", i.e. a disk.}. By setting $\Delta(r)=0$, for $0 \leq a \leq M$ we find the inner and outer event horizons of spinning BHs, where the metric shows a coordinate singularity,
\begin{equation}
r_\pm=M\pm\sqrt{M^2 -a^2}\,,
\end{equation}
while for $a>M$ we get a naked singularity. For $a=M$ the horizons $r_+$ and $r_-$ coincide and we get an extremal BH, while for $a=0$ the radius of the outer horizon becomes the Schwarzschild radius $r_s=2M$ and the inner horizon collapses into the curvature singularity at $r=0$. We can consider the outer horizon as the "surface" of the BH, thus the angular velocity $\Omega_H$ of the compact object is defined as the angular velocity of a ZAMO computed at $r=r_+$:
\begin{equation}\label{BH_angular_velocity}
\Omega_H=-\left.\frac{g_{t\phi}}{g_{\phi\phi}}\right|_{r=r_+}=\frac{a }{2 M r_+}
\end{equation}

Outside the event horizon $r_+$, Kerr BHs feature an ellipsoidal surface, called ergosphere, where $g_{00}$ changes its sign, which is distinct from the event horizon only for $a>0$:
\begin{equation}
r_{ergo}(\theta)=M+\sqrt{M^2 -a^2\cos^2(\theta)}\,,
\end{equation}
The volume $r_+<r<r_{ergo}(\theta)$ is the ergoregion: observers in this volume are allowed to acquire also negative energies, but they cannot be static due to four-vectors proportional to $(1,0,0,0)$ being space-like inside the ergoregion. In the next sections we will see the role played by the ergoregion and the event horizon in BH energy extraction mechanisms like superradiance.

\section{The Penrose process}

Some years before the discovery of BH superradiance, Roger Penrose in 1969 noticed how the presence of an event horizon and the possibility of having negative energy classical states, due to the ergoregion, allow mechanisms of energy extraction from BHs\cite{Penrose_grav_collapse,Penrose_extraction_energy,Brito_SR}. He considered the following "thought experiment", involving a particle following a geodesic trajectory in the Kerr metric before decaying into two particles while travelling inside the ergoregion.

Point particle's free motion in GR is described through the geodesic equation\cite{Einstein_foundation_GR,carroll_spacetime_geometry,Wald_General_relativity,Ferrari_GR},
\begin{equation}\label{geodesic_eq}
\frac{d^2 x^\sigma}{d\tau^2}+\Gamma^\sigma_{~\mu\nu}\frac{dx^\mu}{d\tau}\frac{dx^\nu}{d\tau}=0\,,
\end{equation}
where 
\begin{equation}
    \Gamma^\sigma_{~\mu\nu}=\frac{1}{2}g^{\sigma\beta}\left(\partial_\mu g_{\beta\nu}+\partial_\nu g_{\beta\mu}-\partial_\beta g_{\mu\nu} \right)
\end{equation}
is the Levi-Civita connection of the spacetime. Equation \ref{geodesic_eq} can arise from the following lagrangian,
\begin{equation}
\mathcal{L}=\frac{1}{2} \mu_0 g_{\mu\nu}\frac{dx^\mu}{d\tau}\frac{dx^\nu}{d\tau}\,,
\end{equation}
where $\mu_0$ is some parameter having the dimension of a mass. If we are a considering a particle having mass, $\mu_0$ can  be its mass and consequently $\tau$ becomes its proper time. By using this lagrangian, in the equatorial plane we can find the following equations for geodesic motion of particles with mass in Kerr spacetime\cite{Bardeen_rotating_BHs,Chandrasekhar1983,carroll_spacetime_geometry,Wald_General_relativity,Ferrari_GR}:
\begin{subequations}
\begin{align}
 &\frac{dt}{d\tau}=\frac{1}{\Delta(r)}\left[\left(r^2+a^2+\frac{2 a^2 M}{r}\right)E-\frac{2 a M}{r}L \right]\,,\\
&\frac{d\phi}{d\tau}=\frac{1}{\Delta(r)}\left[\frac{2 a M}{r}E+\left(1-\frac{2 M}{r}\right)L \right]\,,\\
&\label{radial_eq}
\left(\frac{dr}{d\tau} \right)^2=E^2+\frac{2 M}{r^3}\left(a E-L\right)^2+\frac{1}{r^2}\left.-(a^2 E^2-L^2\right)-\frac{\Delta(r)}{r^2}\,,
\end{align}
\end{subequations}
where $L=g_{\phi\mu}\frac{dx^\mu}{d\tau}$ and $E=g_{t\mu}\frac{dx^\mu}{d\tau}$ respectevely are the conserved angular momentum and energy per unit mass parameters.

The particle considered by Penrose has mass $\mu_0$ and is at rest at infinity, i.e. its energy parameter must be $E_0=\frac{\mathcal{E}_0}{\mu_0}=1$, where $\mathcal{E}_0$ is its energy. If we set $r=r_0$ as the minimum $r$ reached by the particle before decaying, $\frac{dr}{d\tau}$ must be null at that radial position, thus from the condition $\left.\frac{dr}{d\tau}\right|_{r=r_0}=0$ and \ref{radial_eq} we get
\begin{equation}\label{eq_turnpoint}
L=\frac{-2 a M E \pm\sqrt{r_0\Delta(r)\left(2M+r E^2-r \right)}}{r_0-2M}\,.
\end{equation}
From the decay we will get two particles having mass $\mu_f$, energies $\mathcal{E}_1=\mu_f E_1$ and $\mathcal{E}_2=\mu_f E_2$, angular momenta $\mathcal{L}_1=\mu_f L_1$ and $\mathcal{L}_2=\mu_f L_2$. By imposing the conservation of total energy and total angular momentum and applying equation \ref{eq_turnpoint}, we get the energy for the products\cite{Brito_SR}:
\begin{equation}
\mathcal{E}_1=\frac{\mu_0}{2}\left(1\pm\sqrt{\frac{2 M\left(1-\frac{4\mu_f^2}{\mu_0^2}\right)}{r_0}} \right)~,~\mathcal{E}_2=\frac{\mu_0}{2}\left(1\mp\sqrt{\frac{2 M\left(1-\frac{4\mu_f^2}{\mu_0^2}\right)}{r_0}} \right)
\end{equation}
This result shows how one of the two resulting particles will have energy higher than the one of the starting particle coming from infinity if $r_0<2M \left(1-\frac{4\mu_f^2}{\mu_0^2}\right)$ and negative energies are allowed, i.e. if the decay happens in the ergoregion. If the decay happens at the horizon we get the maximum gain of energy, while in the case with $\mu_f=0$ we get maximum efficiency and the photon having negative energy is doomed to fall into the BH \footnote{Roger Penrose originally considered this case.}\cite{Penrose_extraction_energy,Brito_SR,orbits_ergosphere}. By absorbing the negative-energy particle, the BH loses part of its rotational energy, which is emitted through the other particle escaping at infinity.

What happens with the Penrose process gives us a first picture of what role is played by the ergoregion and the event horizon in energy extraction processes. The ergoregion acts like a material medium giving friction by allowing negative energy states, consequently it is technically sufficient for the extraction process. In fact the extraction of energy happens even when the negative energy particle does not cross the event horizon, thus it is possible even if the event horizon is not present, i.e. in the case of exotic compact objects. What the event horizon does, instead, is just dumping the negative energy states, therefore having a stabilizing effect on the system.

\section{Massless bosonic perturbations of Kerr black holes}\label{Mass-less_bosonic_BH_perturbs}

In this section we briefly introduce how first order massless bosonic perturbations of Kerr BHs are treated, so that we have enough instruments for introducing the superradiant scattering in the next section.

In GR the dynamics of non-gravitational fields in curved spacetime is determined by the following coupling prescription, which transforms special-relativistic field theory's lagrangians/actions into ones featuring general covariance. Given a generic special-relativistic field $\Psi$, we get its generalization in curved spacetime by substituting in its action flat spacetime $\eta_{\mu\nu}$ with $g_{\mu\nu}$, standard derivatives $\partial_\mu$ with covariant ones $\nabla_\mu$ and the volume element $d^4 x$ with its covariant counterpart $\sqrt{-g} d^4 x$\cite{carroll_spacetime_geometry,Wald_General_relativity,Ferrari_GR}. Thus, if its lagrangian density in special-relativistic field theory is $\mathcal{L}=\mathcal{L}(\Psi,\partial_\mu\Psi,\eta_{\mu\nu})$, its general-relativistic action will be
\begin{equation}
S_m[\Psi,g_{\mu\nu}]=\int d^4 x \sqrt{-g} \mathcal{L}(\Psi,\nabla_\mu\Psi,g_{\mu\nu})\,.
\end{equation}
Therefore, by following this coupling prescription we can easily get the Einstein-Hilbert action for the specific case of gravitation coupled with a massless scalar and a U(1) gauge vector\footnote{We are keeping it generic, but this can be the electromagnetic radiation.} fields:
\begin{equation}\label{EH_bosonic_action}
S_{EH}[g_{\mu\nu}, A_{\mu}, \Phi]=\int d^4 x \sqrt{-g} \left(\frac{1}{16\pi}R-\frac{1}{4}F_{\mu\nu}F^{\mu\nu}-\frac{1}{2}\nabla^\mu \Phi \nabla_\mu \Phi \right)\,,
\end{equation}
where $F_{\mu\nu}=\partial_\mu A_\nu-\partial_\nu A_\mu$ \footnote{Due to the absence of torsion in GR, $F_{\mu\nu}=\nabla_\mu A_\nu-\nabla_\nu A_\mu=\partial_\mu A_\nu-\partial_\nu A_\mu$. Conversely, this is not true in theories extending GR having non-zero torsion, which makes the standard coupling prescription problematic. In fact in that case we would have $F_{\mu\nu}=\nabla_\mu A_\nu-\nabla_\nu A_\mu=\partial_\mu A_\nu-\partial_\nu A_\mu+T^\sigma_{~\mu\nu}A_\sigma$, where the torsion breaks the U(1) symmetry.} is the field strenght of a U(1) gauge vector field $A_\mu$ and $\Phi$ is a scalar field. From \ref{EH_bosonic_action} we get the (massless) Klein-Gordon and Maxwell (vacuum) equations in curved spacetime coupled with the Einstein-Klein-Gordon-Maxwell equations for the metric:
\begin{equation}\label{mass-less_KG}
\nabla_\mu \nabla^\mu \Phi=0\,,
\end{equation}
\begin{equation}\label{Maxwell}
\nabla_\mu F^{\mu\nu}=0\,,
\end{equation}
\begin{equation}\label{GR-scalar-vector}
R_{\mu\nu}-\frac{1}{2}g_{\mu\nu} R=8\pi \left(\partial_\mu \Phi \partial_\nu \Phi -\frac{1}{2}g_{\mu\nu}\nabla_\mu\Phi \nabla^\mu \Phi +F_\mu^{~\alpha}F_{\nu\alpha}-\frac{1}{4}g_{\mu\nu}F_{\alpha\beta}F^{\alpha\beta}\right)\,.
\end{equation}
By considering first order perturbations of a vacuum metric $\bar{g}_{\mu\nu}$ we neglect the backreaction on the spacetime, which is second order's in the perturbations and sourced by the energy-momentum currents of the involved fields. In the case of $\Phi$ and $A_\mu$ it means neglecting their stress-energy tensor on the right-hand side of equation \ref{GR-scalar-vector}, thus decoupling them from the metric, resulting in the scalar and Maxwell equations on a fixed curved background:
\begin{equation}\label{mass-less_KG_perturb}
\bar\nabla_\mu \bar\nabla^\mu \Phi=0\,,
\end{equation}
\begin{equation}\label{Maxwell_perturb}
\bar\nabla_\mu F^{\mu\nu}=0\,,
\end{equation}
where $\bar\nabla_\mu$ is the covariant derivative of the background. In the case of the metric perturbations, instead, the backreaction comes from the non-linearities of the Ricci tensor, which physically arise as the effect of spacetime interacting with its own energy-momentum\footnote{Gravitational energy-momentum in GR cannot be expressed as a tensor, thus there are some subleties and ambiguities in how to define it and difficulties in visualizing it as the source of self-interaction in the Einstein Field Equations. The most used formalizations are the Landau-Lifshitz\cite{Landau_classic_field_theory} and the Einstein\cite{Einstein_energy_GR_1,Einstein_energy_GR_2} stress-energy pseudo-tensors, which cannot be distinguished in the weak field limit but are rather different in the strong-gravity regime. There is an example of alternative formulation of the same classical gravitational physics described by GR based on torsion instead of curvature, i.e. the Teleparallel Equivalent of General Relativity (TEGR), which, instead, has a clear tensor definition of gravitational energy\cite{Teleparallel_gravity}. Thus we may deduce that the problem of gravitational energy-momentum in GR arises exclusively from the mathematical formulation and first principles on which the theory is based.}\cite{Deser_self-interaction,GR_from_QG}. By perturbing the Ricci tensor, we get the first order field equations of the metric perturbations,
\begin{equation}\label{GW_curved_eqs}
\bar\nabla^{\sigma}\left(\bar\nabla_\sigma h_{\mu\nu}-\bar\nabla_\mu h_{\sigma\nu}-\bar\nabla_\nu h_{\sigma\mu} \right)+\bar\nabla_\mu\bar\nabla_\nu h^\sigma_{~\sigma}=0
\end{equation}
where $h_{\mu\nu}=g_{\mu\nu}-\bar{g}_{\mu\nu}$ is the perturbation, describing gravitational waves on a fixed curved vacuum background. Equations \ref{GW_curved_eqs} are invariant under the following gauge transformation,
\begin{equation}
h_{\mu\nu}\rightarrow h_{\mu\nu}+\bar\nabla_\mu \epsilon_\nu+\bar\nabla_\nu \epsilon_\mu\,,
\end{equation}
corresponding to an infinitesimal coordinate transformation of $g_{\mu\nu}$. Equation \ref{mass-less_KG_perturb} on a Kerr background is easily separable\cite{scalar_separability}, while solving \ref{Maxwell_perturb} and \ref{GW_curved_eqs} is not an easy task due to the complicate structure of the metric.

In 1972 Teukolsky discovered that by applying the Newman-Penrose (NP) formalism\cite{Newman-Penrose} the perturbations of Kerr BHs are all described by the following master equation\cite{Teukolsky_1972,Teukolsky_perturb_1,Chandrasekhar1983}:
\begin{small}
\begin{equation}
\begin{aligned}
&\left[\frac{(r^2+a^2)^2}{\Delta(r)}-a^2 \sin^2\theta \right] \frac{\partial^2 \psi}{\partial t^2}+\frac{4 M a r}{\Delta(r)}\frac{\partial^2\psi}{\partial t\partial\phi}+\left[\frac{a^2}{\Delta(r)}-\frac{1}{\sin^2\theta} \right]\frac{\partial^2 \psi}{\partial \phi^2}\\
&-\Delta^{-s}(r)\frac{\partial}{\partial r}\left(\Delta^{s+1}(r) \frac{\partial\psi}{\partial r}\right)-\frac{1}{\sin\theta}\frac{\partial}{\partial\theta}\left(\sin\theta\frac{\partial\psi}{\partial\theta}\right)-2 s \left[\frac{a(r-M)}{\Delta(r)}+\frac{i \cos\theta}{\sin^2\theta} \right]\frac{\partial\psi}{\partial\phi}\\
&-2 s \left[\frac{M(r^2-a^2)}{\Delta(r)}-r-i a \cos\theta \right]\frac{\partial\psi}{\partial t}+\left(s^2 \cot^2\theta-s \right)\psi=0\,.
\end{aligned}
\end{equation}
\end{small}
By choosing the ansatz $\psi=e^{i(m\phi- \omega t)}\prescript{}{s}S_{lm}(\theta)\prescript{}{s}R_{lm}(r)$, Teukolsky separated the wave equation into radial and angular ordinary differential equations:
\begin{equation}\label{Teukolsky_radial}
\small\left[\Delta^{-s}(r) \frac{d}{dr}\left(\Delta^{s+1}(r)\frac{d}{dr} \right)+\frac{K^2(r)-2 i s (r-M)K(r)}{\Delta(r)}+4 i s \omega r -\lambda \right]\prescript{}{s}R_{lm}(r)=0\,,
\end{equation}
\begin{small}
\begin{equation}\label{spin-weighted-spheroidal}
\begin{aligned}
&\left[\frac{1}{\sin\theta}\frac{d}{d\theta}\left(\sin\theta\frac{d}{d\theta}\right)+\left(a^2\omega^2\cos^2\theta-\frac{m^2}{\sin^2\theta}-2 a \omega s \cos\theta\right.\right.~~~~~~~~~~~~~~~~~~~~~\\
&~~~~~~~~~~~~~~~~~~~~~~~~~~~~~~~~~~~~~~~~-\left.\left.\frac{2 m s \cos\theta}{\sin^2\theta}-s^2\cot^2\theta+s+\prescript{}{s}A_{lm}\right)\right]\prescript{}{s}S_{lm}(\theta)=0\,,
\end{aligned}
\end{equation}
\end{small}
where $K(r)=(r^2+a^2)\omega-a m$, $\lambda=A_{slm}+a^2\omega^2-2 a m \omega$ and $\prescript{}{s}A_{sm}$ is a separation constant. The parameter $s$ is the spin weight of the perturbation: it is equal to $0$ for scalar fields, $\pm1/2$ for spin 1/2 fermions, $\pm1$ for electromagnetic perturbations and $\pm2$ for gravitational waves.

By imposing the orthonormality condition $\int_0^{2\pi}\int_0^\pi |\prescript{}{s}S_{lm}(\theta)|^2 \sin\theta d\theta d\phi=1$ and regularity in all the domain, we can numerically find the eigenvalue $\prescript{}{s}A_{lm}$ and the solutions to the angular equation, i.e. the spin-weighted spheroidal harmonics $\prescript{}{s}S_{lm}(\theta,\phi)=e^{i\phi}\prescript{}{s}S_{lm}(\theta)$\cite{Berti_spheroidal_harmonics}. For $a\omega=0$ $\prescript{}{s}S_{lm}(\theta)$ reduces to the spin-weighted spherical harmonics $\prescript{}{s}Y_{lm}(\theta)$, thus for $a\omega\ll 1$ we get $\prescript{}{s}A_{lm}=l(l+1)-s(s+1)+O(a^2\omega^2)$. The resolution of the radial equation, instead, needs some boundary conditions at the BH horizon and at infinity for the solutions $\prescript{}{s}R_{lm}(r)$. If we define the tortoise coordinate $r_*$ such that $\frac{dr}{dr_*}=\frac{\Delta(r)}{r^2+a^2}$, we get the following asymptotic solutions\cite{Berti_BH_QNMs,Brito_SR}:
\begin{equation}
r\rightarrow r_+~~
\implies~~\prescript{}{s}R_{lm}\sim\mathcal{T}\frac{e^{-i k_H r_*}}{\Delta^{s}(r)}+\mathcal{O} e^{i k_Hr_*}\,,
\end{equation}
\begin{equation}
r\rightarrow \infty~~
\implies~~\prescript{}{s}R_{lm}\sim\mathcal{I}\frac{e^{-i\omega r_*}}{r}+\mathcal{R}\frac{e^{i\omega r_*}}{r^{2s+1}}\,,~~
\end{equation}
where $k_H=\omega-m\Omega_H$ and $\mathcal{T}$, $\mathcal{O}$, $\mathcal{I}$ and $\mathcal{R}$ are constants, while the expression for the BH angular velocity $\Omega_H$ can be found in equation \ref{BH_angular_velocity}. Regularity of the solution at $r=r_+$ imposes $\mathcal{O}=0$, which excludes radiation outgoing from the horizon, in perfect accordance with the nature of the surface of a BH. At infinity the condition to be imposed depends on what we are looking for. If we need to compute the quasi-normal modes of the system, we impose the radiation to be purely out-going by fixing $\mathcal{I}=0$. If we are computing a scattering, instead, we keep both $\mathcal{I}$ and $\mathcal{R}$ as free parameters, respectevely representing the amplitude of incident and radiated waves.

For each value of the spin weight the scalar $\psi$ describes the perturbation for a specific spin and polarization. For $s=0$ Teukolsky's equation is just Klein-Gordon (equation \ref{mass-less_KG}) on a Kerr background, therefore $\psi=\Phi$. In the case of electromagnetic perturbations, instead, $\psi$ is related with the components of $F_{\mu\nu}$:
\begin{equation}
s=+1~\rightarrow~\psi=\phi_0~,~~~~~~~\phi_0=F_{\mu\nu}l^\mu m^\nu\,,
\end{equation}
\begin{equation}
s=-1~\rightarrow~\psi=\rho^{-2}\phi_2~,~~\phi_2=F_{\mu\nu}\bar{m}^\mu n^\nu\,,
\end{equation}
where 
\begin{equation}
\rho=m^\mu \bar{m}^\nu \nabla_\nu l_\mu=-\frac{1}{r-i a \cos\theta}
\end{equation}
 is a spin coefficient of the NP formalism and
\begin{subequations}
\begin{align}
&l^\mu=\left[\frac{r^2+a^2}{\Delta(r)},1,0,\frac{a}{\Delta(r)}\right]\,,\\
&n^\mu=\frac{1}{2\Sigma(r,\theta)}\left[r^2+a^2,-\Delta(r),0,a\right]\,,\\
&m^\mu=\frac{1}{\sqrt{2}\left(r+i a\cos\theta\right)}\left[i a \sin\theta,0,1,\frac{i}{\sin\theta}\right]
\end{align}
\end{subequations}
is the Kinnersley tetrad basis of the Kerr metric (the fourth tetrad $\bar{m}^\mu$ is the complex conjugate of $m^\mu$)\cite{Newman-Penrose,Chandrasekhar1983,Brito_SR}. Finally, gravitational radiation is encoded through Weyl scalars,
\begin{equation}\label{s=+2}
s=+2~\rightarrow~\psi=\Psi_0~,~~~~~~\Psi_0=-C_{\alpha\beta\mu\nu}l^\alpha m^\beta l^\mu m^\nu\,,
\end{equation}
\begin{equation}\label{s=-2}
s=-2~\rightarrow~\psi=\rho^{-4}\Psi_4~,~\Psi_4=-C_{\alpha\beta\mu\nu}l^\alpha \bar{m}^\beta l^\mu \bar{m}^\nu\,,
\end{equation}
where $C_{\alpha\beta\mu\nu}$ is the Weyl curvature tensor of $g_{\mu\nu}=\bar{g}_{\mu\nu}+h_{\mu\nu}$.

\section{Superradiant scattering in Kerr black holes}\label{section:SR_scattering}

In this section we will introduce superradiant scattering in spinning BHs, which can be considered as the wave equivalent of the Penrose process. This effect is possible for any bosonic wave, i.e. scalar, vector or tensor fields, scattered off a Kerr BH\cite{Teukolsky_perturb_3,Starobinski_GW_EM_SR,Bekenstein_energy_extraction,Brito_SR}. By applying what we introduced in the previous section about BH perturbation theory, we can analytically demonstrate the amplification of radiation in the scattering process.
It can be demonstrated that, through a redefinition of the perturbation and by using the tortoise coordinate defined in the previous section, the radial Teukolsky equation \ref{Teukolsky_radial} for bosonic fields can be re-written in a Schrödinger-like form with a real-valued potential\cite{Detweiler_resonant_BH, Ferrari_GR},
\begin{equation}\label{Schrodinger-like}
\left[\frac{d^2}{dr_*^2}+V_{\text{eff}}(r_*) \right]\psi=0\,,
\end{equation}
where $V_{\text{eff}}$ has the following behaviour at the horizon and at infinity:
\begin{equation}\label{potetial_boundary}
V_{\text{eff}}\sim\left\lbrace
\begin{matrix}
 k_H ~~~~r_*\rightarrow-\infty
\\
\omega ~~~~~~r_*\rightarrow +\infty
\end{matrix}\right.\,.
\end{equation}
Thus, the boundary conditions we previously defined can be rewritten in the following form:
\begin{equation}\label{boundary_SR}
\psi\sim\left\lbrace
\begin{matrix}
\displaystyle\mathcal{T}e^{-i k_H r_*}~~~~~~~~~~~~~~~~~r_*\rightarrow-\infty
\\
\displaystyle\mathcal{I}e^{-i\omega r_*}+\mathcal{R}e^{i\omega r_*}~~~~~~r_*\rightarrow +\infty
\end{matrix}\right.\,.
\end{equation}
Because of $V_{\text{eff}}\in\mathbf{R}$, the complex conjugate of the solution $\bar\psi$ is a solution of \ref{Schrodinger-like}. In addition, $\psi$ and $\bar\psi$ are also linearly independent, hence we can use them for computing the Wronskian of the radial equation, which must be independent of $r_*$:
\begin{equation}
W=\frac{d\psi}{dr_*}\bar\psi-\frac{d\bar\psi}{dr_*}\psi\,.
\end{equation}
Thus, the computations of $W$ at the horizon and at infinity must coincide:
\begin{equation}
-2 i k_H |\mathcal{T}|^2=\left.W\right|_{r_*=-\infty}=\left.W\right|_{r_*=+\infty}=2 i\omega\left(|\mathcal{R}|^2-|\mathcal{I}|^2\right).
\end{equation}
We rewrite this last relationship in the following way,
\begin{equation}
|\mathcal{R}|^2=|\mathcal{I}|^2-\frac{\omega-m\Omega_H}{\omega}|\mathcal{T}|^2\,,
\end{equation}
which implies we have wave amplification, i.e. $|\mathcal{R}|^2>|\mathcal{I}|^2$, for any integer value of the spin weight if
\begin{equation}
0<\omega<m \Omega_H\,,
\end{equation}
which is called superradiant condition\footnote{The ansatz chosen for the wave solutions implies positive frequency. Negative frequency solutions are just waves moving in the opposite direction, or back in time.}\cite{Brito_SR, Ferrari_GR}. This condition gives the frequency window in which the amplification mechanism is active but it also specifies that there is superradiance only if $m\Omega_H>0$, i.e. there is amplification if the field "co-rotates" with the BH. 

The boundary condition at the horizon may induce us in thinking that superradiance might not be possible if we consider horizon-less compact objects\footnote{The exotic compact objects are proposals of stars mimicking BHs in beyond SM and GR scenarios\cite{Cardoso_ECOs}.}, but that is not the case. Exactly like in the Penrose process, the ergoregion is sufficient for triggering superradiance, while the horizon just dumps negative energy states. The hypothetical absence of an horizon would have a destabilizing effect for the system due to the development of negative energy region, triggering what are called ergoregion instabilities\cite{Maggio_2019}, but superradiance would still be present\cite{superradiance_ergoregion_instability,Eskin_2016,Brito_SR}.

In 1973 Bekenstein pointed out that superradiance is also implied by the Laws of BH Thermodynamics\cite{Bekenstein_energy_extraction, Brito_SR}. In any infinitesimal classical transformation involving spinning BHs, the relationship between mass/energy of BHs, area and angular momentum $J$ is the following\cite{Wald_BH_thermo,Damour_BH_entropy,Page_BH_thermodynamics}:
\begin{equation}\label{BH_first_law}
\delta M=\frac{k_H}{8\pi}\delta A_H+\Omega_H\delta J\,,
\end{equation}
where $A_H$ is the area of the event horizon and $k_H=\frac{\sqrt{M^2-a^2}}{r_+^2+a^2}$ is called surface gravity. Moreover, during a classical BH transformation the area cannot decrease, showing irreversibility in BHs in an explicit form\cite{Wald_BH_thermo,Damour_BH_entropy,Page_BH_thermodynamics}:
\begin{equation}\label{BH_second_law}
\delta A_H\geq 0.
\end{equation}
BH area plays the same role of entropy, consequently \ref{BH_first_law} and \ref{BH_second_law} are the BH equivalent of the First and Second Laws of Thermodynamics\cite{Wald_BH_thermo,Damour_BH_entropy,Page_BH_thermodynamics}. When a perturbing field is involved, the ratio $\frac{\delta J}{\delta M}$ is equal to the ratio of the angular momentum and energy fluxes of the interacting wave at the horizon, which is $\frac{\delta J}{\delta M}=\frac{m}{\omega}$ for monocromatic radiation having definite azimuthal number $m$ \footnote{This result does not depend on the spin of the boson and can be computed by integrating the stress-energy tensor of the involved field. But it can easily be deduced by considering the absorbtion of one particle having energy $\mathcal{E}=\hbar \omega$ and total azimuthal angular momentum $J_z=\hbar m$.}\cite{Bekenstein_energy_extraction, Brito_SR}. Therefore we can rewrite the First Law \ref{BH_first_law} in the following way,
\begin{equation}
\delta M =\frac{k_H \omega}{8\pi} \frac{ \delta A_H}{\omega-m \Omega_H}\,,
\end{equation}
which, as a consequence of the Second Law \ref{BH_second_law}, gives energy and angular momentum extraction if the superradiant condition $0<\omega<m\Omega_H$ is satisfied\cite{Bekenstein_energy_extraction, Brito_SR}. Important sidenote: the analogy between BH mechanics and thermodynamics is not just formal. In 1972 Bekenstein conjectured BHs should have an entropy $S_H$ proportional to $A_H$\cite{Bekenstein_BH_entropy}, confirmed by Hawking two years later by discovering that quantum field theory in curved spacetime implies thermal radiation emitted by BHs (Hawking's radiation\footnote{The impact of this breakthrough is huge for our understanding of BHs. It implies that from a quantum mechanical point of view they are not really black, i.e. they emit the feeble Hawking's radiation, and consequently the compact object evaporates by emitting all its energy. This process creates the information loss paradox\cite{Hawking_BH_information_paradox,BH_information_paradox}: what happens to all the information of the objects swallowed by BHs if they emit just thermal radiation? The resolution of this problem is one of the big open issues in the theoretical study of BHs and fundamental interactions.})\cite{Hawking_radiation,Wald_BH_thermo,Damour_BH_entropy,Page_BH_thermodynamics}:
\begin{equation}
S_H=\frac{k_B A_H}{4}~,~T_H=\frac{k_H}{2\pi k_B}\,,
\end{equation}
where $T_H$ is the BH temperature, $k_B$ is the Boltzmann constant and the expressions are in geometrized units $G=c=\hbar=1$.

\section{Superradiant instabilities}

\begin{figure}[h]
\centering
\includegraphics[width=1\textwidth]{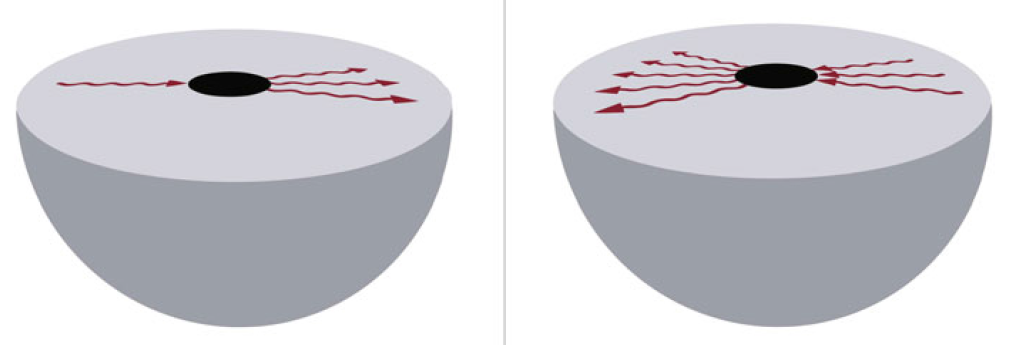}\\
\caption{BH bomb: the perturbation is continuously scattered and amplified because of a confining mirror, thus causing a superradiant instability. Image by Ana Sousa Carvalho, taken from the book \textit{Superradiance: New Frontiers in Black Hole Physics}\cite{Brito_SR}.}
\label{figure_BH_bomb}
\end{figure}

Superradiance can trigger instabilities in BHs if there is some confinement acting on the scattered waves\cite{Press_BH_bomb,Damour_BH_instability,Detweiler_scalar_instability,Dolan_spin-1_instability,Brito_magnetic_SR,Cardoso_AdS-Kerr_unstable,Cardoso_superradiant_instability,Brito_spin-2_SR_slow_rot,Brito_hydrogenic_spin-2_SR,Spectra_grav_atom, Brito_SR}. The confinement forces the perturbation into getting continuously scattered on the compact object, getting gradually amplified until the superradiant condition is satisfied, thus triggering a superradiant instability. A simple example is given by a BH enclosed in a reflecting mirror, first considered by Press and Teukolsky in 1972\cite{Press_BH_bomb}, represented in Figure \ref{figure_BH_bomb}.
This unphysical setting can be considered as a toy-model simplifying more realistic scenarios. 

A natural way for having confinement is by considering a mass term for the bosonic perturbation\cite{Damour_BH_instability,Detweiler_scalar_instability,Zouros_scalar_instability,Dolan_spin-1_instability,Cardoso_superradiant_instability,Brito_spin-2_SR_slow_rot,Brito_hydrogenic_spin-2_SR,Spectra_grav_atom, Brito_SR}. What happens in this case is that the mass changes the effective potential for the perturbation, generating a potential well which under certain conditions can work as a confinement: the resulting field will thus have a Yukawa-like behaviour $e^{-\mu r}/r$, where $\mu$ is the mass of the boson. Isolating the effective potential by expressing the wave equation in a Schrödinger-like form can be done easily only in the case of scalar fields, while for massive vector and tensor perturbations Teukolsky's formalism breaks and consequently things get way more convoluted\footnote{In 2018 \textit{Frolov et al}\cite{Spin-1_separability} separated the massive vector perturbations for a wide class of background spacetimes which include also the Kerr metric. By using Frolov's formalism we could in principle rewrite also the spin-1 radial equation in a Schrödinger-like form. The separability of massive spin-2 perturbations, instead, is still an open problem, hence there is no known way of isolating the effective potential for such case.}. This type of confinement is the wave equivalent of what happens with time-like particles\footnote{In fact geodesic motion is the eikonal limit of wave propagation on a fixed curved spacetime.}: their mass generates a potential well in their effective potential and thus stable orbits are possible, otherwise particles escape at infinity or get swallowed by the BH\footnote{Photons do have a possible orbit, called photon sphere, but it is unstable.}. \begin{figure}[h]
\centering
\includegraphics[width=0.85\textwidth]{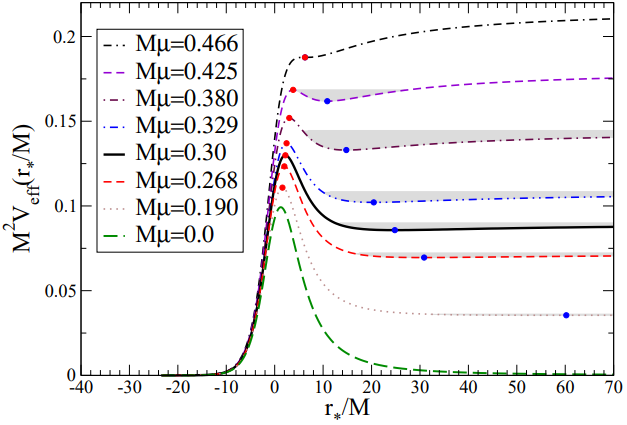}\\
\caption{The effective potential of a scalar field in a Schwarzschild BH. The potential well generated by the mass $\mu$ can be clearly noticed. Image taken from \textit{Barranco et al} \cite{Barranco_2011}.}
\end{figure}
Superradiant instabilities triggered by massive bosonic perturbations have a potential astrophysical observable impact if the boson mass is ultralight, i.e. $\mu\ll eV$, thus this phenomenon can play an important role in the potential detection of hypotethical ultralight bosons \cite{Constraint_bosons_BH_spin,BH_spin_constraint_axion,Brito_hydrogenic_spin-2_SR,Binary_BHs_bosons_1,Binary_BHs_bosons_2,Binary_BHs_bosons_3,GWs_signatures,Stochastic_GW_1,Stochastic_GW_2,GW_searches_ultralight_bosons,Ultralight_vector_SR_signatures,Cardoso_constrsaint_dark_photon_SR,EHT_constraint_DM,discovering_axion_BHs,string_axiverse_BH,kodama_bh_axiverse,axion_cloud,GW_probes_Dark_Matter,string_compactification_superradiance}. We will go deep into the perturbative numerical treatment of this type of instability in Chapters \ref{chapter:spherical_harmonic} and \ref{chapter:numerical_SR_GR}.

Another example of superradiant instability we will describe in great detail comes from non-linear interactions between the bosonic field and matter in theories extending GR. In such scenario a massless boson can acquire an effective mass: if the accreting matter around the BH generates a cavity, the effectve mass generates a potential well able to trigger superradiant instabilities\cite{cardoso_plasma_tests,Dima_2020,Scalar_plasma_SR}. In Chapter \ref{chapter:plasma_SR} we will show how this happens in scalar-tensor theories and how this phenomenon can be used for constraining the modification to GR.

In the next chapter we will introduce the ultralight exotic bosons, where they come from, their perturbative field equations and their actual observational constraints, so that we can see what role can superradiance have in astrophysical detections.

\chapter{Ultralight exotic bosons}\label{chapter:ultralight_bosons}

Several BH \cite{Peccei_Quinn_axion,axion_dark_matter_1,axion_dark_matter_2,string_axiverse,hidden_photons_string,Low_energy_particle_frontier,discovering_axion_BHs,string_axiverse_BH,kodama_bh_axiverse,axion_cloud,GW_probes_Dark_Matter,string_compactification_superradiance,ALPs,improved_bound_ultralight_scalars,scalar&vector_ultralight_DM,dark_photon_1,dark_photon_2,dark_photon_3,fundamental_physics_intensity_frontier,photon_graviton_mass,bi-metric_gravity_1,bi-metric_gravity_2,bi-metric_gravity_5} and GR's 
\cite{modified_gravity,photon_graviton_mass,massive_gravity_1,massive_gravity_2,massive_gravity_braneworld, Brito_spin-2_SR_slow_rot,BH_scalar_tensor,testing_GR,fourth_order_gravity_kerr_instability,Horndeski_modified_gravity,superradiance_horndeski,bi-metric_gravity_1,bi-metric_gravity_2,bi-metric_gravity_3,bi-metric_gravity_4,bi-metric_gravity_5,bi-metric_gravity_6} extensions predict the existence of exotic bosons of arbitrary spin having ultralight masses much smaller than the electronvolt scale. The proposed extra bosons are scalar particles, e.g. QCD and string theory axions\cite{Peccei_Quinn_axion,axion_dark_matter_1,axion_dark_matter_2,string_axiverse,Low_energy_particle_frontier,discovering_axion_BHs,string_axiverse_BH,kodama_bh_axiverse,axion_cloud,GW_probes_Dark_Matter,string_compactification_superradiance,ALPs,improved_bound_ultralight_scalars,scalar&vector_ultralight_DM}, or "dark photons", i.e. vector particles with ultralight masses\cite{Low_energy_particle_frontier,fundamental_physics_intensity_frontier,dark_photon_1,dark_photon_2,dark_photon_3,scalar&vector_ultralight_DM,photon_graviton_mass}, or ultralight tensor particles, arising from massive gravity and bi-metric gravity theories\cite{photon_graviton_mass,massive_gravity_1,massive_gravity_2, massive_gravity_braneworld,Brito_spin-2_SR_slow_rot,bi-metric_gravity_1,bi-metric_gravity_2,bi-metric_gravity_3,bi-metric_gravity_4,bi-metric_gravity_5,bi-metric_gravity_6}. 

Their coupling to ordinary matter is feeble, making them ideal candidates for dark matter, but also difficult to be tested through direct detection. Though, if such hypothetical particles exist, they are expected to fall in the interval of particle masses being able to trigger superradiant instabilities in BHs. In fact, as we will see in Chapter \ref{chapter:numerical_SR_GR}, in a BH having mass $M$ the instability triggered by a boson having mass $\mu$ is relevant, i.e. features a time-scale of astrophysical relevance, if $\mu M \lesssim 1$: if we consider BH masses of at least the order of the solar mass, i.e. the only ones whose existence has been proved, the boson particle has to be ultralight ($\mu\lesssim 10^{-10} eV$). 

The recent emergence of gravitational-wave astronomy \cite{GW_discovery, GW_detect_1, GW_detect_2, GW_detect_3} 
and the growth of high-energy astrophysics observations \cite{high_energy_astro_1, high_energy_astro_2, 
high_energy_astro_3, high_energy_astro_4} give the opportunity of testing these exotic bosons through gravitational interactions by exploiting the superradiant instabilities\cite{Constraint_bosons_BH_spin,BH_spin_constraint_axion,Brito_hydrogenic_spin-2_SR,Binary_BHs_bosons_1,Binary_BHs_bosons_2,Binary_BHs_bosons_3,GWs_signatures,Stochastic_GW_1,Stochastic_GW_2,GW_searches_ultralight_bosons,Ultralight_vector_SR_signatures,Cardoso_constrsaint_dark_photon_SR,EHT_constraint_DM,discovering_axion_BHs,string_axiverse_BH,kodama_bh_axiverse,axion_cloud,GW_probes_Dark_Matter,string_compactification_superradiance}, therefore suggesting the idea of using BHs as particle-physics laboratories and detectors \cite{Brito_evolution_SR_instabilities}.

In this chapter we review the different types of ultralight bosons, their perturbative equations in vacuum backgrounds and their observational constraints.

\section{Scalar ultralight bosons}

The prototype of scalar light particle is the QCD axion proposed by Peccei and Quinn\cite{Peccei_Quinn_axion,kodama_bh_axiverse} as a possible solution to the strong CP problem in Quantum Chromodynamics (QCD). The problem consists in the detected suppression of Charge-Parity symmetry violations not expected by the BH's QCD theory, which, instead, allows large violations.
Axions and Axion-Like Particles (ALPs)\cite{ALPs} also emerge in the so-called "string axiverse", i.e. in the context of String Theory models\cite{axion_dark_matter_1,axion_dark_matter_2,string_axiverse,discovering_axion_BHs,string_axiverse_BH,kodama_bh_axiverse,ALPs,improved_bound_ultralight_scalars,scalar&vector_ultralight_DM}. This proposal of unified theory of fundamental interactions predicts the existence of extra dimensions of spacetime, which in some models are compactified in extremely small scales in order to be consistent with the fact that they are not present in our every-day experience (i.e. Kaluza-Klein mechanism\cite{Kaluza-Klein_theories}). The effective field theories (EFT) of such compactified extra-dimensional models give a plethora of axion fields emerging in the low-energy scale as "leaks" of the extra dimensions in the standard 4-dimensional spacetime, with masses ranging from $10^{-10} eV$ down to the Planck scale. Detecting a large amount of axion fields would therefore be a possible signature of extra-dimensional compactification.

The general action for a massive scalar field minimally coupled \footnote{i.e. featuring the standard GR coupling we introduced in section \ref{Mass-less_bosonic_BH_perturbs}, which is consistent with the Strong Equivalence Principle (SEP)\cite{testing_GR}: gravitational and inertial mass are equivalent and any local gravitational or non-gravitational experiment in any inertial reference frame is equivalent.} with GR is\cite{carroll_spacetime_geometry,Wald_General_relativity,Ferrari_GR}
\begin{equation}
    S[g_{\mu\nu},\Phi]=\int d^4 x \sqrt{-g}\left[\frac{1}{16\pi}R-\frac{1}{2}\nabla_\mu\Phi \nabla^\mu\Phi+V(\Phi)\right]\,,
\end{equation}
where $V(\Phi)$ is the self-potential, representing the self-interactions of the scalar field. Consequently, the field equation of the scalar field will be
\begin{equation}
    \nabla_\mu \nabla^\mu \Phi+V'(\Phi)=0\,.
\end{equation}
If we consider the first order perturbation in the scalar field of a vacuum metric $\bar g_{\mu\nu}$ in models having $V'(0)=0$\footnote{Obviously we always have $V(0)=0$, because $V(0)\neq 0$ cannot have any impact on the dynamics of the scalar field.}, we get the following perturbative equation for $V''(0)>0$:
\begin{equation}\label{Klein-Gordon_Perturb_eq}
    \bar\nabla_\mu \bar\nabla^\mu \Phi+\mu^2\Phi=0\,,
\end{equation}
where $\mu=\sqrt{2 V''(0)}$ is the mass of the scalar perturbation and $\bar\nabla_\mu$ is the covariant derivative of the background metric. Conversely, if $V''(0)<0$ we get a field featuring tachyonic condensation.

There are also theories having a non-minimal coupling between the scalar field and the scalar curvature, i.e. scalar-tensor theories\cite{scalar-tensor_theory_GWs,scalar-tensor_comment,testing_GR}, but such scenarios are beyond GR and will be introduced in Chapter \ref{chapter:plasma_SR}.

\section{Dark photons}\label{dark_photons}

 Massive vector bosons arise from the so-called "hidden U(1) sector", which is, again, a feature of String Theory models\cite{hidden_photons_string,Low_energy_particle_frontier,dark_photon_1,dark_photon_2,dark_photon_3,fundamental_physics_intensity_frontier,scalar&vector_ultralight_DM,photon_graviton_mass}. The Kaluza-Klein mechanism arising from the compactification of extra dimensions, in fact, gives also U(1) gauge fields in the low-energy EFT\cite{Kaluza-Klein_theories}, which can acquire an ultra-light mass through spontaneous symmetry breakings from Higgs or Stückelberg mechanisms\cite{hidden_photons_string,Low_energy_particle_frontier,dark_photon_1,dark_photon_2,dark_photon_3,fundamental_physics_intensity_frontier,photon_graviton_mass}.

The action for a massive vector field $A_\mu$ minimally coupled with GR is just the action of the Einstein-Maxwell theory plus a vector mass term:
\begin{equation}
    S[g_{\mu\nu},A_\mu]=\int d^4 x \sqrt{-g}\left(\frac{1}{16\pi}R-\frac{1}{4}F_{\mu\nu}F^{\mu\nu}+\frac{1}{2}\mu^2 A_\mu A^\mu\right)\,,
\end{equation}
where $F_{\mu\nu}=\partial_\mu A_\nu-\partial_\nu A_\mu$ is the U(1) field strength and $\mu$ is the mass of the boson. The resulting field equations for the massive vector are the Proca equations in curved space-time\cite{Proca_eqs}:
\begin{equation}
    \nabla_\mu F^{\mu\nu}+\mu^2A^\nu=0\,,
\end{equation}
which can also be rewritten in the following form:
\begin{equation}
    \nabla_\mu\nabla^\mu A_\nu-R^\mu_{~\nu}A_\mu+\mu^2 A_\nu=0~,~\nabla^\mu A_\mu=0\,.
\end{equation}
The presence of the mass term adds a dynamical degree of freedom to the ones we usually have in Maxwell theory, thus these field equations give 3 polarizations.
We might be tempted to add some general self-interaction through a potential $V(A_\mu A^\mu)$ for non-linear evolutions like in the case of scalar fields, but such class of models would be plagued by ghosts and tachyonic instabilities\cite{proca_self-interact_problem_1,proca_self-interact_problem_2,proca_self-interact_problem_3}. These pathologies can be solved by including some Higgs-like mechanism\cite{non-linear_dark-photon_SR,non-linear_proca_star}:
\begin{equation}\label{non-linear-Proca_UV-complete}
    S[g_{\mu\nu},A_\mu,\Phi]=\int d^4 x \sqrt{-g}\left[\frac{1}{16\pi}R-\frac{1}{4}F_{\mu\nu}F^{\mu\nu}-\frac{1}{2}\left(D_\mu\Phi\right)^* D^\mu\Phi+V(|\Phi|^2)\right]\,,
\end{equation}
where $D_\mu\Phi=\left(\partial_\mu -i g_0 A_\mu \right)\Phi$ is the U(1) covariant derivative of the Higgs-like complex scalar $\Phi$ and $g_0$ is the coupling constant of the U(1) field. The low energy EFT of this type of theories are exactly self-interacting Proca models, thus the pathologies are just the effect of the breakdown of the EFT framework\cite{non-linear_proca_star}. Consequently, for the non-linear study of superradiance one is obliged to use models like \ref{non-linear-Proca_UV-complete}, e.g. in ref.\cite{non-linear_dark-photon_SR}. The equations for the perturbative study on a background $\bar g_{\mu\nu}$ are
\begin{equation}
    \bar\nabla_\mu F^{\mu\nu}+\mu^2 A^\nu=0\,,
\end{equation}
which can be rewritten in the following form when the background is vacuum:
\begin{equation}
    \bar\nabla_\mu\bar\nabla^\mu A_\nu+\mu^2 A_\nu=0~,~\bar\nabla^\mu A_\mu=0\,.
\end{equation}

\section{Ultralight spin-2 fields}

Considering the hypothesis of a massive gravity might raise eyebrows. Why shall we consider such "obscenity"? The answer could be: why not? As a matter of fact, the most banal reason for considering such "wild" proposal concerns the foundations of classical field theory itself, i.e. understanding if a predictive massive tensor field theory is mathematically possible. This is useful both for theorizing massive gravity itself\cite{massive_gravity_1,massive_gravity_2,theoretical_observational_consistency_massive_gravity,photon_graviton_mass} and/or considering some beyond SM spin-2 massive particles interacting with gravity, i.e. massive bi-metric theories\cite{bi-metric_gravity_1,bi-metric_gravity_2,bi-metric_gravity_3,bi-metric_gravity_4,bi-metric_gravity_5,bi-metric_gravity_6}. Beyond-SM non-gravitational massive spin-2 fields have been considered in particle physics as possible mesons fixing issues in the hadronic confinement\cite{bi-metric_gravity_1,bi-metric_gravity_2} and in the context of String Theory\cite{bi-metric_gravity_5}. Moreover, brane-world scenarios\cite{braneworld_1,braneworld_2} can make gravity acquire mass, as it is demonstrated by the DGP model\cite{massive_gravity_braneworld}: in such models the extra dimensions are large and our 4D universe is confined in a sub-space while gravity travels freely in all dimensions, thus the effective 4D gravity can acquire a mass. Massive gravitational degrees of freedom can also arise from theories involving a Riemann-Cartan geometry in the framework of higher-curvature Poincaré Gauge Theory\cite{ghost-free_higher_curvature}, which have a close mathematical relationship with massive bi-metric theory. From a phenomenological point of view, instead, massive gravity and massive bi-metric theories could be an explanation for the acceleration of the expansion of the universe\cite{massive_gravity_cosmology_1,massive_bi-gravity_cosmology_1,massive_bi-gravity_cosmology_2,theoretical_observational_consistency_massive_gravity,cosmology_generalized_massive_gravity}, thus they are considered among possible modifications of GR through an ultralight tensor mass\cite{testing_GR}. The reason for such cosmological effect is simple: the Yukawa-like profile $\sim e^{-\mu r}/r$ for the weak-field limit of the massive graviton makes gravity weaker on a cosmological scale for values of the mass $\mu$ of the order of the Hubble constant, thus producing an effective acceleration of the expansion of the Universe without the need of any "dark energy".

\subsection{The special-relativistic linear theory}

The dynamics of a massive tensor field is much more involved than the vector and scalar cases. In flat spacetime, at the linear level, the only Lorentz-invariant theory free of ghosts and tachyons is the one found by Fierz and Pauli in 1939\cite{Fierz-Pauli,massive_gravity_1,massive_gravity_2,Brito_spin-2_SR_slow_rot,photon_graviton_mass},
\begin{equation}\label{Fierz-Pauli_action}
    S_{F-P}[h_{\mu\nu}]=\int d^4 x \left[-\frac{1}{2}\partial_\lambda h_{\mu\nu}\partial^\lambda h^{\mu\nu}+\partial_\lambda h_{\mu\nu}\partial^\mu h^{\lambda\nu}-\partial_\mu h^{\mu\nu}\partial_\nu h^\lambda_{~\lambda}\right.~~~
\end{equation}
\[~~~~~~~~~~~~~~~~~~~~~~~~~~~~+\left.\frac{1}{2}\partial_\mu h^\lambda_{~\lambda}\partial^\mu h^\nu_{~\nu}+\frac{\mu^2}{2}\left(h_{\mu\nu}h^{\mu\nu}-h^{\mu}_{~\mu}h^{\nu}_{~\nu} \right) \right]\,, \]
where $h_{\mu\nu}$ is the spin-2 field and $\mu$ is its mass, propagating five dynamical degrees of freedom. Other choices for the mass term give a sixth degree of freedom, i.e. a scalar ghost having negative energy. By varying the action, we get the following field equations\cite{Fierz-Pauli,massive_gravity_1,massive_gravity_2,Brito_spin-2_SR_slow_rot,photon_graviton_mass}:
\begin{equation}\label{Fierz-Pauli_eqs}
    G^F_{\mu\nu}-\frac{\mu^2}{2} \left(h_{\mu\nu}-\eta_{\mu\nu}h^{\lambda}_{~\lambda}\right)=0\,,
\end{equation}
where
\begin{equation}
\small G^F_{\mu\nu}=\frac{1}{2}\left[\partial^\alpha\partial_\mu\left(h_{\alpha\nu}-\eta_{\alpha\nu}h^\lambda_{~\lambda} \right)-\partial_\lambda \partial^\lambda \left(h_{\mu\nu}-\eta_{\mu\nu}h^\lambda_{~\lambda} \right)-\eta_{\mu\nu}\partial_\alpha\partial_\beta h^{\alpha\beta}+\partial^\alpha\partial_\nu h_{\alpha\mu}\right]
\end{equation}
is the linearization of the Einstein tensor $G_{\mu\nu}=R_{\mu\nu}-\frac{1}{2}g_{\mu\nu} R$ on a flat background. By taking the divergence and the trace of equations \ref{Fierz-Pauli_eqs}, we can rewrite them in the following form\cite{Fierz-Pauli,massive_gravity_1,massive_gravity_2,Brito_spin-2_SR_slow_rot,photon_graviton_mass}:
\begin{equation}
    \left(\partial_\lambda \partial^\lambda +\mu^2 \right) h_{\mu\nu}=0~,~\partial_\mu h^\mu_{~\nu}=0~,~h^\lambda_{~\lambda}=0\,.
\end{equation}
Because of Weinberg's Soft Graviton Theorem\cite{soft_theorem_1,soft_theorem_2}, massless spin-2 fields must be gauge fields transforming like
\begin{equation}
    h'_{\mu\nu}=h_{\mu\nu}+\partial_\mu \epsilon_\nu+\partial_\nu \epsilon_\mu
\end{equation}
in order to have a Lorentz invariant S-Matrix. The only kinetic term for $h_{\mu\nu}$ invariant under such gauge transformation is $G^F_{\mu\nu}$, which consequently is the only possible kinetic term for a spin-2 field\cite{GR_from_QG}.

The Soft Graviton Theorem also implies that the Lorentz invariance of the S-Matrix must impose the universal coupling between spin-2 particles and other fields through a term $h_{\mu\nu}T^{\mu\nu}$ in the action, where $T^{\mu\nu}$ is the stress-energy tensor of the interacting fields\cite{soft_theorem_1,soft_theorem_2}\footnote{Soft Theorem thus naturally implies the Equivalence Principle, without having to introduce it as a first principle.}. This coupling makes the linear spin-2 theory inconsistent when other interacting fields are present, both in the massive and massless cases, because the total energy-momentum of all the fields is not conserved. This contradiction hints at the non-linear extension of the theory, given that the energy to be conserved has to be the one of the other fields plus the energy of the spin-2 fields, which means including the self-interactions of $h_{\mu\nu}$ through its own $T^{\mu\nu}$\footnote{This is exactly how we can recover GR from its linearization: see \cite{Deser_self-interaction} for further details.}.

By carrying the limit for $\mu\rightarrow 0$, the predictions of Fierz-Pauli theory do not reduce to the ones of linearized GR\cite{vDMZ_discontinuity_1,vDMZ_discontinuity_2}, thus apparently making the theory fail any observational test. As a matter of fact the extra degrees of freedom introduced by the mass term do not decouple in this limit, giving rise to what is known as van-Dam-Veltmann-Zakharov (vDVZ) discontinuity. Though, this is an artifact of the linear theory that cannot be present in its non-linear completion\cite{Vainshtein_1972}: around any massive object having mass $M$ for distances $r\leq r_V\sim M/(\mu^4 M^2_P)$ ($r_V$ is called Vainshtein radius) the non-linearities dominate\footnote{This mechanism is called Vainshtein screening and it is present also in other modified theories of gravity, e.g. some of the scalar-tensor models, playing a crucial role in hiding modified gravity effects in the Solar System regime of gravity\cite{testing_GR}.}, thus making the linear theory invalid.

The correct massless limit of the linear theory can be recovered by applying the so-called "Stückelberg trick"\cite{Stueckelberg_1957,smooth_mass-less_limit,massive_gravity_1}. In fact, by making the substitution $h_{\mu\nu}\rightarrow h_{\mu\nu}+\left(\partial_\mu A_\nu+\partial_\nu A_\mu\right)/\mu$ in the spin-2 action, followed by the substitution $A_\mu\rightarrow A_\mu+\partial_\mu \Phi/\mu^2$, we can carry the limit $\mu\rightarrow 0$ without ambiguities on the degrees of freedom: when the sources are conserved we get massless spin-2 plus uncoupled massless spin-1 and spin-0\cite{smooth_mass-less_limit}.

\subsection{The non-linear theories}

While the non-linear completion of a massless spin-2 field arising from the resummation of all the orders of self-interaction generated by the coupling with its own stress-energy is unique\cite{Deser_self-interaction,GR_from_QG}, i.e. GR, that is not the case when we also have the mass term\cite{massive_gravity_1,massive_gravity_2,Brito_spin-2_SR_slow_rot,photon_graviton_mass}.

If we simply add the Fierz-Pauli mass term (defined on curved spacetime) to the action of non-linear massless gravitons, which is the Einstein-Hilbert term, we have to depend on some background $\bar g_{\mu\nu}$ (called "absolute metric")\cite{massive_gravity_1,massive_gravity_2,EFT_massive_spin-2},
\begin{equation}
    S[h_{\mu\nu}]=\int d^4 x \sqrt{-\bar g} \left[R(h_{\mu\nu},\bar g_{\mu\nu})+\frac{\mu^2}{4}\bar g^{\mu\alpha}\bar g^{\nu\beta}\left(h_{\mu\nu}h_{\alpha\beta}-h_{\mu\alpha}h_{\mu\beta} \right) \right]\,,
\end{equation}
where $\bar g$ is the determinant of $\bar g_{\mu\nu}$ and $R(h_{\mu\nu},\bar g_{\mu\nu})$ is the Ricci scalar of the metric $g_{\mu\nu}=\bar g_{\mu\nu}+h_{\mu\nu}$. We can further extend it by introducing some non-linear generalization of Fierz-Pauli $V(h_{\mu\nu},\bar g_{\mu\nu})$, but in general the non-linear completions of Fierz-Pauli theory depend on some absolute metric and are plagued by the presence of an extra sixth mode having negative energy, i.e. the Boulware-Deser ghost\cite{Boulware-Deser_ghost,massive_gravity_1,massive_gravity_2,Brito_spin-2_SR_slow_rot}. What happens is that the non-linear interactions re-introduce the ghost mode which had been cancelled by the Fierz-Pauli choice of the mass term in the linear theory. The mass of the BD ghost depends on the absolute metric: in particular, it becomes infinite and decouples in the case of flat background, thus this is why it disappears in the linear Fierz-Pauli theory\cite{Boulware-Deser_ghost,massive_gravity_1,massive_gravity_2,Brito_spin-2_SR_slow_rot}. 

This ghost issue halted research on massive gravity for some decades, until the formulation of the de-Rham-Gabadadze-Tolley two-parameter family of theories (dRGT) in 2010, the first non-linear massive gravity models claimed to be ghost-free\cite{massive_gravity_1,massive_gravity_2,massive_gravity_general_refence_1,massive_gravity_general_refence_2,massive_gravity_general_refence_3,Brito_spin-2_SR_slow_rot}\footnote{dRGT is widely considered ghost-free in the scientific community, though there are some minority counter-arguments: see ref.\cite{hidden_ghost_massive_gravity}}. This theory was found by analizing models of 5D GR giving an effective graviton mass in 4D subspace, e.g. DGP model\cite{massive_gravity_braneworld}, given that 5D gravity propagates five degrees of freedom, the exact number of polarizations expected for a 4D ghost-free massive gravity\cite{massive_gravity_2}. The dRGT theory features a "reference metric" $f_{\mu\nu}$, i.e. a non-dynamical tensor other than the metric $g_{\mu\nu}$ which generalizes the concept of background metric but can differ drastically from the background of some perturbation of $g_{\mu\nu}$. This tensor is required for building the mass term, because of the impossibility of doing so by using just the metric tensor. The dRGT theory is described by the following action for a generic reference metric\cite{massive_gravity_1,massive_gravity_2,massive_gravity_general_refence_1,massive_gravity_general_refence_2,massive_gravity_general_refence_3}:
\begin{equation}\label{dRGT_action}
    S_{dRGT}[g_{\mu\nu}]=\int d^4 x \sqrt{-g} \left[R+2\mu^2 \sum\limits_{n=0}^3 \beta_n e_n \left(\sqrt{\mathbf{g^{-1} f}} \right)\right]\,,
\end{equation}
where $g$ is the determinant of the metric, $\mu$ is the mass of the graviton, $R$ is the Ricci scalar curvature of the metric tensor, $\mathbf{g^{-1}}$ and $\mathbf f$ respectively represent the matrix of the inverse metric tensor and the matrix of the reference metric. The coefficients $\beta_n=(-1)^n \left[\frac{1}{2}(4-n)(3-n)-(4-n)\alpha_3+\alpha_4 \right]$ are defined by free parameters $\alpha_3$ and $\alpha_4$\footnote{The four parameters are reduced to two by fixing a zero cosmological constant and imposing $\mu$ to be the mass of the graviton\cite{massive_gravity_general_refence_1}.}, while $e_n$ are polynomials defined in the following way for a generic matrix $\mathbf{X}$:
\begin{equation}
    e_0\left(\mathbf{X}\right)=1\,,~~~~~~~~~~~~~~~~~~~~~~~~~~~~
\end{equation}
\[e_1\left(\mathbf{X}\right)=\text{Tr}\left(\mathbf{X}\right)\,,~~~~~~~~~~~~~~~~~~~~~\]
\[~~e_2\left(\mathbf{X}\right)=\frac{1}{2}\left[\text{Tr}\left(\mathbf{X}^2\right)-\left(\text{Tr}\left(\mathbf{X}\right)\right)^2\right]\,,\]
\[~~~~~~~~~~~~~~~~~~~~~~~~~~~~e_3\left(\mathbf{X}\right)=\frac{1}{6}\left[\left(\text{Tr}\left(\mathbf{X}\right)\right)^3-3\left(\text{Tr}\mathbf{X}\right) \text{Tr}\left(\mathbf{X}^2\right)+2\text{Tr}\left(\mathbf{X}^3\right)\right]\,,\]
\[e_4\left(\mathbf{X}\right)=\text{det}\left(\mathbf{X}\right)\,.~~~~~~~~~~~~~~~~~~~~\]
The variations of the dRGT action gives the following vacuum field equations:
\begin{equation}
    R_{\mu\nu}-\frac{1}{2}g_{\mu\nu} R+\mu^2 V^f_{\mu\nu}(g_{\mu\nu})=0\,,
\end{equation}
where $V^f_{\mu\nu}(g_{\mu\nu})$ is the tensor potential of $g_{\mu\nu}$ defined on the reference metric $f_{\mu\nu}$\cite{bi-metric_gravity_3,bi-metric_gravity_4}:
\begin{equation}\label{tensor_potential_dRGT}
 V^f_{\mu\nu}(g_{\mu\nu})=\sum\limits_{n=0}^3 (-1)^n \beta_n g_{\lambda\mu}Y^\lambda_{(n)\nu}\left(\sqrt{\mathbf{g^{-1} f}} \right)\,,
\end{equation}
and the matrix functions $Y^\lambda_{(n)\nu}$ are defined in the following way for a generic matrix $\mathbf{X}$,
\begin{equation}
    \mathbf{Y}_{(n)} (\mathbf{X})=\sum\limits_{k=0}^n (-1)^k \mathbf{X}^{n-k} e_k (\mathbf{X})\,.
\end{equation}
The other class of non-linear theories is the generalization of dRGT theory which promotes the reference metric $f_{\mu\nu}$ to a dynamical field, i.e. bi-metric gravity\cite{bi-metric_gravity_1,bi-metric_gravity_2,bi-metric_gravity_3,bi-metric_gravity_4,bi-metric_gravity_5,bi-metric_gravity_6}, first conceived in 1971 in order to describe the interaction between gravity and some hypothetical spin-2 meson\cite{bi-metric_gravity_1}. Bi-metric gravity suffers from the very same BD ghost issue, consequently the resolution of the problem in massive gravity through dRGT paved the way for defining the class of ghost-free bi-metric theories. The action of this family of models is\cite{bi-metric_gravity_3,bi-metric_gravity_4}
\begin{equation}\label{bi-metric_action}
 \small{S_{gf}[g_{\mu\nu},f_{\mu\nu}]=\int d^4 x \sqrt{|g|}\left[ R_g+\frac{M^2_f}{M^2_g}\sqrt{\frac{|f|}{|g|}} R_f+\frac{2 M^4_{v}}{M^2_g} \sum\limits_{n=0}^4 \beta_n e_n \left(\sqrt{\mathbf{g^{-1} f}} \right)\right]}\,,
\end{equation}
where $R_g=R(g_{\mu\nu})$ and $R_f=R(f_{\mu\nu})$ are the curvature scalars of the propagating tensor fields $g_{\mu\nu}$ and $f_{\mu\nu}$ respectively, $M_g$ and $M_f$ are their respective coupling constants, $g$ and $f$ are their respective determinants, $M_{v}$ is the coupling of the potential and $\mu$ is the mass parameter of the theory. Bi-metric gravity propagates seven degrees of freedom, i.e. two massless modes plus five massive polarizations, while its field equations in vacuum are the following\cite{bi-metric_gravity_3,bi-metric_gravity_4,Brito_spin-2_SR_slow_rot}:
\begin{equation}\label{bi-metric_field_eqs_g}
    R^g_{\mu\nu}-\frac{1}{2}g_{\mu\nu} R_g+\frac{ M^4_{v}}{M^2_g} V^f_{\mu\nu}(g_{\mu\nu})=0\,,
\end{equation}
\begin{equation}\label{bi-metric_field_eqs_f}
    R^f_{\mu\nu}-\frac{1}{2}f_{\mu\nu} R_f+\frac{ M^4_{v}}{M^2_f} V^g_{\mu\nu}(f_{\mu\nu})=0\,,
\end{equation}
where the tensor potential $V^f_{\mu\nu}(g_{\mu\nu})$ is the one defined in \ref{tensor_potential_dRGT}\footnote{The fourth term in the sum of the bi-metric potential in \ref{bi-metric_action} becomes non-dynamical if $f_{\mu\nu}$ is kept fixed as a reference metric\cite{bi-metric_gravity_3}, thus it can be neglected in the case of the potential of dRGT gravity in \ref{dRGT_action}. This is why we can use the same expression for $V^f_{\mu\nu}(g_{\mu\nu})$ in both theories despite the fourth term is missing in the action of dRGT.}, while the other potential is 

\begin{equation}\label{tensor_potential_f}
 V^g_{\mu\nu}(f_{\mu\nu})=\sum\limits_{n=0}^3 (-1)^n \beta_{4-n} f_{\lambda\mu}Y^\lambda_{(n)\nu}\left(\sqrt{\mathbf{f^{-1} g}} \right)\,.
\end{equation}

\subsection{First order perturbations on a curved background}\label{spin-2_curved_background}

The only consistent equations on a curved background $\bar g_{\mu\nu}$ governing the first order perturbations of massive spin-2 modes $h_{\mu\nu}$ of a non-linear theory are the following\cite{consistency_spin-2_massive_perturbs}:
\begin{equation}\label{massive_spin-2_perturbs_wave}
    \bar\nabla_\sigma \bar\nabla^\sigma h_{\mu\nu}+2 \bar R^{~\alpha~\beta}_{\mu~\nu}h_{\alpha\beta}-\bar R^\sigma_{~\mu}h_{\sigma\nu}-\bar R^\sigma_{~\nu}h_{\sigma\mu}+h_{\mu\nu} \bar R-\bar g_{\mu\nu} h_{\alpha\beta}\bar R^{\alpha\beta}+\mu^2 h_{\mu\nu}=0\,,
\end{equation}
\begin{equation}\label{massive_spin-2_costraint_eqs}
\bar\nabla^\mu h_{\mu\nu}=0~,~\bar g^{\mu\nu}h_{\mu\nu}=0\,,~~~~~~~~~~~~~~~~~~~~~~~~~~~~~~~~~~~~~~~~~~~~~~~~~~~~~~~~~~~~~~~~~~~~~~~~~~~~~~~~~~
\end{equation}
where $\bar R^{~\alpha~\beta}_{\mu~\nu}$ and $\bar R^\sigma_{~\nu}=\bar R_{~~~\alpha\nu}^{\alpha\sigma}$ are the Riemann and Ricci curvature tensor of $\bar g_{\mu\nu}$ respectively, $\bar R=\bar R_{~~~\alpha\beta}^{\alpha\beta}$ is the background Ricci scalar, $\bar\nabla_\mu$ is the background covariant derivative and $\mu$ is the mass of the perturbation. In the case of a Ricci-flat background, e.g. a BH in GR, we get these simplified expressions for the perturbation:
\begin{equation}
    \bar\nabla_\sigma \bar\nabla^\sigma h_{\mu\nu}+2 \bar R^{~\alpha~\beta}_{\mu~\nu}h_{\alpha\beta}+\mu^2 h_{\mu\nu}=0\,,
\end{equation}
\begin{equation}
    \bar\nabla^\mu h_{\mu\nu}=0~,~\bar g^{\mu\nu}h_{\mu\nu}=0\,.
\end{equation}
Perturbative equations equivalent to \ref{massive_spin-2_perturbs_wave} and \ref{massive_spin-2_costraint_eqs} can be deduced from the linear theory in flat spacetime by making the substitution ($\partial_\sigma\rightarrow \bar\nabla_\sigma$, $\eta_{\mu\nu}\rightarrow \bar g_{\mu\nu}$, $d^4 x \rightarrow d^4 x \sqrt{-\bar g}$) to the Fierz-Pauli action \ref{Fierz-Pauli_action} and varying the resulting action:
\begin{equation}\label{massive_graviton_eqs}
    \bar{\mathcal{E}}^{\alpha\beta}_{\mu\nu}h_{\alpha\beta}-\frac{\mu^2}{2} \left(h_{\mu\nu}-\bar g_{\mu\nu}\bar g^{\alpha\beta}h_{\alpha\beta}\right)=0\,,
\end{equation}
where $\bar{\mathcal{E}}^{\alpha\beta}_{\mu\nu}$ is the operator associated with the linearization of the Einstein tensor on a background $\bar g_{\mu\nu}$:
\begin{equation}
\footnotesize\bar{\mathcal{E}}^{\alpha\beta}_{\mu\nu}=\frac{1}{2}\left(\bar g^{\alpha\beta}\bar \nabla_\mu \bar \nabla_\nu-\delta^\alpha_\mu \delta^\beta_\nu \bar \nabla_\sigma \bar\nabla^\sigma-\bar g_{\mu\nu}\bar g^{\alpha\beta} \bar \nabla_\sigma \bar\nabla^\sigma-\delta^\alpha_\mu \bar \nabla^\beta\bar\nabla_\nu-\delta^\alpha_\nu \bar \nabla^\beta\bar\nabla_\mu+\bar g_{\mu\nu}\bar \nabla^\alpha \bar \nabla^\beta\right)\,.
\end{equation}
In the case of dRGT theory the presence of a non-dynamical reference metric $f_{\mu\nu}$ complicates the situation: in general we get more involved equations, but the case $f_{\mu\nu}=\bar g_{\mu\nu}$ gives us exactly \ref{massive_graviton_eqs} \cite{Linearized_dRGT}.

The linearization of bi-metric gravity is even more complex, but it can be greatly simplified for a specific class of solutions. Among its solutions in fact there exist a subset having proportional metrics $\bar g_{\mu\nu}=C^2 \bar f_{\mu\nu}$ for some constant $C$, which surprisingly coincide with GR solutions with a cosmological costant due to field equations \ref{bi-metric_field_eqs_g} and \ref{bi-metric_field_eqs_f} reducing to the following expressions\cite{bi-metric_solutions, Brito_spin-2_SR_slow_rot}:
\begin{equation}
    \bar R^g_{\mu\nu}-\frac{1}{2}\bar g_{\mu\nu} \bar R_g+\Lambda_g \bar g_{\mu\nu}=0\,,
\end{equation}
\begin{equation}
    \bar R^f_{\mu\nu}-\frac{1}{2}\bar f_{\mu\nu} \bar R_f+\Lambda_f \bar f_{\mu\nu}=0\,,
\end{equation}
where $\bar R^g_{\mu\nu}$ and $\bar R^f_{\mu\nu}$ are the Ricci curvature tensors of $\bar g_{\mu\nu}$ and $\bar f_{\mu\nu}$ respectively, $\bar R_g=\bar R^g_{\mu\nu}\bar g^{\mu\nu}$ and $\bar R_f=\bar R^f_{\mu\nu}\bar f^{\mu\nu}$ are their curvature scalars and $\Lambda_f=\Lambda_g$ are the effective cosmological constant(s). Thus, we consider the perturbation of such proportional-metric solutions, i.e. $g_{\mu\nu}=\bar g_{\mu\nu}+\delta g_{\mu\nu}$ and $f_{\mu\nu}=C^2 \bar g_{\mu\nu}+\delta f_{\mu\nu}$, in the case of a BH background, therefore we also neglect the cosmological constant. The resulting perturbation equations give two massless and five massive modes\cite{bi-metric_gravity_3,Brito_spin-2_SR_slow_rot},
\begin{equation}
    \bar{\mathcal{E}}^{\alpha\beta}_{\mu\nu}h_{\alpha\beta}=0\,,~~~~~~~~~~~~~~~~~~~~~~~~~~~~~~~~
\end{equation}
\begin{equation}
    \bar{\mathcal{E}}^{\alpha\beta}_{\mu\nu}\mathcal{h}_{\alpha\beta}-\frac{\mu^2}{2}\left(\mathcal{h}_{\mu\nu}-\bar g_{\mu\nu}\bar g^{\alpha\beta}\mathcal{h}_{\alpha\beta} \right)=0\,,
\end{equation}
where $\mu^2=M_v^4 (C \beta_1+2 C^2 \beta_2+C^3 \beta_3 )[(C^2 M_f^2 )^{-1}+M_g^{-2} ]$ is the mass (squared) of the massive modes, while $h_{\mu\nu}$ and $\mathcal{h}_{\mu\nu}$ are the following linear combinations of $\delta g_{\mu\nu}$ and $\delta f_{\mu\nu}$:
\begin{equation}
    h_{\mu\nu}=\frac{M_g \delta g_{\mu\nu}+C M_f\delta f_{\mu\nu}}{\sqrt{C^2 M_f^2+M_g^2}}~,~\mathcal{h}_{\mu\nu}=\frac{M_g \delta f_{\mu\nu}-C M_f\delta g_{\mu\nu}}{\sqrt{C^2 M_f^2+M_g^2}}\,.
\end{equation}

\section{Observational bounds on ultralight boson masses}

In this section we analize the current and future possible bounds on ultralight boson masses imposed by astrophysical observations and (the very few possible) experiments. We consider both constraints arising from BH superradiant instabilities and the ones coming from other phenomenology. The interactions considered here are both those currently formalized as non-massive in the standard theory, i.e. gravity and electromagnetism, and proposals of new fields, e.g. scalar interactions, dark photons, new tensor particles.

\subsection{Constraints on the photon mass}

Though there are robust theoretical arguments excluding the possibility of a non-zero mass for the photon\cite{photon_graviton_mass}, e.g. the main one is how this hypothesis would break the conservation of electric charge, its hypothetical mass has been continuously constrained since early 20th century. The current best bound arising from laboratory testing of Coulomb's Law is still the one found by Williams, Faller and Hill in 1971\cite{photon_mass_lab_test,photon_graviton_mass}, i.e. $\mu_\gamma\lesssim 10^{-14} eV$, which is slightly better than the one found in the same year by Kroll\cite{photon_mass_schumann_resonance_test,photon_graviton_mass} by exploiting the Schumann resonances of Earth's atmosphere ($\mu_\gamma\lesssim 3 \times 10^{-13} eV$). The measurements of the magnetic field associated with the solar wind in our solar system, instead, gave us the strongest bound directly measurable, i.e. $\mu_\gamma\lesssim 10^{-18}eV$ (Ryutov, 2007 \cite{photon_mass_test_solar_wind,photon_graviton_mass}). The strongest constraints on the photon mass, though, come from observations on galactic and extra-galactic scales. From the observations of the Crab Nebula, in fact, the computed upper bound of the mass of the photon is $\mu_\gamma\lesssim 10^{-26} eV$\cite{photon_mass_test_crab,Chibisov_1976,photon_graviton_mass}, i.e. the so-called Yamaguchi-Chibisov limit. For such mass scales, we would have $\mu_\gamma M\lesssim 10^{-6}$ for BHs having mass $M$, with the upper limit coming from the theoretical maximum possible mass $M=5\times 10^{10}M_\odot$\footnote{$M_\odot\simeq 2\times 10^{30} kg$ is the mass of the Sun.} for super-massive BHs (SMBHs)\cite{maximum_BH_mass}: for such values we expect superradiant instabilities that would be negligible for astrophysical time scales, thus superradiance cannot be used for constraining more tightly the photon mass\cite{Brito_SR} (see sub-section \ref{ultralight_bounds_superradiance} about bounds from BH superradiance). Therefore any spin-1 ultralight field detected through BH superradiance would necessarily be some new beyond-SM particle.

\subsection{Constraints on the graviton mass}

In the previous section we saw how the hypothesis of a non-zero mass for the graviton is, instead, much more complex. Despite for a long time massive gravity has been considered not viable from a theoretical point of view, the hypothetical non-zero mass of the graviton has been widely tested, but the bounds found are less stringent than the ones of the photon\cite{photon_graviton_mass,graviton_mass_bound_review}. From the data of the decay rate of the Hulse-Taylor binary pulsar system Finn and Sutton (2002, \cite{graviton_mass_test_binary_pulsar,photon_graviton_mass,graviton_mass_bound_review}) found the bound $\mu_g\lesssim 7.6\times 10^{-20}eV$. Instead, the best observational constraint found by the LIGO-Virgo-Kagra Collaboration through the detection of gravitational waves is $\mu_g\lesssim 1.27\times 10^{-23} eV$\cite{GW_detect_1,graviton_mass_bound_review,LIGOScientific:2021sio}. Another very good model-independent bound on the graviton mass comes from the analysis of planetary motion in the solar system, which gives $\mu_g\lesssim 7.2\times 10^{-23} eV$ from the analysis of Mars (Will, 1997 \cite{C_Will_bound_graviton_mass,graviton_mass_bound_review}). With such bounds in this case, given the limit for the mass of SMBHs\cite{maximum_BH_mass} shown in the previous sub-section, we get $\mu_g M\lesssim 10^{-2}$ for a BH having mass $M$\cite{Brito_SR}, which is on the limit of the detection window of SMBH superradiance\footnote{Though this is valid under the current numerical knowledge of spin-2 BH superradiant instability, which excludes the polar dipolar mode, for which only a slow-rotation approximation is known\cite{Brito_spin-2_SR_slow_rot}. This mode is expected to trigger even stronger instabilities than the ones used for current bounds, thus a more precise numerical result is expected to give tighter constraints. See Section \ref{section:spin-2_SR} for further details.}. There are even tighter bounds that can be computed down to orders $\mu_g\lesssim 10^{-33}eV$ (i.e. Hubble scale mass), e.g. by exploiting also phenomena on the galactic or cosmological scale, but they are model-dependent and thus cannot give absolute constraints\cite{graviton_mass_bound_review}. Nevertheless a graviton mass of the order of the Hubble scale is a quite natural option for explaining the acceleration of the expansion of the Universe\cite{massive_gravity_cosmology_1,massive_bi-gravity_cosmology_1,massive_bi-gravity_cosmology_2,theoretical_observational_consistency_massive_gravity,cosmology_generalized_massive_gravity,graviton_mass_bound_review}. Even though such order of magnitude is far from the BH superradiance mass-window, this value could be constrained by taking into account that Kerr BHs are unstable under monopolar massive tensor perturbations if $\mu_g M\lesssim 0.438$, even in the non-spinning case, due to massive gravity BHs featuring a massive graviton hair\cite{massive_spin-2_instability_babichev,Brito_spin-2_SR_slow_rot,massive_gravity_hairy_BHs,testing_GR}.

\subsection{Bounds on ultralight bosons from BH superradiance}\label{ultralight_bounds_superradiance}

The key feature of spinning BH superradiant instabilities is that BHs are spun down by the energy extraction process, consequently what we expect is that the existence of ultralight bosons would exclude from the observations some BH mass-spin configurations depending on the mass and spin of the involved bosons\cite{Brito_SR,string_axiverse,string_axiverse_BH,discovering_axion_BHs,Cardoso_superradiant_instability,Brito_spin-2_SR_slow_rot}. This would cause holes in the Regge plane, i.e. exclusion regions in the BH mass-spin plots, that can be exploited for constraining the existence of ultralight bosons by using reliable BH mass and spin measurements and the theoretical estimates of the instability time scales\cite{Brito_SR,string_axiverse,string_axiverse_BH,discovering_axion_BHs,Cardoso_superradiant_instability,Brito_spin-2_SR_slow_rot,Axion_BH_AdvLIGO,GW_searches_ultralight_bosons,Ultralight_vector_SR_signatures,BH_spin_constraints_axion,Constraint_bosons_BH_spin}. By assuming an approximately continuous BH mass spectrum $M_\odot\lesssim M\lesssim 10^{10} M_\odot$\cite{maximum_BH_mass,BH_spin_theory&observ}, assuming all of them can be spinning, and by taking into account that superradiant instability is relevant if $\mu M\lesssim 1$ we can constrain an approximate range $10^{-21}eV\lesssim \mu\lesssim 10^{-10} eV $ for the boson mass $\mu$, where the most massive BHs constrain the lightest particles and vice-versa\cite{Brito_SR}. 

\begin{figure}[H]
\centering
\includegraphics[width=0.62\textwidth]{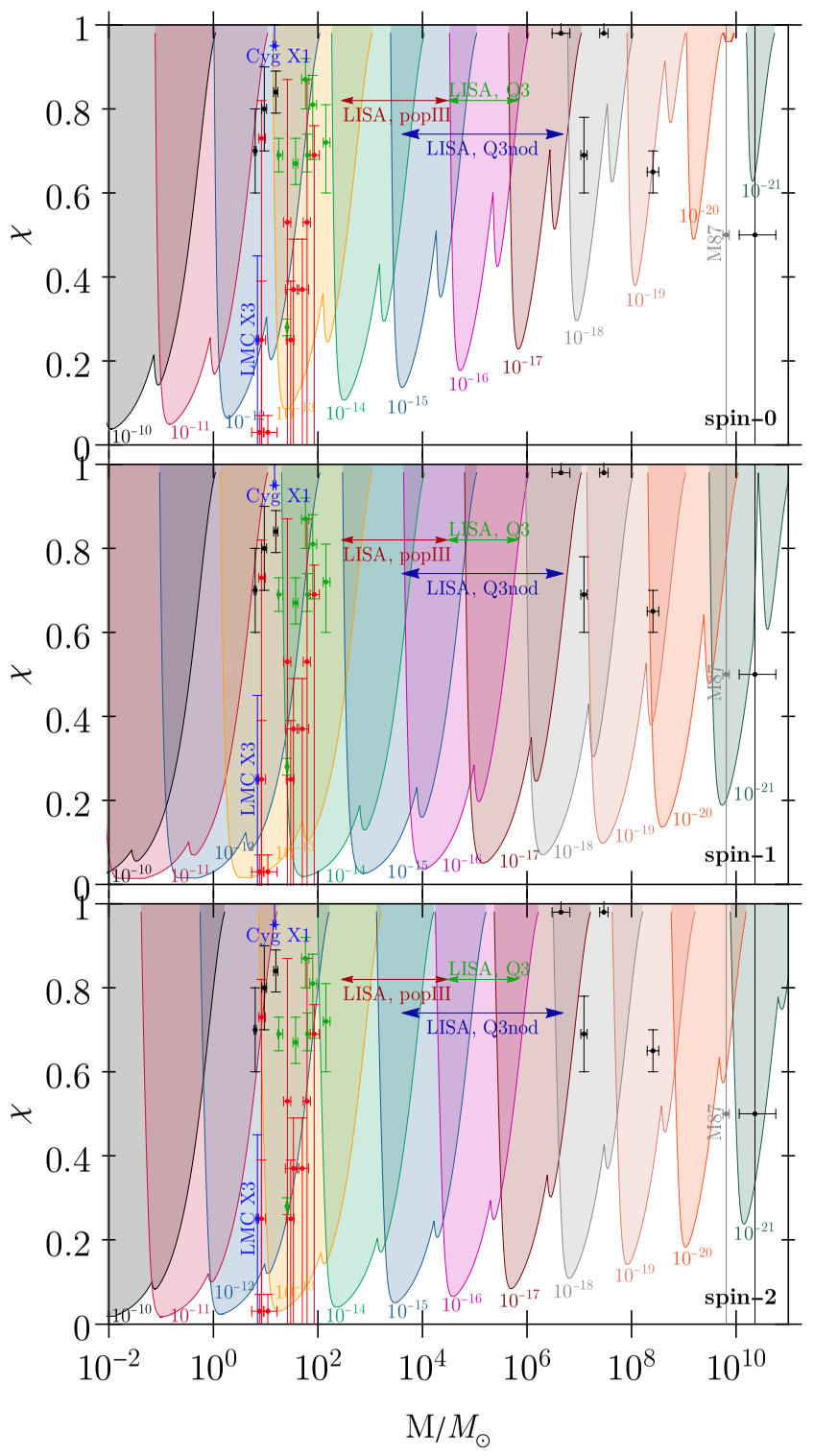}\\
\caption{The exclusion regions in the BH mass ($M$) vs adimensional spin parameter ($\chi=a/M)$ Regge plane produced by BH superradiant instability in the case of scalar, vector and tensor ultralight fields, for different values (see colors) of the boson mass. For each mass the separatrix corresponds to an instability time scale being equal to the Salpeter time $\tau=2\times 10^7 yr$. The plotted dots refer to observations excluding specific values of the boson mass. In this plot the spin-2 polar dipolar mode, which could provide stronger constraints, is not considered because its instability time is currently confirmed only for the small spin approximation: given that this mode has a unique behaviour, its computation is more involved and thus its exact value has been computed only recently \cite{dias2023black} and is still under further investigation (see subsection \ref{tensor_numerical_results} for more details). Image and data elaboration by \textit{Brito et al.}, from the book \textit{Superradiance: New Frontiers in Black Hole Physics}\cite{Brito_SR}.}
\label{regge_plane_ultralight_boson_constraints}
\end{figure}

In Figure \ref{regge_plane_ultralight_boson_constraints} the exclusion zones for scalar (spin-0), vector (spin-1) and tensor (spin-2) fields for different values of the boson masses are plotted in the Regge plane. The plotted dots refer to observed BH configurations: the black ones are stellar or supermassive BH spins
estimated through measurements of K$\alpha$ iron line or the continuum fitting method, red
dots are data from LIGO-Virgo primary and secondary BHs in merger events, while the remaining green one are LIGO-Virgo data for the remnants of the events (further info can be found in section 6.2.1 of \cite{Brito_SR}). 
The existence in nature of some observed BH mass-spin configurations implies the exclusion of the boson masses associated to the exclusion regions occupied by the observed BHs, thus exploiting superradiant instabilities for constraining the ultralight bosons. The refinement of observations through the introduction of new and more precise gravitational wave detectors\cite{extreme_gravity_detections_fundamental_physics}, e.g. the Laser Interferometer Space Antenna (LISA)\cite{LISA_GW_detector}, together with improvements in numerical results from the predictions can give us tighter bounds in future\cite{Brito_SR}.

\begin{figure}[H]
\centering
\includegraphics[width=0.57\textwidth]{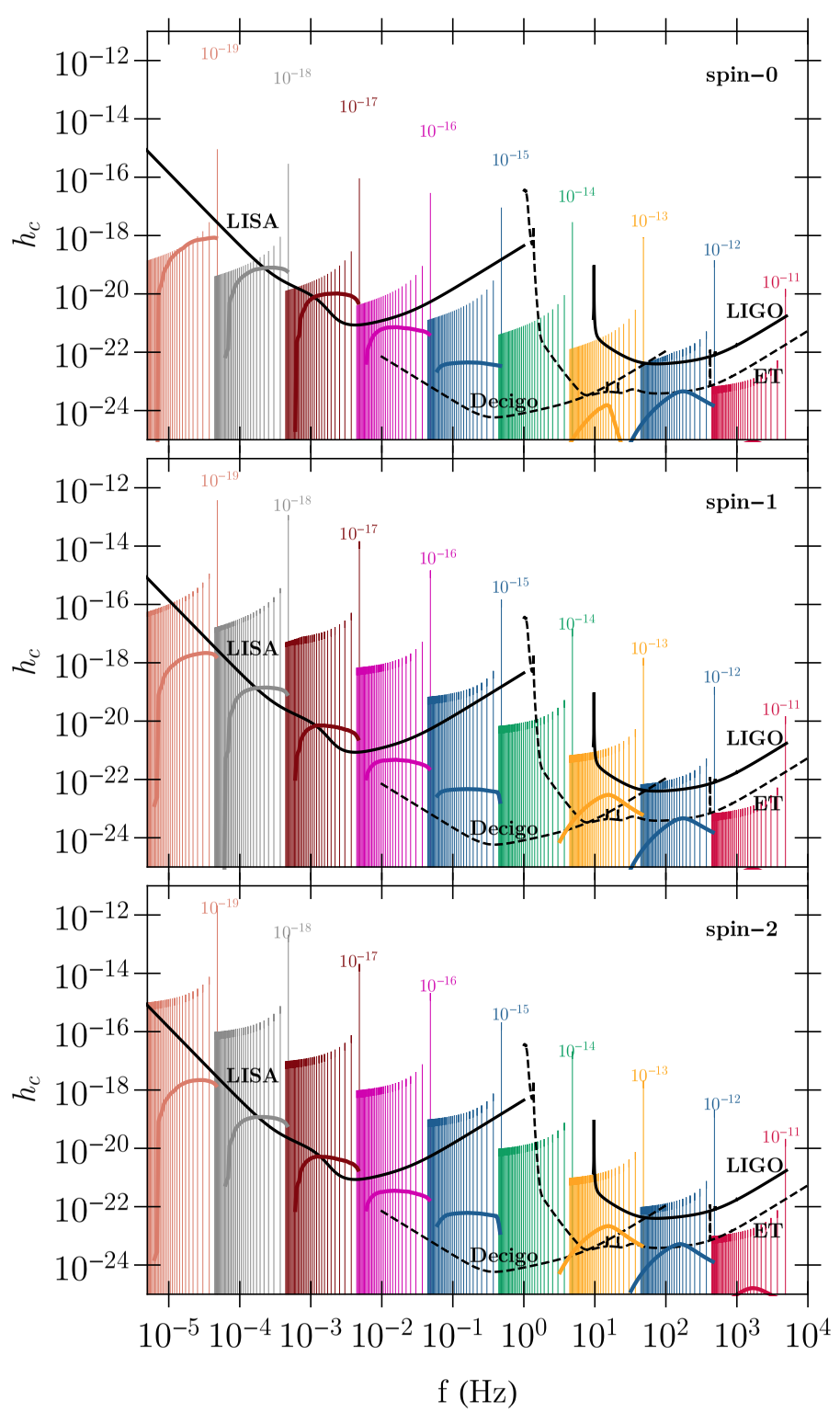}\\
\caption{GW strain amplitudes of ultralight bosonic condensates vs their frequency, compared with the noise curves of LISA, DECIGO, Einstein Telescope and Adv. LIGO. The amplitude
is computed at its peak value for an observation time of 4 yrs and an initial BH spin of $\chi=0.9$, while the considered emission is from the dominant unstable hydrogenic mode (see Sections \ref{section:spin-0_SR}, \ref{section:spin-1_SR} and \ref{section:spin-2_SR} for details about the hydrogenic approximation.). Image and data elaboration by \textit{Brito et al.}, from the book \textit{Superradiance: New Frontiers in Black Hole Physics}\cite{Brito_SR}.}
\label{continuous_GW_detectability}
\end{figure}

BH superradiant instabilities are expected to cause the condensation of bosonic clouds enveloping the compact objects\cite{East_non-linear_SR_1,East_non-linear_SR_2,Herdeiro_Kerr_Hair,Yoshino_axion_cloud_SR,Brito_evolution_SR_instabilities}, which would emit continuous GWs due the generation of non-zero quadrupoles arising from the asymmetries of the clouds\cite{East_non-linear_SR_1,East_non-linear_SR_2,Herdeiro_Kerr_Hair,Yoshino_axion_cloud_SR}. With the activation of the space-based detector LISA and third generation ground-based GW detectors\cite{einstein_telescope_GW_detector,next_gen_GW_detectors,freq_response_3rd_gen_GW_detectors,DECIGO_GW_detector}, e.g. Einstein Telescope and Cosmic Explorer, the continuous emissions from bosonic clouds in the mass interval $10^{-19} eV\lesssim\mu\lesssim 10^{-11} eV$ could be detectable\cite{Brito_SR}. This can be clearly seen in Figure \ref{continuous_GW_detectability}, where the expected strain amplitudes of the continuous GW emissions from bosonic condensates around BHs is plotted in comparison with the expected noise curves of LISA, DECIGO, Einstein Telescope and Advanced LIGO with respect to the frequency\cite{Brito_SR}.

\begin{figure}[H]
\centering
\includegraphics[width=1\textwidth]{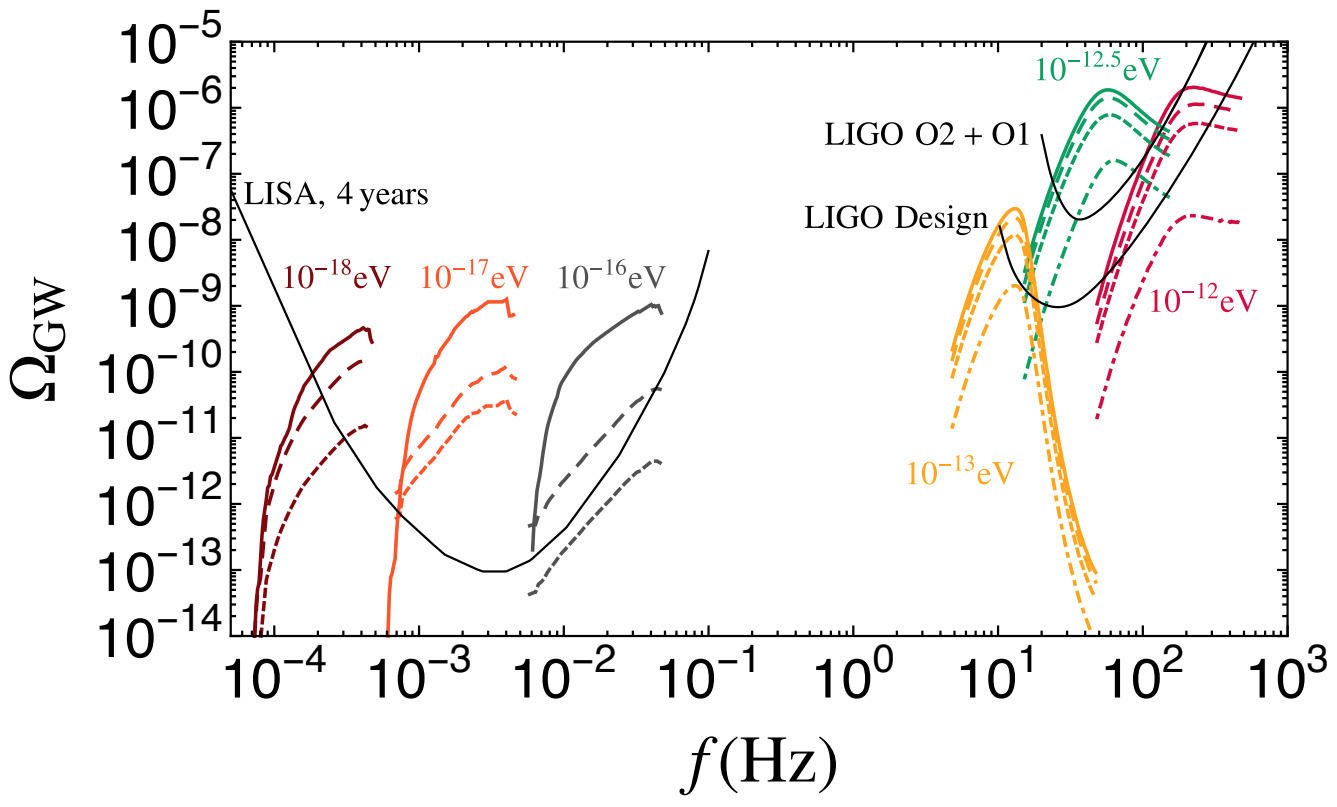}\\
\caption{Stochastic GW signals from scalar BH condensates in the LISA and LIGO bands, compared with the noise curves. In the case of LISA, the three different plots for each scalar mass value correspond to “optimistic”
(top), “less optimistic” (middle) and “pessimistic” (bottom) choices of astrophysical models by the authors. For LIGO, instead, the different signals for the boson masses shown
are associated to a uniform initial spin distribution with, from top to bottom, $\chi \in$ $[0.8, 1]$, $[0.5, 1]$, $[0, 1]$ and $[0, 0.5]$. Image and data elaboration by \textit{Brito et al.}, from Ref. \cite{Stochastic_GW_2}.}
\label{stochastic_boson_GW_emissions}
\end{figure}

In addition to individual GW sources, we expect a great number of emissions from clouds that are too faint to be individually detected that would generate a stochastic GW background\cite{Brito_SR,Stochastic_GW_2}. This effect was first computed in \cite{Stochastic_GW_2} in the case of emissions from scalar clouds, whose results are shown in Figure \ref{stochastic_boson_GW_emissions}. The GW background is characterized by its adimensional energy spectrum $\displaystyle\Omega_{GW}=\frac{1}{\rho_c}\frac{d\rho_{GW}}{d\ln f}$, where $\rho_{GW}$ is the energy density of the background, $\rho_c$ is the critical density of the Universe at present time and $f$ is the GW frequency measured by the detector. In the case of the most optimistic astrophysical models, LIGO could costrain scalar masses in the range $2\times 10^{-13} eV\lesssim\mu\lesssim 10^{-12} eV$, while LISA would be able to detect in the range $5\times 10^{-19} eV\lesssim\mu\lesssim 5\times 10^{-16} eV$\cite{Stochastic_GW_2,Brito_SR}. Similar ranges should be expected also for the vector and tensor cases\cite{Brito_SR}.

\chapter{Spherical harmonic decomposition of fields in Kerr spacetimes}\label{chapter:spherical_harmonic}

In this chapter we will start outlining a numerical technique for the resolution of bosonic field equations in BH spacetimes based on the spherical harmonic decomposition of perturbations, set up for computing the quasi-bound states (QBSs) of massive fields. This approach does not rely on any separation ansätze, thus it can potentially be applied in any BH perturbation theory problem, even non-separable ones. The technique has been shown for the first time by \textit{Baumann et al.} in Ref. \cite{Spectra_grav_atom}, where it was applied for solving Klein-Gordon and Proca equations in a Kerr background. In this chapter we will show the formalism of the tensor spherical harmonic decomposition used in \cite{Spectra_grav_atom} and extend it also to the case of spin-2 fields.

While in the case of scalar fields the harmonic decomposition is straightforward, the vector and tensor cases need some careful extra manipulation. In the following sections we will show how the vector and tensor problems can be reduced to the case of coupled Klein-Gordon fields by stripping their spinful nature through spin-eigenstate decompositions. After such operation, the coupled fields can be consequently decomposed with scalar spherical harmonics like in the standard Klein-Gordon case.

\section{Decomposition of Klein-Gordon fields}

We start with the simplest case, i.e. scalar fields. The explicit expression of the scalar field perturbative equation \ref{Klein-Gordon_Perturb_eq} in the case of Kerr background in Boyer-Lindquist coordinates (metric \ref{Kerr_metric}) is the following:
\begin{equation}\label{Klein-Gordon-Kerr}
\left[\frac{(r^2+a^2)^2}{\Delta(r)}-a^2 \sin^2\theta \right] \frac{\partial^2 \Phi}{\partial t^2}+\frac{4 M a r}{\Delta(r)}\frac{\partial^2\Phi}{\partial t\partial\phi}+\left[\frac{a^2}{\Delta(r)}-\frac{1}{\sin^2\theta} \right]\frac{\partial^2 \Phi}{\partial \phi^2}
\end{equation}
\[-\frac{\partial}{\partial r}\left(\Delta(r) \frac{\partial\Phi}{\partial r}\right)-\frac{1}{\sin\theta}\frac{\partial}{\partial\theta}\left(\sin\theta\frac{\partial\Phi}{\partial\theta}\right)+\mu_S^2\Phi=0\,,\]
where $\Delta(r)=r^2-2 M r+a^2$, $M$ is the mass of the BH, $a$ is its spin parameter, $\mu_S$ is the mass of the scalar perturbation $\Phi$ and $x^\mu=\left(t,r,\theta,\phi\right)$ are the Boyer-Lindquist coordinates. 

We define the following angular operators,
\begin{equation}\label{angular_momentum_op_scalar}
   \mathbfcal{L}^2:= -\frac{1}{\sin\theta}\frac{\partial}{\partial\theta}\left(\sin\theta\frac{\partial}{\partial\theta}\right)-\frac{1}{\sin^2\theta} \frac{\partial^2 }{\partial\phi^2}\,,
\end{equation}
\begin{equation}\label{azimuth_angular_momentum_op_scalar}
    \mathbfcal{L}_{\boldsymbol z}=-i\frac{\partial}{\partial\phi}\,,
\end{equation}
whose common eigen-functions are the spherical harmonics $Y_{l,m}(\theta,\phi)$\cite{rose_angular_momentum}:
\begin{equation}
    \mathbfcal{L}^2 Y_{l,m}=l(l+1)Y_{l,m}~,~\mathbfcal{L}_{\boldsymbol z} Y_{l,m}=m Y_{l,m}~,~|m|\leq l~,~l\in\mathbb{N}~,~m\in\mathbb{Z}\,.
\end{equation}
Given that $\Phi$ describes particles, we can associate the spherical harmonics $Y_{l,m}$ to states of particles having angular momentum $l(l+1)$ with azimuthal component equal to $m$, in perfect accordance with first quantization\cite{Sakurai_quantum_mechanics}. Thus we call $\mathbfcal{L}$ and $\mathbfcal{L}_{\boldsymbol z}$ angular momentum and azimuthal angular momentum operators respectively: in the next sections we will see how expressions \ref{angular_momentum_op_scalar} and \ref{azimuth_angular_momentum_op_scalar} are the scalar case representations of angular momentum operators that can be applied to tensor fields of any rank. 

For the resolution of \ref{Klein-Gordon-Kerr} we exploit the stationarity and the axisymmetry of the background by considering solutions having definite frequency $\omega$ and azimuthal angular momentum $m$, i.e. $\partial_t \Phi=-i\omega \Phi$ and $\mathbfcal{L}_{\boldsymbol z} \Phi=m \Phi$.
Thus we rewrite equation \ref{angular_momentum_op_scalar} in the following form, which exploits the angular momentum operators and is more suitable for its asymptotic study (we will see it in Section \ref{section:boundary_conditions}):
\begin{equation}\label{klein-gordon_manipulation}
    \frac{\Sigma(r,\theta)}{\Delta(r)}\left(\bar \Box+\mu_S^2\right)\Phi=\frac{1}{\Delta(r)}\left\lbrace\frac{\Sigma(r,\theta)}{\Delta(r)}(\mu_S^2-\omega^2)-\frac{m^2 a^2}{\Delta(r)}+\mathbfcal{L}^2\right.~~~~~~~~~
\end{equation}
\[\left.-\frac{\partial}{\partial r}\left[\Delta(r)\frac{\partial}{\partial r} \right]+\frac{4 m a\omega M r}{\Delta(r)}-\omega^2\frac{2 M r(a^2+r^2)}{\Delta(r)}\right\rbrace\Phi=~~~~~~~~~~~~~~~~~~\]
\[=\left\lbrace\displaystyle\frac{1}{\Delta(r)}[\mathbfcal{L}^2+\left(\mu_S^2-\omega^2\right) a^2\cos^2\theta]-\frac{1}{\Delta(r)}\frac{\partial}{\partial r}\left[\Delta(r)\frac{\partial}{\partial r}\right]+\mu_S^2-\omega^2\right.\]
\[~~~~~\left.-\frac{P_+^2}{(r-r_+)^2}-\frac{P_-^2}{(r-r_-)^2}+\frac{A_+}{(r_+-r_-)(r-r_+)}-\frac{A_-}{(r_+-r_-)(r-r_-)}\right\rbrace\Phi=\]
\[=0\,,~~~~~~~~~~~~~~~~~~~~~~~~~~~~~~~~~~~~~~~~~~~~~~~~~~~~~~~~~~~~~~~~~~~~~~~~~~~~~~~~~~~\]
where $\Sigma(r,\theta)=r^2+a^2\cos^2\theta$, $\bar\Box=\bar\nabla_\sigma \bar \nabla^\sigma$ is the D'Alembert operator of the background metric, $\displaystyle r_\pm=M\pm\sqrt{M^2-a^2}$ are the roots of $\Delta(r)$ and $P_\pm$ and $A_\pm$ are parameters defined by 
\begin{equation}
    \small P_\pm=\frac{m a-2\omega M r_\pm}{r_+-r_-}=\left(m \Omega_H-\omega\right)\frac{2 M r_+}{r_+-r_-}\,,~~~~~~~~~~~~~~~~~~~~~~~~~~~~~~~~~~~~~~~~~~~~~~~~~~~~
\end{equation}
\begin{equation}
    \small A_\pm=P_+^2+P_-^2 + M^2(\mu_S^2-7\omega^2)\pm(\mu_S^2-2\omega^2)M(r_+-r_-)+(M^2-a^2)(\mu_S^2-\omega^2)\,.
\end{equation}
We now proceed with the spherical harmonic decomposition, thus we express the solution in the following form,
\begin{equation}\label{phi_ansatz}
\Phi=e^{-i \omega t} F(r)\sum\limits_{l\geq|m|}^\infty B_{l}(\zeta(r))Y_{l,m}(\theta,\phi)\,,
\end{equation}
where $m$ is fixed. $F(r)$ is a function we will define in Section \ref{section:boundary_conditions} when we will manage the asymptotic behaviour of the field and its boundary conditions, while $B_{l}(\zeta(r))$ are the radial functions of the solution to be computed, depending on an auxiliary radial coordinate $\zeta(r)$ that will be fixed in Section \ref{section:chebyshev_interpol}. Because of the absence of spherical symmetry in the background metric, the decomposition does not separate the radial and angular parts, consequently what we get is an infinite cascade of coupled radial equations:

\begin{equation}
    \small\left\lbrace\frac{\partial^2}{\partial\zeta^2}+\left[\left(\frac{1}{r-r_+}+\frac{1}{r-r_-} \right)\frac{1}{\zeta'(r)}+\frac{2 F'(r)}{F(r)}\frac{1}{\zeta'(r)}+\frac{\zeta''(r)}{\zeta'^2(r)}\right]\frac{\partial}{\partial\zeta}+\right.
\end{equation}
\begin{small}
\[+\frac{1}{\zeta'^2(r)}\frac{ F''(r)}{F(r)}+\left(\frac{1}{r-r_+}+\frac{1}{r-r_-} \right)\frac{1}{\zeta'^2(r)}\frac{F'(r)}{F(r)}+\frac{1}{\zeta'^2(r)}\left[\frac{P_+^2}{(r-r_+)^2}+\frac{P_-^2}{(r-r_-)^2}-\right.\]
\[\left.\left.-\frac{A_+}{(r_+-r_-)(r-r_+)}+\frac{A_-}{(r_+-r_-)(r-r_-)}-(\mu_S^2-\omega^2)-\frac{l(l+1)}{\Delta(r)}\right]\right\rbrace B_{l}(\zeta(r))\]
\[-\frac{1}{\zeta'^2(r)}\frac{a^2 \left(\mu_S^2-\omega^2 \right)}{\Delta(r)}\sum\limits_{l'\geq|m|}^\infty c^m_{l l'} B_{l'}(\zeta(r))=0\,,\]
\end{small}
where the couplings $c^m_{l l'}$ are spherical harmonic bra-ket overlap integrals\cite{Dolan_scalar_SR},
\begin{equation}\label{eq:c_ll^m}
\begin{aligned}
    c^m_{l l'}&=\left\langle l,m\right|\cos^2\theta \left|l', m  \right\rangle=\int\limits_0^{2\pi}d\phi \int\limits_0^\pi d\theta \sin\theta Y_{l,m}^*(\theta,\phi)\cos^2\theta Y_{l',m}(\theta,\phi)\\
    &=\frac{1}{3}\delta_{l l'}+\frac{2}{3}\sqrt{\frac{2 l'+1}{2 l+1}}\left\langle l' , m , 2 , 0 \right| \left. l , m\right\rangle\left\langle l' , 0 , 2 , 0 \right| \left. l , 0\right\rangle\,,
\end{aligned}
\end{equation} $\left| l_1 , m_1, l_2 , m_2\right\rangle=\left| l_1 , m_1\right\rangle \otimes \left| l_2 , m_2\right\rangle$ , while $\left\langle l_1 , m_1 , l_2, m_2 \right| \left. l , m\right\rangle$, instead, are Clebsch-Gordan coefficients for the sum of angular momentum states $\left| l_1 , m_1\right\rangle$  and $\left| l_2 , m_2\right\rangle$\footnote{They are the coefficients of the linear combination\\$\displaystyle\left| l , m\right\rangle=\sum\limits_{m_1=-l_1}^{l_1}\sum\limits_{m_2=-l_2}^{l_2}\left\langle l_1 , m_1 , l_2, m_2 \right| \left. l , m\right\rangle \left| l_1 , m_1\right\rangle \otimes \left| l_2 , m_2\right\rangle$, where the resulting total angular momentum $l$ must satisfy the triangle inequality $|l_1-l_2|\leq l\leq|l_1+l_2|$\cite{rose_angular_momentum}.}. These radial equations can be put  in the following more compact form,
\begin{equation}\label{radial_wave_eqs}
\left[\frac{\partial^2 }{\partial\zeta^2}+C_{1}(r(\zeta))\frac{\partial}{\partial\zeta}+ C_{2}(r(\zeta))\right]B_{l}(\zeta)+\sum\limits_{l'\geq|m|}^\infty C_{3 , l}^{l'}(r(\zeta))B_{l'}(\zeta)=0\,,
\end{equation}
where we have the following radial functions:
\begin{equation}
    \small C_1(r)=\left(\frac{1}{r-r_+}+\frac{1}{r-r_-} \right)\frac{1}{\zeta'(r)}+\frac{2 F'(r)}{F(r)}\frac{1}{\zeta'(r)}+\frac{\zeta''(r)}{\zeta'^2(r)}\,,
\end{equation}
\begin{equation}
    \small C_2(r)=\frac{1}{\zeta'^2(r)}\frac{ F''(r)}{F(r)}+\left(\frac{1}{r-r_+}+\frac{1}{r-r_-} \right)\frac{1}{\zeta'^2(r)}\frac{F'(r)}{F(r)}~~~~
\end{equation}
\begin{small}
\[~+\frac{1}{\zeta'^2(r)}\left[\frac{P_+^2}{(r-r_+)^2}+\frac{P_-^2}{(r-r_-)^2}-\frac{A_+}{(r_+-r_-)(r-r_+)}\right.\]
\[\left.+\frac{A_-}{(r_+-r_-)(r-r_-)}-(\mu_S^2-\omega^2)-\frac{l(l+1)}{\Delta(r)}\right]\,,~~~~~~~~~\]
\end{small}
\begin{equation}
    \small C_{3 , l}^{l'}(r)=-\frac{1}{\zeta'^2(r)}\frac{a^2 \left(\mu_S^2-\omega^2 \right)}{\Delta(r)}c^m_{l l'}\,. ~~~~~~~~~~~~~~~~~~~~~~~~~~~~~~~~~~~~~
\end{equation}

\section{Tensor spin-eigenstate decomposition of spin-1 and spin-2 fields}\label{section:spin-eigenstate_formalism}

For the decomposition of spin-1 and spin-2 fields we need a mathematical framework generalizing to the vector and tensor cases what we did with scalar fields. \textit{Baumann et al.} \cite{Spectra_grav_atom} extended the angular momentum operator $\mathbfcal{L}$ to tensor representations of any rank by exploiting the properties of Killing vectors and Lie derivatives. 

They defined the following vectors parallel to the directions $t$, $x$, $y$, $z$ of cartesian spacetime, 
\begin{small}
\begin{equation}\label{killing}
k_t^\mu=-i\left[\begin{matrix}
1\\0\\0\\0
\end{matrix}\right]
~,~
k_x^\mu=i\left[\begin{matrix}
0\\0\\\sin\phi\\ \cot\theta\cos\phi
\end{matrix}\right]
~,~
k_y^\mu=i\left[\begin{matrix}
0\\0\\ -\cos\phi\\ \cot\theta\sin\phi
\end{matrix}\right]
~,~
k_z^\mu=-i\left[\begin{matrix}
0\\0\\0\\1
\end{matrix}\right]\,,
\end{equation}
\end{small}which are Killing vectors for spacetimes featuring stationarity and spherical symmetry\footnote{They are Schwarzschild metric's Killing vectors. Only $k^\mu_t$ and $k^\mu_z$ are also Kerr's.}. They used these vectors for building the following associated differential operators,
\begin{equation}
    \boldsymbol{k_i}=k_i^\mu \partial_\mu~,~ i\in\left\lbrace t, x , y, z \right\rbrace\,,
\end{equation}
among which the ones from space-like vectors happen to be generators of the $SO(3)$ algebra:
\begin{equation}
    \left[\boldsymbol{k_j},\boldsymbol{k_k}\right]=i\epsilon_{jkl}\boldsymbol{k_l}~,~ j,k,l\in\left\lbrace x , y, z \right\rbrace\,.
\end{equation}
This is a consequence of the spherical symmetry associated with the chosen Killing vectors in \ref{killing}: each one of them describes a rotation axis, therefore they collectively represent spacetime symmetry under any space rotation, i.e. $SO(3)$ symmetry. 

The next ingredient comes from Lie derivatives\cite{Yano_Lie_derivatives}, which enable us to evaluate the derivative $\mathbfcal{L}_{\boldsymbol X}({\boldsymbol T})$ of tensor fields ${\boldsymbol T}$ along the flow defined by a vector field $X^\mu$ (with $\boldsymbol{X}=X^\mu\partial_\mu$). By taking into account that $\mathbfcal{L}_{\boldsymbol X}\mathbfcal{L}_{\boldsymbol Y}=\mathbfcal{L}_{\boldsymbol X \boldsymbol Y}$, we can write Lie derivatives commutators as  $[\mathbfcal{L}_{\boldsymbol X},\mathbfcal{L}_{\boldsymbol Y}]=\mathbfcal{L}_{[\boldsymbol X , \boldsymbol Y]}$. This last property is the key to finding the angular momentum operators. In fact, by defining the operators $\mathbfcal{L}_{\boldsymbol j}$ as the Lie derivatives along the flow of Killing vectors $\boldsymbol{k_j}$ we find that 
\begin{equation}
   \mathbfcal{L}_{\boldsymbol j}=\mathbfcal{L}_{\boldsymbol k_j} ~,~[\mathbfcal{L}_{\boldsymbol j}, \mathbfcal{L}_{\boldsymbol k}]=i\epsilon_{jkl}\mathbfcal{L}_{\boldsymbol l}~,~ j,k,l\in\left\lbrace x , y, z \right\rbrace\,,
\end{equation}
which means $\mathbfcal{L}_{\boldsymbol j}$ are angular momentum operators\cite{Sakurai_quantum_mechanics,rose_angular_momentum}. This algebra has a quadratic Casimir invariant $\mathbfcal{L}^2=\mathbfcal{L}_{\boldsymbol x}^2+\mathbfcal{L}_{\boldsymbol y}^2+\mathbfcal{L}_{\boldsymbol z}^2$, which can be identified as the total angular momentum squared, commuting with all the operators $\mathbfcal{L}_{\boldsymbol j}$\cite{Sakurai_quantum_mechanics,rose_angular_momentum}. Therefore, in analogy with quantum mechanics, it is possible to find tensor irreducible representations of SO(3) of any rank identified by their eigenvalues with respect to $\mathbfcal{L}^2$ and a chosen $\mathbfcal{L}_{\boldsymbol i}$\cite{Sakurai_quantum_mechanics,rose_angular_momentum,Spectra_grav_atom}. By making the standard choice $\mathbfcal{L}_{\boldsymbol i}=\mathbfcal{L}_{\boldsymbol z}$, we will have the representation \footnote{It has an arbitrary chosen tensor rank.} $\left| j,j_z \right\rangle$ carrying total angular momentum $j$ and total azimuthal angular momentum $j_z$, with $|j_z|\leq j$, $j\in\mathbb{N}$ and $j_z\in\mathbb{Z}$. Hence we have
\begin{equation}\label{L_eigenvalue_eqs}
\mathbfcal{L}^2  \left| j,j_z \right\rangle=j(j+1) \left| j,j_z \right\rangle~,~\mathbfcal{L}_{\boldsymbol z}\left| j,j_z \right\rangle=j_z \left| j,j_z \right\rangle\,,
\end{equation}
and we can define the operators $\mathbfcal{L}_{\boldsymbol +}=\mathbfcal{L}_{\boldsymbol x}+i \mathbfcal{L}_{\boldsymbol y}$ and $\mathbfcal{L}_{\boldsymbol -}=\mathbfcal{L}_{\boldsymbol x}-i \mathbfcal{L}_{\boldsymbol y}$, which give
\begin{equation}\label{L+L-}
\mathbfcal{L}_{\boldsymbol \pm} \left| j,j_z \right\rangle=\sqrt{(j\mp j_z)(j\pm j_z+1)}\left| j,j_z\pm 1\right\rangle\,.
\end{equation}
Now we have all the required instruments for generating tensor eigenstates of any rank. 

We start by looking for spin-1 eigenstates, i.e. a basis of elements $\chi^\mu_{s,s_z}$ ($s=1$, $s_z\in\left \lbrace -1,0,1\right\rbrace$) for the vector representation having angular momentum arising from the vector nature only (i.e. spin). The general rule is "spin $\leq$ tensor rank", thus in the vector representation angular momentum higher than 1 cannot be intrinsic and consequently requires the coupling of the vector eigenstates with some scalar spherical harmonics $Y_{l,m}$, as we shall see in the following paragraphs when we will build the vector and tensor harmonics. From \ref{L_eigenvalue_eqs} and \ref{L+L-} we get the equations to be solved for finding the spin-up vector eigenstate,
\begin{equation}\label{spinupeq}
\mathbfcal{L}_{\boldsymbol +}  \chi^\mu_{1,1}=0~,~\mathbfcal{L}_{\boldsymbol z}\chi^\mu_{1,1}=\chi^\mu_{1,1}\,,
\end{equation}
where $\chi^\mu_{1,1}$ is the vector eigenstate with total spin $s=1$ and azimuthal spin $s_z=1$. The general vector solution to \ref{spinupeq} is
\begin{equation}\label{chi1_1}
 \chi^\mu_{1,1}\partial_\mu=\frac{e^{i\phi}}{\sqrt{2}}\left[F_r(r)\sin\theta\partial_r+F_\theta(r)\left(\cos\theta \partial_\theta+\frac{i}{\sin\theta}\partial_\phi \right)\right]\,,
\end{equation}
where $F_r(r)$ and $F_\theta(r)$ are free functions that can be chosen arbitrarily\footnote{The absence of any radial partial derivative inside the angular momentum operators introduces these free radial functions. The general solution includes also a third function $F_{\phi}(r)$ coupled with pseudo-vector components, therefore to be set equal to zero in the vector case \cite{Spectra_grav_atom}.}.
From \ref{L+L-} we get $\mathbfcal{L}_{\boldsymbol -}  \chi^\mu_{1,1}=\sqrt{2}\chi^\mu_{1,0}$ and $\mathbfcal{L}_{\boldsymbol -}  \chi^\mu_{1,0}=\sqrt{2}\chi^\mu_{1,-1}$, hence for $s_z=0$ and $s_z=-1$ respectevely we get
\begin{equation}\label{chi1_0}
 \chi^\mu_{1,0}\partial_\mu=-F_r(r)\cos\theta\partial_r+F_\theta(r)\sin\theta \partial_\theta\,,
\end{equation}
\begin{equation}\label{chi1_-1}
 \chi^\mu_{1,-1}\partial_\mu=\frac{e^{-i\phi}}{\sqrt{2}}\left[-F_r(r)\sin\theta\partial_r+F_\theta(r)\left(-\cos\theta \partial_\theta+\frac{i}{\sin\theta}\partial_\phi \right)\right]\,.
\end{equation}
These vector eigenstates form a complete basis just for space dimensions, therefore we need an extra vector for covering also the temporal one, thus carrying spin $s=0$\footnote{The temporal component is invariant under space rotations.}:
\begin{equation}\label{tau_vector}
\tau^\mu=F_t(r)\delta^\mu_0\,.
\end{equation}

Computations involving projections on spin-eigenstates and objects having covariant indices imply we need to compute also the associated covector eigenstates. The complex conjugation of these covectors to be found must be $\left\langle s , s_z \right|$ eigenstates, therefore, because of the orthogonality condition $\left\langle s , s_z \right.\left| s',s_z'\right\rangle=\delta_{s,s'}\delta_{s_z , s_z'}$, we must have
\begin{equation}\label{vec_ortho}
\left(\tilde{\chi}^{s,s_z}_\mu\right)^* \chi_{s',s_z'}^\mu=\delta_{s,s'}\delta_{s_z , s_z'}~,~\tilde{\tau}_\mu\tau^\mu=1~,~\tilde{\tau}_\mu\chi_{s',s_z'}^\mu=0~,~\left(\tilde{\chi}^{s,s_z}_\mu\right)^*\tau^\mu=0
\end{equation}
where $\tilde{\tau}_\mu$ and $\tilde{\chi}^{s,s_z}_\mu$ respectevely are the duals of $\tau^\mu$ and $\chi_{s,s_z}^\mu$. The covector spin eigenstates therefore will be \footnote{Partial derivatives and $dx^\mu$ differentials respectevely behave like a coordinate basis for vectors and covectors, with $\partial_\mu$ and $dx^\mu$ being dual to each other. This is why we can always express vectors and covectors as differential operators and differential 1-forms respectevely.}:
\begin{subequations}
\begin{align}
&\label{tau_covector}
\tilde{\tau}_\mu=\frac{1}{F_t(r)}\delta_\mu^0\,,\\
&\label{dualchi1_1}
 \tilde{\chi}_\mu^{1,1}dx^\mu=\frac{e^{i\phi}}{\sqrt{2}}\left[\frac{1}{F_r(r)}\sin\theta dr+\frac{1}{F_\theta(r)}\left(\cos\theta d\theta+i \sin\theta d\phi \right)\right]\,,\\
&\label{dualchi1_0}
 \tilde{\chi}_\mu^{1,0}dx^\mu=-\frac{1}{F_r(r)}\cos\theta dr+\frac{1}{F_\theta(r)}\sin\theta d\theta\,,\\
&\label{dualchi1_-1}
\tilde{\chi}_\mu^{1,-1}dx^\mu=\frac{e^{-i\phi}}{\sqrt{2}}\left[-\frac{1}{F_r(r)}\sin\theta dr+\frac{1}{F_\theta(r)}\left(-\cos\theta d\theta+i\sin\theta d\phi \right)\right]\,.
\end{align}
\end{subequations}
We re-label the spin-1 eigenstates in the following way,
\begin{equation}\label{vector_eigenstate_basis}
\theta^\mu_1=\tau^\mu~,~\theta^\mu_2=\chi^\mu_{1,1}~,~\theta^\mu_3=\chi^\mu_{1,0}~,~\theta^\mu_4=\chi^\mu_{1,-1}\,,
\end{equation}
\begin{equation}\label{covector_eigenstate_basis}
\tilde\theta_\mu^1=\tilde\tau_\mu~,~\tilde\theta_\mu^2=\tilde\chi_\mu^{1,1}~,~\tilde\theta_\mu^3=\tilde\chi_\mu^{1,0}~,~\tilde\theta_\mu^4=\tilde\chi_\mu^{1,-1}\,,
\end{equation}
in order to proceed with the spin-eigenstate decomposition of the Proca field, which requires the dual representation so that we can evade the non-regularity of eigenstates \ref{chi1_1}, \ref{chi1_0}, \ref{chi1_-1}:
\begin{equation}\label{proca_decomposed}
    A_\mu=\sum\limits_{J=1}^4 \psi_J \tilde\theta^J_\mu\,.
\end{equation}
The scalars $\psi_J$ encode the dynamics of the field: in Section \ref{section:spin-eigenstate_formalism} we will proceed with their spherical harmonic decomposition.

From the vector basis defined by \ref{chi1_1} \ref{chi1_0} \ref{chi1_-1} \ref{tau_vector} we can build a rank 2 tensor spin-eigenstate basis by using Clebsch-Gordan coefficients\cite{rose_angular_momentum,Sakurai_quantum_mechanics}. Because of the properties of these coefficients, in general we expect to get eigenstates having spin $0\leq s \leq 2$, but not all of them are useful for the decomposition of spin-2 fields $h_{\mu\nu}$. In fact we need to take into account that $h_{\mu\nu}$ must be symmetric, thus we will not consider the antisymmetric rank-2 eigenstates\footnote{The rank-2 tensor product between the elements of the vector eigenstate basis give rise also to antisymmetric spin-1 eigenstates: this is why the electromagnetic field $F_{\mu\nu}$, which is antisymmetric, is spin-1.}.
The space vector eigenstates \ref{chi1_1} \ref{chi1_0} \ref{chi1_-1} give five spin-2 eigenstates plus one carrying zero spin,\begin{subequations}
\begin{align}
 &\label{chi2_2}
\chi^{\mu\nu}_{2,2}=\chi^\mu_{1,1} \chi^\nu_{1,1}\,,\\
&\label{chi2_1}
    \chi^{\mu\nu}_{2,1}=\frac{1}{\sqrt{2}}\left(\chi^\mu_{1,0} \chi^\nu_{1,1}+\chi^\mu_{1,1} \chi^\nu_{1,0}\right)\,,\\
&\label{chi2_0}
\chi^{\mu\nu}_{2,0}=\frac{1}{\sqrt{6}}\chi^\mu_{1,1} \chi^\nu_{1,-1}+\sqrt{\frac{2}{3}}\chi^\mu_{1,0} \chi^\nu_{1,0}+\frac{1}{\sqrt{6}}\chi^\mu_{1,-1} \chi^\nu_{1,1}\,,\\
&\label{chi2_-1}
\chi^{\mu\nu}_{2,-1}=\frac{1}{\sqrt{2}}\left(\chi^\mu_{1,0} \chi^\nu_{1,-1}+\chi^\mu_{1,-1} \chi^\nu_{1,0}\right)\,,\\
&\label{chi2_-2}
    \chi^{\mu\nu}_{2,-2}=\chi^\mu_{1,-1} \chi^\nu_{1,-1}\,,\\
&\label{chi0_0}
\chi^{\mu\nu}_{0,0}=\frac{1}{\sqrt{3}}\left(\chi^\mu_{1,1} \chi^\nu_{1,-1}-\chi^\mu_{1,0} \chi^\nu_{1,0}+\chi^\mu_{1,-1} \chi^\nu_{1,1}\right)\,.
\end{align}
\end{subequations}
By combining $\tau^\mu$ with vectors $\chi^\mu_{1,s_z}$ we get the temporal elements of the rank-2 basis, composed by a spin-0 and three spin-1 eigenstates:
\begin{subequations}
\begin{align}
&\label{tau00}
\tau^{\mu\nu}_{0,0}=\tau^\mu \tau^\nu\\
&\label{tau11}
\tau^{\mu\nu}_{1,1}=\frac{1}{\sqrt{2}}\left(\tau^\mu \chi^\nu_{1,1}+\tau^\nu \chi^\mu_{1,1} \right)\,,\\
&\label{tau10}
\tau^{\mu\nu}_{1,0}=\frac{1}{\sqrt{2}}\left(\tau^\mu \chi^\nu_{1,0}+\tau^\nu \chi^\mu_{1,0} \right)\,,\\
&\label{tau1-1}
\tau^{\mu\nu}_{1,-1}=\frac{1}{\sqrt{2}}\left(\tau^\mu \chi^\nu_{1,-1}+\tau^\nu \chi^\mu_{1,-1} \right)\,.
\end{align}
\end{subequations}
The procedure for building the rank-2 dual basis starting from \ref{tau_covector}, \ref{dualchi1_1}, \ref{dualchi1_0}, \ref{dualchi1_-1} is identical to the one applied for the vector case, thus here follow the resulting "dualized" expressions for the eigenstates:
\begin{subequations}
    \begin{align}
        &\tilde{\chi}_{\mu\nu}^{2,2}=\tilde{\chi}_\mu^{1,1} \tilde{\chi}_\nu^{1,1}\,,\\
        &\tilde{\chi}_{\mu\nu}^{2,1}=\frac{1}{\sqrt{2}}\left(\tilde{\chi}_\mu^{1,0} \tilde{\chi}_\nu^{1,1}+\tilde{\chi}_\mu^{1,1} \tilde{\chi}_\nu^{1,0}\right)\,,\\
        &\tilde{\chi}_{\mu\nu}^{2,0}=\frac{1}{\sqrt{6}}\tilde{\chi}_\mu^{1,1} \tilde{\chi}_\nu^{1,-1}+\sqrt{\frac{2}{3}}\tilde{\chi}_\mu^{1,0} \tilde{\chi}_\nu^{1,0}+\frac{1}{\sqrt{6}}\tilde{\chi}_\mu^{1,-1} \tilde{\chi}_\nu^{1,1}\,,\\
        &\tilde{\chi}_{\mu\nu}^{2,-1}=\frac{1}{\sqrt{2}}\left(\tilde{\chi}_\mu^{1,0} \tilde{\chi}_\nu^{1,-1}+\tilde{\chi}_\mu^{1,-1} \tilde{\chi}_\nu^{1,0}\right)\,,\\
        &\tilde{\chi}_{\mu\nu}^{2,-2}=\tilde{\chi}_\mu^{1,-1} \tilde{\chi}_\nu^{1,-1}\,,\\
        &\tilde{\chi}_{\mu\nu}^{0,0}=\frac{1}{\sqrt{3}}\left(\tilde{\chi}_\mu^{1,1} \tilde{\chi}_\nu^{1,-1}-\tilde{\chi}_\mu^{1,0} \tilde{\chi}_\nu^{1,0}+\tilde{\chi}_\mu^{1,-1} \tilde{\chi}_\nu^{1,1}\right)\,,\\
        &\tilde{\tau}_{\mu\nu}^{0,0}=\tilde{\tau}_\mu \tilde{\tau}_\nu\,,\\
        &\tilde{\tau}_{\mu\nu}^{1,1}=\frac{1}{\sqrt{2}}\left(\tilde{\tau}_\mu \tilde{\chi}_\nu^{1,1}+\tilde{\tau}_\nu \tilde{\chi}_\mu^{1,1} \right)\,,\\
        &\tilde{\tau}_{\mu\nu}^{1,0}=\frac{1}{\sqrt{2}}\left(\tilde{\tau}_\mu \tilde{\chi}_\nu^{1,0}+\tilde{\tau}_\nu \tilde{\chi}_\mu^{1,0} \right)\,,\\
        &\tilde{\tau}_{\mu\nu}^{1,-1}=\frac{1}{\sqrt{2}}\left(\tilde{\tau}_\mu \tilde{\chi}_\nu^{1,-1}+\tilde{\tau}_\nu \tilde{\chi}_\mu^{1,-1} \right)\,.
    \end{align}
\end{subequations}
We now re-order the rank-2 spin-eigenstate basis \footnote{Like in the vector case, for each choice of $F_t(r)$, $F_r(r)$, $F_\theta(r)$ we get a different set of spin-eigenstates: we will fix these functions later when we will need to address the issue of boundary conditions for $h_{\mu\nu}$.} in the following way,
\begin{equation}
\begin{aligned}
&\Theta_0^{\mu\nu}=\tau^{\mu\nu}_{0,0}~,~\Theta_1^{\mu\nu}=\tau^{\mu\nu}_{1,1}~,~\Theta_2^{\mu\nu}=\tau^{\mu\nu}_{1,0}~,~\Theta_3^{\mu\nu}=\tau^{\mu\nu}_{1,-1}~,~\Theta_4^{\mu\nu}=\chi^{\mu\nu}_{0,0}\,,\\
&\Theta_5^{\mu\nu}=\chi^{\mu\nu}_{2,2}~,~\Theta_6^{\mu\nu}=\chi^{\mu\nu}_{2,1}~,~\Theta_7^{\mu\nu}=\chi^{\mu\nu}_{2,0}~,~\Theta_8^{\mu\nu}=\chi^{\mu\nu}_{2,-1}~,~\Theta_{9}^{\mu\nu}=\chi^{\mu\nu}_{2,-2}\,,
\end{aligned}
\end{equation}
\begin{equation}
\begin{aligned}
&\tilde{\Theta}^0_{\mu\nu}=\tilde{\tau}_{\mu\nu}^{0,0}~,~\tilde{\Theta}^1_{\mu\nu}=\tilde{\tau}_{\mu\nu}^{1,1}~,~\tilde{\Theta}^2_{\mu\nu}=\tilde{\tau}_{\mu\nu}^{1,0}~,~\tilde{\Theta}^3_{\mu\nu}=\tilde{\tau}_{\mu\nu}^{1,-1}~,~\tilde{\Theta}^4_{\mu\nu}=\tilde{\chi}_{\mu\nu}^{0,0}\,,\\
&\tilde{\Theta}^5_{\mu\nu}=\tilde{\chi}_{\mu\nu}^{2,2}~,~\tilde{\Theta}^6_{\mu\nu}=\tilde{\chi}_{\mu\nu}^{2,1}~,~\tilde{\Theta}^7_{\mu\nu}=\tilde{\chi}_{\mu\nu}^{2,0}~,~\tilde{\Theta}^8_{\mu\nu}=\tilde{\chi}_{\mu\nu}^{2,-1}~,~\tilde{\Theta}^{9}_{\mu\nu}=\tilde{\chi}_{\mu\nu}^{2,-2}\,,
\end{aligned}
\end{equation}
so that we can express the spin-2 field as
\begin{equation}\label{h_spin_decomposed}
h_{\mu\nu}=\sum\limits_{J=0}^{9}\Psi_J \tilde{\Theta}^J_{\mu\nu}\,,
\end{equation}
where $\Psi_J$ are scalar functions and the tensor basis satisfies the following orthonormality condition:
\begin{equation}\label{ortho_basis}
\left(\tilde{\Theta}^J_{\mu\nu}\right)^*\Theta_K^{\mu\nu}=\delta^J_K\,.
\end{equation}

The functions $\psi_J$ and $\Psi_J$ can be considered as components of a 4-dimensional and a 10-dimensional spinors $\boldsymbol{\psi}$ and $\boldsymbol{\Psi}$ respectevely, each function describing the evolution of a specific spin-eigenstate of $A_\mu$ and $h_{\mu\nu}$ respectively. In the rest of this text we will apply the Einstein summation convention to the uppercase latin spinor indices.

\section{Spin decomposition of spin-1 and spin-2\\field equations}

In this section we will apply the spin-eigenstate decomposition defined in the previous section to the field equations of spin-1 and spin-2 massive perturbations in Kerr spacetime. The results we will get are the first step for carrying the spherical harmonic decomposition of these field equations. We recall the perturbative field equations of spin-1 and spin-2 massive perturbations on a Kerr spacetime (see sections \ref{dark_photons} and \ref{spin-2_curved_background} for details about these equations):
\begin{subequations}
\begin{align}
&\label{proca_wave_eqs}\left(\bar\Box +\mu_V^2\right)A_{\mu}=0\,,\\
&\label{divergence-less_spin1}\bar\nabla^\mu A_\mu=0\,,\\
&\label{massive_graviton_wave_eqs}\bar\Box h_{\mu\nu}+2 \bar R^{~\alpha~\beta}_{\mu~\nu}h_{\alpha\beta}+\mu_T^2 h_{\mu\nu}=0\,,\\
&\label{divergence-less_spin2}\bar\nabla^\mu h_{\mu\nu}=0\,,\\
&\label{trace-less_spin2}h_{\mu\nu}\bar g^{\mu\nu}=0\,,
\end{align}
\end{subequations}
where $\bar g^{\mu\nu}$ is the inverse Kerr metric tensor (see equation \ref{Kerr_metric}), $\bar\nabla_\sigma$ is its covariant derivative, $\bar R^{~\alpha~\beta}_{\mu~\nu}$ is its Riemann curvature tensor, $\bar\Box=\bar\nabla_\sigma\bar\nabla^\sigma$ is its D'Alembertian operator and $\mu_V$ and $\mu_T$ are the masses of $A_\mu$ and $h_{\mu\nu}$ respectively.
Like in the scalar case, we impose $A_\mu$ and $h_{\mu\nu}$ to be monochromatic waves having some frequency $\omega$ and some definite total azimuthal angular momentum $j_z$\footnote{Obviously $A_{\mu}$ and $h_{\mu\nu}$ in general have differing frequencies $\omega$ and azimuthal angular momentum $j_z$: we are using the same characters for these parameters just for simplifying the notation, but they are generic.}:
\begin{equation}\label{A_omega_m}
i\partial_t A_{\mu}=\omega A_{\mu}~,~\mathbfcal{L}_{\boldsymbol z} A_{\mu}=j_z A_{\mu}\,,
\end{equation}
\begin{equation}\label{h_omega_m}
i\partial_t h_{\mu\nu}=\omega h_{\mu\nu}~,~\mathbfcal{L}_{\boldsymbol z} h_{\mu\nu}=j_z h_{\mu\nu}\,.
\end{equation}
First, we decompose the wave equations \ref{proca_wave_eqs} and \ref{massive_graviton_wave_eqs}:
\begin{small}
\begin{subequations}
\begin{align}
&\label{vector_wave_op+mass_decomposed}\begin{aligned}
        &\left(\theta^{\mu}_I \right)^*\left(\bar\Box +\mu_V^2\right)A_{\mu}=\left(\theta^{\mu}_I \right)^*\left(\bar\Box+\mu_V^2 \right)\left(\psi_J \tilde{\theta}^J_{\mu}\right)=\\
        &=(\bar \Box+\mu_V^2)\psi_I+\left[2 \left(\theta^{\mu}_I \right)^*\bar\nabla^\sigma \tilde{\theta}^J_{\mu} \partial_\sigma+\left(\theta^{\mu}_I \right)^*\bar\Box \tilde{\theta}^J_{\mu}\right]\psi_J=0\,,
    \end{aligned}\\
&\label{tensor_wave_op+mass_decomposed}\begin{aligned}
        &\left(\Theta^{\mu\nu}_I \right)^*\left(\bar\Box h_{\mu\nu}+2 \bar R^{~\alpha~\beta}_{\mu~\nu}h_{\alpha\beta}+\mu_T^2 h_{\mu\nu}\right)=\\
        &=\left(\Theta^{\mu\nu}_I \right)^*\left(\delta^\alpha_\mu\delta^\beta_\nu\bar\Box+2 \bar R^{~\alpha~\beta}_{\mu~\nu}+\mu_T^2\delta^\alpha_\mu\delta^\beta_\nu\right)\left(\Psi_J \tilde{\Theta}^J_{\alpha\beta}\right)=\\
        &=(\bar \Box+\mu_T^2)\Psi_I+\left[2 \left(\Theta^{\mu\nu}_I \right)^*\bar\nabla^\sigma \tilde{\Theta}^J_{\mu\nu} \partial_\sigma+\left(\Theta^{\mu\nu}_I \right)^*\bar\Box \tilde{\Theta}^J_{\mu\nu}+2 \left(\Theta^{\mu\nu}_I \right)^*\bar R^{~\alpha~\beta}_{\mu~\nu}\tilde{\Theta}^J_{\alpha\beta}\right]\Psi_J=0\,.
    \end{aligned}
\end{align}
\end{subequations}\end{small}
If we exploit \ref{A_omega_m} and \ref{h_omega_m} we can rewrite the Klein-Gordon-like part of the decomposed wave equations in the following forms, resembling the one of the scalar case:
\begin{subequations}
\begin{align}
&\label{klein-gordon-like_vector_case}\begin{aligned}
        &\frac{\Sigma(r,\theta)}{\Delta(r)}\left(\bar\Box+\mu_V^2 \right)\psi_I=\frac{a^2 \left(\theta^{\mu}_I \right)^*}{\Delta^2(r)}\left(\mathbfcal{L}_{\boldsymbol z} ^2\tilde{\theta}^J_{\mu}+2 \mathbfcal{L}_{\boldsymbol z}\tilde{\theta}^J_{\mu} \mathbfcal{L}_{\boldsymbol z} \right)\psi_J\\
        &+\frac{1}{\Delta(r)}\left\lbrace \mathbfcal{L}^2-\frac{\partial}{\partial r}\left[\Delta(r) \frac{\partial}{\partial r}\right]+\frac{4 j_z a \omega M r}{\Delta(r)} -\frac{2 M \omega^2 r(a^2+r^2)}{\Delta(r)}-\frac{a^2 j_z^2}{\Delta(r)}\right\rbrace \psi_I\\
        &+\frac{ \Sigma(r,\theta)}{\Delta(r)}(\mu_V^2-\omega^2)\psi_I-\frac{4 a \omega M r}{\Delta^2(r)} \left(\theta^{\mu}_I \right)^*\mathbfcal{L}_{\boldsymbol z} \tilde{\theta}^J_{\mu}\psi_J\,,
\end{aligned}\\
&\label{klein-gordon-like_tensor_case}\begin{aligned}
        &\frac{\Sigma(r,\theta)}{\Delta(r)}\left(\bar\Box+\mu_T^2 \right)\Psi_I=\frac{a^2 \left(\Theta^{\mu\nu}_I \right)^*}{\Delta^2(r)}\left(\mathbfcal{L}_{\boldsymbol z} ^2\tilde{\Theta}^J_{\mu\nu}+2 \mathbfcal{L}_{\boldsymbol z}\tilde{\Theta}^J_{\mu\nu} \mathbfcal{L}_{\boldsymbol z} \right)\Psi_J\\
        &+\frac{1}{\Delta(r)}\left\lbrace \mathbfcal{L}^2-\frac{\partial}{\partial r}\left[\Delta(r) \frac{\partial}{\partial r}\right]+\frac{4 j_z a \omega M r}{\Delta(r)} -\frac{2 M \omega^2 r(a^2+r^2)}{\Delta(r)}-\frac{a^2 j_z^2}{\Delta(r)}\right\rbrace \Psi_I\\
        &+\frac{ \Sigma(r,\theta)}{\Delta(r)}(\mu_T^2-\omega^2)\Psi_I-\frac{4 a \omega M r}{\Delta^2(r)} \left(\Theta^{\mu\nu}_I \right)^*\mathbfcal{L}_{\boldsymbol z} \tilde{\Theta}^J_{\mu\nu}\Psi_J\,.
    \end{aligned}
\end{align}
\end{subequations}
From formulae \ref{chi1_1} \ref{chi1_0} \ref{chi1_-1} of the vector eigenstates we can define the following angular operators,
\begin{equation}\label{angular_operators}
\boldsymbol{D_\pm}=\frac{e^{\pm i\phi}}{\sqrt{2}}\left(\pm\cos\theta \partial_\theta+\frac{i}{\sin\theta}\partial_\phi \right)~,~\boldsymbol{D_0}=\sin\theta \partial_\theta\,,
\end{equation}
acting on spherical harmonics $Y_{l,m}(\theta,\phi)$ in the following way:
\begin{small}\begin{subequations}
\begin{align}
&\label{D+-}
\begin{aligned}
    &\boldsymbol{D_\pm}Y_{l,m}=\\
    &=\sqrt{\frac{(l+1)^2(l\mp m)(l\mp m-1)}{2(2 l+1)(2 l-1)}}Y_{l-1 ,m\pm 1}+\sqrt{\frac{l^2(l\pm m+1)(l\pm m+2)}{2(2 l+3)(2 l+1)}}Y_{l+1 ,m\pm 1}\,,
\end{aligned}\\
&\label{D0}
\begin{aligned}
&\boldsymbol{D_0}Y_{l,m}=\\
&=\sqrt{\frac{l^2(l+ m+1)(l-m+1)}{(2 l+1)(2 l+3)}}Y_{l+1 ,m}-\sqrt{\frac{(l+1)^2(l+ m)(l- m)}{(2 l+1)(2 l-1)}}Y_{l-1 ,m}\,.
\end{aligned}
\end{align}
\end{subequations}
\end{small}We can use these new operators, the vector eigenstates \ref{dualchi1_1}, \ref{dualchi1_0}, \ref{dualchi1_-1} and the temporal dependency of the fields for rewriting the partial derivatives of the scalar fields $\psi_J$ and $\Psi_J$ as
\begin{equation}\label{partial_deriv}
\partial_\sigma=-i\omega\delta^0_\sigma+\delta^1_\sigma \partial_r+ F_\theta(r)\left[\left(\tilde{\chi}^{1,1}_\sigma\right)^*\boldsymbol{D_+}+\left(\tilde{\chi}^{1,0}_\sigma\right)^*\boldsymbol{D_0}+\left(\tilde{\chi}^{1,-1}_\sigma\right)^*\boldsymbol{D_-}\right]\,,
\end{equation}
which is a way of expressing derivatives more suitable for computing the spherical harmonic decomposition. By exploiting all the mathematical manipulations we did in \ref{vector_wave_op+mass_decomposed}, \ref{tensor_wave_op+mass_decomposed},\ref{klein-gordon-like_vector_case}, \ref{klein-gordon-like_tensor_case}, \ref{partial_deriv} and in the Klein-Gordon case in \ref{klein-gordon_manipulation}, we can rewrite the spin-decomposed wave equations \ref{vector_wave_op+mass_decomposed} and \ref{tensor_wave_op+mass_decomposed} in the following form\cite{Spectra_grav_atom},
\begin{equation}\label{wave_eqs}
\begin{aligned}
&\left\lbrace\frac{1}{\Delta(r)}[\mathbfcal{L}^2+a^2\cos^2\theta(\mu^2-\omega^2)]-\frac{1}{\Delta(r)}\frac{\partial}{\partial r}\left[\Delta(r)\frac{\partial}{\partial r}\right]+\mu^2-\omega^2\right.\\
&\left.-\frac{P_+^2}{(r-r_+)^2}-\frac{P_-^2}{(r-r_-)^2}+\frac{A_+}{(r_+-r_-)(r-r_+)}-\frac{A_-}{(r_+-r_-)(r-r_-)}\right\rbrace\Psi^{(S)}_I\\
&+\left(S_I^{~J}+Q_I^{~J}\mathbfcal{L}_{\boldsymbol z}+R_I^{~J}\partial_r+P_I^{~J}\boldsymbol{D_+}+Z_I^{~J}\boldsymbol{D_0}+M_I^{~J}\boldsymbol{D_-}\right)\Psi^{(S)}_J=0\,,
\end{aligned}
\end{equation}
where $\displaystyle r_\pm=M\pm\sqrt{M^2-a^2}$, $\displaystyle P_\pm=\frac{j_z a-2\omega M r_\pm}{r_+-r_-}$, $\displaystyle A_\pm=P_+^2+P_-^2 + M^2(\mu^2--7\omega^2)\pm(\mu^2-2\omega^2)M(r_+-r_-)+(M^2-a^2)(\mu^2-\omega^2)$, $\Psi^{(S)}_J$ can be either the vector or the tensor field ($\Psi^{(1)}_J=\psi_J$, $\Psi^{(2)}_J=\Psi_J$) and $\mu$ is the mass of the perturbation. In equations \ref{wave_eqs} we can see several Klein-Gordon-like equations, modified and coupled by spin-mixing matrices. In the case of spin-1 perturbations the spin-mixing matrices have the following expressions\cite{Spectra_grav_atom}:
\begin{subequations}
    \begin{align}
        &\label{Smatrix_s1}
        S_I^{~J}=\frac{a^2}{\Delta^2(r)}\left(\theta^{\mu}_I \right)^* \mathbfcal{L}_{\boldsymbol z}^2\tilde{\theta}^J_{\mu}-\frac{4 a \omega M r}{\Delta^2(r)}\left(\theta^{\mu}_I \right)^* \mathbfcal{L}_{\boldsymbol z}\tilde{\theta}^J_{\mu}+\frac{\Sigma(r,\theta)}{\Delta(r)}\left(\theta^{\mu}_I \right)^* \left(\bar\Box-2 i \omega \bar\nabla^0 \right) \tilde{\theta}^J_{\mu}\,,\\
        &\label{Rmatrix_s1}
        R_I^{~J}=\frac{2\Sigma(r,\theta)}{\Delta(r)} \left(\theta^{\mu}_I \right)^*\bar\nabla^1 \tilde{\theta}^J_{\mu}\,,\\
        &\label{Qmatrix_s1}
        Q_I^{~J}=\frac{2 a^2}{\Delta^2(r)}\left(\theta^{\mu}_I \right)^*\mathbfcal{L}_{\boldsymbol z}\tilde{\theta}^J_{\mu}\,,\\
        &\label{Pmatrix_s1}
        P_I^{~J}=\frac{2\Sigma(r,\theta)}{\Delta(r)}F_\theta(r)\left(\tilde{\chi}^{1,1}_\sigma\right)^* \left(\theta^{\mu}_I \right)^* \bar\nabla^\sigma \tilde{\theta}^J_{\mu}\,,\\
        &\label{Zmatrix_s1}
        Z_I^{~J}=\frac{2\Sigma(r,\theta)}{\Delta(r)}F_\theta(r)\left(\tilde{\chi}^{1,0}_\sigma\right)^* \left(\theta^{\mu}_I \right)^* \bar\nabla^\sigma \tilde{\theta}^J_{\mu}\,,\\
        &\label{Mmatrix_s1}
        M_I^{~J}=\frac{2\Sigma(r,\theta)}{\Delta(r)}F_\theta(r)\left(\tilde{\chi}^{1,-1}_\sigma\right)^* \left(\theta^{\mu}_I \right)^* \bar\nabla^\sigma \tilde{\theta}^J_{\mu}\,.
    \end{align}
\end{subequations}
The spin-mixing matrices' structure in the spin-2 case is almost identical to the one of spin-1 perturbations, differing only because of the involved spin-eigenstate basis and the presence of the Riemann curvature:
\begin{subequations}
{\allowdisplaybreaks[4]
\begin{align}
    &\begin{aligned}
        \label{Smatrix_s2}
        S_I^{~J}= &\frac{a^2}{\Delta^2(r)}\left(\Theta^{\mu\nu}_I \right)^*  \mathbfcal{L}_{\boldsymbol z}^2\tilde{\Theta}^J_{\mu\nu}-\frac{4 a \omega M r}{\Delta^2(r)}\left(\Theta^{\mu\nu}_I \right)^* \mathbfcal{L}_{\boldsymbol z}\tilde{\Theta}^J_{\mu\nu}+\frac{\Sigma(r,\theta)}{\Delta(r)}\left(\Theta^{\mu\nu}_I \right)^* \left(\bar\Box\right.\\
        &\left.-2 i \omega \bar\nabla^0 \right) \tilde{\Theta}^J_{\mu\nu}+\frac{2\Sigma(r,\theta)}{\Delta(r)} \left(\Theta^{\mu\nu}_I \right)^*\bar R^{~\alpha~\beta}_{\mu~\nu}\tilde{\Theta}^J_{\alpha\beta}\,,
    \end{aligned}\\
    &\label{Rmatrix_s2}
    R_I^{~J}=\frac{2\Sigma(r,\theta)}{\Delta(r)} \left(\Theta^{\mu\nu}_I \right)^*\bar\nabla^1 \tilde{\Theta}^J_{\mu\nu}\,\\
    &\label{Qmatrix_s2}
    Q_I^{~J}=\frac{2 a^2}{\Delta^2(r)}\left(\Theta^{\mu\nu}_I \right)^*\mathbfcal{L}_{\boldsymbol z}\tilde{\Theta}^J_{\mu\nu}\,,\\
    &\label{Pmatrix_s2}
    P_I^{~J}=\frac{2\Sigma(r,\theta)}{\Delta(r)}F_\theta(r)\left(\tilde{\chi}^{1,1}_\sigma\right)^* \left(\Theta^{\mu\nu}_I \right)^* \bar\nabla^\sigma \tilde{\Theta}^J_{\mu\nu}\,,\\
    &\label{Zmatrix_s2}
    Z_I^{~J}=\frac{2\Sigma(r,\theta)}{\Delta(r)}F_\theta(r)\left(\tilde{\chi}^{1,0}_\sigma\right)^* \left(\Theta^{\mu\nu}_I \right)^* \bar\nabla^\sigma \tilde{\Theta}^J_{\mu\nu}\,,\\
    &\label{Mmatrix_s2}
    M_I^{~J}=\frac{2\Sigma(r,\theta)}{\Delta(r)}F_\theta(r)\left(\tilde{\chi}^{1,-1}_\sigma\right)^* \left(\Theta^{\mu\nu}_I \right)^* \bar\nabla^\sigma \tilde{\Theta}^J_{\mu\nu}\,.
\end{align}}
\end{subequations}The spin-decomposed wave equations \ref{wave_eqs} are a generalization of the ones computed in \cite{Spectra_grav_atom} for spin-1 perturbations, now extended also for the spin-2 case, showing a general common structure.

The field equations left to be decomposed are the Lorenz constraints \ref{divergence-less_spin1} and \ref{divergence-less_spin2} and the trace-less constraint \ref{trace-less_spin2}. By substituting the decomposed expressions \ref{proca_decomposed}, \ref{h_spin_decomposed} and \ref{partial_deriv} into the constraints and projecting the vector eigenstate basis on the spin-2 Lorenz constraint, we get the spin-decomposed expressions for these constraints:
\begin{subequations}
{\allowdisplaybreaks[4]
\begin{align}
    &\label{Lorenz_s1}
    \left(\rho^{J}\partial_r +\pi^{J}\boldsymbol{D_+}+\zeta^{J}\boldsymbol{D_0}+\mu^{J}\boldsymbol{D_-}+\sigma^{J}\right) \psi_J=0\,,\\
    &\label{Lorenz_s2}
    \left(\rho_I^{~J}\partial_r +\pi_I^{~J}\boldsymbol{D_+}+\zeta_I^{~J}\boldsymbol{D_0}+\mu_I^{~J}\boldsymbol{D_-}+\sigma_I^{~J}\right) \Psi_J=0\,,\\
    &\label{trace-less}
    T^J\Psi_J =0\,.
\end{align}}
\end{subequations}
The expressions for the spin-constraint vectors and matrices are the following:
\begin{subequations}
{\allowdisplaybreaks[4]
\begin{align}
    &\label{sigmavector}
    \sigma^{J}=-i\omega\bar g^{0\mu} \tilde{\theta}^J_{\mu}+ \bar g^{\mu\sigma}\bar\nabla_\sigma\tilde{\theta}^J_{\mu}\,,\\
    &\label{rhovector}
    \rho_{J}= \bar g^{1\mu} \tilde{\theta}^J_{\mu}\,,\\
    &\label{pivector}
    \pi^{J}=F_\theta(r)\left(\tilde{\chi}_\sigma^{1,1}\right)^*\bar g^{\sigma\mu} \tilde{\theta}^J_{\mu}\,,\\
    &\label{zetavector}
    \zeta^{J}=F_\theta(r)\left(\tilde{\chi}_\sigma^{1,0}\right)^*  \bar g^{\sigma\mu} \tilde{\theta}^J_{\mu}\,,\\
    &\label{muvector}
    \mu^{J}=F_\theta(r)\left(\tilde{\chi}_\sigma^{1,-1}\right)^* \bar g^{\sigma\mu} \tilde{\theta}^J_{\mu}\,,\\ 
    &\label{sigmamatrix}
    \sigma_I^{~J}=-i\omega\left(\theta^\nu_I\right)^* \bar g^{0\mu} \tilde{\Theta}^J_{\mu\nu}+ \left(\theta^\nu_I\right)^*\bar g^{\mu\sigma}\bar\nabla_\sigma\tilde{\Theta}^J_{\mu\nu}\,,\\
    &\label{rhomatrix}
    \rho_I^{~J}=\left(\theta^\nu_I\right)^*  \bar g^{1\mu} \tilde{\Theta}^J_{\mu\nu}\,,\\
    &\label{pimatrix}
    \pi_I^{~J}=F_\theta(r)\left(\theta^\nu_I\right)^*\left(\tilde{\chi}_\sigma^{1,1}\right)^*\bar g^{\sigma\mu} \tilde{\Theta}^J_{\mu\nu}\,,\\
    &\label{zetamatrix}
    \zeta_I^{~J}=F_\theta(r)\left(\theta^\nu_I\right)^*\left(\tilde{\chi}_\sigma^{1,0}\right)^*  \bar g^{\sigma\mu} \tilde{\Theta}^J_{\mu\nu}\,,\\
    &\label{mumatrix}
    \mu_I^{~J}=F_\theta(r)\left(\theta^\nu_I\right)^*\left(\tilde{\chi}_\sigma^{1,-1}\right)^* \bar g^{\sigma\mu} \tilde{\Theta}^J_{\mu\nu}\,,\\
    &\label{Tvector}
    T^J=\bar g^{\mu\nu}\tilde{\Theta}^J_{\mu\nu}\,.
\end{align}}
\end{subequations}
Also in this case the expressions found are the ones computed by \textit{Baumann et al.} in \cite{Spectra_grav_atom} and their spin-2 generalization. The $I$ index appearing in \ref{Lorenz_s2}, \ref{sigmamatrix}, \ref {rhomatrix}, \ref{pimatrix}, \ref{zetamatrix} and \ref{mumatrix} runs from $I=0$ to $I=3$, while $J$ runs from $J=0$ to $J=9$, consequently the constraint matrices are rectangular.
The spin-1 Lorenz and the spin-2 trace-less constraints, on the contrary, are just scalar equations, therefore their decomposition consist just in inserting the decomposed field in their expressions.

The field equations of spin-2 fields, though, can be reduced by substituting the trace-less constraint \ref{trace-less} into the other field equations\footnote{We found out that without this substitution finding solutions is practically impossible.}. In particular, for this substitution we exploit the following decomposition for the total angular-momentum squared operator\cite{rose_angular_momentum,Sakurai_quantum_mechanics},
\begin{equation}
    \mathbfcal{L}^2=\frac{1}{2}\left(\mathbfcal{L}_+ \mathbfcal{L}_- +\mathbfcal{L}_- \mathbfcal{L}_+\right)+\mathbfcal{L}_z^2\,,
\end{equation}
where we defined the operators $\mathbfcal{L}_\pm$ in \ref{L+L-}, which in the scalar representation reduce to 
\begin{equation}
    \mathbfcal{L}_\pm=\pm e^{\pm i\phi}\left(\partial_\theta\pm i\cot\theta\partial_\phi\right)\,.
\end{equation}
Thus, we eliminate the purely temporal component $\Psi_0$ from the equations by substituting  $\Psi_0=\left(T^0\right)^{-1}\sum\limits_{J=1}^9 T^J \Psi_J$, and we exploit the algebra of the angular momentum operators. The resulting wave equations for the remaining components of the spin-2 field have the same structure of \ref{wave_eqs},
\begin{equation}\label{spin2_wave_eqs}
\begin{aligned}
&\left\lbrace\frac{1}{\Delta(r)}[\mathbfcal{L}^2+a^2\cos^2\theta(\mu^2-\omega^2)]-\frac{1}{\Delta(r)}\frac{\partial}{\partial r}\left[\Delta(r)\frac{\partial}{\partial r}\right]+\mu^2-\omega^2\right.\\
&\left.-\frac{P_+^2}{(r-r_+)^2}-\frac{P_-^2}{(r-r_-)^2}+\frac{A_+}{(r_+-r_-)(r-r_+)}-\frac{A_-}{(r_+-r_-)(r-r_-)}\right\rbrace\Psi_I\\
&+\left(\tilde{S}_I^{~J}+\tilde{Q}_I^{~J}\mathbfcal{L}_{\boldsymbol z}+\tilde{R}_I^{~J}\partial_r+\tilde{P}_I^{~J}\boldsymbol{D_+}+\tilde{Z}_I^{~J}\boldsymbol{D_0}+\tilde{M}_I^{~J}\boldsymbol{D_-}\right)\Psi_J=0\,,
\end{aligned}
\end{equation}
where $I , J\in\left\lbrace 1 , 2 , 3 , 4 , 5 , 6 , 7 , 8 , 9\right\rbrace$\footnote{It will be applied for all the following equations regarding the spin-2 perturbation.} and the spin-mixing matrices have been transformed by the substitution:
\begin{subequations}
{\allowdisplaybreaks[4]
\begin{align}
    &\begin{aligned}
        \label{S_tilde_matrix}
        \tilde{S}_I^{~J}=&S_I^{~J}-S_I^{~0}\left(T^0\right)^{-1}T^J-Q_I^{~0}\mathbfcal{L}_z\left(\left(T^0\right)^{-1}T^J\right)-R_I^{~0}\partial_r\left(\left(T^0\right)^{-1}T^J\right)\\&-P_I^{~0}\boldsymbol{D_+}\left(\left(T^0\right)^{-1}T^J\right)-Z_I^{~0}\boldsymbol{D_0}\left(\left(T^0\right)^{-1}T^J\right)-M_I^{~0}\boldsymbol{D_-}\left(\left(T^0\right)^{-1}T^J\right)\,,
    \end{aligned}\\
    &\label{Q_tilde_matrix}
    \tilde{Q}_I^{~J}=Q_I^{~J}-Q_I^{~0}\left(T^0\right)^{-1}T^J\,,\\
    &\label{R_tilde_matrix}
    \tilde{R}_I^{~J}=R_I^{~J}-R_I^{~0}\left(T^0\right)^{-1}T^J\,,\\
    &\label{P_tilde_matrix}
    \tilde{P}_I^{~J}=P_I^{~J}-R_I^{~0}\left(T^0\right)^{-1}T^J\,,\\
    &\label{Z_tilde_matrix}
    \tilde{Z}_I^{~J}=Z_I^{~J}-R_I^{~0}\left(T^0\right)^{-1}T^J\,,\\
    &\label{M_tilde_matrix}
    \tilde{M}_I^{~J}=M_I^{~J}-R_I^{~0}\left(T^0\right)^{-1}T^J\,.
\end{align}}
\end{subequations}
Also the Lorenz constraint keeps the same structure,
\begin{equation}\label{Lorenz_s2_new}
    \left(\tilde\rho_I^{~J}\partial_r +\tilde\pi_I^{~J}\boldsymbol{D_+}+\tilde\zeta_I^{~J}\boldsymbol{D_0}+\tilde\mu_I^{~J}\boldsymbol{D_-}+\tilde\sigma_I^{~J}\right) \Psi_J=0\,,
\end{equation}
and its constraint matrices have experienced the same transformation:
\begin{subequations}
{\allowdisplaybreaks[4]
\begin{align}
    &\begin{aligned}
        \label{sigma_tilde_matrix}
        \tilde{\sigma}_I^{~J}=&\sigma_I^{~J}-\sigma_I^{~0}\left(T^0\right)^{-1}T^J-\rho_I^{~0}\partial_r\left(\left(T^0\right)^{-1}T^J\right)\\&-\pi_I^{~0}\boldsymbol{D_+}\left(\left(T^0\right)^{-1}T^J\right)-\zeta_I^{~0}\boldsymbol{D_0}\left(\left(T^0\right)^{-1}T^J\right)-\mu_I^{~0}\boldsymbol{D_-}\left(\left(T^0\right)^{-1}T^J\right)\,,
    \end{aligned}\\
    &\label{rho_tilde_matrix}
    \tilde{\rho}_I^{~J}=\rho_I^{~J}-\rho_I^{~0}\left(T^0\right)^{-1}T^J\,,\\
    &\label{pi_tilde_matrix}
    \tilde{\pi}_I^{~J}=\pi_I^{~J}-\pi_I^{~0}\left(T^0\right)^{-1}T^J\,,\\
    &\label{zeta_tilde_matrix}
    \tilde{\zeta}_I^{~J}=\zeta_I^{~J}-\zeta_I^{~0}\left(T^0\right)^{-1}T^J\,,\\
    &\label{mu_tilde_matrix}
    \tilde{\mu}_I^{~J}=M_I^{~J}-\mu_I^{~0}\left(T^0\right)^{-1}T^J\,.
\end{align}}
\end{subequations}
The 0th wave equation, instead, becomes a second-order constraint,
\begin{equation}\label{second_order_constraint}
\begin{aligned}
\left(A^J\mathbfcal{L}^2+R^J \partial_r^2+\tilde{S}^{J}+\tilde{Q}^{J}\mathbfcal{L}_{\boldsymbol z}+\tilde{R}^{J}\partial_r\right.&+\tilde{P}^{J}\mathbfcal{L}_++\tilde{M}^{J}\mathbfcal{L}_-\\
&\left.+\tilde{P}_0^{~J}\boldsymbol{D_+}+\tilde{Z}_0^{~J}\boldsymbol{D_0}+\tilde{M}_0^{~J}\boldsymbol{D_-}\right)\Psi_J=0\,,
\end{aligned}
\end{equation}
where the constraint vectors have the following expressions:
\begin{subequations}
{\allowdisplaybreaks[4]
\begin{align}
    &\begin{aligned}
        \label{S_tilde_vector}
        &\tilde{S}^{J}=\tilde{S}_0^{~J}-\left\lbrace\frac{\mathbfcal{L}^2+a^2\cos^2\theta(\mu^2-\omega^2)}{\Delta(r)}-\frac{1}{\Delta(r)}\frac{\partial}{\partial r}\left[\Delta(r)\frac{\partial}{\partial r}\right]+\mu^2-\omega^2\right.\\
        &\left.-\frac{P_+^2}{(r-r_+)^2}-\frac{P_-^2}{(r-r_-)^2}+\frac{A_+}{(r_+-r_-)(r-r_+)}-\frac{A_-}{(r_+-r_-)(r-r_-)}\right\rbrace\left(\frac{T^J}{T^0}\right)\,,
    \end{aligned}\\
    &A^J=-\frac{\left(T^0\right)^{-1}T^J}{\Delta(r)}\,,\\
    &R^J=\left(T^0\right)^{-1}T^J\,,\\
    &\tilde{R}^J=\tilde{R}_0^{~J}+2\frac{\partial}{\partial r}\left[\left(T^0\right)^{-1}T^J\right]-\frac{\partial\ln\Delta(r)}{\partial r}\,,\\
    &\tilde{Q}^J=\tilde{Q}_0^{~J}-\frac{2}{\Delta(r)}\mathbfcal{L}_z\left[\left(T^0\right)^{-1}T^J\right]\,,\\
    &\tilde{P}^J=-\frac{1}{\Delta(r)}\mathbfcal{L}_-\left[\left(T^0\right)^{-1}T^J\right]\,,\\
    &\tilde{M}^J=-\frac{1}{\Delta(r)}\mathbfcal{L}_+\left[\left(T^0\right)^{-1}T^J\right]\,.
\end{align}}
\end{subequations}
\section{Spherical harmonic decomposition of spin-1 and spin-2 fields}\label{section:vector_tensor_harmonics}

Having defined a formalism for dealing with the vector and tensor nature of fields in curved spacetime, we can now proceed with the spherical harmonic decomposition. We introduce the following vector and tensor spherical harmonics, based on Clebsch-Gordan summation of angular momentum and the spin-eigenstates we introduced in section \ref{section:spin-eigenstate_formalism}:
\begin{subequations}
{\allowdisplaybreaks[4]
\begin{align}
    &\label{vector_spherical}
    Y^{\mu}_{l,s,j,j_z}=\sum\limits_{m=-l}^l\sum\limits_{s_z=-s}^s \left\langle l,m,s,s_z \right.\left|j,j_z\right\rangle Y_{l,m} \left[\delta_{0,s} \tau^{\mu}+\left(1-\delta_{0,s}\right) \chi^{\mu}_{s,s_z} \right]\,,\\
    &\label{tensor_spherical}
    Y^{\mu\nu}_{l,s,j,j_z}=\sum\limits_{m=-l}^l\sum\limits_{s_z=-s}^s \left\langle l,m,s,s_z \right.\left|j,j_z\right\rangle Y_{l,m} \left[\delta_{s,1} \tau^{\mu\nu}_{s,s_z}+\left( 1-\delta_{s,1}\right)\delta_1^i \chi^{\mu\nu}_{s,s_z} \right]\,,\\
    &\label{dual_vector_spherical}
    \tilde{Y}_{\mu}^{l,s,j,j_z}=\sum\limits_{m=-l}^l\sum\limits_{s_z=-s}^s \left\langle l,m,s,s_z\right.\left|j,j_z\right\rangle Y_{l,m} \left[\delta_{0,s} \tilde{\tau}_{\mu}+\left(1-\delta_{0,s}\right) \tilde{\chi}_{\mu}^{s,s_z} \right]\,,\\
    &\label{dual_tensor_spherical}
    \tilde{Y}_{\mu\nu}^{l,s,j,j_z}=\sum\limits_{m=-l}^l\sum\limits_{s_z=-s}^s \left\langle l,m,s,s_z\right.\left|j,j_z\right\rangle Y_{l,m} \left[\delta_{s,1} \tilde{\tau}_{\mu\nu}^{s,s_z}+\left( 1-\delta_{s,1}\right) \tilde{\chi}_{\mu\nu}^{s,s_z} \right]\,.
\end{align}}
\end{subequations}
The purely temporal component $\Psi_0$ of the spin-2 field can be decomposed separately through the following spin-0 harmonics,
\begin{equation}
Y^{\mu\nu}_{j,j_z}=Y_{j,j_z}\tau^{\mu\nu}_{0,0}~,~\tilde{Y}_{\mu\nu}^{j,j_z}=Y_{j,j_z}\tilde{\tau}_{\mu\nu}^{0,0}\,,
\end{equation}
though, given we have eliminated it from the equations by substituting the trace-less constraint, it will not enter in our computations. These vector and tensor spherical harmonics are orthonormal states having definite total angular momentum $j$, total azimuthal angular momentum $j_z$, total spin $s$ and total non-spin angular momentum $l$. We generalize to spin-1 and spin-2 the ansatz \ref{phi_ansatz} we used for the scalar case 
\begin{subequations}
\begin{align}
    &\label{A_ansatz}
    A_{\mu}=e^{-i \omega t} F(r)\sum\limits_{s=0}^1\sum\limits_{l=0}^{\infty}\sum\limits_{j=\text{max}\left\lbrace |j_z|,|l-s|\right\rbrace}^{l+s}B_{l,s,j}(\zeta(r))\tilde{Y}_{\mu}^{l,s,j,j_z}\,,\\
    &\label{h_ansatz}
    \begin{aligned}
        h_{\mu\nu}=e^{-i \omega t} F(r)\sum\limits_{j=|j_z|}^\infty&\tilde{B}_{j}(\zeta(r))\tilde{Y}_{\mu\nu}^{j,j_z}\\
        &+e^{-i \omega t} F(r)\sum\limits_{s=0}^{2}\sum\limits_{l=0}^{\infty}\sum\limits_{\scriptsize\begin{matrix}j=\text{max}\left\lbrace |j_z|\right.,\\\left.|l-s|\right\rbrace    
        \end{matrix}}^{l+s}\tilde{B}_{l,s,j}(\zeta(r))\tilde{Y}_{\mu\nu}^{l,s,j,j_z}\,.
    \end{aligned}
\end{align}
\end{subequations}
where $j_z$ is kept fixed and also in this case the asymptotic behaviour of the solutions at the horizon and at infinity is expressed through some radial function $F(r)$. The radial functions $B_{l,s,j}(\zeta(r))$ describe $A_\mu$, while $\tilde{B}_{j}(\zeta(r))$ and $\tilde{B}_{l,s,j}(\zeta(r))$ are radial functions for pure-temporal space states of $h_{\mu\nu}$ respectively, in both cases depending on the auxiliary radial coordinate $\zeta(r)$.
Because of the orthonormality of the vector and tensor spin-eigenstate basis, the inverse of expressions \ref{proca_decomposed} and \ref{h_spin_decomposed} are $\psi_J=A_{\mu}\left(\theta^{\mu}_J \right)^*$ and $\Psi_J=h_{\mu\nu}\left(\Theta^{\mu\nu}_J \right)^*$ respectively, thus giving the following spherical harmonics expansions,
\begin{subequations}
\begin{align}
    &\label{psi_spherical_harmonics}
    \psi_J=e^{-i \omega t} F(r)\sum\limits_{s=0}^1\sum\limits_{l=0}^{\infty}\sum\limits_{j=\text{max}\left\lbrace |j_z|,|l-s|\right\rbrace}^{l+s}B_{l,s,j}(\zeta(r))\tilde{Y}_{(1)J}^{l,s,j,j_z}(\theta,\phi)\,,\\
    &\label{Psi_spherical_harmonics}
    \Psi_J=e^{-i \omega t} F(r)\sum\limits_{s=0}^{2}\sum\limits_{l=0}^{\infty}\sum\limits_{j=\text{max}\left\lbrace |j_z|,|l-s|\right\rbrace}^{l+s}\tilde B_{l,s,j}(\zeta(r))\tilde{Y}_{(2)J}^{l,s,j,j_z}(\theta,\phi)\,,\\
    &\label{Psi0_spherical_harmonics}
    \Psi_0=e^{-i \omega t} F(r)\sum\limits_{j=|j_z|}^\infty\tilde{B}_{j}(\zeta(r))Y_{j,j_z}(\theta,\phi)\,,
\end{align}
\end{subequations}
where $\tilde{Y}_{(1)J}^{l,s,j,j_z}=\tilde{Y}_{\mu}^{l,s,j,j_z}\left(\theta^{\mu}_J \right)^*$ and $\tilde{Y}_{(2)J}^{l,s,j,j_z}=\tilde{Y}_{\mu\nu}^{l,s,j,j_z}\left(\Theta^{\mu\nu}_J \right)^*$ are the spinor components of the dual vector and tensor spherical harmonics respectively. The vector and the tensor spherical harmonics are orthonormal on the sphere, hence their spinor components will satisfy the orthonormality relationships
\begin{subequations}
\begin{align}
    &\label{ortho_vector_harmonics}\int d\Omega\left(Y^{(1)J}_{l',s',j',j_z'}\right)^*\tilde{Y}_{(1)J}^{l,s,j,j_z}=\delta_{l'}^l\delta_{s'}^s\delta_{j'}^j\delta_{j_z'}^{j_z}\,,\\
    &\label{ortho_tensor_harmonics}\int d\Omega\left(Y^{(2)J}_{l',s',j',j_z'}\right)^*\tilde{Y}_{(2)J}^{l,s,j,j_z}=\delta_{l'}^l\delta_{s'}^s\delta_{j'}^j\delta_{j_z'}^{j_z}\,,
\end{align}
\end{subequations}
where $\int d\Omega=\int_0^{2\pi}d\phi\int^\pi_0 d\theta\sin\theta$ is the integration on the sphere. We must be careful when we use the spinor components of the vector and tensor harmonics, because they do not have the spin-eigenstates in their expression, consequently they give
\begin{subequations}
\begin{align}
    &\mathbfcal{L}^2 Y^{(S)J}_{l,s,j,j_z}=l(l+1) Y^{(S)J}_{l,s,j,j_z}\,,\\
    &\mathbfcal{L}^2 Y_{(S)J}^{l,s,j,j_z}=l(l+1) Y_{(S)J}^{l,s,j,j_z}\,.
\end{align}
\end{subequations}
and are not eigenfunctions of $\mathbfcal{L}_{\boldsymbol z}$. 

By substituting \ref{psi_spherical_harmonics} and \ref{Psi_spherical_harmonics} in the wave equations \ref{wave_eqs} and then projecting on the vector and tensor harmonics respectively we get the infinite cascade of coupled radial equations for spin-1 and spin-2 perturbations,
\begin{equation}
\begin{aligned}\label{radial_wave_eqs_s1_s2}
    \frac{\partial^2 B^{(S)}_{l,s,j}(\zeta)}{\partial\zeta^2}+&\sum\limits_{s'=0}^S\sum\limits_{l'=0}^{\infty}\sum\limits_{j'=\text{max}\left\lbrace |j_z|,|l'-s'|\right\rbrace}^{l'+s'} \left[C_{1,l,s,j}^{l',s',j'}(r)\frac{\partial}{\partial\zeta}+ C_{2,l,s,j}^{l',s',j'}(r)\right.\\
    &+\Gamma_{l,s,j}^{l',s',j'}(r)+D_{l,s,j}^{l',s',j'}(r)+\Lambda_{l,s,j}^{l',s',j'}(r)\bigg]B^{(S)}_{l',s',j'}(\zeta)=0\,,
\end{aligned}
\end{equation}
where $B^{(S)}_{l,s,j}=\delta_{S,1}B_{l,s,j}+\delta_{S,2}\tilde B_{l,s,j}$  and the expressions for the mixing functions can be found in Appendix \ref{couplings_wave_eqs}. When computing these radial equations we must be careful in not introducing unphysical modes and/or equations, i.e. always respect the rule $s,s'\in\left[0,S\right]$ for the radial wave equations. $\Lambda_{l,s,j}^{l',s',j'}(r)$ are the only eigenstate-mixing arising from the Klein-Gordon-like part of the wave equations \footnote{$C_{2,l,s,j}^{l',s',j'}(r)$ also come from the Klein-Gordon-like part, but they are not really mixing functions. Their indices in fact are all from Kronecker deltas, thus it is just the identity matrix multiplied by a function. $C_{1,l,s,j}^{l',s',j'}(r)$, instead, are sums of a non-mixing part like $C_{2,l,s,j}^{l',s',j'}(r)$ and a part arising from spin-mixing matrices.}, and in fact they are generalizations of the only mixing present in the case of scalar perturbations (see Appendix \ref{couplings_wave_eqs}). 

The expressions for the decomposed Lorenz constraints are, instead, less involving,
\begin{subequations}
\begin{align}
&\label{radial_lorenz_s1}
\sum\limits_{s'=0}^1\sum\limits_{l'=0}^{\infty}\sum\limits_{\scriptsize\begin{matrix}j'=\text{max}\left\lbrace |j_z|\right.,\\\left.|l'-s'|\right\rbrace    
\end{matrix}}^{l'+s'} \left[C_{3,j}^{l',s',j'}(r)\frac{\partial}{\partial\zeta}+ C_{4,j}^{l',s',j'}(r)+\tilde{D}_{j}^{l',s',j'}(r)\right]B_{l',s',j'}(\zeta)=0\,,\\
&\label{radial_lorenz_s2}
\sum\limits_{s'=0}^2\sum\limits_{l'=0}^{\infty}\sum\limits_{\scriptsize\begin{matrix}j'=\text{max}\left\lbrace |j_z|\right.,\\\left.|l'-s'|\right\rbrace    
\end{matrix}}^{l'+s'} \left[C_{3,l,s,j}^{l',s',j'}(r)\frac{\partial}{\partial\zeta}+ C_{4,l,s,j}^{l',s',j'}(r)+\tilde{D}_{l,s,j}^{l',s',j'}(r)\right]\tilde B_{l',s',j'}(\zeta)=0\,,
\end{align}
\end{subequations}
while the expression of the decomposed second-order constraint for the spin-2 field is close to the one of the wave equations:
\begin{equation}\label{decomposed_second-order_constraint}
   \sum\limits_{s'=0}^2\sum\limits_{l'=0}^{\infty}\sum\limits_{\scriptsize\begin{matrix}j'=\text{max}\left\lbrace |j_z|\right.,\\\left.|l'-s'|\right\rbrace    
\end{matrix}}^{l'+s'}\left[C^{l',s',j'}_{5,j}\frac{\partial^2}{\partial\zeta^2}+C^{l',s',j'}_{6,j}\frac{\partial}{\partial\zeta}+C^{l',s',j'}_{7,j}+L^{l',s',j'}_{j}\right]\tilde B_{l',s',j'}=0\,.
\end{equation}
Also in the case of the constraints, we could be tricked by the involved formalism and thus introduce unphysical modes: in Equation \ref{radial_lorenz_s1} $s'\in\left \lbrace 0 , 1 \right\rbrace$, while in \ref{radial_lorenz_s2} and \ref{decomposed_second-order_constraint} $0\leq s'\leq 2$, but in \ref{radial_lorenz_s2} we have $s\in\left \lbrace 0 , 1 \right\rbrace$\footnote{This is a consequence of the fact that the spin-2 Lorenz constraint is a 4-vector, hence we projected it on the vector eigenstate basis.}.
The full expressions of the involved coupling functions for the Lorenz and second-order constraints can be found in Appendices \ref{couplings_lorenz} and \ref{couplings_second-order} respectively.

The couplings between different radial equations, as well as the number of equations itself, is infinite, consequently we will need to apply some truncation for $j$ and $l$. The number of spherical harmonics required for good results depends on how close to $M$ is the BH spin parameter $a$, i.e. computing perturbations of nearly extremal BHs requires many modes. In the next section we will address the issue of computing mixing matrices and therefore analyze the relationship between $a$ and spherical harmonics truncation.

\section{Computation of the spherical harmonic coupling terms of spin-1 and spin-2 fields}\label{section:overlap_parametrization}

The coupling terms we found in the previous section are bra-ket integrals of rational functions $X$ arising from the Kerr metric and its Christoffel symbols, functions whose dependence on $\theta$ is through trigonometric functions $\cos\theta$ and $\sin\theta$ while they depend on $\phi$ through some phase functions. We can generically label such functions as $X(r, \theta, e^{i\phi}; a, M)$, where $a<M$ is the BH spin parameter, $M$ is the BH mass. For $a=0$ these functions are linear combinations of trigonometric functions in $\theta$, while for $a\neq 0$ they are cumbersome rational functions involving $\cos\theta$ and $\sin\theta$. Brute force computation of all the integrals $\left\langle l_1, m_1 \right|X(r, \theta, e^{i\phi}; a, M)\left| l_2 , m_2\right\rangle$ is very time-consuming and must be performed by fixing the parameters of the system\footnote{In general, in the spin-1 case carrying all the integrations analytically is extremely slow for given values of the parameters, while the spin-2 case is totally unmanageable. Brute force numerical integration can even be more problematic, given the extreme behaviour of the involved functions in some cases.}, which implies that any change in the parameters requires recomputing those integrals from scratch. Furthermore, the number of relevant overlapping integrals can be very high, especially for highly spinning BHs. Indeed, if we truncate our spherical harmonic expansion such that $l_1, l_2\leq L$ for some value $L>0$, the number of overlap integrals to be computed will be proportional to $L^2 (L+2)^2$\footnote{If we just consider $\left|l,m \right\rangle$ for $0\leq l\leq L$, we have $\sum\limits_{l=0}^L (2 l+1)=L(L+2)$ possible spherical harmonics. Each integral involves two spherical harmonics, thus the number of integrals is proportional to $L^2(L+2)^2$}. Moreover, the higher the spin of the perturbation is, the higher the number of integrals will be. In fact, on one hand the spin-1 case mixing matrices have $16$ components and Lorenz constraint vectors are $4$-dimensional. On the other hand, in the spin-2 case we have $9^2$ dimensional spin-mixing matrices, $4\times9$-dimensional Lorenz spin-constraint matrices and a $4$-dimensional vectors for the second-order constraint.

A way to bypass such problems is to Fourier-expand $X(r, \theta, e^{i\phi}; a, M)$,
\begin{equation}\label{Fourier-expand}
\begin{small}\begin{aligned}
    \left\langle l_1, m_1 \right| X(r, \theta, e^{i\phi}; a, M) \left| l_2 , m_2 \right\rangle&=\sum\limits_{n=0 , k}^\infty \tilde{X}_{n,k}(M,a; r)\left\langle l_1, m_1 \right| e^{i k \phi} \cos(n \theta) \left| l_2 , m_2\right\rangle\\
    &+\sum\limits_{n=1 , k}^\infty \tilde{X}_{-n,k}(M,a; r)\left\langle l_1, m_1 \right| e^{i k \phi} \sin(n \theta) \left| l_2 , m_2 \right\rangle\,,
\end{aligned}\end{small}
\end{equation}
in which case the problem is reduced to computing integrals of the following type:
\begin{subequations}
\begin{small}\begin{align}
    &\label{bra_ket1}
    \left\langle l_1, m_1 \right| e^{i k \phi} \cos(n \theta) \left| l_2 , m_2\right\rangle=\int\limits_0^{2 \pi}d\phi\int\limits_0^\pi d\theta \sin\theta Y^*_{l_1 ,m_1}(\theta,\phi) e^{i k \phi} \cos(n \theta)Y_{l_2 ,m_2}(\theta,\phi)\,,\\
    &\label{bra_ket2}
    \left\langle l_1, m_1 \right| e^{i k \phi} \sin(n \theta) \left| l_2 , m_2\right\rangle=\int\limits_0^{2 \pi}d\phi\int\limits_0^\pi d\theta \sin\theta Y^*_{l_1 ,m_1}(\theta,\phi) e^{i k \phi} \sin(n \theta)Y_{l_2 ,m_2}(\theta,\phi)\,.
\end{align}\end{small}
\end{subequations} 
In order to compute the Fourier coefficients $\tilde{X}_{n,k}$ without carrying any integration, one can resort to a Taylor expansion of $X(r, \theta, e^{i\phi}; a, M)$ in the black-hole spin parameter at arbitrarily high order. Due to how the spin parameter is coupled to the trigonometric functions, once the integrals (\ref{bra_ket1}) and (\ref{bra_ket2}) are computed analytically it is possible to derive analytical expressions for $\tilde{X}_{n,k}$ at arbitrarily high order in the black-hole spin. In fact $X(r, \theta, e^{i\phi}; a, M)$ is built from the Kerr metric and the spin-eigenstate basis, thus it will already be a complex Fourier expansion in $\phi$ and its dependency in $\theta$ comes from trigonometric functions. In addition, almost all the sines and cosines appearing in the expression of $X(r, \theta, e^{i\phi}; a, M)$ are multiplied with the spin parameter $a$. All these properties give the important result that Fourier-expanding in $\theta$ and $\phi$ or Taylor-expanding in $a$ is mathematically almost equivalent if the order is high: in both cases we get a Fourier expansion. But Taylor-expanding is easier, quicker and enables us to always find analytic parametric coefficients. Though, given that our goal is getting a Fourier expansion and that the spin parameter is not always coupled with trigonometric functions, what we really need is expanding only specific portions of $X$. Specifically, through Taylor expansion we need to move to the numerators the angular dependencies appearing in the denominators, i.e. expanding functions $1/\Sigma(r,\theta)$ and the denominator of $1/T^0$ appearing in the expressions of $X$\footnote{For accurate results, one needs to expand $1/\Sigma(r,\theta)$ functions at least for order equal to $2 L$ in the spin expansion. At the end of this section we explain why.}. Therefore, by computing this "surgical" expansion in $a$ we get the analytical expression of "partial" Taylor expansions of all the $2 N+1$ coefficients of the $N$-th order Fourier expansion in $\theta$ \footnote{There are also some extra terms of higher Fourier order that need to be discarded.}. Integrals \ref{bra_ket1} and \ref{bra_ket2} can be computed analytically by exploiting the properties of spherical harmonics and Legendre polynomials, thus getting parametrized expressions which result in saving computation time.

We recall that spherical harmonics are defined as\cite{rose_angular_momentum} 
\begin{equation}\label{spherical_harmonics}
Y_{l,m}(\theta,\phi)=\sqrt{\frac{(2 l+1)(l-m)!}{4\pi(l+m)!}}P_l^m(\cos\theta)e^{i m \phi}\,,
\end{equation}
where $P_l^m$ are the Legendre associated functions, and consequently their complex conjugate is $Y_{l , m}^*=(-1)^m Y_{l,-m}$. 
Moreover, we can express their products in the following way\cite{rose_angular_momentum}:
\begin{equation}\label{harmonics_product}
\begin{aligned}
    Y_{l_1 , m_1}Y_{l_2 , m_2}&=\sum\limits_{l=|l_1-l_2|}^{l_1+l_2}\sqrt{\frac{(2 l_1+1)(2 l_2+1)}{4\pi(2 l+1)}}\left\langle l_1 , 0 , l_2 , 0 \right| \left. l , 0\right\rangle\\
    &\times\left\langle l_1 , m_1 , l_2 , m_2 \right| \left. l , m_1+m_2\right\rangle Y_{l , m_1+m_2}\,,
\end{aligned}
\end{equation}
where $\left\langle l_1 , m_1 , l_2 , m_2 \right| \left. l , m\right\rangle$ are the Clebsch-Gordan coefficients.
By exploiting these properties, bra-kets (\ref{bra_ket1}) and (\ref{bra_ket2}) can be written as
\begin{subequations}
\begin{small}
\begin{align}
    &\label{cos_braket0}\begin{aligned}
    \left\langle l_1, m_1 \right| e^{i k \phi} \cos(n \theta) \left| l_2 , m_2\right\rangle&=\sum\limits_{l=|l_1-l_2|}^{l_1+l_2}(-1)^{m_1}\left\langle l_1 , 0 , l_2 , 0 \right| \left. l , 0\right\rangle \left\langle l_1 , -m_1 , l_2 , m_2 \right| \left. l , -k\right\rangle\\
    &\times \sqrt{\frac{\left( l_1+\frac{1}{2}\right)\left( l_2+\frac{1}{2}\right)(l+k)!}{(l-k)!}}\int_0^\pi d\theta \sin\theta \cos(n\theta) P_l^{-k}(\cos\theta)\,,   
    \end{aligned}\\
    &\label{sin_braket0}\begin{aligned}
    \left\langle l_1, m_1 \right| e^{i k \phi} \sin(n \theta) \left| l_2 , m_2\right\rangle&=\sum\limits_{l=|l_1-l_2|}^{l_1+l_2}(-1)^{m_1}\left\langle l_1 , 0 , l_2 , 0 \right| \left. l , 0\right\rangle\left\langle l_1 , -m_1 , l_2 , m_2 \right| \left. l , -k\right\rangle\\
    &\times\sqrt{\frac{\left( l_1+\frac{1}{2}\right)\left(l_2+\frac{1}{2}\right)(l+k)!}{(l-k)!}} \int_0^\pi d\theta \sin\theta \sin(n\theta) P_l^{-k}(\cos\theta)\,.
    \end{aligned}
\end{align}
\end{small}
\end{subequations}Functions $\cos(n\theta)$ and $\sin(n\theta)$ can be expressed as finite linear combinations of associated Legendre functions:
\begin{equation}\label{cos-sin-legendre0}
\cos(n\theta)=\sum\limits_{l=0}^n a^n_l P_l^0(\cos\theta)\,,\qquad\sin(n\theta)=\sum\limits_{l=0}^n b^n_l P_l^1(\cos\theta)\,.
\end{equation}
From the definition of the associated Legendre functions~\cite{math_meth}, we have $P_l^0(\cos\theta)=P_l(\cos\theta)$ and $P_l^1(\cos\theta)=\frac{d}{d\theta}\left[P_l(\cos\theta)\right]$, where $P_l$ are the Legendre polynomials. Thus, by taking the derivative of the first equation in~(\ref{cos-sin-legendre0}) with respect to $\theta$ and comparing it with the second equation, we get
\begin{equation}\label{cos-sin-coeff}
b^n_l=-\frac{1}{n}a^n_l\,.
\end{equation}
One can therefore focus only on the first equation in~(\ref{cos-sin-legendre0}). Functions $\cos(n\theta)$ are Čebyšëv polynomials $T_n(\cos\theta)=\cos(n\theta)$~\cite{special_polynomials}, hence we can exploit this property for finding the $a^n_l$ coefficients. Functions $T_n(x)$ and $P_l(x)$ are in fact special cases of ultraspherical polynomials $C^\gamma_n(x)$~\cite{special_polynomials}:
\begin{equation}\label{ultraspherical}
T_n(x)=\lim\limits_{\gamma\rightarrow 0}\frac{n+2\gamma}{2\gamma}C^\gamma_n(x)\,,\qquad P_l(x)=C^{1/2}_l(x)\,,
\end{equation}
Different types of ultraspherical polynomials can be related through the following expression~\cite{special_polynomials}, which is a special case of the connection relation for Jacobi polynomials\footnote{Ultraspherical polynomials are a special case of Jacobi polynomials~\cite{special_polynomials}.}:
\begin{equation}\label{connection_jacobi}
C^\gamma_n(x)=\sum\limits_{j=0}^{\floor{n/2}}\frac{(\gamma-\beta)_j\left(\gamma\right)_{n-j}}{j!\left(\beta+1\right)_{n-j}}\left(\frac{\beta+n-2 j}{\beta} \right) C^\beta_{n-2 j}(x)\, ,
\end{equation}
where $\displaystyle\left(x \right)_y=\frac{\Gamma(x+y)}{\Gamma(x)}$ is the Pochhammer's symbol for $x,y\in\mathbb{C}$ and $\Gamma(x)$ is Euler's gamma function. From Eqs.~(\ref{ultraspherical}) and~(\ref{connection_jacobi}) we get the expressions for Čebyšëv polynomials as linear combinations of Legendre polynomials
\begin{equation}\label{chebyshev-legendre}
T_n(x)=-\sum\limits_{j=0}^{\floor{n/2}}\frac{n\Gamma\left(j-1/2\right)\Gamma\left(n-j\right)}{8 j!\Gamma\left(3/2+n-j\right)}(1+2n-4j)P_{n-2 j}(x)\,,
\end{equation}
and consequently, by using Eq.~(\ref{cos-sin-coeff}), we can rewrite Eqs.~(\ref{cos-sin-legendre0}) as
\begin{subequations}
\begin{align}
    &\label{cos-legendre}
    \cos(n\theta)=-\sum\limits_{j=0}^{\floor{n/2}}\frac{n\Gamma\left(j-1/2\right)\Gamma\left(n-j\right)}{8 j!\Gamma\left(3/2+n-j\right)}\left(1+2 n-4 j \right)P^0_{n-2j}(\cos\theta)\,,\\
    &\label{sin-legendre}
    \sin(n\theta)=\sum\limits_{j=0}^{\floor{n/2}}\frac{\Gamma\left(j-1/2\right)\Gamma\left(n-j\right)}{8 j!\Gamma\left(3/2+n-j\right)}\left(1+2 n-4 j \right)P^1_{n-2j}(\cos\theta)\,.
\end{align}
\end{subequations}
By using these last results, we can write the general parametric expressions for the bra-kets~(\ref{cos_braket0}) and~(\ref{sin_braket0}) in closed form:
\begin{subequations}
\begin{align}
    &\label{cos_braket}\begin{aligned}
        &\left\langle l_1, m_1 \right| e^{i k \phi} \cos(n \theta) \left| l_2 , m_2\right\rangle=\\
        &=\sum\limits_{l=|l_1-l_2|}^{l_1+l_2} \sum\limits_{j=0}^{\floor{n/2}}\frac{(-1)^{m_1+1}}{2}\frac{n\Gamma\left(j-1/2\right)\Gamma\left(n-j\right)}{8 j!\Gamma\left(3/2+n-j\right)}\sqrt{\frac{(2 l_1+1)(2 l_2+1)(l+k)!}{(l-k)!}}\\
        &\times  \left(1+2 n-4 j \right) \left\langle l_1 , -m_1 , l_2 , m_2 \right| \left. l , -k\right\rangle \left\langle l_1 , 0 , l_2 , 0 \right| \left. l , 0\right\rangle I(n-2 j, 0, l,-k)\,,
    \end{aligned}\\
    &\label{sin_braket}\begin{aligned}
        &\left\langle l_1, m_1 \right| e^{i k \phi} \sin(n \theta) \left| l_2 , m_2\right\rangle=\\
        &=\sum\limits_{l=|l_1-l_2|}^{l_1+l_2} \sum\limits_{j=0}^{\floor{n/2}}\frac{(-1)^{m_1}}{2}\frac{\Gamma\left(j-1/2\right)\Gamma\left(n-j\right)}{8 j!\Gamma\left(3/2+n-j\right)}\sqrt{\frac{(2 l_1+1)(2 l_2+1)(l+k)!}{(l-k)!}}\\
        &\times \left(1+2 n-4 j \right) \left\langle l_1 , -m_1 , l_2 , m_2 \right| \left. l , -k\right\rangle \left\langle l_1 , 0 , l_2 , 0 \right| \left. l , 0\right\rangle I(n-2 j, 1, l,-k)\,,
    \end{aligned}
\end{align}
\end{subequations}
where we have $\displaystyle I(l,m,l',m')=\int_{-1}^1 dx P_l^m(x) P_{l'}^{m'}(x)$. This last integral can be explicitly performed~\cite{legendre_integral}:
\begin{equation*}
\begin{aligned}
    I(l,m,l',m')&=\sqrt{\frac{(l+m)!(l'+m')!}{(l-m)!(l'-m')!}}\sum\limits_{j=|l-l'|}^{l+l'}\sqrt{\frac{(j-m-m')!}{(j+m+m')!}}\\
    &\times\left\langle l , m , l', m' \right| \left. j , m+m'\right\rangle\left\langle l , 0 , l', 0 \right| \left. j , 0\right\rangle I_0(j,m+m')\,,
\end{aligned}
\end{equation*}
where for $m>0$ we have
\begin{equation*}
I_0(l,m)=\int_{-1}^1 dx P_l^m(x)=\frac{[(-1)^m+(-1)^l]2^{m-2}m\Gamma\left(l/2 \right)\Gamma\left((l+m+1)/2\right)}{\left((l-m)/2 \right)!\Gamma\left((l+3)/2 \right)}.
\end{equation*}
Because of $P_0^0(\cos\theta)=1$, for $l,m=0$ we have $I_0(l,m)=2$, while for $m<0$ we can exploit $\displaystyle I_0(l,m)=(-1)^m\frac{(l+m)!}{(l-m)!}I_0(l,-m)$ from the properties of $P_l^m$. Expression~(\ref{cos_braket}) gives an indeterminate result for $n=0$, therefore, by taking into account that $\cos(0 \theta)=1=P_0^0(\cos\theta)$, for this case we need to use the following expression:
\begin{equation}\label{n0cos_braket}
\begin{aligned}
    \left\langle l_1, m_1 \right| e^{i k \phi} \left| l_2 , m_2\right\rangle&=\sum\limits_{l=|l_1-l_2|}^{l_1+l_2}\frac{(-1)^{m_1}}{2} \sqrt{\frac{(2 l_1+1)(2 l_2+1)(l+k)!}{(l-k)!}}\\
    &\times\left\langle l_1 , -m_1 , l_2 , m_2 \right| \left. l , -k\right\rangle \left\langle l_1 , 0 , l_2 , 0 \right| \left. l , 0\right\rangle I_0(l,-k)\,.
\end{aligned}
\end{equation}
In Appendix \ref{referee_alternative} we report also an alternative, though computationally less efficient, result for integrals \ref{bra_ket1} and \ref{bra_ket2}, while in Appendix \ref{simplified_axisym_braket} we give a simplified result for the $k=0$ case. All these results have been published in \cite{Lingetti_spherical_overlaps}.

With the method we defined in these pages we can also estimate the relationship between spherical harmonics truncation and the BH spin parameter $a$. If we consider truncation at $l=L$ we notice that in \ref{cos_braket0} and \ref{sin_braket0} the highest degree Legendre associated functions involved will have $l=2 L$ \footnote{This is a consequence of the properties of Clebsch-Gordan coefficients.}. Moreover, the dominating Legendre associated functions in \ref{cos-legendre} and \ref{sin-legendre} are the ones having highest degree, and for higher $n$ values this property is even more pronounced. Thus, we can say that the most important contribution at $l=L$ is given by Fourier terms of order $n\approx 2 L$. Given the almost exact equivalence between $n$-th order Taylor expansions in $a$ and $n$-th order Fourier expansions in $\theta$, we conclude that truncating at $j,l=L$ is approximately equivalent to expanding in $a$ until order $2 L$. 

\chapter{Numerical computation of superradiant instabilities in GR black holes}\label{chapter:numerical_SR_GR}

In this chapter we will give numerical results for the computation of unstable superradiant states in GR spinning BHs, i.e. quasi-bound states (QBSs) of ultralight bosonic fields satisfying the superradiant condition. We will apply the spherical harmonic decomposition we defined in the previous chapter, then discretize the radial equations through a Čebyšëv interpolation, in order to transform them in a matrix non-linear eigenvalue problem. We will attack the matrix equations with a non-linear inverse iteration algorithm, through which we will reproduce the results by \textit{Baumann at al.}\cite{Spectra_grav_atom} for spin-0 and spin-1. We will apply the technique also to the spin-2 case and show the problems arising. This numerical approach requires some approximated solution as starting guess for the iterations, therefore we will review the previous results we used as guess. In this chapter we will finally fix the auxiliary radial coordinate $\zeta$, which is required for applying the Čebyšëv interpolation, and the function $F(r)$ we introduced previously, in accordance with the geometry of Kerr metric and the boundary conditions of the problem.

\section{Boundary conditions and field ansätze}\label{section:boundary_conditions}

When the mass of a test field propagating in a BH spacetime is non-zero, the particles of the field will experience an effective radial potential well, thus they will admit quasi-bound states (QBSs). These type of states are found by imposing the following boundary conditions:
\begin{itemize}
\item purely in-going waves at the event horizon, because in classical physics nothing can escape from a BH;
\item non radiating waves at infinity, i.e. waves asymptotically decreasing to zero at infinity.
\end{itemize}
The other possible states are the quasi-normal modes (QNMs), having same boundary condition at the horizon, but radiating at infinity\footnote{This is the reason why QBSs are also called "non-radiative QNMs".}. QBSs in BH spacetimes describe massive particles orbiting around a BH, thus they are states of fields confined in the proximity of the central compact object: this is why we expect to find superradiantly unstable states among QBSs. They are called "quasi-bound" instead of "bound" because these states are not stationary due to the presence of the BH event horizon. This type of systems present many characteristics in common with atoms, therefore, not surprisingly, they are also called "gravitational atoms"\cite{Spectra_grav_atom}. Obviously there also differences, e.g. the most important one is being classical bosonic states.

\subsection{Field ansatz for scalar quasi-bound states}

In the previous chapter in the field ansätze we introduced the function $F(r)$ with the objective of extracting the asymptotic behaviour of the solution, thus now we will define its expression according to the boundary conditions of QBSs. In the scalar field case the asymptotic study of the field equation \ref{klein-gordon_manipulation} gives the following expressions:
\begin{subequations}
\begin{align}
    &\left[\frac{\partial^2}{\partial r^2}+\frac{1}{r-r_+}\frac{\partial}{\partial r}+\frac{P_+^2}{(r-r_+)^2}\right]\Phi \sim 0~~~~~~~~~~~~~~~~~~~~r\rightarrow r_+\,,\\
    &\left[\frac{\partial^2}{\partial r^2}+\frac{2}{r}\frac{\partial}{\partial r}+\frac{(4\omega^2-2\mu^2) M}{r}+\omega^2-\mu^2 \right]\Phi\sim 0~~~~~~~r\rightarrow\infty\,,    
\end{align}
\end{subequations}
Therefore we get the following asymptotic solutions at the horizon and at infinity,
\begin{equation}\label{asymptotic_Phi}
\Phi\sim\left\lbrace
\begin{matrix}
A_{in} \displaystyle(r-r_+)^{ i P_+}+A_{out} \displaystyle(r-r_+)^{- i P_+}~~~~~~~~~~~~~~~~~~~~~~~~~~~~r\rightarrow r_+
\\
\displaystyle \mathcal{R}~ r^{-1-\nu+\frac{2 M^2\mu^2}{\nu}}e^{\frac{\mu^2 M}{\nu}r}+\mathcal{B}~ r^{-1+\nu-\frac{2 M^2\mu^2}{\nu}}e^{-\frac{\mu^2 M}{\nu}r}~~~~~~~~~~~r\rightarrow \infty
\end{matrix}\right.\,,
\end{equation}
which describe QBSs if we impose $A_{out}=0$ and $\mathcal{R}=0$, where we simplified the expressions by introducing the new parameter $\nu=\frac{\mu^2 M}{\sqrt{\mu^2-\omega^2}}$:
\begin{equation}\label{freq}
\omega=\mu\sqrt{1-\frac{\mu^2 M^2}{\nu^2}}\,.
\end{equation}
\textit{Baumann et al.}\cite{Spectra_grav_atom} chose the following expression for $F(r)$,
\begin{equation}\label{asymptotic_F_function}
    F(r)=\left(\frac{r-r_+}{r-r_-} \right)^{i P_+}(r-r_-)^{-1+\nu-\frac{2\mu^2 M^2}{\nu}}e^{-\frac{\mu^2 M}{\nu}(r-r_+)}\,,
\end{equation}
which is the product of the solution at the horizon with the one at infinity, including some sub-leading terms, e.g. the behaviour $\Phi\sim(r-r_-)^{- i P_+}$ near $r=r_-$ .

\subsection{Field ansatz for vector quasi-bound states}

Fixing the ansatz in the vector case, instead, means also choosing the vector basis. In fact, we recall that the vector and (co)vector basis we used for the decomposition of spin-1 field equations feature 3 free function $F_t(r)$, $F_r(r)$, $F_\theta(r)$ that need to be fixed. In \cite{Spectra_grav_atom}, \textit{Baumann et al.} fix them by imposing that all the spinor components of the Proca field $A_\mu$ must scale the same way both at the horizon and at infinity, thus choosing
\begin{equation}\label{basis_choice}
    F_t(r)=1~,~F_r(r)=\frac{r-r_+}{r-r_-}~,~F_\theta(r)=\displaystyle\frac{1}{r-r_-}\,.
\end{equation}
They found this functions by carrying a brute-force asymptotic analysis of the Proce equations in Kerr spacetime before carrying any decomposition of any type. Such approach is extremely challenging, especially if we look also for some tensor extension for fixing also the ansatz of spin-2 fields, thus we look for some more manageable alternative.

Let us consider the vacuum Maxwell equations in Kerr spacetime in the Lorenz gauge,
\begin{equation}
    \bar\Box A_\mu=0~,~\bar\nabla^\mu A_\mu=0
\end{equation}
where $A_\mu$ is the electromagnetic four-potential. Such equations are invariant under the transformation $A_\mu \rightarrow A_\mu+\partial_\mu \Phi$ if the scalar field $\Phi$ satisfies the massless Klein-Gordon equation $\bar\Box \Phi=0$. This implies that solutions $A_\mu=\partial_\mu \Phi$ do satisfy the field equations but are not physical, i.e. $\partial_\mu \Phi$ gives electromagnetic gauge modes. If we consider scalar solutions having azimuthal angular momentum $j_z$, what we get is just the massless case of what we computed in \ref{klein-gordon_manipulation} :
\begin{equation}
    \begin{aligned}
        &\frac{\Sigma(r,\theta)}{\Delta(r)}\bar\Box\Phi=\left\lbrace\displaystyle\frac{1}{\Delta(r)}[\boldsymbol{\mathbfcal{L}}^2-\omega^2 a^2\cos^2\theta]-\frac{1}{\Delta(r)}\frac{\partial}{\partial r}\left[\Delta(r)\frac{\partial}{\partial r}\right]-\omega^2-\right.\\
        &\left.-\frac{P_+^2}{(r-r_+)^2}-\frac{P_-^2}{(r-r_-)^2}+\frac{A_+}{(r_+-r_-)(r-r_+)}-\frac{A_-}{(r_+-r_-)(r-r_-)}\right\rbrace\Phi=0\,,
    \end{aligned}
\end{equation}
which gives $\Phi\sim (r-r_+)^{\pm i P_+}$ for $r\rightarrow r_+$. In spite of the fact that solutions $A_\mu=\partial_\mu \Phi$ are just gauge modes, we can exploit them and their asymptotic behaviour for the computation of the general behaviour of $A_\mu$ at the horizon. In fact, second order wave equations give just two possible asymptotic behaviours at the horizon, in-going or out-going wave, and therefore the difference between gauge modes and physical modes must be negligible when $r\rightarrow r_+$. Thus, from the asymptotic behaviour of $\Phi$ we can deduce the one of $A_\mu$:
\begin{equation}
    A_\mu\sim\partial_\mu\Phi\sim\left[\begin{matrix}
 (r-r_+)^{\pm i P_+}~~~\\ (r-r_+)^{\pm i P_+ -1}\\ (r-r_+)^{\pm i P_+}~~~\\ (r-r_+)^{\pm i P_+}~~~
\end{matrix}\right]\,.
\end{equation}
This result is valid also for the case of Proca fields, due to the mass term being negligible at the horizon. We can, therefore, compare it with the one computed by \textit{Baumann et al.}, and easily find out that their result and the choice they made for the spin-eigenstate basis are compatible with what we got. This is quite remarkable and hence worth being extended to the rank-2 tensor case, given that we managed to compute the asymptotic result without having to carry all the cumbersome manipulations \textit{Baumann et al.} had to (see sub-section \ref{QBSs_ansatz_spin2}).

The choice made in \ref{basis_choice} is also compatible with negligible spin-mixing matrices when studying the asymptotic behaviour of the Proca equations at infinity\footnote{They scale at least as $1/r^2$.}\cite{Spectra_grav_atom}, thus the result for the asymptotic solutions of $\psi_J$ is the same as the one of the scalar case:
\begin{equation}\label{asymptotic_A}
\psi_J\sim\left\lbrace
\begin{matrix}
A_{in} \displaystyle(r-r_+)^{ i P_+}+A_{out} \displaystyle(r-r_+)^{- i P_+}~~~~~~~~~~~~~~~~~~~~~~~~~~~~r\rightarrow r_+
\\
\displaystyle \mathcal{R}~ r^{-1-\nu+\frac{2 M^2\mu^2}{\nu}}e^{\frac{\mu^2 M}{\nu}r}+\mathcal{B}~ r^{-1+\nu-\frac{2 M^2\mu^2}{\nu}}e^{-\frac{\mu^2 M}{\nu}r}~~~~~~~~~~~r\rightarrow \infty
\end{matrix}\right.\,.
\end{equation}
Consequently $F(r)$ will have the same expression \ref{asymptotic_F_function} of the spin-0 case\cite{Spectra_grav_atom}. 

\subsection{Field ansatz for tensor quasi-bound states}\label{QBSs_ansatz_spin2}

We expect that in the asymptotic study of spin-2 fields something similar to what we saw in the spin-1 case should happen, taking into account that rank-2 tensor fields can be considered as linear combinations of tensor products between vectors $h_{\mu\nu}= \sum_j c_j V^j_\mu V^j_\nu$. This property suggests that having all $\Psi_J$ scalars scale in the same way should imply the scaling of the rank-2 tensor basis to be the tensor product of the scalings of the vector basis chosen by \textit{Baumann et al} for the spin-1 case. But this property does not fully address the structure of the field equations of massive spin-2 perturbations, hence we need some more rigorous procedure for finding the scalings of the components of the field. We apply the method based on gauge modes we used in the spin-1 case, in order to avoid carrying a brute-force asymptotic analysis of the field equations at the horizon.

Thus, let us consider the linearized field equations of gravitational waves $h_{\mu\nu}$ in the trace-less Lorenz gauge in Kerr spacetime:
\begin{equation}
    \bar\Box h_{\mu\nu}+2 \bar R^{~\alpha~\beta}_{\mu~\nu}h_{\alpha\beta}=0~,~\bar\nabla^\mu h_{\mu\nu}=0~,~h_{\mu\nu}\bar g^{\mu\nu}=0
\end{equation}
These equations are invariant under the following gauge transformation,
\begin{equation}
    h_{\mu\nu}\rightarrow h_{\mu\nu}+\epsilon \left(\bar\nabla_\mu A_\nu+\bar\nabla_\nu A_\mu \right)\,,
\end{equation}
(associated to the $\epsilon$-infinitesimal coordinate transformation $x^\mu\rightarrow x^\mu+\epsilon A^\mu$) if $A_\mu$ satisfies the field equations of electromagnetic waves in the Lorenz gauge. Therefore, at the horizon we must have \[h_{\mu\nu}\sim \left(\bar\nabla_\mu A_\nu+\bar\nabla_\nu A_\mu \right)\] and consequently, like in the spin-1 case, we can compute the (massive) spin-2 asymptotic solution from the asymptotic solution of electromagnetic waves we found previously:
\begin{equation}
    h_{\mu\nu}\sim\left[\begin{matrix}
 (r-r_+)^{\pm i P_+}& (r-r_+)^{\pm i P_+-1}& (r-r_+)^{\pm i P_+}& (r-r_+)^{\pm i P_+}\\ (r-r_+)^{\pm i P_+-1}& (r-r_+)^{\pm i P_+-2}& (r-r_+)^{\pm i P_+-1}& (r-r_+)^{\pm i P_+-1}\\ (r-r_+)^{\pm i P_+}& (r-r_+)^{\pm i P_+-1}& (r-r_+)^{\pm i P_+}& (r-r_+)^{\pm i P_+}\\ (r-r_+)^{\pm i P_+}& (r-r_+)^{\pm i P_+-1}& (r-r_+)^{\pm i P_+}& (r-r_+)^{\pm i P_+}\end{matrix}\right]\,,
\end{equation}
If we take into account that the rank-2 tensor basis is built through linear combinations of tensor products of vector spin-eigenstates, we deduce from the result we computed that the conjecture we made about the scalings of the components of $h_{\mu\nu}$ is correct. This result proves that the choice made in \ref{basis_choice} for the free functions of spin-1 case works also for the case of spin-2 perturbations, making all $\Psi_J$ scalars scale at the horizon like Klein-Gordon fields $\Psi_J\sim (r-r_+)^{\pm i P_+}$.

This choice works well also for the asymptotic analysis of $h_{\mu\nu}$ at infinity. For this case, in fact, we can fully exploit the spin-eigenstate decomposition of the field equations, by taking advantage of the similarities in the structure of the spin-mixing matrices between Proca and massive spin-2 fields. The biggest difference is in the presence of the Riemann curvature in matrix $S_I^{~J}$, but we still have $\lim\limits_{r\rightarrow\infty}r S_I^{~J}=0$ like in the Proca case \cite{Spectra_grav_atom} . The structure of the other mixing and constraint matrices is the same as the one of the Proca case: in general all the matrices are $O(1/r^2)$ at least and consequently do not contribute at infinity, exactly like in \cite{Spectra_grav_atom}. Thus the asymptotic behaviour is the same as Klein-Gordon's one, i.e. $\Psi_J\sim r^{-1\mp\nu\pm\frac{2 M^2\mu^2}{\nu}}e^{\pm\frac{\mu^2 M}{\nu}r}$. The results we got show that the spin-mixing matrices in \ref{wave_eqs} do not contribute to the asymptotic analysis at the horizon and at infinity, thus implying that also in the spin-2 case for QBSs we can use the expression \ref{basis_choice} for $F(r)$.

\section{Radial matrix equation}\label{section:chebyshev_interpol}

In the previous chapter we described how, by applying and extending the procedure originally developed by \textit{Baumann et al.} \cite{Spectra_grav_atom}, the equations of BH perturbations of any spin can be reduced to an infinite cascade of coupled radial differential equations. The numerical technique they developed for attacking such equations is rather universal and extremely powerful, thus, once all the mathematical framework for building the cascade of radial equations is set up, we can just follow \textit{Baumann et al.} for the next steps. After having imposed some truncation for $j$ and $l$, what they did is trasforming the set of coupled equations into a matrix equation by discretizing the radial coordinate through Čebyšëv interpolations.

The Čebyšëv polynomial of order $k$ is the polynomial such that $T_k(\cos\theta)=$ $=\cos(k\theta)$, with $k\in\mathbb{N}$ \cite{special_polynomials, approximation_theory,chebyshev_and_fourier_spectral_methods}. These polynomials form a complete functional basis, thus they can be used for expanding any smooth function $f\in C^\infty[-1,1]$:
\begin{equation}\label{chebyshev_exp}
f(\zeta)=\displaystyle\sum\limits_{k=0}^\infty a_k T_k(\zeta)~,~\displaystyle a_k=\frac{2-\delta^0_k}{\pi}\int\limits_{-1}^1 d\zeta \frac{T_k(\zeta)f(\zeta)}{\sqrt{1-\zeta^2}}\,.
\end{equation}
Let us consider some function $G$ defined on the complex unit circle such that $G(z)=G(1/z)=f(\zeta)$ and $\zeta=\frac{z+z^{-1}}{2}$. This function will be analytic in some annulus $\rho^{-1}\leq|z|\leq\rho$ because of the smoothness of $f$, and its Čebyšëv expansion will coincide with its Laurent series. Consequently, if we consider a truncation of \ref{chebyshev_exp} and ellipses $\zeta(\theta)=\frac{1}{2}\left(\rho+\frac{1}{\rho} \right)\cos\theta +\frac{i}{2}\left(\rho-\frac{1}{\rho} \right)\sin\theta$ (Bernstein ellipses) we will have
\begin{equation}\label{chebyshev_conv}
 \tilde{f}_N(\zeta)=\sum\limits_{k=0}^N a_k T_k(\zeta)~,~|f(\zeta)-\tilde{f}_N(\zeta)|\leq\frac{\mathcal{C}\rho^{-N}}{\rho-1}
\end{equation}
for some constant $\mathcal{C}$, where $\rho$ defines the biggest possible Bernstein ellipse inside which $f(\zeta)$ is analytic, i.e. Čebyšëv polynomial expansions feature exponential convergence \cite{approximation_theory}. What we will use for our computation, Čebyšëv interpolations, is a slightly different expansion. An interpolating polynomial function $f_N(\zeta)$ approximating $f(\zeta)$ is a linear combination of polynomials $p_k(\zeta)$ which, for a given set of $N+1$ points $\zeta_k$ (interpolation points), assumes values $f_N(\zeta_k)=f(\zeta_k)$. If $p_n(\zeta_k)=\delta_{nk}$ the interpolating polynomial function is defined as
\begin{equation}\label{interpolation}
f_N(\zeta)=\displaystyle\sum\limits_{k=0}^N f(\zeta_k) p_k(\zeta)~,~p_n(\zeta)=\displaystyle\frac{\prod_{k\neq n}(\zeta-\zeta_k)}{\prod_{k\neq n}(\zeta_n-\zeta_k)}=\frac{p(\zeta)w_n}{\zeta-\zeta_n}\,,
\end{equation}
where $p_n(\zeta)$ are called Lagrange polynomials, $w_n=1/p'(\zeta_n)$ are their corresponding weights and $p(\zeta)=\displaystyle\prod\limits_{k=0}^N(\zeta-\zeta_k)$ is the ($N+1$)-node polynomial. The interpolation error can be computed exactly, i.e. \cite{approximation_theory}
\begin{equation}\label{interpolation_error}
f(\zeta)-f_N(\zeta)=\frac{f^{(N+1)}(\zeta)p(\zeta)}{(N+1)!}\,,
\end{equation}
and is under control if $\lim\limits_{N\rightarrow\infty}|p(\zeta)|<\infty$ $\forall\zeta$, thus we choose $p(\zeta)=\displaystyle\frac{T_{N+1}(\zeta)}{2^N}$ because Čebyšëv polynomials are bounded. Our interpolation will consequently be a Čebyšëv expansion and have the Čebyšëv nodes as interpolation points \cite{approximation_theory,Spectra_grav_atom}:
\begin{equation}\label{nodes}
\zeta_k=\cos\left(\frac{\pi(2k+1)}{2(N+1)}\right)~,~w_k=(-1)^k\sin\left(\frac{\pi(2k+1)}{2(N+1)}\right)
\end{equation}
The interpolation of the derivatives gives terrible results if we apply the definition in \ref{interpolation}, but this problem can be solved by using the second barycentric form of the Lagrange polynomials \cite{barycentric_lagrange1, barycentric_lagrange2,myths_poynomial_interpol}:
\begin{equation}\label{second_barycentric_form}
p_n(\zeta)=\frac{w_n}{\zeta-\zeta_n}\left(\sum\limits_{k=0}^N\frac{w_k}{\zeta-\zeta_k} \right)^{-1}\,.
\end{equation}
This formula, in fact, can be used for the computation of the differentiation matrices $p_k'(\zeta_n)$ and  $p_k''(\zeta_n)$ when $n\neq k$, while the diagonal elements can be found by imposing the annihilation of the constant functions by the matrices \cite{diff_matrix}:
\begin{equation}\label{diff_matrix}
 p_k'(\zeta_n)=\left\lbrace \begin{matrix}
~~~~~~\frac{w_k/w_n}{\zeta_n-\zeta_k}~~~~~~~~~~~n\neq k~~~~~~~~~~~~~~~~~~\\~\\
-\sum\limits_{m,m\neq n}^N p_m'(\zeta_n)~~~~~n=k~~~~~~~~~~~~~~~~~~
\end{matrix} \right.
\end{equation}
\begin{equation}\label{ddiff_matrix}
p_k''(\zeta_n)=\left\lbrace \begin{matrix}
~~~~2 p_k'(\zeta_n)\left(p_n'(\zeta_n)-\frac{1}{\zeta_n-\zeta_k} \right)~~~~~~n\neq k\\~\\
2 p_k'(\zeta_n)p_n'(\zeta_n)+\sum\limits_{m,m\neq n}^N \frac{2 p_m'(\zeta_n)}{\zeta_n-\zeta_m}~~~~n=k
\end{matrix} \right.
\end{equation}
Čebyšëv interpolation works only if the coordinate is bounded in the interval $(-1,1)$, thus we cannot directly use the radial coordinate $r\in\left[r_+ , \infty\right)$ for our computations. This is the reason why we previously introduced the auxiliary radial coordinate $\zeta$ and expressed all the radial equations without using $r$, but we have not fixed yet any mapping between $\zeta$ and $r$. \textit{Baumann et al.} defined two possible mappings, both of them having the horizon at $\zeta=-1$, $r=\infty$ at $\zeta=1$ and $r_-$ sufficiently far from the dominion of integration \footnote{By mapping $r_-$ far we avoid numerical issues regarding singularities arising from the inner horizon.}; we will use the one mapping better both near and far regions \cite{Spectra_grav_atom}:
\begin{equation}\label{mapping}
\zeta(r)=\frac{r-\sqrt{4 r_+ (r-r_-)+r_+^2}}{r-r_-}~,~r(\zeta)=\frac{4 r_++r_- (\zeta^2-1)}{(\zeta-1)^2}\,.
\end{equation}
By applying the Čebyšëv interpolation we can proceed with the discretization of the radial equations \ref{radial_wave_eqs}, \ref{radial_wave_eqs_s1_s2}, \ref{radial_lorenz_s1}, \ref{radial_lorenz_s2} and \ref{decomposed_second-order_constraint}:
\begin{subequations}
\begin{align}
&\label{discr_radial_wave_eqs}
\begin{aligned}
    \sum\limits_{k=0}^N \left[p_k''(\zeta_n)+C_{1}(r(\zeta_n))p_k'(\zeta_n)\right]B_{l}(\zeta_k)&+C_{2}(r(\zeta_n))B_{l}(\zeta_n)\\
    &+\sum\limits_{l'\geq|m|}^\infty C_{3 , l}^{l'}(r(\zeta_n))B_{l'}(\zeta_n)=0\,,
\end{aligned}\\
&\label{discr_radial_wave_eqs_s1_s2}
\begin{aligned}
    &\sum\limits_{k=0}^N \Bigg\lbrace p_k''(\zeta_n)B^{(S)}_{l,s,j}(\zeta_k)+ \sum\limits_{s'=0}^S\sum\limits_{l'=0}^{\infty}\sum\limits_{\scriptsize\begin{matrix}j'=\text{max}\left\lbrace |j_z|\right.,\\\left.|l'-s'|\right\rbrace    
\end{matrix}}^{l'+s'} B^{(S)}_{l',s',j'}(\zeta_k)\Bigg[C_{1,l,s,j}^{l',s',j'}(r(\zeta_n))p_k'(\zeta_n)+\\
    & \delta_{kn}\left(C_{2,l,s,j}^{l',s',j'}(r(\zeta_n))+\Gamma_{l,s,j}^{l',s',j'}(r(\zeta_n))+D_{l,s,j}^{l',s',j'}(r(\zeta_n))+\Lambda_{l,s,j}^{l',s',j'}(r(\zeta_n))\right)\bigg]\Bigg\rbrace =0\,,
\end{aligned}\\   
&\label{discr_lorenz_constraint_s1}
    \begin{aligned}
        \sum\limits_{k=0}^N\sum\limits_{s'=0}^1\sum\limits_{l'=0}^{\infty}\sum\limits_{\scriptsize\begin{matrix}j'=\text{max}\left\lbrace |j_z|\right.,\\\left.|l'-s'|\right\rbrace    
\end{matrix}}^{l'+s'} \Big[C_{3,j}^{l',s',j'}&(r(\zeta_n))p_k'(\zeta_n)+\left(C_{4,j}^{l',s',j'}(r(\zeta_n))\right.\\
        &\left.\left.+\tilde{D}_{j}^{l',s',j'}(r(\zeta_n))\right)\delta_{kn}\right]B_{l',s',j'}(\zeta_k)=0\,,
    \end{aligned}\\
&\label{discr_lorenz_constraint_s2}
    \begin{aligned}
        \sum\limits_{k=0}^N\sum\limits_{s'=0}^2\sum\limits_{l'=0}^{\infty}\sum\limits_{\scriptsize\begin{matrix}j'=\text{max}\left\lbrace |j_z|\right.,\\\left.|l'-s'|\right\rbrace    
\end{matrix}}^{l'+s'} \Big[C_{3,l,s,j}^{l',s',j'}(r(\zeta_n))&p_k'(\zeta_n)+\left(C_{4,l,s,j}^{l',s',j'}(r(\zeta_n))\right.\\
        &\left.\left.+\tilde{D}_{l,s,j}^{l',s',j'}(r(\zeta_n))\right)\delta_{kn}\right]\tilde B_{l',s',j'}(\zeta_k)=0\,,
    \end{aligned}\\
&\label{second-order_constraint_s2}
    \begin{aligned}
     \sum\limits_{k=0}^N\sum\limits_{s'=0}^2\sum\limits_{l'=0}^{\infty}&\sum\limits_{\scriptsize\begin{matrix}j'=\text{max}\left\lbrace |j_z|\right.,\\\left.|l'-s'|\right\rbrace    
\end{matrix}}^{l'+s'}\left[C^{l',s',j'}_{5,j}(r(\zeta_n))p_k''(\zeta_n)+C^{l',s',j'}_{6,j}(r(\zeta_n))p_k'(\zeta_n)\right.\\
     &+\left.\left(C^{l',s',j'}_{7,j}(r(\zeta_n))+L^{l',s',j'}_{j}(r(\zeta_n))\right) \delta_{kn}\right]\tilde B_{l',s',j'}(\zeta_k)=0\,.
    \end{aligned}
\end{align}
\end{subequations}
For each perturbation problem, in either $B_{l}(\zeta_n)$ or $B_{l,s,j}(\zeta_n)$ or $\tilde{B}_{l,s,j}(\zeta_n)$ variables, the corresponding linear algebraic equations among \ref{discr_radial_wave_eqs}, \ref{discr_radial_wave_eqs_s1_s2}, \ref{discr_lorenz_constraint_s1}, \ref{discr_lorenz_constraint_s2} and \ref{second-order_constraint_s2} can be collectively considered as a matrix equation
\begin{equation}\label{matrix_eq}
\mathbfcal{M}(\nu)\underline{B}=\underline 0
\end{equation}
representing a non-linear eigenvalue problem in the $\nu$ frequency parameter, where $\underline{B}$ is the super-vector built from the variables $B_{l}(\zeta_n)$ or $B_{l,s,j}(\zeta_n)$ or $\tilde B_{l,s,j}(\zeta_n)$. In theory we could find the frequency by just imposing the matrix to be singular, i.e. $\det \mathbfcal{M}(\nu)=0$, but the high number of components of the matrix does not allow us to pursue such path. What we need is a technique called non-linear inverse iteration.

\section{Non-linear inverse iteration}\label{section:non-linear_inverse_iteration}

The non-linear inverse iteration algorithm is based on Newton's method of the tangents \cite{non-linear-eigenvalue}, which we briefly describe here. Let us consider the one-dimensional non-linear eigenvalue problem, i.e. finding the roots of a function:
\[f(\lambda)=0~,~f\in C^1(\mathbb{C})~,~\lambda\in\mathbb{C}\,.\]
If we have some good guess $\lambda_0$ for a solution, we can linearize the equation, i.e.
\[f(\lambda_0)+f'(\lambda_0)(\lambda-\lambda_0)=O((\lambda-\lambda_0)^2)\,,\]
and invert it in order to find a better approximation of the solution:
\[\lambda=\lambda_0-\frac{f(\lambda_0)}{f'(\lambda_0)}+O((\lambda-\lambda_0)^2)\,.\]
Newton's method consists in iterating this procedure. From the guess $\lambda_0$ we get an approximation $\lambda_1$, then we use $\lambda_1$ as the linearization point for finding an approximation $\lambda_2$ and so on until convergence on the exact result:
\begin{theo}
\textbf{Newton's method of the tangents}\\~\\
Given the non-linear equation
\[f(\lambda)=0~,~f\in C^1(\mathbb{C})~,~\lambda\in\mathbb{C}\,,\]
choose some guess solution $\lambda_0$, iterate until convergence the following sequence for $k=\lbrace 0,1,2,... \rbrace$:
\[\lambda_{k+1}=\lambda_k-\frac{f(\lambda_k)}{f'(\lambda_k)}\,.\]
\end{theo}
\begin{figure}[H]
\centering
\includegraphics[width=1\textwidth]{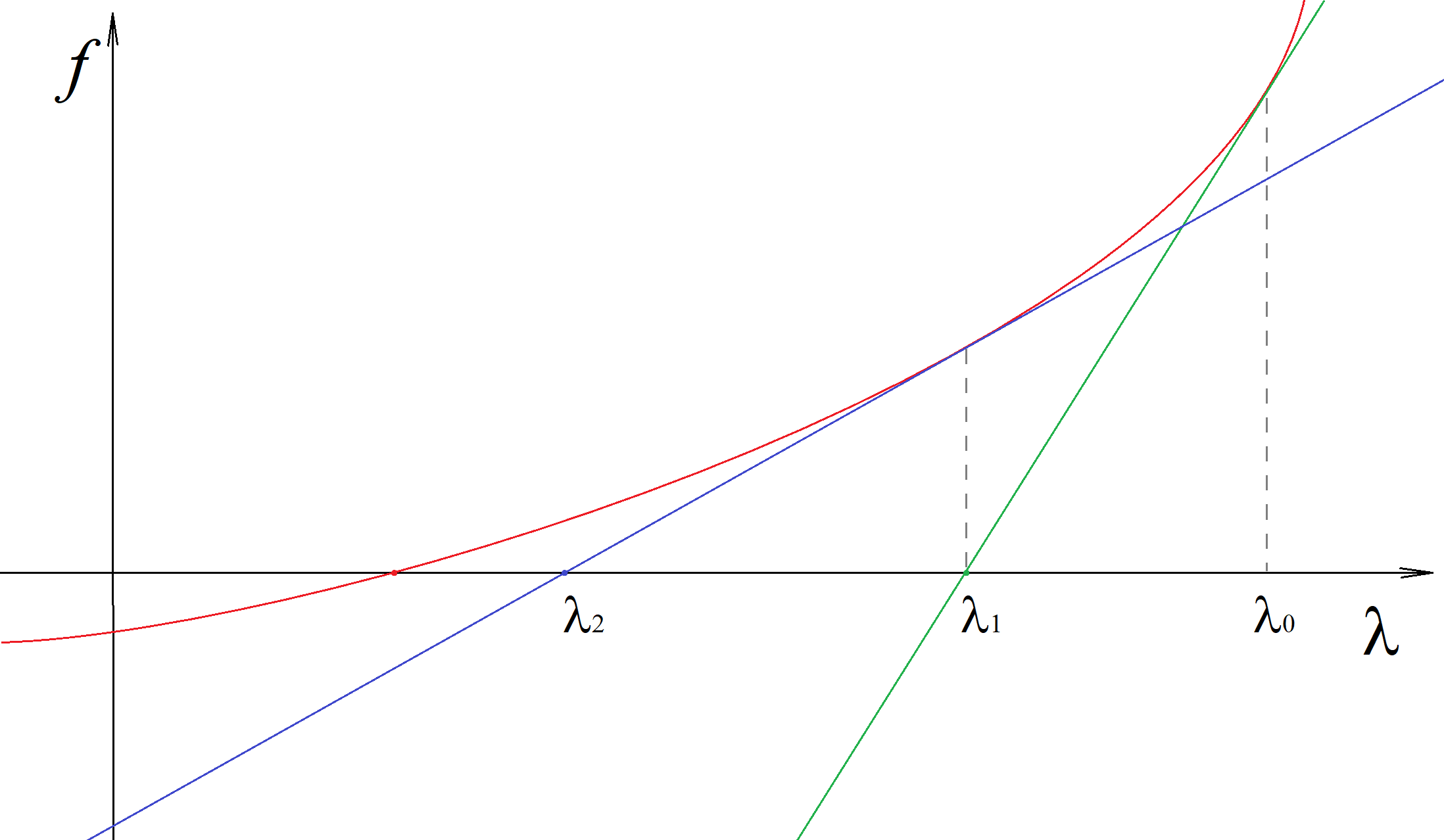}\\
\caption{Newton's method of the tangents: the linearization of the function $f(\lambda)$ around some approximation $\lambda=\lambda_0$ of its root gives a result $\lambda=\lambda_1$ which is a better approximation than the starting guess if we are close enough to the exact result. By iterating the procedure we converge on the root.}
\end{figure}

Let us now extend this method to the D-dimensional case, i.e. a non-linear eigenvalue problem involving a matrix $\bold{A}(\lambda)$ and vectors $\underline{x}$:
\[\bold{A}(\lambda) \underline{x}=\underline{0}~,~\bold{A}\in\mathbb{M}(m,n)~,~A_{ij}\in C^1(\mathbb{C})~,~\lambda\in\mathbb{C}~,~\underline{x}\in\mathbb{C}^n\,.\]
In this case we also need to impose some normalization to  $\underline{x}$, i.e.
\[\displaystyle\left\langle \underline{y},\displaystyle\underline{x}\right\rangle=\underline{y}^{*T}\underline{x}=1\]
for some choice $\underline{y}\in\mathbb{C}^n$,  because the matrix $\bold{A}(\lambda)$ has to be singular in order to have a non-null kernel. We proceed with the linearization around some guess $(\lambda_0,\underline{x}_0)$,
\begin{small}
\[\displaystyle\begin{bmatrix}\textbf{A}(\lambda_0)\underline{x}_0\\\left\langle \underline{y}, \underline{x}_0\right\rangle-1\end{bmatrix}+\textbf{J}(\underline{x}_0,\lambda_0)\begin{bmatrix}\underline{x}-\underline{x}_0\\\lambda-\lambda_0\end{bmatrix}=O\left(\left\|\begin{bmatrix}\underline{x}-\underline{x}_0\\\lambda-\lambda_0\end{bmatrix}\right\|^2\right)~,~\textbf{J}(\underline{x},\lambda)=\begin{bmatrix}\textbf{A}(\lambda)&\textbf{A}'(\lambda)\underline{x}\\\underline{y}^{*T}&0 \end{bmatrix}\,,\]
\end{small}
thus the D-dimensional extension will be:
\begin{theo}
\textbf{Newton's method (D-dimensional)}\\~\\
Given the non-linear eigenvalue problem 
\[\bold{A}(\lambda) \underline{x}=\underline{0}~,~\bold{A}\in\mathbb{M}(m,n)~,~A_{ij}\in C^1(\mathbb{C})~,~\lambda\in\mathbb{C}~,~\underline{x}\in\mathbb{C}^n\,,\]
choose some guess solution ($\underline{x}_0$,$\lambda_0$) and normalization $\underline{y}$, iterate the following sequence for $k=\lbrace 0,1,2,... \rbrace$ until convergence:
\[\begin{bmatrix}\underline{x}_{k+1}\\\lambda_{k+1}\end{bmatrix}=\begin{bmatrix}\underline{x}_k\\\lambda_k\end{bmatrix}-\textbf{J}^{-1}(\underline{x}_k,\lambda_k)\begin{bmatrix}\textbf{A}(\lambda_k)\underline{x}_k\\\left\langle \underline{y}, \underline{x}_k\right\rangle-1\end{bmatrix}\,.\]
\end{theo}
This method can be reorganized as follows \cite{non-linear-eigenvalue}:
\begin{theo}
\textbf{Non-linear inverse iteration} (Unger, 1950)\\~\\
Given the non-linear eigenvalue problem 
\[\bold{A}(\lambda) \underline{x}=\underline{0}~,~\bold{A}\in\mathbb{M}(m,n)~,~A_{ij}\in C^1(\mathbb{C})~,~\lambda\in\mathbb{C}~,~\underline{x}\in\mathbb{C}^n\,,\]
choose some guess solution ($\underline{x}_0$,$\lambda_0$) $| \left\|\underline{x}_0\right\|=1$ and normalization $\underline{y}$, iterate the following steps for $k=\lbrace 0,1,2,... \rbrace$ until convergence:
\begin{enumerate}
\item{solve linear problem $\displaystyle\textbf{A}(\lambda_k)\underline{\tilde{x}}_{k+1}=\textbf{A}'(\lambda_k)\underline{x}_{k}$ for $\underline{\tilde{x}}_{k+1}$};
\item{set $\displaystyle\lambda_{k+1}=\lambda_{k}-\frac{\left\langle\underline{y},\underline{x}_k\right\rangle}{\left\langle\underline{y},\underline{\tilde{x}}_{k+1}\right\rangle}$ and normalize $\displaystyle\underline{x}_{k+1}=\frac{\underline{\tilde{x}}_{k+1}}{\left\|\underline{\tilde{x}}_{k+1}\right\|}$}.\\~\\
\end{enumerate}
\end{theo}
In the non-linear inverse iteration algorithm most of the error arises in the norm $\left\|\underline{x}\right\|$ of the solution, while keeping the quadratic convergence of Newton's method: this is a great property, as we are interested only in the direction of $\underline{x}$. With this technique we can solve our non-linear eigenvalue problem $\mathbfcal{M}(\nu)\underline{B}=\underline 0$.

\section{Scalar superradiant instabilities}\label{section:spin-0_SR}

In this section we will show the numerical results for the scalar case we computed by applying the numerical technique described in the previous sections. The QBSs of the scalar field can be computed analytically if we apply either small mass $\mu M \ll 1$\footnote{When we are not comparing scalar, vector and tensor bosons we drop the subscript specifying the boson type. Thus in this specific section when we write $\mu$ we mean $\mu_S$.} \cite{Detweiler_scalar_instability, Spectra_grav_atom, Brito_SR} (i.e. the so-called "hydrogenic" approximation) or WKB $\mu M \gg 1$ \cite{WKB_ZOUROS_scalar_kerr,Brito_SR} approximations, therefore these results can be potentially used as guess solutions for the exact numerical computation. Given that we are interested in superradiantly unstable states and, as we shall see in a few lines, given that such configurations appear when $\mu M\lesssim 1$, for our results we used the small mass approximation as a starting guess.

\begin{figure}[H]
\centering
\includegraphics[width=0.67\textwidth]{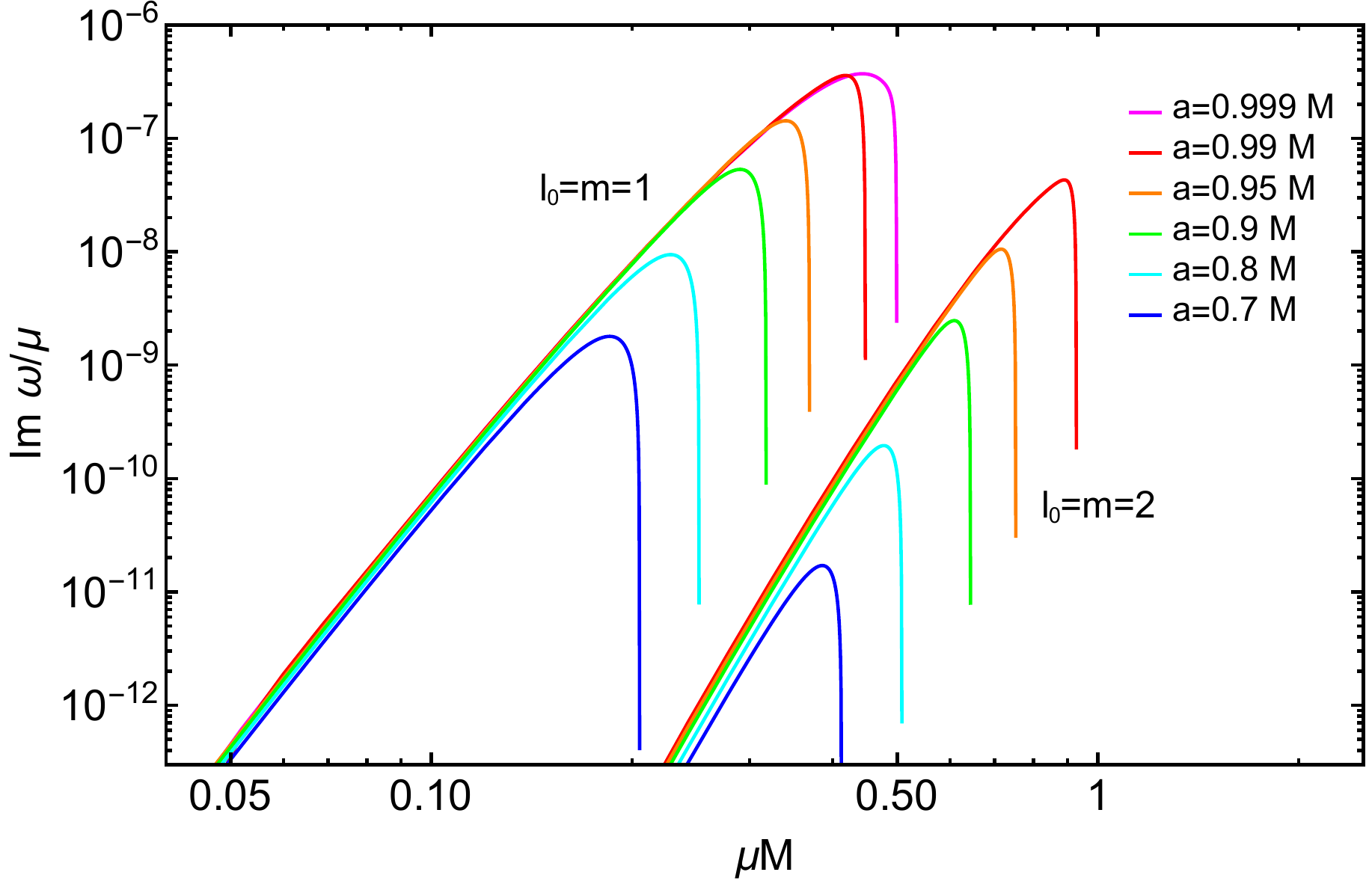}\\
\caption{Imaginary part of the frequency: dimensionless plots of unstable states, for different values of the BH spin, of the imaginary part of the frequency vs boson mass for the most unstable states $l=1$ and $l=2$ ($n=0$). The lower the value of $l$ is, the higher the instability will be. In particular, for each value of $l$ the most unstable state is the one having $l=m$. These results show that the superradiant instabilities are relevant for $\mu M\lesssim 1$ and confirm the validity of the superradiant condition.}
\label{plot_instability_peaks}
\end{figure}

\begin{figure}[H]
\centering
\includegraphics[width=0.7\textwidth]{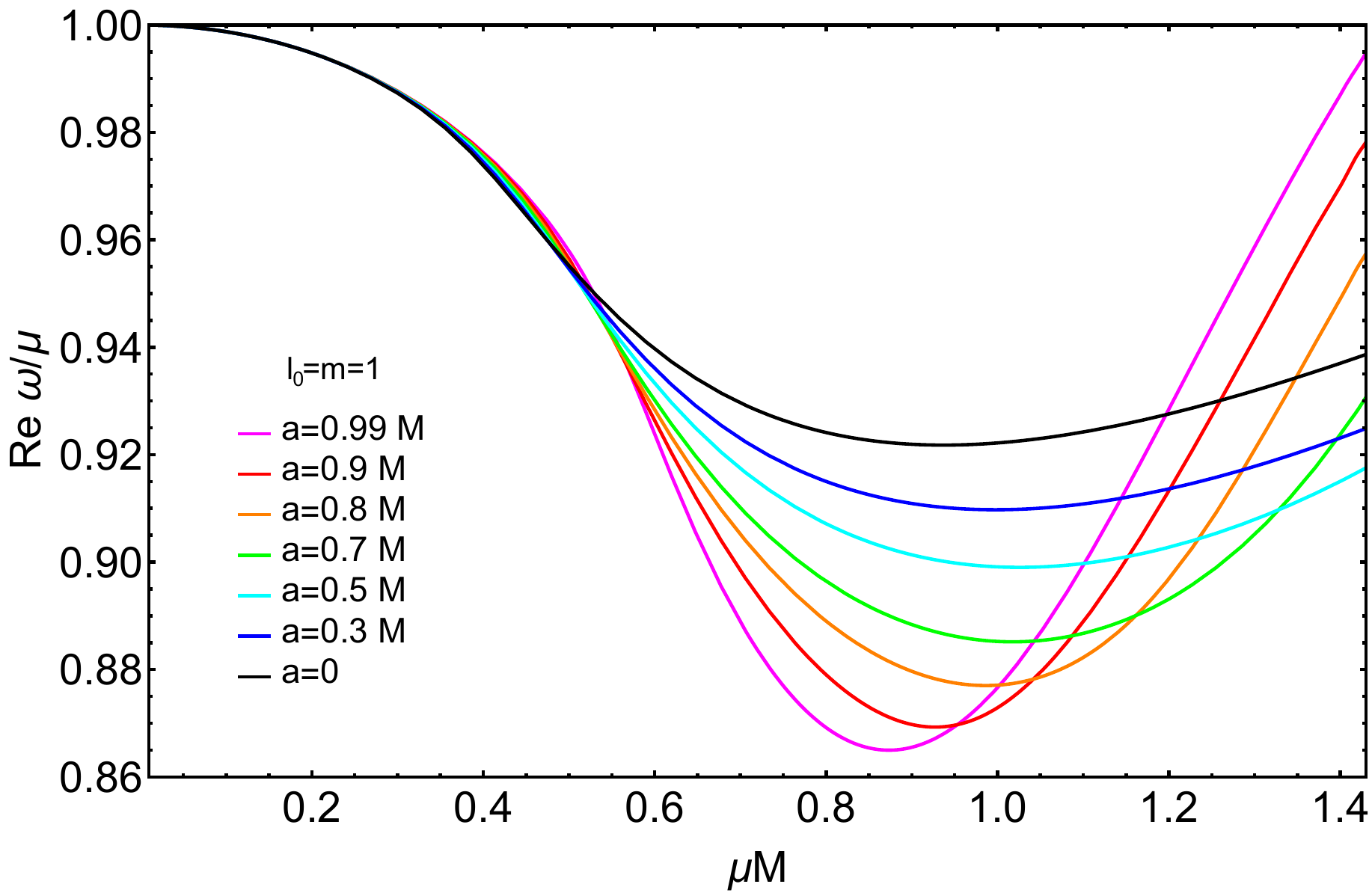}\\
\includegraphics[width=0.7\textwidth]{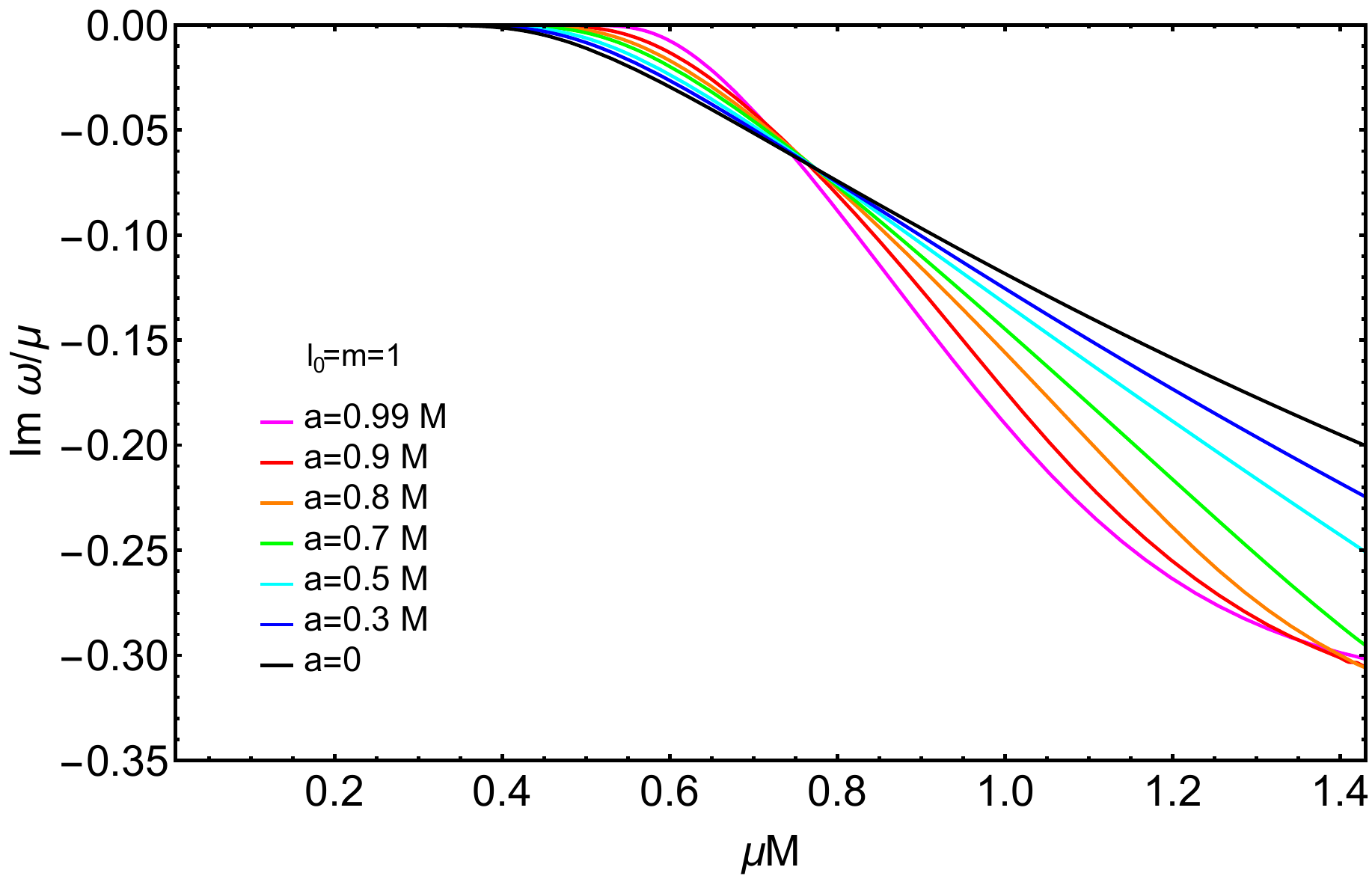}\\
\caption{Frequency of the $l=m=1$ ($n=0$) state vs boson mass for different values of the BH spin. We notice that around $\mu M\simeq 0.4-0.5$ the plots of $\operatorname{Re}\omega$ experience an inflection point: that is exactly where the hydrogenic approximation breaks down completely.}
\label{l=m=1_plots}
\end{figure}

In the small mass approximation $\omega\sim\mu$, consequently the angular and radial parts of the Klein-Gordon equation \ref{klein-gordon_manipulation} can be separated by using spherical harmonics. The radial solutions in the frequency domain and the real part of the associated frequencies have the same expressions of the quantum states of the non-relativistic hydrogen atom, though the solution differ in the presence of a non-zero imaginary part of the frequency, and $\mu M$ works as a "gravitational fine structure" constant\footnote{This is why this approximation is called "hydrogenic".}, \cite{Detweiler_scalar_instability, Spectra_grav_atom, Brito_SR}:
\begin{equation}
    \Phi\approx e^{-i \omega_{nlm}t}R_{nl}(r)Y_{l,m}(\theta,\phi)~,~n\in\mathbb{N}
\end{equation}
\begin{equation}\label{scalar_hydrogenic_radial_function}
    R_{nl}(r)=\tilde{r}^l e^{-\tilde{r}/2}L^{(2l+1)}_{n}\left(\tilde{r} \right)~~,~~~~\tilde{r}=\frac{2 r M \mu^2}{n+l+1}\,,
\end{equation}
\begin{equation}\label{scalar_hydrogenic_freq}
    \omega_{nlm}=\mu \left[1-\frac{\mu^2 M^2}{2 (n+l+1)^2} \right]+i \Gamma_{nlm}+O(\mu^3 M^3)\,,
\end{equation}
\begin{equation}
\begin{aligned}
   \Gamma_{nlm}=&2 r_+ \mu (m\Omega_H-\operatorname{Re}\omega_{nlm})(\mu M)^{4 l+4}\\
   &\times\frac{2^{4l+2}(2l+n+1)!}{(n+l+1)^{2l+4}n!}\left[\frac{l!}{(2l)!(2l+1)!}\right]^2\\
   &\times\prod\limits_{k=1}^l\left[k^2 \left(1-\frac{a^2}{M^2} \right)+\left(\frac{m a}{M}-2 r_+\mu\right) \right]\,,
\end{aligned}
\end{equation}
where $L^{(2l+1)}_{n}(\tilde r)$ are associated Laguerre polynomials. The sign of $\Gamma_{nlm}$, i.e. the imaginary part of the frequency, is determined by the factor $(m\Omega_H-\operatorname{Re}\omega_{nlm})$, which gives unstable states when the superradiant condition is satisfied. Given that the exact solution at the horizon is $\Phi\sim(r-r_+)^{iP_+}$, the expression of the radial function \ref{scalar_hydrogenic_radial_function} shows explicitly that the hydrogenic approximation neglects what happens in the immediate proximity of the BH event horizon, i.e. it is a non-relativistic approximation. The analytical computation through matched asymptotics of the following orders in the $\mu M\ll1$ expansion give the equivalent of hyperfine corrections in the hydrogen atoms, thus opening the possibility to hyperfine transitions in the gravitational atom \cite{Spectra_grav_atom}.
In Figure \ref{plot_instability_peaks}, we give the results we computed for the imaginary part of the frequency of the most unstable modes as a function of $\mu M$, which are in excellent agreement with the known literature\cite{dolan_scalar_freq_domain,Spectra_grav_atom,Brito_SR}.
In Figure \ref{l=m=1_plots}, instead, we show $\operatorname{Re}\omega$ and $\operatorname{Im}\omega$ for $l=m=1$ ($n=0$) for different values of the BH spin, which coincide with the ones computed in \cite{dolan_scalar_freq_domain}. For $\mu M\lesssim0.4$, the curves of the real part are degenerate in the BH spin and parabolic, in accordance with the hydrogenic approximation.
In Fig. \ref{l=m=1_plots} we notice that for $\mu M\lesssim 0.4$ the curves of $\operatorname{Re}\omega$ are approximately parabolic and degenerate in the BH spin, i.e. the exact values are just corrections of the hydrogenic results, while for higher values they deviate completely from the parabolic behaviour. Thus for $\mu M\gtrsim 0.4$ in general we could not use the hydrogenic guess for the computation of valid results and had to use an iterative guess instead: every result for a given boson mass value has been used for the computation of a result for a slightly higher mass.

\begin{figure}[H]
\centering
\includegraphics[width=0.7\textwidth]{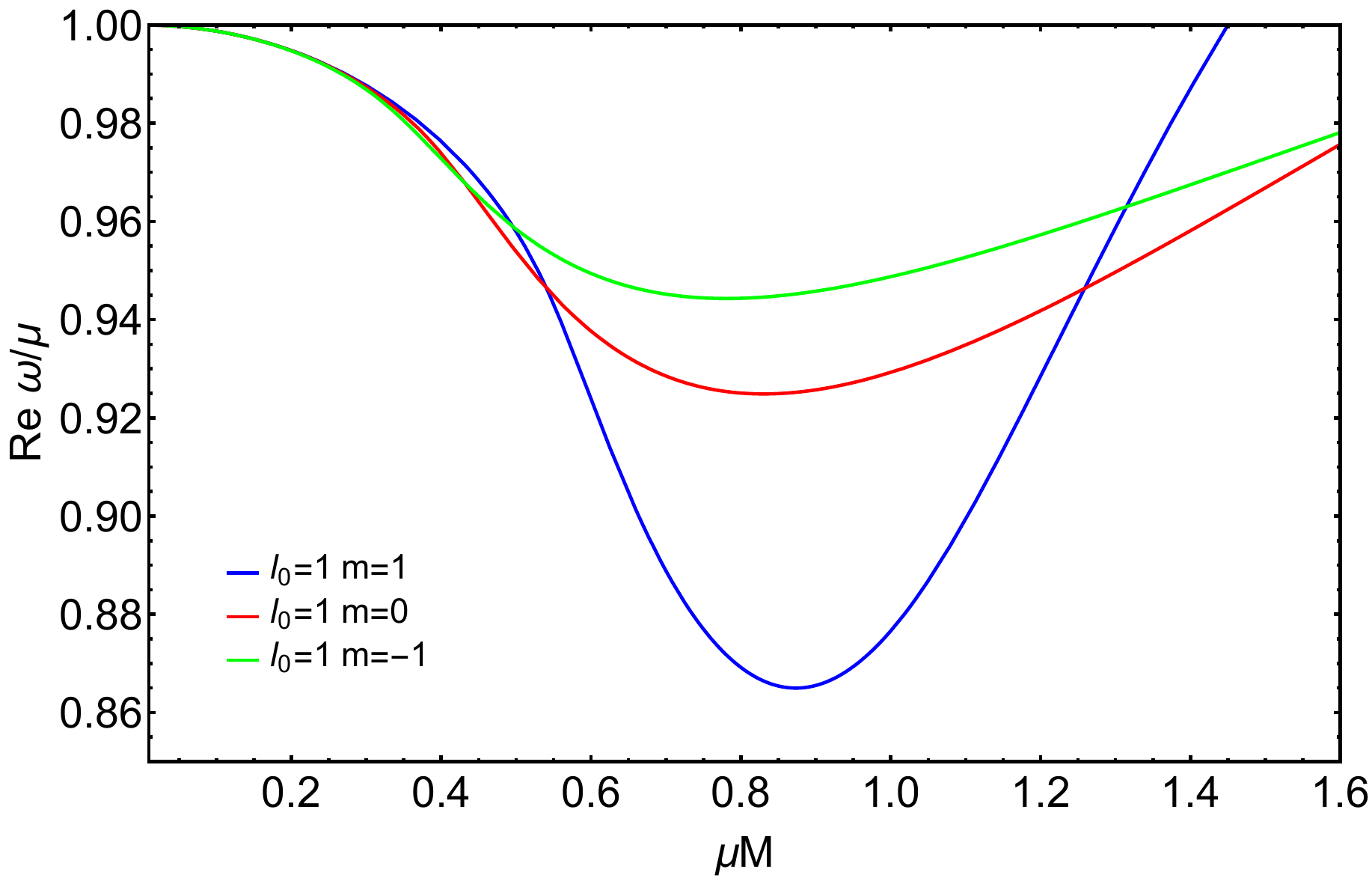}\\
\includegraphics[width=0.7\textwidth]{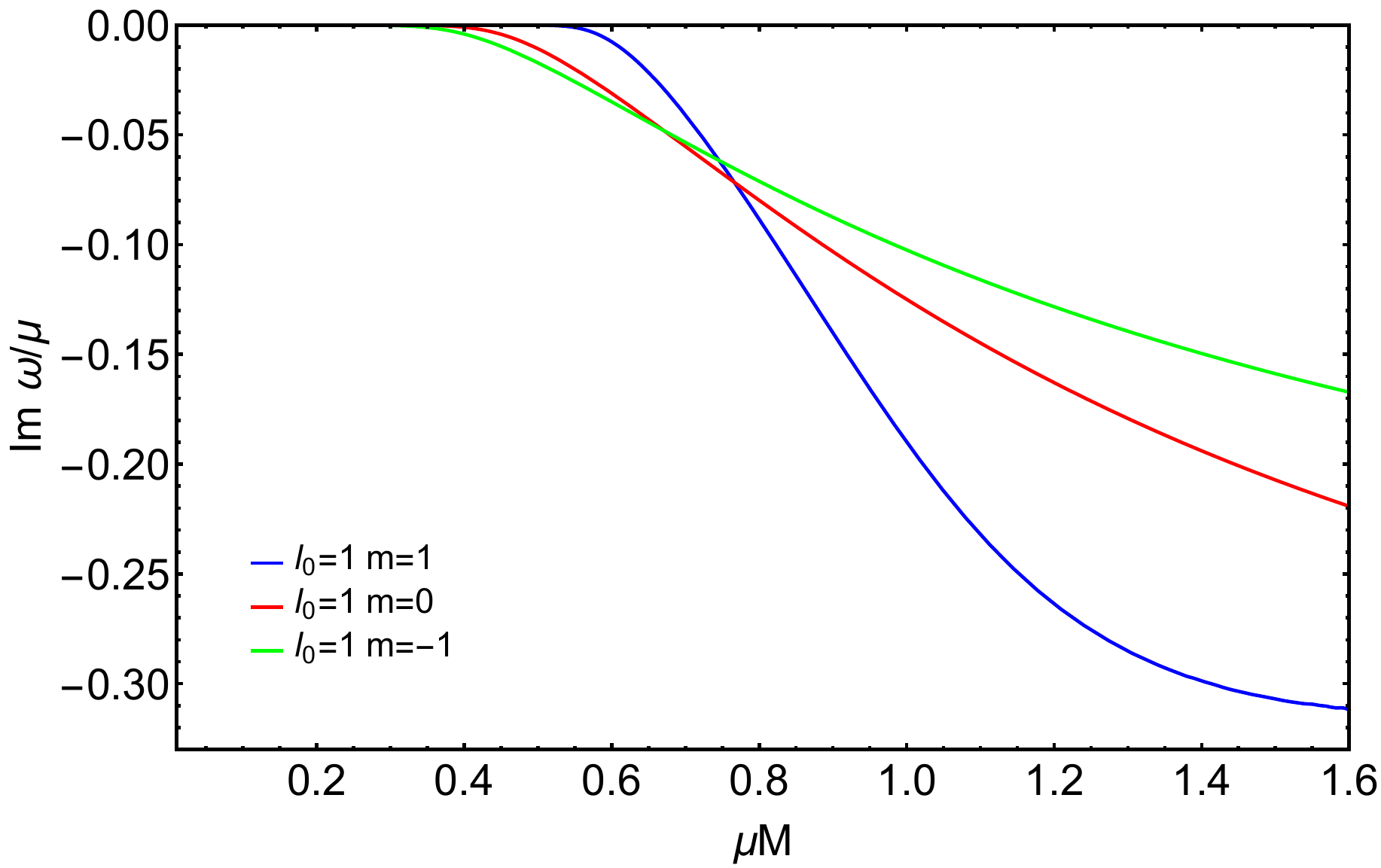}\\
\caption{Frequencies of the $l=1$ ($n=0$) states vs boson mass for different values of the azimuthal angular momentum $m$. Also in this plot we notice the hydrogenic approximation breaking down around $\mu M\simeq 0.4-0.5$: for $\mu M\lesssim 0.4$ the states have real frequency approximately degenerate in $m$, while for $\mu M \gtrsim 0.4$ they deviate completely from the parabolic hydrogenic behaviour. Also in this case we had to use an iterative guess for the higher boson masses.}
\label{l=1_a=0.99_plots}
\end{figure}

Something similar happens also if we consider different values of the azimuthal angular momentum $m$. This is clearly shown by Figure \ref{l=1_a=0.99_plots}, where we compare states ($n=0$, $l=1$) having same BH spin $a=0.99 M$ but having different $m$. Again, for $\mu M\lesssim 0.4$ the curves of the real part are parabolic in accordance with the hydrogenic approximation, i.e. in this case they are degenerate in the azimuthal angular momentum. Also this plots are in complete agreement with \cite{dolan_scalar_freq_domain}.

\section{Vector superradiant instabilities}\label{section:spin-1_SR}

\begin{figure}[H]
\centering
\includegraphics[width=1\textwidth]{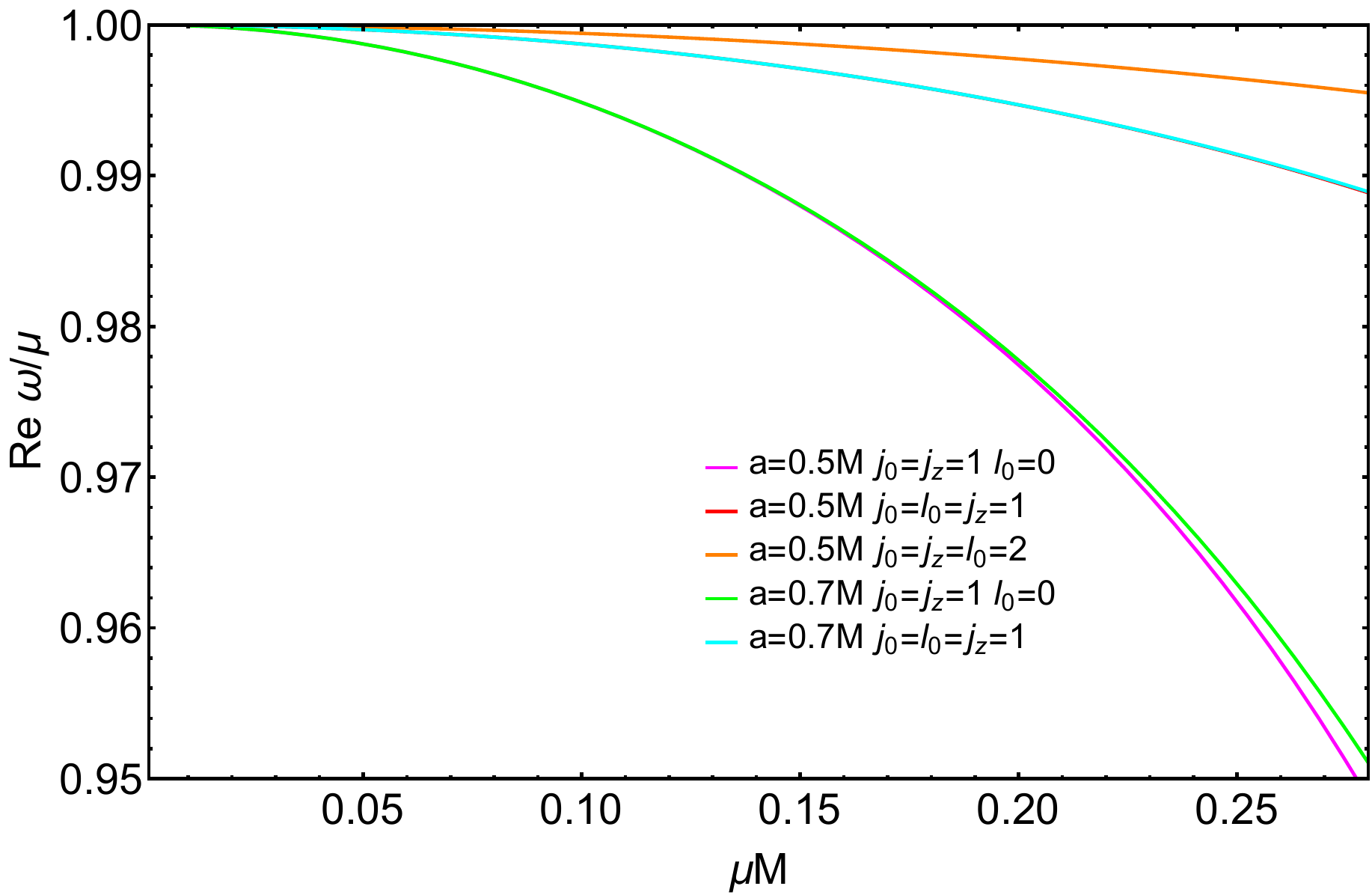}\\
\caption{Real part of the frequency vs boson mass for different configurations ($n=0$). We notice how configurations differing only in $a/M$ almost overlap each other, though for higher values of $\mu M$ the difference increases. As a matter of fact, in the small mass approximation the real part depends only in $\mu$, $l$ and $n$, thus all the other dependencies arise as relativistic corrections to the hydrogenic result. For higher values of $\mu M$ we expect a total departure from the parabolic hydrogenic behaviour after some inflection point, like in the scalar case.}
\label{Re_spin-1_plots}
\end{figure}

The small mass approximation ($\mu M\ll 1$, $\omega\sim\mu$) is the only known analytical result for the massive vector perturbations of the Kerr metric\cite{Ultralight_vector_SR_signatures, Spectra_grav_atom}, thus also for the results of spin-1 superradiant instabilities we used that framework as a starting guess. This result differs from the scalar one just in the use of vector spherical harmonics for the separation of the angular part and in the presence of a temporal component $A_0$, while the radial part $R_{nl}$ and $\operatorname{Re}\omega$ have exactly the same expressions of the scalar case, i.e. the ones in \ref{scalar_hydrogenic_radial_function} and \ref{scalar_hydrogenic_freq} respectively\cite{Ultralight_vector_SR_signatures, Spectra_grav_atom}:
\begin{subequations}
    \begin{align}
    &A_0\approx e^{-i\omega_{n l j j_z} t}R^{(0)}_{n j j_z}(r)Y_{j,j_z}(\theta,\phi)\,,\\
    &A_i\approx e^{-i\omega_{n l j j_z} t} R_{nl}(r)Y^{l, j j_z}_i (\theta,\phi)\,,
    \end{align}
\end{subequations}
where $Y^{l, j j_z}_i (\theta,\phi)$ are the pure-orbital vector spherical harmonics as defined in \cite{Thorne_multipole_expansions,Maggiore_GWs:vol_1}, $l$ and $j$ are the orbital and total angular momentum parameters respectively, $j_z$ is the total azimuthal angular momentum. The temporal radial function $R^{(0)}_{n j j_z}(r)$ is fixed by the space components of the field through the Lorenz constraint $\bar\nabla^\mu A_\mu=0$, i.e. $\partial_t A_t\simeq \partial_r A_r+\frac{1}{r^2} \partial_\theta A_\theta+\frac{1}{r^2 \sin^2\theta}\partial_\phi A_\phi$ in this approximation. The pure-orbital vector spherical harmonics, instead, are defined as angular vector solutions to the secular equations of the laplacian operator in flat spacetime $\nabla_F^2$, i.e. $-r^2 \nabla_F^2 Y^{l, j j_z}_i=l(l+1)Y^{l, j j_z}_i$:
\begin{equation}\label{Thorne_vector_harmonics}
    Y^{l, j j_z}_i=\sum\limits_{m=-l}^l \sum\limits_{s_z=-1}^1 \left\langle  l,m,s,s_z\right. \left|j, j_z \right\rangle Y_{l,m} \xi^{(s_z)}_{i}\,,
\end{equation}
where $\xi^{(s_z)}_{i}$ are 3-vectors built from the orthonormal coordinate cartesian basis $\left\lbrace \vec{e}_{(x)}, \vec{e}_{(y)}, \vec{e}_{(z)}\right\rbrace$\footnote{In this approximation we can raise and lower tensor indices with the flat metric because of the non-relativistic nature of the framework.}:
\begin{subequations}\label{ortho_cartesian_basis}
\begin{align}
&\xi_i^{(\pm 1)}=\mp\frac{1}{\sqrt{2}}\left(e_i^{(x)}\pm i e^{(y)}_i \right)\,,\\
&\xi_i^{(0)}=e^{(z)}_i\,.
\end{align}
\end{subequations}
The expression \ref{Thorne_vector_harmonics} shows striking similarities with the vector harmonics \ref{vector_spherical} and \ref{dual_vector_spherical} defined by \textit{Baumann et al.}\cite{Spectra_grav_atom}. As a matter of fact $\xi_i^{(s_z)}$ describes spin-1 eigenstates, thus \ref{Thorne_vector_harmonics} coincides with \ref{vector_spherical} and \ref{dual_vector_spherical} if we set $F_r(r)=1$ and $F_\theta(r)=\frac{1}{r}$. We can thus consider the vector harmonics used for the small boson mass limit as the non-relativistic (or weak-field) equivalent of the ones defined by \textit{Baumann et al.}. The expressions defining the frequency of spin-1 QBSs are the following\cite{Ultralight_vector_SR_signatures, Spectra_grav_atom}:
\begin{subequations}
\begin{align}
    &\omega_{n l j j_z}=\mu\left[1-\frac{\mu^2 M^2}{(n+l+1)^2}\right]+i \Gamma_{n l j j_z}+O(\mu^3 M^3)\,,\\
    &\begin{aligned}
    \Gamma_{n l j j_z}=&2 r_+ \mu (j_z\Omega_H-\operatorname{Re}\omega_{nljj_z})(\mu M)^{2 l+2 j+4}\\
   &\times\frac{2^{2l+2j+2}(2l+n+1)!}{(n+l+1)^{2l+4}n!}\left[\frac{l!}{(l+j)!(l+j+1)!}\right]^2\\
   &\times\left[1+\frac{2(1+l-j)(1-l+j)}{l+j}\right] \,.
   \end{aligned}
\end{align}    
\end{subequations}
The spin-1 case can therefore be considered as a simple generalization of the scalar solution. In order to use this result as a starting guess, we need to transform the small-mass solution so that it is written in the vector spherical harmonic basis defined by \textit{Baumann et al.} in \cite{Spectra_grav_atom} that we showed in Section \ref{section:vector_tensor_harmonics}. Thus, what follows is the discretized expression for a starting guess having orbital angular momentum $l_0$, total angular momentum $j_0$ and total azimuthal angular momentum $j_z$:
\begin{equation}
B_{l,s,j}(\zeta_k)=\left[\int d\Omega\frac{R_{nl_0}}{F(r)}\sum_i Y^{l_0,j_0 jz}_{i}\left(Y^{(1)i}_{l,s,j,j_z} \right)^*+\delta_{s,0}\frac{R^{(0)}_{n j_0 j_z}(r)}{F(r)}\right]_{r=r(\zeta_k)}.
\end{equation}
Here follow the numerical results we computed, the first ones which highly rely on the parametrization of the overlap integrals we introduced in \ref{section:overlap_parametrization}. 
\begin{figure}[H]
\centering
\includegraphics[width=1\textwidth]{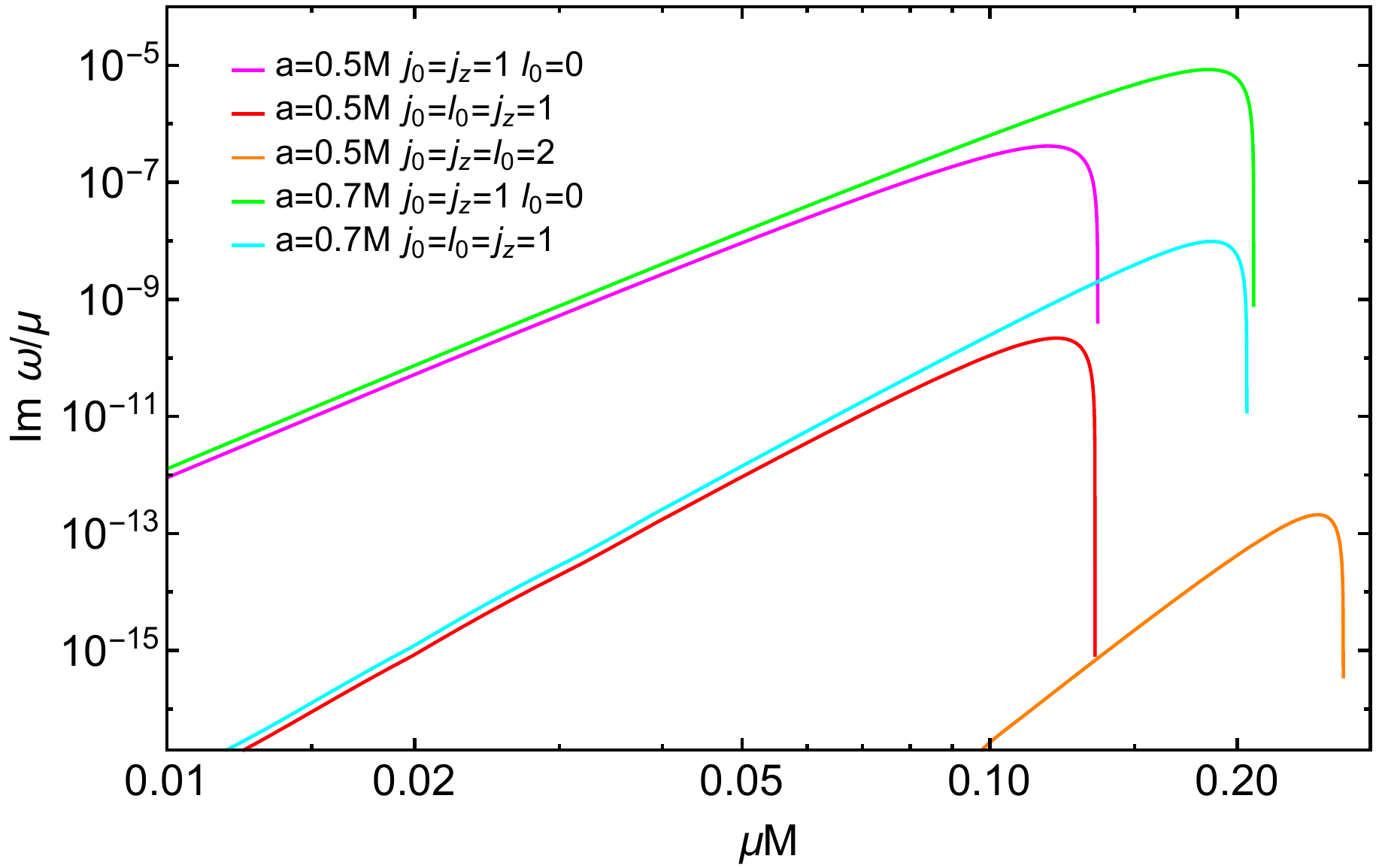}\\
\caption{Instability peaks vs boson mass for different configurations ($n=0$). In this plots we have the results for $a/M=0.5$ and $a/M=0.7$ for the corresponding most unstable states (purple and green respectively), i.e. $l_0=0$ $j_0=j_z=1$. In addition, there are some other examples plotted which show the general rules we find in literature \cite{Dolan_spin-1_instability,Spectra_grav_atom}. Again, like in the scalar case, the most unstable states are the ones having $j_0=j_z$, while the lower $l_0$ is the higher the instability will be. Also in this case, higher values for $j_z$ and/or $a/M$ give wider instability windows due to the superradiant condition $0\leq \omega_R\leq j_z \Omega_H$.}
\label{Im_spin-1_plots}
\end{figure}
The results shown in Fig. \ref{Im_spin-1_plots} and \ref{Re_spin-1_plots} are in agreement with what is recorded in literature about the topic \cite{Dolan_spin-1_instability, Spectra_grav_atom}, thus can be considered as a proof of principle of how the parametrization of the overlap integrals introduced in \ref{section:overlap_parametrization} is effective. By increasing the value of $\mu M$ we face higher difficulties in finding valid results, given that a higher precision is required. The same applies for angular momentum parameters, therefore, given that we are dealing with a spinful field, the spin-1 case is more challenging than the scalar one when we look for higher multipole modes.

\section{Tensor superradiant instabilities}\label{section:spin-2_SR}

The computation of massive spin-2 QBSs in BH spacetimes is way more involved than the spin-0 and spin-1 cases. The unknown functions to be found are ten, the field equations are 15 and the couplings arising from the geometry of Kerr spacetime are extremely involved.
The only known solutions for massive spin-2 BH perturbations have been computed in the small BH spin\cite{Brito_spin-2_SR_slow_rot} and hydrogenic \cite{Brito_hydrogenic_spin-2_SR} approximations, where the latter is the only analytical result. These results can be used as guess solutions for the computation of the exact numerical results of the linearized spin-2 field equations. Both tensor bases used in these works, i.e. Regge-Wheeler's\cite{Regge-Wheeler} and the so-called "pure orbital" one defined in\cite{Thorne_multipole_expansions}\cite{Maggiore_GWs:vol_1}, differ from the basis we use in our numerical technique: we consequently need to compute some change of basis for each case, as we did in the previous section. 

In the tensor case, the small boson mass approximation is able to capture only states whose radial functions and frequency spectrum show a hydrogenic behaviour, i.e. $s=2$ pure-space states, thus missing monopolar and dipolar solutions. For the computation of the latters we therefore need to use also the non-spinning / small spin solutions as starting guess for small spin results and iterate the procedure up to higher spins.

\subsection{Hydrogenic guess solutions}

In the $M \mu\ll 1$ limit the pure-space components of $h_{\mu\nu}$ dominate over the temporal ones, i.e. the field is well approximed by linear combinations of $s=2$ states\footnote{Among pure-space states there is also $s=0$, but the trace-less constraint imposes it to be negligible in this regime.}. In \cite{Brito_hydrogenic_spin-2_SR} \textit{Brito et al.} found the following analytical expressions for the dominant contribution to the solution, which are the tensor generalization of the hydrogenic spin-1 result:
\begin{equation}\label{hydrogenic_guess}
h_{\alpha\beta}\approx\left[\begin{matrix}0&0\\0&\sum\limits_{n,l,j,j_z} e^{-i \omega_{nlj j_z} t} R_{nl}(r)Y^{l,j jz}_{ab} \end{matrix} \right]~,~~~a,b\in\lbrace 1,2,3 \rbrace\,,
\end{equation}
\begin{equation}
    \omega_{nlj j_z}=\mu \left[1-\frac{\mu^2 M^2}{2 (n+l+1)^2} \right]+i \Gamma_{nlj j_z}\,,
\end{equation}
\begin{equation}
    \Gamma_{nljj_z}\propto -(M\mu)^{2(l+j)+5}(\operatorname{Re}\omega_{nljj_z}-j_z \Omega_H)\,,
\end{equation}
where $R_{nl}(r)$ have the same expression of the scalar case \ref{scalar_hydrogenic_radial_function}, $\Omega_H$ is the angular velocity of the event horizon, $n$ is the hypertone number. The angular functions $Y^{l,j jz}_{ab}$ are the pure-orbital tensor harmonics \cite{Thorne_multipole_expansions,Maggiore_GWs:vol_1}, which satisfy the secular equation of the laplacian $\nabla_F^2$ of flat spacetime, i.e.
\begin{equation}\label{orbital_secular_eq}
-r^2\nabla_F^2 Y^{l,j jz}_{ab}=l(l+1) Y^{l,j jz}_{ab}\,,
\end{equation}
and feature the following expressions:
\begin{equation}\label{orbital_harmonics}
Y^{l,j jz}_{ab}=\sum\limits_{m=-l}^l \sum\limits_{s_z=-2}^2 \left\langle  l,m,s,s_z\right. \left|j, j_z \right\rangle Y_{l,m} t^{(s_z)}_{ab}
\end{equation}
\begin{equation}\label{spin2_eigenstate}
t^{(s_z)}_{ab}=\sum\limits_{s_z',s_z''=-1}^1\left\langle 1, s_z' , 1, s_z''\right. \left|2 , s_z \right\rangle \xi_a^{(s_z')}\xi_b^{(s_z'')}
\end{equation}
where $\xi_a^{(s_z)}$ are the 3-vector orthonormal basis defined in \ref{ortho_cartesian_basis}. The tensor basis here shown is the non-relativistic equivalent of the tensor basis we defined in Section \ref{section:vector_tensor_harmonics}: the two tensor harmonic bases coincide if we set $F_r(r)=1$ and $F_\theta(r)=\frac{1}{r}$, like in the vector case.

For a given hydrogenic $(l_0,j_0,j_z)$ mode in the non-relativistic basis, we have the following expression for its corresponding $\underline{B}$ super-vector\footnote{Please note that the transformation is not diagonal: in fact $\tilde{B}_{l,s,j}$ has non-null components also for $s\neq 2$. States having a definite spin are just an approximation valid in the small boson mass limit, but they are not in general possible because of spin-mixing.}:
\begin{equation}
\tilde{B}_{l,s,j}(\zeta_k)=\int d\Omega\left[\frac{R^{nl}}{F(r)}Y^{l_0,j_0 jz}_{ab}\left(Y^{(2)ab}_{l,s,j,j_z} \right)^*\right]_{r=r(\zeta_k)}
\end{equation}
Also in this case we get linear combinations of spherical harmonic bra-ket integrals of trigonometric functions, whose results are given by \ref{cos_braket}, \ref{sin_braket}, \ref{n0cos_braket}.

\subsection{Slow-rotation guess solutions}

In\cite{Brito_spin-2_SR_slow_rot} the perturbation in the (first order) $a\ll M$ approximation is decomposed into polar and axial $(j,j_z)$ modes\footnote{Under parity transformations, polar (even) and axial (odd) modes are respectively multiplied by $(-1)^j$ and $(-1)^{j+1}$ }
\begin{equation}\label{Regge-Wheeler}
h_{\mu\nu}(\omega,t,r,\theta,\phi)=\sum\limits_{j,j_z}  e^{-i\omega t}\left[h_{\mu\nu}^{pol, j j_z}(\omega,r,\theta,\phi)+h_{\mu\nu}^{ax, j j_z}(\omega,r,\theta,\phi) \right]
\end{equation}
\[h_{\mu\nu}^{ax, j j_z}=\left[\begin{matrix}
 0&0& h_0^{j j_z}\csc\theta\partial_\phi Y_{j,j_z}& - h_0^{j j_z}\sin\theta\partial_\theta Y_{j,j_z}\\ 
*&0& h_1^{j j_z}\csc\theta\partial_\phi Y_{j,j_z}& - h_1^{j j_z}\sin\theta\partial_\theta Y_{j,j_z}\\ 
*& *&- h_2^{j j_z}\csc\theta X_{j,j_z}&h_2^{j j_z}\sin\theta W_{j,j_z}\\
 *& *& *&h_2^{j j_z}\sin\theta X_{j,j_z}\end{matrix}\right]\]
\[h_{\mu\nu}^{pol, j j_z}=\left[\begin{matrix}
 f(r)H_0^{j j_z}Y_{j,j_z}&H_1^{j j_z}Y_{j,j_z}& \eta_0^{j j_z}\partial_\theta Y_{j,j_z}& \eta_0^{j j_z}\partial_\phi Y_{j,j_z}\\ 
*&f^{-1}(r)H_2^{j j_z}Y_{j,j_z}& \eta_1^{j j_z}\partial_\theta Y_{j,j_z}& \eta_1^{j j_z}\partial_\phi Y_{j,j_z}\\ 
*& *&\begin{matrix} r^2(K^{j j_z}Y_{j,j_z}+\\+G^{j j_z} W_{j,j_z})\end{matrix} &r^2 G^{j j_z} X_{j,j_z}\\
 *& *& *&\begin{matrix} r^2 (K^{j j_z}Y_{j,j_z}-\\-G^{j j_z} W_{j,j_z})\sin^2\theta\end{matrix}\end{matrix}\right]\]
where $\displaystyle f(r)=1-\frac{2 M}{r}$, $\displaystyle X_{j,j_z}=2 \partial_\phi \left(\partial_\theta Y_{j,j_z}-\cot\theta Y_{j, j_z} \right)$, $W_{j,j_z}=\partial_\theta^2 Y_{j,j_z}- -\cot\theta\partial_\theta Y_{j,j_z}-\csc^2\theta\partial_\phi^2 Y_{j,j_z}$, while $h_0^{j j_z}$, $h_1^{j j_z}$, $h_2^{j j_z}$, $H_0^{j j_z}$, $H_1^{j j_z}$, $\eta_0^{j j_z}$, $\eta_1^{j j_z}$, $K^{j j_z}$ and $G^{j j_z}$ are frequency-dependent radial functions.

\begin{figure}[H]
\centering
\includegraphics[width=1\textwidth]{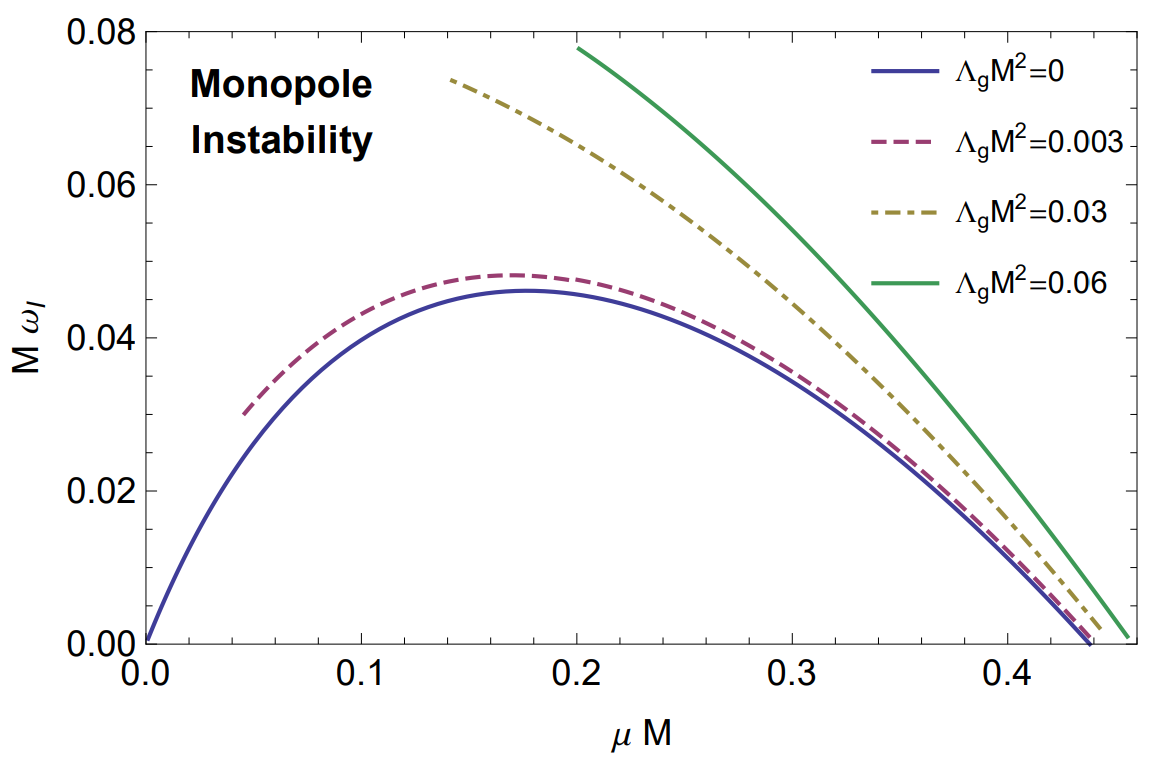}\\
\caption{Monopole instability vs boson mass in Schwarzschild and Schwarzschild-De-Sitter BHs, by \textit{Brito et al.} \cite{Brito_spin-2_SR_slow_rot}.}
\label{fig:monopole_instability}
\end{figure}

Through this framework \cite{Brito_spin-2_SR_slow_rot} \textit{Brito et al.} got both the hydrogenic modes (that can be computed also in the small mass limit \cite{Brito_hydrogenic_spin-2_SR}) and the monopolar $j=0$ and dipolar $j=1$ solutions. GR BHs are found to be unstable against monopolar perturbations, even in the non-rotating case: the instability is associated to the growth of a tensor hair, due to GR BHs being just a subset of the possible massive gravity's BH solutions\cite{massive_gravity_hairy_BHs,Brito_spin-2_SR_slow_rot}. The monopolar instability is active for small masses $0<\mu M \lesssim 0.47$, is characterized by an imaginary part $\omega_I=\operatorname{Im}\omega\sim 0.7 \mu$ in the small mass limit and has maximum $M \omega_I \sim 0.046$ \cite{Brito_spin-2_SR_slow_rot} (Fig. \ref{fig:monopole_instability}). Through the numerical technique described in this work, we have been able to confirm the monopole instability for $a=0$.
In the dipolar sector, instead, they found superradiant states. The axial dipolar mode features an extremely weak instability (almost five orders weaker than its equivalent in the vector case), whose analytical expression for the fundamental mode for $a j_z /(M^2\mu)\lesssim j$ is the following\cite{Brito_spin-2_SR_slow_rot}:
\begin{equation}
    M\omega_I\approx\frac{40}{19683}(a/M-2 r_+ \mu)(M\mu)^{11}\,.
\end{equation}
The polar dipolar mode, instead, experiences a strong superradiant instability, higher than the one of the hydrogenic sector\cite{Brito_spin-2_SR_slow_rot}, whose expression for the dominant mode is what follows:
\begin{equation}
    M\omega_I \sim (j_z a/M-2 r_+ \omega_R) (M\mu)^3~,~~\omega_R\approx 0.72 (1-M\mu)+O(a/M)\,.
\end{equation}
Computing the exact value of this instability is of the highest importance in the superradiance program, given that confirming its high value implies tighter constraints on ultralight spin-2 bosons (see subsection \ref{tensor_numerical_results}). We will give a description of the actual state of the art of this computation in the next sub-section.

By inverting \ref{h_ansatz} and carrying the Čebyšëv interpolation, we can get the formula for the computation of the $\underline{B}$ super-vector of ansatz \ref{Regge-Wheeler}:
\begin{equation}\label{basis_transform}
\tilde{B}_{l,s,j}(\zeta_k)=\int d\Omega\left[\frac{h_{\mu\nu}}{e^{- i \omega t} F(r)} \left(Y^{(2)\mu\nu}_{l,s,j,j_z}\right)^*\right]_{r=r(\zeta_k)}
\end{equation}
Such transformation can be simplified by exploiting operators $\boldsymbol{D_\pm}$ and $\boldsymbol{D_0}$ in \ref{angular_operators} and angular momentum raising/lowering operators in \ref{L+L-}, given that in the scalar case the latters can be expressed as 
\begin{equation}\label{L+L-_spherical_coordinates}
\mathbfcal{L}_\pm=\pm e^{\pm i\phi}\left(\partial_\theta\pm i\cot\theta\partial_\phi\right)
\end{equation}
What we get are the decomposed expressions of the formulas appearing in the Regge-Wheeler basis, whose expressions can be found in Appendix \ref{appendix_Regge-Wheeler}.

Those cumbersome formulas can be exploited for reducing \ref{basis_transform} to some linear combinations of spherical harmonic bra-kets of trigonometric functions, i.e. linear combinations of integrals whose analytical results are given by \ref{cos_braket}, \ref{sin_braket}, \ref{n0cos_braket}.
In the case of the non-spinning monopole, it can be easily demonstrated that ansatz \ref{Regge-Wheeler} reduces just to the following spin-decomposed expression\footnote{The total angular momentum of the monopole must be $j=0$, thus we get to the result by using the Clebsch-Gordan coefficients for the summation of the orbital angular momentum with the spin one.}
\begin{equation}
\frac{h_{\mu\nu}}{e^{-i\omega t}}=f(r) H^{j=0,j_z=0}_0 \sqrt{4\pi} F^2_t(r) \tilde{\tau}^{0,0}_{\mu\nu} Y_{0,0}+H_1^{j=0,j_z=0}\sqrt{\frac{8\pi}{3}}F_t(r)F_r(r) \times~~~~~~~~~~~~~~
\end{equation}
\[\times\left(\tilde{\tau}^{1,1}_{\mu\nu}Y_{1,-1}-\tilde{\tau}^{1,0}_{\mu\nu}Y_{1,0}+\tilde{\tau}^{1,-1}_{\mu\nu}Y_{1,1} \right)+f^{-1}(r) H^{j=0,j_z=0}_2 F^2_r(r)\left[\frac{2}{3}\sqrt{\frac{6\pi}{5}}\times\right.~~~~~~~~~~~~~~\]
\[\left.\times\left(\tilde{\chi}^{2,2}_{\mu\nu}Y_{2,-2}+\tilde{\chi}^{2,-2}_{\mu\nu}Y_{2,2}-\tilde{\chi}^{2,1}_{\mu\nu}Y_{2,-1}-\tilde{\chi}^{2,-1}_{\mu\nu}Y_{2,1}+\tilde{\chi}^{2,0}_{\mu\nu}Y_{2,0} \right)-\sqrt{\frac{4\pi}{3}}\tilde{\chi}^{0,0}_{\mu\nu}Y_{0,0} \right]-\]\[-K^{j=0,j_z=0}r^2 F^2_\theta(r)\left[\frac{2}{3}\sqrt{\frac{6\pi}{5}}\left(\tilde{\chi}^{2,2}_{\mu\nu}Y_{2,-2}+\tilde{\chi}^{2,-2}_{\mu\nu}Y_{2,2}-\tilde{\chi}^{2,1}_{\mu\nu}Y_{2,-1}-\tilde{\chi}^{2,-1}_{\mu\nu}Y_{2,1}+\right.\right.~~~~~~~~~~~~~~~~~~~~~~\]\[\left.\left.+\tilde{\chi}^{2,0}_{\mu\nu}Y_{2,0} \right)+\sqrt{\frac{16\pi}{3}}\tilde{\chi}^{0,0}_{\mu\nu}Y_{0,0} \right]~~~~~~~~~~~~~~~~~~~~~~~~~~~~~~~~~~~~~~~~~~~~~~~~~~~~~~~~~~~~~~~~~~~~~~~~~~~~~~~
\]
whose associated $\underline{B}$ super-vector can be straightforwardly computed. Extensions to this expression can be computed also for the spinning case both for the monopolar and dipolar modes.

\subsection{Numerical results}\label{tensor_numerical_results}
The computation of tensor QBSs through the numerical technique described in Chapters \ref{chapter:spherical_harmonic} and \ref{chapter:numerical_SR_GR} has not been completed yet. In the first round of runs our code was designed to carry the computation without applying the substitution of the trace-less constraint $h_{\mu\nu}{\bar g}^{\mu\nu}=0$ into the other tensor perturbative equations, thus considering it as a field equation to be included into the matrix equation to be solved. Through such framework we have been able to confirm the non-spinning monopolar instability, thus getting a consistency check for our technique, though we quickly noticed that the spinning BH states could not be computed with sufficient precision. Therefore the technique has been modified such that the trace-less constraint is eliminated by substitution, i.e. getting what we show in Chapters \ref{chapter:spherical_harmonic} and \ref{chapter:numerical_SR_GR}: we expect this modification to be pivotal in getting precise results. In such way the number of elements of the matrix describing the field equations is highly reduced, but we get an important increase in the complexity of the expressions of the couplings among angular momentum states, thus greatly expanding the time needed for the computation.

\begin{figure}[H]
\centering
\includegraphics[width=1\textwidth]{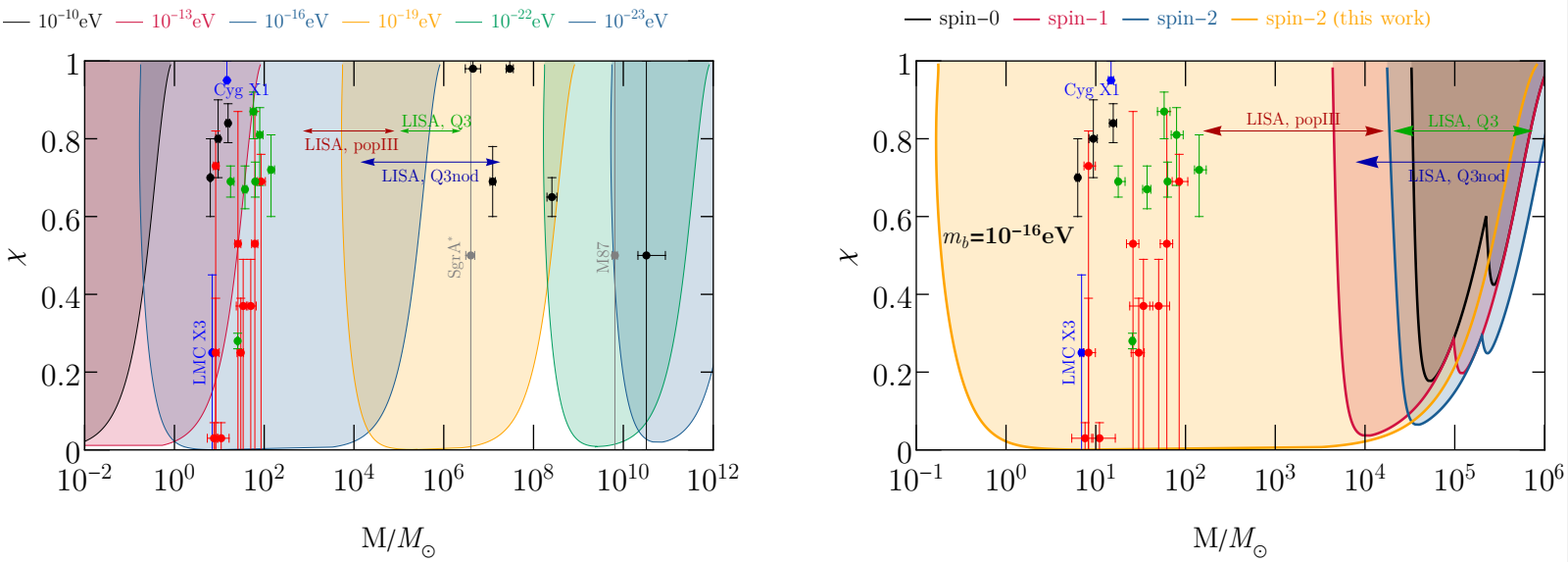}\\
\caption{On the left the exclusion regions in the BH spin-mass diagram are shown, arising from the BH superradiance of the dominant polar dipolar mode. For each mass of the field $m_b$, the considered separatrix for each value of the mass corresponds to an instability
timescale $\tau_S= 4.5\times 10^7 yrs$ (Salpeter time). On the right, instead, the exclusion regions for scalar, vector and tensor bosons having mass $m_b=10^{-16}eV$ arising from the dominant modes are compared. Image from \cite{dias2023black}, where all details about data points can be found.}
\label{fig:updated_spin2_constraints}
\end{figure}

Although at the moment the results of our method are not available yet, we are to show those we computed through an alternative technique in \cite{dias2023black}. This other method, developed by O. J. Dias and J. E. Santos, differs from the one we are adopting, given that the field equations are discretized both in $r$ and $\theta$ coordinates through a pseudospectral collocation grid on Gauss-Chebyshev-Lobbato
points\cite{dias2023black,dias_ultraspinning_BHs,Dias_2016}. The overall result is a confirmation of the slow-rotation computation of the dominant polar dipolar mode\cite{Brito_spin-2_SR_slow_rot} and an extension up to $\chi=a/M=0.8$, thus proving more solidly that it is the most superradiantly unstable mode. Though, the full computation of the monopole is still missing.
\begin{figure}[H]
\centering
\includegraphics[width=0.8\textwidth]{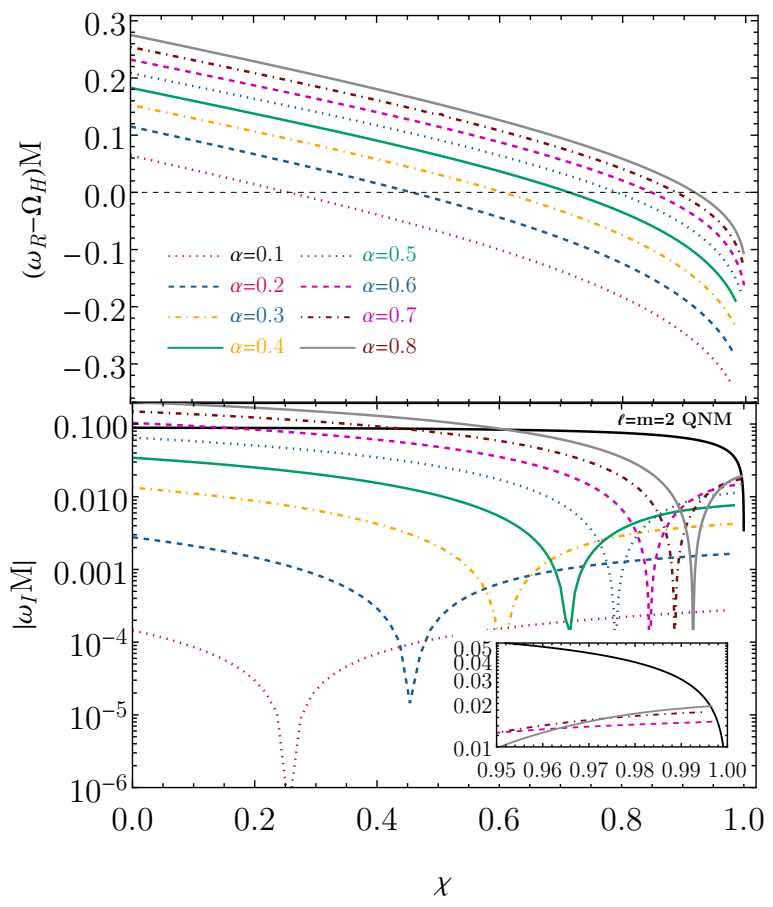}\\
\caption{Real and imaginary parts of $\omega$ of the dominant polar dipolar QBSs vs BH adimensional spin $\chi=a/M$ for different values of the gravitational fine structure constant $\alpha=\mu M$, by \textit{Dias et al.} \cite{dias2023black} (the unstable states are the ones having $\omega_R<\Omega_H$). The instability can be as large as $M\omega_I=0.019$ for $\alpha=0.8$, corresponding to an instability time $\tau=2.6\times 10^{-4}(M/M_\odot) s $. The polar dipolar modes are compared with the dominant $l=m=2$ Kerr QNMs (small bottom panel).}
\label{fig:polar_dipolar_instability}
\end{figure}
The results shown in Fig. \ref{fig:polar_dipolar_instability} are well described by the following polynomial fit\cite{dias2023black},
\begin{subequations}
\begin{align}
    &\omega=\omega_R+i \omega_I\,,\\
    &\frac{\omega_R}{\mu}\approx\left(\sum\limits_{i=0}^3 a_i \chi^i \right)\left(1+\alpha \sum\limits_{i=0}^3 b_i \chi^i+\alpha^2 \sum\limits_{i=0}^2 c_i \chi^i\right)\,,\\
    &\omega_I\approx-\alpha^3 (\omega_R-\Omega_H)\sum\limits_{i=0}^2 d_i \chi^i\,,
\end{align}
\end{subequations}
where\\$a_i\approx(0.73,-0.05,0.15,-0.12)$, $b_i=(-1.21,0.68,-0.55,0.61)$, $c_i=(0.69,-0.58,-0.11)$ and $d_i=(1.47,1.86,-2.75)$, $\alpha=\mu M$ and $\chi= a/M$. When the instability is active this fit is accurate within 2\% and 80\%, respectively, for $\alpha\in[0.05,0.8]$ and $\chi\in[0,\approx 0.99]$.
By including the polar dipole, we get tighter the constraints on spin-2 ultralight bosons arising from BH superradiance (Fig. \ref{fig:updated_spin2_constraints}), thus upgrading the ones shown in subsection \ref{ultralight_bounds_superradiance}.
If we consider current observations both in the EM and GW bands, these results on the spin-2 Regge plane exclude the region $\mu\in [5\times 10^{-17}
,5\times 10^{-11}] eV$, thus severely constraining the observability of dipolar radiation from BH binaries in massive gravity models\cite{Cardoso_massive_GWs,cardoso2023dipolar}. Moreover, a few BH mass-spin measurements for BH masses $M/M_\odot\in[1,10^{10}]$ would be sufficient for probing ultralight spin-2 fields in a range $\mu\in[10^{-23},10^{-10}]eV$, including also the value $\mu\sim 10^{-22}eV$ which corresponds to valid dark matter candidates\cite{Ultralight_cosmo_DM}.

\chapter{Superradiant instabilities triggered by plasma in scalar-tensor theories}\label{chapter:plasma_SR}

    This chapter is dedicated to the application of the numerical technique developed by \textit{Baumann et al.} \cite{Spectra_grav_atom}, previously shown in Chapters \ref{chapter:spherical_harmonic} and \ref{chapter:numerical_SR_GR}, to a beyond-GR scenario giving superradiant instabilities from non-linear interactions between gravity, matter and an extra field. We consider scalar-tensor extensions of GR\cite{Fujii:2003pa}, i.e. a family of theories in which the gravitational sector features one or more scalar fields non-minimally coupled with the metric tensor. This is the simplest class of extensions of GR\cite{testing_GR}, thus they can be considered among the most interesting and well-studied modified theories of gravity. If we consider just one scalar field extending GR, a quite general action of scalar-tensor theories in the so-called "Jordan frame" is the following \cite{Fujii:2003pa}:
\begin{equation}
\label{eq:jordanaction}
    S= \frac{1}{16\pi}\int d^4x \sqrt{-g}[\mathcal{F}(\Phi)R-\mathcal{Z}(\Phi)g^{\mu\nu}\partial_\mu \Phi \partial_\nu \Phi+\mathcal{U}(\Phi)]+S_m(\psi_m, g_{\mu\nu})\,,
\end{equation}
where $g_{\mu\nu}$ is the metric, $R$ is its Ricci scalar, $\Phi$ is a scalar field, while the last term $S_m(\psi_m, g_{\mu\nu})$ is the action of some generic matter fields $\psi_m$ minimally coupled with the metric. The functions $\mathcal{F}$, $\mathcal{Z}$ and $\mathcal{U}$ model how the scalar field self-interacts and is non-minimally coupled with the metric sector, thus different choices recover different theories. For instance, if $\mathcal{F}(\Phi)=\Phi$, $\mathcal{Z}(\Phi)\propto\Phi^{-1}$, and $\mathcal{U}(\Phi)=0$, the action \eqref{eq:jordanaction} gives Brans-Dicke theory\cite{testing_GR, Fujii:2003pa,Brans-Dicke}. Actions with scalar fields non-minimally coupled to gravity can also arise as EFTs from the compactification of extra dimensions in string theory and Kaluza-Klein-like theories\cite{Kaluza-Klein_theories}, but also from large dimensions in braneworld scenarios\cite{braneworld_1,braneworld_2}. Scalar-tensor theories have been intensively studied in cosmology\cite{Faraoni:2004pi, Clifton:2011jh}, while their astrophysical implications for compact objects have been investigated in detail\cite{testing_GR}.

A fundamental requirement for these theories for being considered viable is their consistency with observations. Given that GR has been extensively tested with great precision in the weak-field regime \cite{Will:2014kxa,testing_GR}, i.e. in the length scales between the micrometer and the astronomical unit, scalar-tensor theories must have the same weak-field limit of GR in order to be considered as possible modifications. Consequently, scalar-tensor theories with interesting cosmological phenomenology must feature some screening mechanism, e.g. Vainshtein screening\cite{Vainshtein_scalar-tensor}, hiding the scalar field on local scales\cite{Hinterbichler:2010es, Khoury:2003aq}. Instead, in the strong gravity regime, i.e. the one of compact objects, the deviations from GR might be more dramatic, therefore potentially giving rise to important phenomenology to be studied.

In this chapter we show a detailed analysis of matter-triggered spinning BH superradiant instabilities in scalar-tensor theories, which we published in \cite{Scalar_plasma_SR}\footnote{Please note that the signature used here is the opposite of the one used in \cite{Scalar_plasma_SR}.}. 

\section{Superradiant instabilities induced by plasma}

This effect was first shown in \cite{Cardoso:2013opa, Cardoso:2013fwa}, where it was unveiled how the presence of matter in the proximity of BHs can trigger either spontaneous scalarization\footnote{Spontaneous scalarization is the process of scalar tachyonic condensation triggered by some effective negative mass arising from non-linear interactions between the metric and the scalar field due to curvature\cite{spontaneous_scalarization}.} or superradiant instabilities, because of the trace of the stress-energy tensor of the surrounding matter giving rise to an effective mass for the scalar field acting as a confinement. In the following sections we will show how these superradiant instabilities can arise in realistic models of accreting BHs, and give results proving that superradiance could in principle be used for constraining scalar-tensor theories. A similar analysis was recently performed by \textit{Dima et al. }in \cite{Dima_2020}, where, through a spin-0 toy-model, the authors investigated plasma-driven \cite{Pani:2013hpa,Conlon:2017hhi} superradiant instabilities of photons in GR for BHs accreting tenuous plasma\footnote{We suggest also reading \cite{Wang:2022hra} for an extension to the Proca case and \cite{Cannizzaro:2020uap, Cannizzaro:2021zbp} for a recent analysis of photon-plasma interactions in curved spacetime.}. The authors of \cite{Dima_2020} have shown that the instability can be significantly quenched by the complex geometry of accretion disks and the high density values of plasma near BHs. Nevertheless, in \cite{Scalar_plasma_SR} we have shown that for realistic accretion-disk configurations this problem can be circumvented in scalar-tensor theories because of the effective mass depending on the scalar-tensor coupling. 

In the context of photon-plasma interactions in GR BH spacetimes, in the cold and collision-less plasma approximation the effective photon mass coincides with the plasma frequency\cite{Pani:2013hpa,Conlon:2017hhi,Cannizzaro:2020uap, Cannizzaro:2021zbp}:
\begin{equation}
    \omega_p=\sqrt{\frac{4\pi e^2 n_e }{m_e}}\approx 10^{-12}\sqrt{\frac{n_e}{10^{-3}{\rm cm}^{-3}}}\,{\rm eV}\,,
\end{equation}
where $n_e$ is the number density of the free electrons in the plasma, $m_e$ and $e$ are their mass and electric charge respectively. BH superradiant instabilities are most effective if $\omega_p M\sim {\cal O}(0.1)$, where $M$ is the BH mass, while they are highly suppressed when $\omega_p M\gg1$, thus implying $\omega_p\lesssim 10^{-11}\,{\rm eV}$ for astrophysical BHs with $M> M_\odot$. Consequently, the effective mass can trigger superradiant instabilities if $n_e\sim 10^{-3}-10^{-2}\,{\rm cm}^{-3}$, i.e. for plasma densities typical of the interstellar medium~\cite{Conlon:2017hhi}, while the plasma density near accreting BHs is several orders of magnitude bigger \cite{Dima_2020}. Hence, in GR astrophysical BHs the effective mass is unable to induce an instability giving a sufficiently short time scale for relevant phenomenology. 

However, in scalar-tensor theories the effective mass squared is\cite{Cardoso:2013opa, Cardoso:2013fwa}
\begin{equation}
    \mu_{\rm eff}^2=2\alpha T\sim 2{\alpha} \rho \,,
\end{equation}
where $T$ is the trace of the stress-energy tensor, $\rho=m_N n_e$ is the matter-energy density of the gas, having nucleon mass $m_N$, ${\alpha}$ is a free parameter related to the non-minimal coupling of the scalar field. The last step above is valid for a non-relativistic disk, as we will discuss in Section \ref{section:effectve_mass}. Thus, in the scalar-tensor case we have the effective mass depending on $n_e^{1/2}$ like in the standard photon-plasma case but we also have the free effective coupling ${\alpha}$. As we shall discuss in Section \ref{sec:results}, depending on the value of ${\alpha}$, the effective mass can be in the optimal range to trigger superradiant instabilities for realistic plasma configurations around BHs.

 Plasma-driven BH superradiant instabilities, though, can also be drastically suppressed by non-linearities\cite{cardoso_plasma_tests}. While transverse modes with frequency $\omega<\omega_p$ do not propagate in a cold plasma in the linear theory, the plasma becomes transparent due to non-linear effects if the electric field is higher than $E_{\rm crit}=\frac{m_e}{e}\sqrt{\omega_p^2-\omega^2}$~\cite{relativistic_nonlinear_propagation_plasma}. This is caused by the fact that a Lorentz boost factor arising from the backreaction of the plasma four-velocity decreases the plasma frequency. This Lorentz factor can be remarkably large during the superradiant growth of the electric field, hence severely limiting the extraction of angular momentum and energy through plasma-driven superradiant instabilities within GR theory \cite{relativistic_nonlinear_propagation_plasma}. Nevertheless, the situation is completely different in the scalar-tensor theories, given that in this case the effective mass is Lorentz-invariant. Also in this case the plasma 4-velocity experiences a change induced by the backreaction but, due to the effective mass depending only on the trace of the stress-energy tensor, it is not suppressed by a Lorentz factor. We will see the details of all of this in Section \ref{sec:nonlinear}.

 \section{Scalar perturbations in the Einstein frame}

In the Jordan frame, in the scalar-tensor action \eqref{eq:jordanaction} the scalar field is non-minimally coupled to the metric while matter is minimally coupled with gravity. If we perform a conformal transformation of the metric and a field redefinition for the scalar field\cite{Fujii:2003pa,Cardoso:2013opa, Cardoso:2013fwa,Scalar_plasma_SR},
\begin{equation}
    \begin{aligned}
    &g_{\mu\nu}^E=\mathcal{F}(\Phi)g_{\mu\nu}\,, \\&
    \Phi_E(\Phi)=\frac{1}{4 \pi}\int d\Phi \Bigg[\frac{3}{4}\frac{\mathcal{F}'(\Phi)^2}{\mathcal{F}(\Phi)^2}+\frac{1}{2}\frac{Z(\Phi)}{\mathcal{F}(\Phi)}\Bigg]^{1/2}\nonumber \,,\\&
    \mathcal{A}(\Phi_E)=\mathcal{F}^{-1/2}(\Phi)\nonumber \,,\\&
    \mathcal{V}(\Phi_E)=\frac{\mathcal{U}(\Phi)}{\mathcal{F}^2(\Phi)}\,, \nonumber
\end{aligned}
\end{equation}
we can describe the system in the so-called Einstein frame, where the action \eqref{eq:jordanaction} takes the following form\cite{Fujii:2003pa,Cardoso:2013opa, Cardoso:2013fwa,Scalar_plasma_SR}:
\begin{equation}
    S=\int d^4x \sqrt{-g^E}\left(\frac{R^E}{16\pi}-\frac{1}{2}g_E^{\mu\nu}\partial_\mu \Phi_E \partial_\nu \Phi_E+\frac{\mathcal{V}(\Phi_E)}{16 \pi}\right)+ S(\psi_m, \mathcal{A}^2(\Phi_E)g^E_{\mu\nu})\,.
\end{equation}
By varying this action we get the following field equations\cite{Fujii:2003pa,Cardoso:2013opa, Cardoso:2013fwa}:,
\begin{subequations}
    \begin{align}
        &G^E_{\mu\nu}=8\pi\left[T^E_{\mu\nu}+\partial_\mu\Phi_E \partial_\nu\Phi_E -\frac{g^E_{\mu\nu}}{2}\left(g_E^{\alpha\beta}\partial_\beta\Phi_E \partial_\beta \Phi_E -\mathcal{V}(\Phi_E)\right)\right]\,,\\
        &\Box_E \Phi_E=-\frac{\mathcal{A}'(\Phi_E)}{\mathcal{A}(\Phi_E)}T^E-\frac{\mathcal{V}'(\Phi_E)}{16\pi}\,,
    \end{align}
\end{subequations}
where $G^E_{\mu\nu}$ is the Einstein tensor of $g^E_{\mu\nu}$, $g_E^{\mu\nu}$ is the inverse metric of the Einstein frame, $T^E_{\mu\nu}$ is the stress-energy tensor of matter in the Einstein frame, $T^E=T^E_{\mu\nu}g_E^{\mu\nu}$, $\Box_E$ is the D'Alembertian operator of $g^E_{\mu\nu}$. The stress-energy tensor of the Einstein frame is related with the one of the physical Jordan frame through the following relationships\cite{Fujii:2003pa,Cardoso:2013opa, Cardoso:2013fwa}:
\begin{equation}
    T^{\mu}_{E~ \nu}=\mathcal{A}^4(\Phi_E)T^{\mu}_{~\nu}~,~T^E_{\mu\nu}=\mathcal{A}^2(\Phi_E)T_{\mu\nu}~,~T^E=\mathcal{A}^4(\Phi_E)T\,.
\end{equation}
In the Einstein frame the scalar field is minimally coupled with the gravity sector, but matter is coupled with the effective metric \textbf{$\mathcal{A}(\Phi)^2g^E_{\mu\nu}$}: the weak equivalence principle is, thus, preserved, while its strong version is violated. We are considering the Einstein frame for our computations, but we should take into account that laboratory clocks and rods refer to the Jordan-frame metric $g_{\mu\nu}= \mathcal{A}^2(\Phi_E) g_{\mu\nu}^E$. Physical asymptotic quantities related to the metric, such as masses and angular momenta, can be computed from their counterpart in the Einstein frame by rescaling the latter with suitable powers of $A(\Phi^{(0)})$. Consequently, recovering GR in the weak-field limit requires $A(\Phi^{(0)})\approx1$, i.e. for our purposes the distinction between asymptotic quantities in the Einstein and Jordan frames is negligible.

In this frame, we assume generic regular behavior for all the involved functions around a GR solution having constant value $\Phi_E^{(0)}$ of the scalar field,
\begin{subequations}
\begin{align}
    \mathcal{V}(\Phi_E)&=\sum_{n=0}^\infty V_n (\Phi_E-\Phi_E^{(0)})^n\, ,\\
    \mathcal{A}(\Phi_E)&=\sum_{n=0}^\infty A_n (\Phi_E-\Phi_E^{(0)})^n\,.
\end{align}
\end{subequations}
By expanding the equations for $\varphi\equiv\Phi_E-\Phi_E^{(0)}\ll1$, we can rearrange the scalar field perturbative equation in a GR background $\bar g^E_{\mu\nu}$ in the following form\cite{Cardoso:2013opa, Cardoso:2013fwa}
\begin{equation}
\label{eq:finalKGeq}
    [\bar\Box_E+\mu_{\rm eff}^2(r, \theta)]\varphi=0\,,
\end{equation}
where an \emph{effective} mass squared term arises,
\begin{equation}
    \mu_{\rm eff}^2(r, \theta)=\frac{V_2}{8\pi}+2{\alpha} T^E(r, \theta)\, ,
\end{equation}
and ${\alpha}:=A_2/A_0$.
By following\cite{Cardoso:2013fwa}, we focus on asymptotically-flat spacetimes, which requires $V_0=V_1=0$, and on theories admitting GR vacuum solutions, which, instead, requires $A_1=0$. For the rest of this analysis we will also fix $V_2=0$, i.e. we assume that the standard bare mass of the scalar field is zero. 

Hence, what we got is a Klein-Gordon equation in GR background with an effective mass squared proportional to the trace of the stress-energy tensor of the surrounding matter. We choose the background to be the Kerr spacetime, given that the backreaction of matter on the metric is typically negligible and owing to the BH no-hair theorems in scalar-tensor theories \cite{Sotiriou:2011dz}.

The sign of the parameter ${\alpha}$ impacts deeply on the phenomenology of the system \cite{Cardoso:2013opa, Cardoso:2013fwa}. As a matter of fact, if ${\alpha}<0$ the effective mass squared in Eq.\eqref{eq:finalKGeq} is negative, thus it can trigger a possible tachyonic condensation, i.e. the scalarization of the BH. In the opposite case ${\alpha}>0$, instead, the effective mass squared is positive and therefore it causes superradiant instability, i.e. what we analyze through our computation.

The physical effect occurring in the presence of a bosonic effective mass is the same we studied  in the previous chapter in the case of bare mass, though now the geometry of the effective-mass term has an important role, as we shall discuss in the following paragraph.

\section{Effective mass}\label{section:effectve_mass}

As we were previously showing, the effective mass-squared term linearly depends on the trace of the stress-energy tensor of the matter fields surrounding the compact object. In this paragraph we give details about this term for realistic accretion disk profiles.

In our computations we have considered different types of effective mass. In general, it is possible to decompose the stress-energy tensor of an accretion disk in four different contributions\cite{Abramowicz2013}:
\begin{equation}
    T^\mu_\nu=(T^\mu_\nu)_{\rm FLU}+(T^\mu_\nu)_{\rm VIS}+(T^\mu_\nu)_{\rm MAX}+(T^\mu_\nu)_{\rm RAD}\, ,
\end{equation}
representing, respectively, the fluid, viscosity, electromagnetic and the radiation components. The majority of models describing accretion disks assume a specific form for the stress-energy tensor. For instance, the so-called "thick" accretion disk models are based on a perfect fluid approximation, which is the assumption we have considered in all our computations, i.e.
\begin{subequations}
\begin{align}
    &(T^\mu_\nu)_{\rm VIS}=(T^\mu_\nu)_{\rm MAX}=(T^\mu_\nu)_{\rm RAD}=0\,,\\
    &(T^\mu_\nu)_{\rm FLU}=(\rho u^\mu)(W u_\nu)-\delta^\mu_\nu P\,,
\end{align}    
\end{subequations}
where $\rho$, $W$, $P$ are respectively the mass-energy density, enthalpy, and pressure. If we neglect the internal energy density of the fluid, the trace of the stress-energy tensor reads $T=\rho-3P$. We shall stress that while the perfect fluid approximation holds for thick disks, in the case we are considering we can apply the same approximation also for the so-called "thin" disks. This type of disks have a non-vanishing stress part, which, for instance, we can describe in the Shakura-Sunyev model by using a nearly-linear viscosity approximation \cite{Shakura-Sunyaev}. However, we can write the stress part as $(T^\mu_\nu)_{\rm VIS}\propto\sigma^{\mu}_{\nu}$, where $\sigma^{\mu\nu}$ is the shear tensor of the four-velocity of the fluid, i.e. a \emph{traceless} tensor, consequently the effective mass does not depend on viscosity. 

In what is following we also neglect the effects of pressure, given that it gives subdominant contributions. As a matter of fact, we are in a non-relativistic regime giving $P\ll\rho$: if we assume the ideal gas' equation of state, we can express pressure as $P=c_s^2 \rho$, where $c_s$ is the fluid's speed of sound, which is at least two orders of magnitude smaller than the speed of light in the case of accretion disks. Thus, the trace of the stress-energy tensor in our models is simply $T\approx \rho$ and we can safely neglect pressure corrections to the effective mass.

\subsection{Accretion disks: truncation, densities and coronae}
In the analysis we performed we considered accretion environments featuring a sharp cut-off at a sufficiently high distance from the BH's event horizon, i.e. a cavity generated by the disk in the proximity of the compact object. The reason for such choice comes from the necessity of having a setting able to potentially trap scalar modes leading to a superradiant instability. A system satisfying this requirement is the one described by truncated disk accretion models, which are commonly used in BH accretion physics. Depending on the accretion rate, truncation can be located close to the Innermost Stable Circular Orbit (ISCO) (in the so-called "high/soft state") or very far from it, even at $200-400M$ or more (in the so-called "low/hard state"). Whenever we have such astrophysical configuration, only a hot coronal flow can exist in the region within the truncation radius and down to the vicinity of the BH\cite{truncation2007, truncation2014, truncation2014b, truncation1997, truncation2013}. According to the high-energy astrophysics community, the comptonization of hot electrons in the coronal medium should explain the hard X-ray tail that follows the black-body-like emission spectrum of the disk. Hence, the model featuring a truncated disk and a corona is successful in explaining the features in the emission spectrum\cite{truncation2014}. Another configuration which can produce sufficiently wide cavities in the density profiles near BHs are counter-rotating disks that extend all the way up to the ISCO. In this setting, the cavity is able to trap modes due to the ISCO being sufficiently far away from the BH horizon, i.e. $6\leq r_{\rm ISCO}/M\leq9$ depending on the BH spin. Finally, another interesting possibility comes from magnetically-arrested disks, in which the cavity is generated by a strong poloidal magnetic field disrupting the disk at a relatively large radius. Also in this model we can have a hot, low-density coronal flow inside the cavity\cite{Narayan:2003by}. In general, these coronal flows are always very tenuous and quasi-spherical, having density way lower than the disk's one by some orders of magnitude \cite{Bisnovatyi-Kogan:1976fbc,DeVilliers:2003gr, DeVilliers:2004zz}. Therefore, in what is following we describe truncated thin and thick disks by also taking into account the presence of a coronal structure. 

\subsection{Plasma profiles}

In our analysis we considered different models of density profiles, which are discussed below.
In all models the time dependence of matter fields is neglected, given that the time scales of interest are much shorter than the typical BH accretion time scales\cite{Cannizzaro:2021zbp}. Moreover, due to the axisymmetry of spinning BHs, we restrict to axisymmetric configurations of the type $\rho=\rho(r,\theta)$. 

\begin{figure}[H]
\centering
\includegraphics[width=1\textwidth]{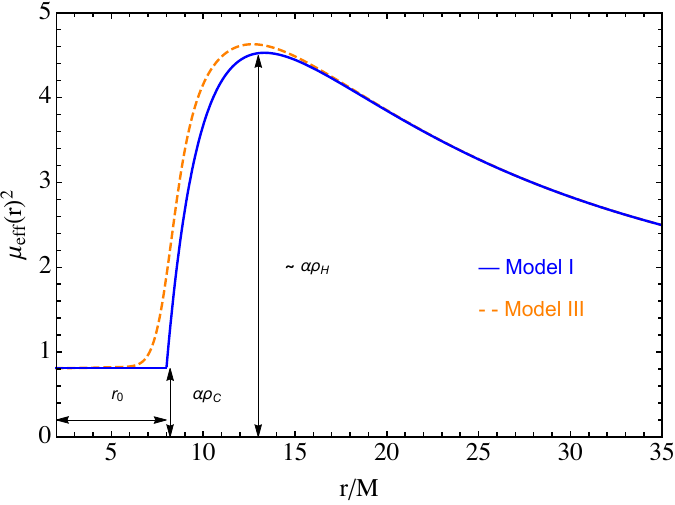}
\caption{Sharp vs smooth cut-offs: comparison between Model I's and Model II's radial profiles of effective mass. The parameter values used for Model I are $\alpha\rho_C M^2=0.9$, $\alpha\rho_H M^2=20$ and $r_0=8M$, while Model III has been plotted for $\beta=500$ and same value for $\alpha\rho_H M^2$. The curves are similar, but Model III features a smoothed cutoff. We have chosen unrealistic values for a matter of convenience, in order to better highlight the three fundamental parameters ($r_0,\alpha\rho_C, \alpha \rho_H$) governing the principal features of the geometry.}
\label{fig:confrontation}
\end{figure}

The first model we considered, Model I, is a thick disc+corona configuration, where the corona is represented through a constant asymptotic term, i.e.
\begin{equation}
     \mu_{\rm eff, I}^2(r,\theta)={\alpha}\left[\rho_H\Theta(r-r_0)\left(1-\frac{r_0}{r}\right)\left(\frac{r_0}{r}\right)^{\frac{3}{2}}+\rho_C\right]\,,
\end{equation}
where $\Theta(x)$ is the Heaviside step function. This model, for $\alpha=1$, coincides with the one studied by \textit{Dima et al.} in \cite{Dima_2020} with a suitable choice of the parameters $\rho_H, \rho_C$, and $r_0$, therefore we used it for a comparation with their analysis. A shortcoming of Model I comes from the constant value introduced for modelling the corona, which impacts also at infinity while the corona should be localized in the cavity near the compact object. Thus, in order to investigate the role of the mass at spatial infinity, we considered a variant of Model I, i.e. Model II, which truncates the corona at $r_0$:
\begin{equation}
     \mu_{\rm eff, II}^2(r,\theta)={\alpha}\left[\rho_H\Theta(r-r_0)\left(1-\frac{r_0}{r}\right)\left(\frac{r_0}{r}\right)^{\frac{3}{2}}+\rho_C\Theta(r_0-r)\right]\,.\label{model2}
\end{equation}
In Model III, instead, we replace sharp cut-off with a sigmoid-like function in order to investigate the effects of Heaviside function of Models I and II:
\begin{equation}
     \mu_{\rm eff, III}^2(r,\theta)=\frac{{\alpha}\rho_H}{1+e^{-2(r-r_0)}}\left[1-\frac{r_0}{r\left(1+\frac{\beta}{r^4}\right)}\right] \left(\frac{r_0}{r}\right)^{\frac{3}{2}}\, .
\end{equation}
Model III, with a suitable choice of $\beta$,  is very close to Model I (Figure \ref{fig:confrontation}), except that the effective mass does not display a sharp cutoff. 

With Model IV we introduce the first realistic profile from an astrophysical point of view, describing a scenario for a standard, truncated thin disk with an additional structure made by an ADAF-type\footnote{ADAF means "Advection-Dominated Accretion Flow".} corona which extends in the inner zones where the disk evaporates \cite{Meyer-Hofmeister:2017ott, Meyer-Hofmeister:2012fol}. Therefore, we use the Shakura-Sunyev solution for modelling the disk, while for the corona we exploit the self-consistent solution described in \cite{Narayan-Yi}. In our analysis, we investigated the effect of the coronal density on the instability by varying it by several orders of magnitude. Moreover, the thickness in thin disks is $H/R\ll1$, thus, in order to try to capture this effect, we follow what \textit{Dima et al.} did \cite{Dima_2020}, i.e. introduced an angular dependency in the form of a factor $\sin^2 \theta$ multiplying the radial Shakura-Sunyev profile. As a matter of fact, we could consider even more thinner profiles, but they would require a higher number of spherical harmonics in the computation of the spectrum (see Section \ref{sec:num}). Like in the case of ADAF-type corona, the geometry is quasi-spherical, thus deviations from spherical symmetry can be safely neglected. The expression of the effective mass for Model IV therefore is the following:
\begin{equation}
     \mu_{\rm eff, IV}^2(r,\theta)={\alpha}\left[\rho_H\Theta(r-r_0)\left(1-\sqrt{\frac{r_0}{r}}\right)^\frac{11}{20}\left(\frac{r_0}{r}\right)^{\frac{15}{8}}     \sin^2 \theta+\rho_C{\left(\frac{1}{r}\right)^{\frac{3}{2}}}\right]\, .
\end{equation}

The final profile we considered, Model V, was studied for exploring differences between the radial geometries of thin and thick disks, by considering a radial profile typical of a thick-disk axisymmetric model with the same corona as in Model IV: 
\begin{multline}
     \mu_{\rm eff, V}^2(r,\theta)={\alpha}\left[\rho_H\Theta(r-r_0)\left(1-\frac{r_0}{r}\right)\left(\frac{r_0}{r}\right)^{\frac{3}{2}}\sin^2 \theta+\rho_C{\left(\frac{1}{r}\right)^{\frac{3}{2}}}\right]\,.
\end{multline}

It is important to highlight how the key properties of these models can be qualitatively captured by three parameters, i.e. $\rho_H$, $r_0$ and $\rho_C$, which should produce the following effects on a physical ground (see also Fig. \ref{fig:confrontation}):  
\begin{itemize}
    \item Parameter $\rho_H$ represents the height of the barrier, which, for sufficiently high values, can naturally confine the scalar modes into a cavity. The higher the $\rho_H$ is, the more efficient the confinement will be. Given that $\rho_H$ represents a potential barrier rather than a bare mass, increasing $\rho_H$ should not stabilize the modes, but only confine them better, thus differing with the standard superradiant instability from massive bosons.
    \item Parameter $r_0$ gives the width of the cavity, which produces an efficient confinement if it is large enough. In particular, the width of the cavity must be greater than, or at least comparable to, the Compton wavelength of the modes \cite{Cardoso_superradiant_instability}. In our study we considered two representative truncation values, i.e. $r_0=8M$ and $r_0=14M$. 
    \item Parameter $\rho_C$, instead, acts as an offset introducing an effective asymptotic mass to the scalar field, which contribute in the stabilization of the modes. Nevertheless, if the barrier is high enough and the modes are strongly confined in it, the effect of $\rho_C$ at infinity should be negligible, because the scalar radiation transmitted at infinity is expected to be extremely small. This effect will be explored by comparing Model I with Model II. 
\end{itemize}

In particular, as we will show in Section \ref{sec:results}, in the disk $\mu_{\rm eff}M\sim \sqrt{\alpha \rho_H} M$ should be sufficiently large for the barrier to confine the mode efficiently, while in the corona $\mu_{\rm eff}M\sim \sqrt{\alpha \rho_C} M$ corresponds to the gravitational coupling that governs the effective mass of the field inside the cavity. As such, the condition for avoiding the suppression of the instability will be $\sqrt{\alpha \rho_C} M\ll {\cal O}(0.1)$.

\section{The numerical method}
\label{sec:num}

In this paragraph we present the numerical techniques we applied for the computation of the spectrum of the scalar perturbation in the astrophysical setting we described in the previous sections.
We assumed the background to be stationary, i.e. Kerr metric, and worked in the frequency domain by considering a $e^{-i \omega t}$ dependence for the scalar field, where $\omega=\omega_R+i\omega_I$ is the complex eigen-frequency of the perturbation. Thus, also in this case solutions having $\omega_I>0$ describe unstable modes, which exponentially grow in time in the perturbative framework\footnote{Obviously this means that perturbation theory works only for an estimate of the instability time: as a matter of fact, the explosive growth of the unstable modes implies the field growing way beyond the limits allowed by perturbation theory, thus breaking the perturbative framework itself. For the full evolution we need non-linear numerical relativity.}.
We recall that in the specific case of superradiant instabilities, this exponential growth is triggered if the mode satisfies the superradiant condition\cite{Brito_SR} (see Section \ref{section:SR_scattering}), i.e. $0<\omega_R<m \Omega_H=\frac{m a}{2 M r_+}$, where $\Omega_H=\frac{a}{2 M r_+}$ is the BH angular momentum, $r_+$ is the radius of its event horizon, and $m$ is the azimuthal number of the mode. 

We used a two-step procedure for the numerical computations, all in the frequency domain. Step 1 consists in finding solutions of Eq. \eqref{eq:finalKGeq} in the the non-spinning BH case by using a direct shooting method\cite{Pani:2013pma}, i.e. in the case of spherical symmetry, when the effective mass profiles depend only on the radial coordinate. Through the imposition of suitable boundary conditions at the horizon and at infinity, the shooting method allows us to solve the eigenvalue problem. Then, in Step 2, we used the wave-functions and eigen-frequencies as starting guess solutions for the computation of the spinning case. We applied the numerical method for non-separable differential equations we described in the previous chapter, i.e. we followed again \textit{Baumann et al.}\cite{Spectra_grav_atom}: we expressed Eq. \eqref{eq:finalKGeq} as a nonlinear eigenvalue problem which we solved with the nonlinear inverse iteration algorithm \cite{non-linear-eigenvalue}. We iteratively solved the problem by gradually increasing the spin, starting from the spherical symmetric case, until we obtained the desired spinning configuration. With this method we could study also quasi-extremal BHs and generic non-separable equations. 

For the case of effective-mass profiles having a $\theta$-dependence through $\sin^2\theta$, the field equations are non-separable even for a non-spinning BH. In this case we added in Step 1 an extra iterative cycle based upon the technique by \textit{Baumann et al.}. Thus, we expressed the generic effective mass of any of the previous models in the following way, 
\begin{equation}
    \mu_{\rm eff}^2(r,\theta)=\mu_r^2(r)(1-k \cos^2\theta)+\mu_0^2(r)\, ,
\end{equation}
where we introduced the fictitious parameter $k$ connecting purely radial profiles ($k=0$) with $\theta$-depending profiles ($k=1$), \textbf{$\lim_{r\to\infty}\mu_r(r)=0$}, while $\mu_0^2(r)$ arises from the BH corona. The extra cycle consists in the application of the nonlinear inverse iteration to the computation of the mode of a non-spinning BH with a non-spherical density profile ($k=1$), using solutions with $k=0$ as starting guess. Thus, at each iteration we gradually increase $k$ and use the previous result as a guess, until we reach the desired configuration with $k=1$ and zero BH spin. Finally, we proceed with Step 2 by using the latter solution as a starting guess in order to find the modes of spinning BHs with $k=1$.

\subsection{Non-spinning BHs with radial density profiles}

The direct shooting method consists in the integration of the separated radial equation of the perturbation from the horizon  to infinity. As a matter of fact, in the non-spinning case when the effective-mass profiles are purely radial the scalar field equation is separable through spherical harmonic decomposition, thus in this case we use the following ansatz:
\begin{equation}
    \varphi(t,r,\theta, \phi)=\sum_{l,m} \frac{\tilde{R}_{lm}(r)}{r}e^{-i\omega t}Y_{lm}(\theta, \phi)\,.
\end{equation}
The Klein-Gordon equation is consequently rearranged in a Schrödinger-like form, i.e.
\begin{equation}
    \mathcal{D} R_{lm}=0\,,
\end{equation}
where $f(r)=1-2M/r$, $M$ is the mass of the BH, and we defined the following differential operator:
\begin{equation}
    \mathcal{D} \equiv   \frac{d^2}{dr_*^{2}}+ \omega^2- f(r) \left[\frac{l(l+1)}{r^2}+\frac{2M}{r^3}+\mu_{\rm eff}^2\right]\,,
\end{equation}
where $r_*$ is the tortoise coordinate given by $dr/dr_*=f(r)$.  Owing to the spherical symmetry of the system, modes with different multipole numbers $l,m$ are decoupled. Then, by imposing suitable boundary conditions, this equation is solved by direct integration. In particular, at the horizon the solution must be a purely in-going wave, given that the horizon behaves as a one-way membrane\cite{membrane_paradigm},
\begin{equation}
\label{eq:boundaryh}
    \tilde{R}_{lm}\sim e^{-i\omega r_*}\sum_n b_{n} (r-2M)^n\,.
\end{equation}
At leading-order at infinity, instead, the general solution has the following asymptotic expression,
\begin{equation}
\tilde{R}_{lm}\sim \mathcal{B} e^{-k_{\infty}r_* }+\mathcal{C} e^{+k_{\infty}r_*}\,,
\end{equation}
where $k_{\infty}=\sqrt{\mu^2_\infty-\omega^2}$ and $\mu_\infty=\lim\limits_{r\to\infty}\mu_{\rm eff}(r,\theta)$. In the standard context of massive boson superradiant instabilities, usually the correct condition for QBSs is $\mathcal{C}=0$, in order to get exponential damping at infinity. Nevertheless, in our case the confinement is provided by a potential barrier in the vicinity of the BH, while the standard case involves an asymptotic mass. Most importantly, in realistic accretion models the effective mass at infinity vanishes. Thus, imposing $\mathcal{C}=0$ in our case would not give damped solutions at infinity, instead the result would be in-going waves from infinity. Obviously this solution cannot be physical, given that it would describe energy injection from infinity. Hence, we must set the opposite conditions $\mathcal{B}=0$, which is the condition that in the standard case corresponds to out-going waves at infinity, i.e. quasi-normal modes (QNMs)\cite{Berti_BH_QNMs}. We can rationalize this choice in the following way. The modes we are looking for are, in some sense, supposed to behave as QNMs if we did not have any effective mass, but the confinement by the barrier in the proximity of the compact object make them become QBSs and thus are prone to the superradiant instability if the BH spins sufficiently fast. The same condition must be applied also when we consider non-vanishing mass at infinity, because we are interested in quasi-bound states featuring $\mu_\infty<|\omega|$.

We also applied a variation of the standard shooting method, characterized by the integration from the horizon to a fixed point and from infinity to the same point, and the imposition of $C^1[r_+,\infty)$ regularity of the wave-function to solve the equations\cite{Pani:2013hpa}. The results we got are independent on the matching point and in general we checked that the two methods give the same results.

\subsection{Axisymmetric configurations}

We now consider the case of non-separable perturbations, which is relevant for spinning BHs but also for non-spinning BHs if the effective mass is axisymmetric, i.e. when it depends on the angular coordinate $\theta$.

We assume Kerr metric background, which makes the perturbations assume a definite azimuthal number $m$. We rewrite Eq. \ref{eq:finalKGeq} in the following form,
\begin{equation}\label{field_eq_baumann}
\begin{aligned}
    &\left\lbrace\frac{1}{\Delta(r)}[\mathbfcal{L}^2+a^2\cos^2\theta(\mu_{\rm eff}^2(r,\theta)-\omega^2)]-\frac{1}{\Delta(r)}\frac{\partial}{\partial r}\left[\Delta(r)\frac{\partial}{\partial r}\right]\right.\\&-\omega^2-\frac{P_+^2}{(r-r_+)^2}-\frac{P_-^2}{(r-r_-)^2}+\frac{A_+}{r-r_+}-\frac{A_-}{r-r_-}\\&+\left.\mu_{\rm eff}^2(r,\theta)\left(1+\frac{B_+}{r-r_+}-\frac{B_-}{r-r_-}\right)\right\rbrace\varphi(t,r,\theta,\phi)=0 \,,
\end{aligned}
\end{equation}
where $\mathbfcal{L}^2=-\frac{1}{\sin\theta}\frac{\partial}{\partial\theta}\left(\sin\theta\frac{\partial}{\partial\theta}\right)-\frac{1}{\sin^2\theta}\frac{\partial^2}{\partial\phi^2}$ is the scalar representation of the angular momentum operator, $ A_\pm=\mp 2\omega^2 M+\frac{P_+^2+P_-^2 -(8 M^2-a^2)\omega^2}{r_+-r_-}$, $ B_\pm=\frac{2M^2-a^2}{r_+-r_-}\pm M$, $ r_\pm=M\pm\sqrt{M^2-a^2}$, $ P_\pm=\frac{m a-2\omega M r_\pm}{r_+-r_-}$ and $\Delta(r)=(r-r_+)(r-r_-)$. 
Note that in the above equation the dependence on $k$ is contained inside $\mu^2_{\rm eff}(r,\theta)$.

At the horizon we must have in-going waves, thus again
\begin{equation}
    \displaystyle\varphi\sim (r-r_+)^{ i P_+}\,,
\end{equation}
while we impose that there are no waves coming from infinity, i.e. 
\begin{equation}
    \varphi\sim r^{-1 - \frac{ M \left(2\omega^2-\mu^2_\infty\right)}{k_\infty}} e^{k_\infty r}\,.
\end{equation} 
Consequently, we apply the following ansatz for the scalar field, which is a simple modification of the one developed by \textit{Baumann et al.}\cite{Spectra_grav_atom}\,,
\begin{equation}\label{S-T_phi_ansatz}
\varphi(t,r,\theta,\phi)= F(r)\sum\limits_{l,m}B_{lm}(\zeta(r))Y_{lm}(\theta,\phi)e^{-i \omega t}\,,
\end{equation}
where 
\begin{equation}
F(r)=\left(\frac{r-r_+}{r-r_-} \right)^{i P_+} (r-r_-)^{-1 - \frac{ M \left(2\omega^2-\mu^2_\infty\right)}{k_\infty}}e^{k_\infty (r-r_+)}
\end{equation}
captures the asymptotic behaviour of the solution.
For simplicity we drop the index $m$ from $B_{lm}$, given that there cannot be mixing in $m$ because of the axisymmetry of the background. In the numerical results presented in the next section we will always consider the case $m=1$. In the above ansatz the radial functions $B_l(\zeta(r))$ depend on the usual auxiliary radial coordinate $\zeta\in(-1,1)$ we previously used in Chapter \ref{chapter:numerical_SR_GR}, defined by the expressions in \ref{mapping}.

The spherical harmonics decomposition gives us the following infinite cascade of radial equations,
\begin{equation}\label{radial_eqs}
    \left[\frac{\partial^2}{\partial\zeta^2}+C^{(1)}_l(\zeta)\frac{\partial}{\partial\zeta}+C^{(2)}_l(\zeta) \right]B_l(\zeta)+\sum\limits_{l'=-4}^4 C^{(3)}_{l,l'}(\zeta)B_{l'}(\zeta)=0\,,
\end{equation}
where we have the following expressions for the couplings,
\begin{subequations}
    \begin{align}
        &\begin{aligned}
        C^{(3)}_{l,l'}(\zeta)=&-\frac{c^{(1)}_{l,l'}}{\zeta'^2(r(\zeta))}\left\lbrace\frac{a^2[\mu^2_r(r(\zeta))+\mu^2_0(r(\zeta))-\omega^2]}{\Delta(r(\zeta))}-k \mu_r^2(r(\zeta))\right.\\
        &\left.\times\left[1+\frac{B_+}{r(\zeta)-r_+}-\frac{B_-}{r(\zeta)-r_-}\right]\right\rbrace+\frac{k c^{(2)}_{l,l'}a^2 \mu_r^2(r(\zeta))}{\zeta'^2(r(\zeta))\Delta(r(\zeta))}\,,
        \end{aligned}\\
        &\begin{aligned}
            c^{(1)}_{l,l'}=&\left\langle l,m\right|\cos^2\theta\left|l',m\right\rangle=\\
            =&\frac{1}{3}\delta_{ll'}+\frac{2}{3}\sqrt{\frac{2l'+1}{2l+1}}\left\langle l',m,2,0\right|\left.l,m\right\rangle\left\langle l',0,2,0\right|\left.l,0\right\rangle\,,
        \end{aligned}\\
        &\begin{aligned}
            c^{(2)}_{l,l'}=&\left\langle l,m\right|\cos^4\theta\left|l',m\right\rangle=\\
            =&\frac{1}{5}\delta_{ll'}+\frac{4}{7}\sqrt{\frac{2l'+1}{2l+1}}\left\langle l',m,2,0\right|\left.l,m\right\rangle\left\langle l',0,2,0\right|\left.l,0\right\rangle\\
            &+\frac{8}{35}\sqrt{\frac{2l'+1}{2l+1}}\left\langle l',m,4,0\right|\left.l,m\right\rangle\left\langle l',0,4,0\right|\left.l,0\right\rangle\,,
        \end{aligned}
    \end{align}
\end{subequations}
and the following expressions for the remaining functions
\begin{multline}
C^{(1)}_l(\zeta)=\left(\frac{1}{r(\zeta)-r_+}+\frac{1}{r(\zeta)-r_-} \right)\frac{1}{\zeta'(r(\zeta))}+\\\frac{1}{\zeta'(r(\zeta))}\frac{2 F'(r(\zeta))}{F(r(\zeta))}+\frac{\zeta''(r(\zeta))}{\zeta'^2(r(\zeta))}\,,
\end{multline}
\begin{multline}
C^{(2)}_l(\zeta)=\frac{1}{\zeta'^2(r(\zeta))}\left\lbrace\frac{ F''(r(\zeta))}{F(r(\zeta))}+\left[\frac{1}{r(\zeta)-r_+}+\right.\right.\\\left.\frac{1}{r(\zeta)-r_-} \right]\frac{F'(r(\zeta))}{F(r(\zeta))}+\frac{P_+^2}{[r(\zeta)-r_+]^2}+\frac{P_-^2}{[r(\zeta)-r_-]^2}\\-\left[\mu_r^2(r(\zeta))+\mu_0^2(r(\zeta))\right]\left[1+\frac{B_+}{r(\zeta)-r_+}-\frac{B_-}{r(\zeta)-r_-}\right]\\-\left.\frac{A_+}{r(\zeta)-r_+}+\frac{A_-}{r(\zeta)-r_-}+\omega^2-\frac{l(l+1)}{\Delta(r(\zeta))}\right\rbrace \,.
\end{multline}
The couplings $c^{(1)}_{l,l'}$ are nonzero for $l'\in\left\lbrace l,l\pm 2 \right\rbrace$, while $c^{(2)}_{l,l'}$ are non-zero for $l'\in\left\lbrace l,l\pm 2, l\pm 4 \right\rbrace$, thus each $l$-mode is coupled with 4 other differing ones. If we set $k=0$ we notice that the only surviving couplings are $c^{(1)}_{l,l'}$, i.e. exactly the ones we had already encountered in \ref{eq:c_ll^m} when studying massive scalar perturbations of Kerr in GR.

Also in this case, we truncate the infinite cascade to some $L$ in order to transform the resulting finite set of radial equations into a matrix form. The radial coordinate is then discretized through the Čebyšëv interpolation we defined in Section \ref{section:chebyshev_interpol}. Consequently the radial functions $B_l$ are described by a set of $(L+1)(N+1)$ coefficients $B_l(\zeta_k)$, where $(N+1)$ is the number of interpolation points, that define a $(L+1)(N+1)$-dimensional array $\underline{B}$, while the radial equations take the following form:
\begin{equation}
\small\sum\limits_{q=0}^N \left[ p_q''(\zeta_n)B_l(\zeta_q)+ C^{(1)}_l(\zeta_n)p_q'(\zeta_n)B_l(\zeta_q)\right]+ C^{(2)}_l(\zeta_n)B_l(\zeta_n)+\sum\limits_{l'=-4}^4 C^{(3)}_{l,l'}(\zeta_n)B_{l'}(\zeta_n)=0\,.
\end{equation}
The differentiation matrices appearing in the expression above are the same we used in Section \ref{section:chebyshev_interpol} defined by \ref{diff_matrix} and \ref{ddiff_matrix}. Therefore we have transformed the problem in a nonlinear eigenvalue problem in $\omega$ and $\underline{B}$, i.e.
\begin{equation}
    \mathbfcal{A}(\omega)\underline{B}=0\,,
\end{equation}
to be solved through nonlinear inverse iteration \cite{non-linear-eigenvalue} (see Section \ref{section:non-linear_inverse_iteration}).

\section{Numerical results}\label{sec:results}

In this paragraph we give an overview of the results we have computed for Model I, II, III, IV and V by using the numerical approach we described in the previous section. Thus, in the following sub-paragraphs we go through the results of each model and make some compared analysis among them. We start by showing the first three models with the same density profiles considered by \textit{Dima et al.} in\cite{Dima_2020}, in order to show that in scalar-tensor theories we can circumvent the obstacles existing in plasma-driven superradiant instabilities. Then, we will consider the results for more realistic models, including axisymmetric ones, so that we can show how superradiant instabilities can be relevant in scalar-tensor theories.

\subsection{Model I: the key requirements for the instability}
\begin{figure*}[h]
\centering
\includegraphics[width=0.46\textwidth]{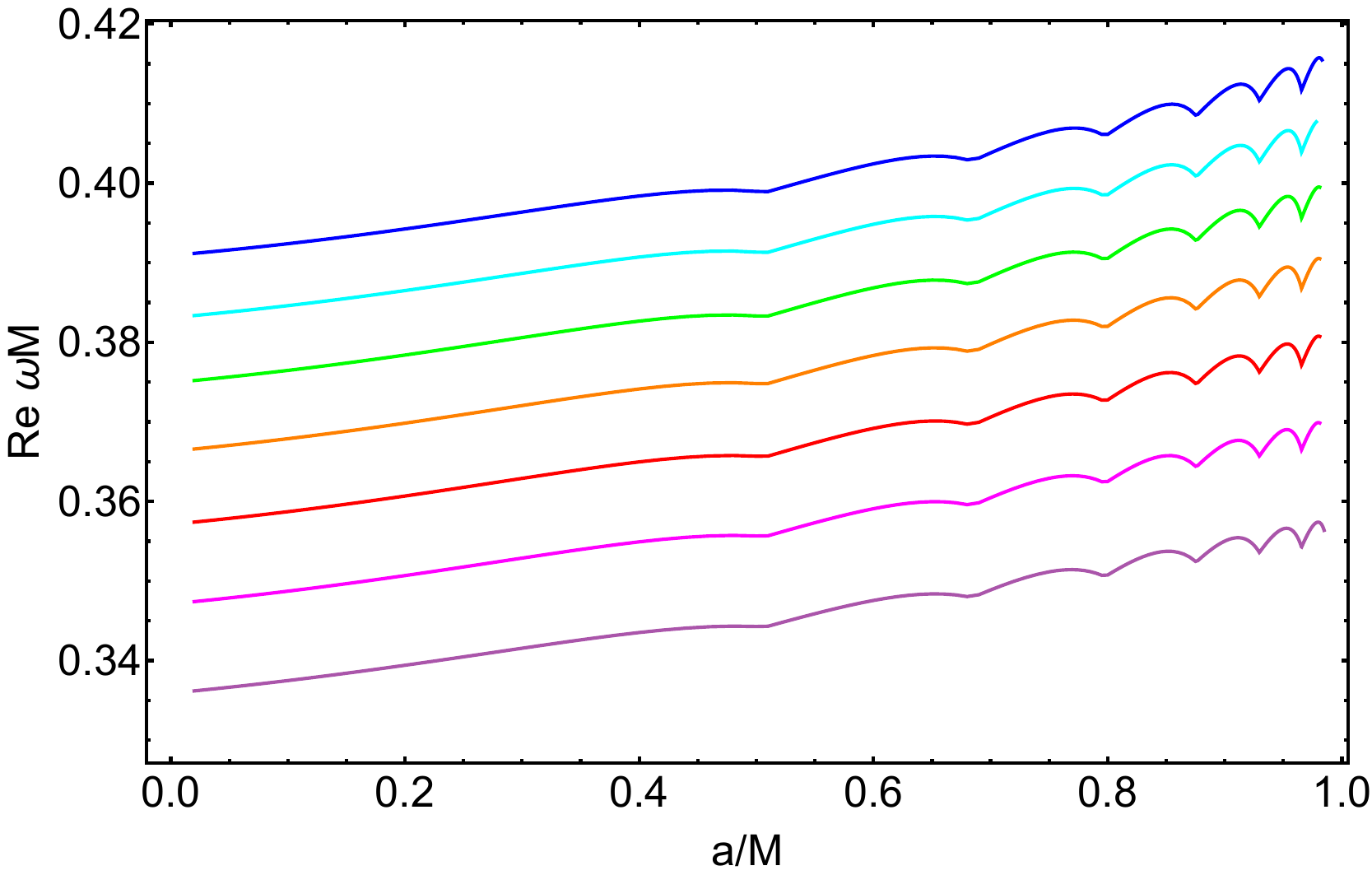}
\includegraphics[width=0.49\textwidth]{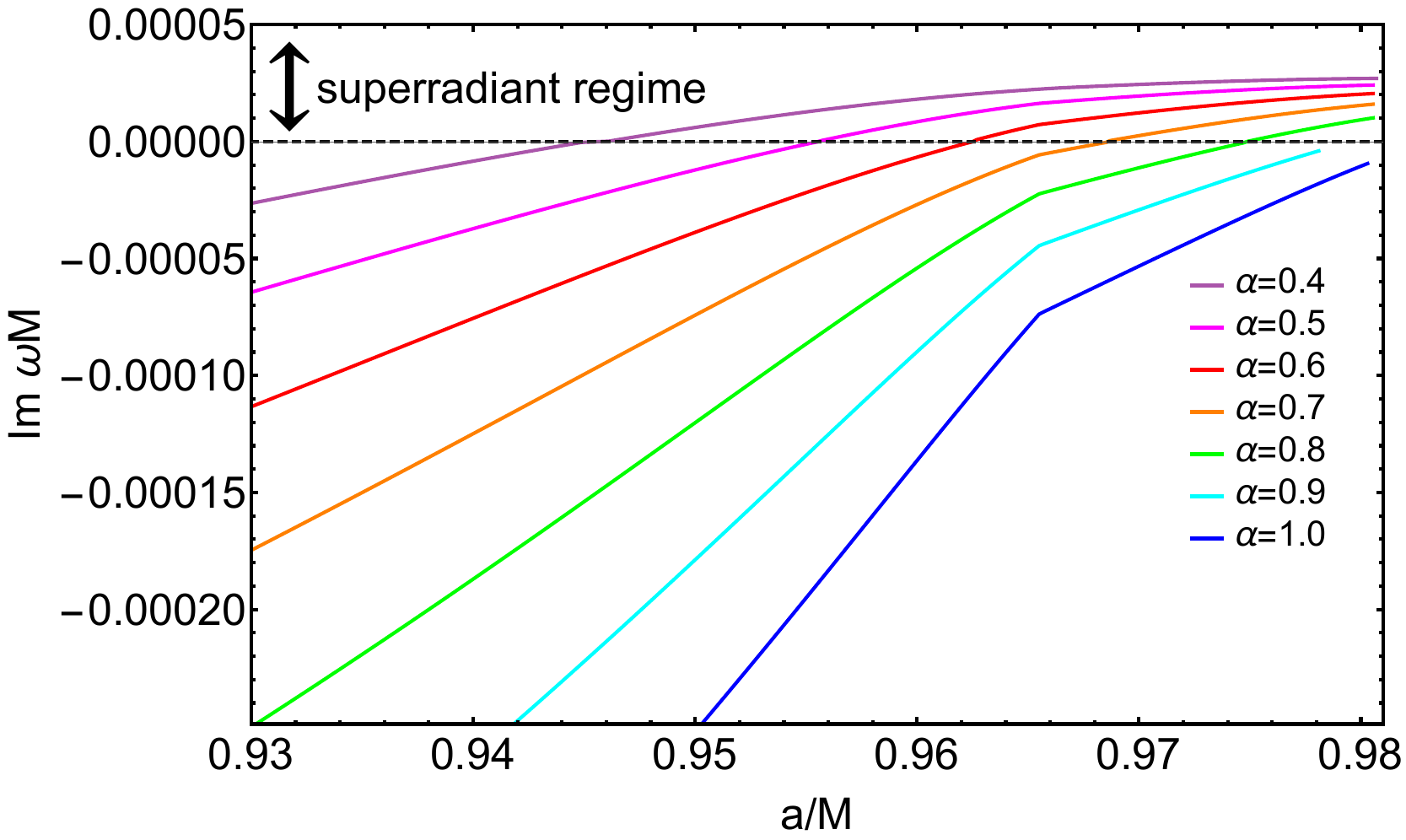}
\caption{On the left panel we show the real part of the mode frequency of Model I, while on the right we focus on the imaginary part for near-extremal BHs in order to show the instability. The curves are functions of the BH spin for different values of the scalar-tensor coupling ${\alpha}$. For lower values of this coupling, $\omega_R$ decreases and we get superradiant instabilities for smaller values of the BH spin. For all the curves $\rho_H=4/M^2,\rho_C=0.09/M^2$, $r_0=8M$, following the same choice made in \cite{Dima_2020}.}
\label{fig:Dima}
\end{figure*}
Figure \ref{fig:Dima} shows the scalar frequencies of Model I for different values of ${\alpha}$ for a specific choice of astrophysical parameters in accordance with \textit{Dima et al.}'s work \cite{Dima_2020}, hence for ${\alpha}=1$ we recover their results. In this specific case, we confirm that the superradiant instability cannot be active for BH spin below $a/M=0.99$. However, by considering lower values of ${\alpha}$ we can make the effective mass of the scalar field decrease, and thus also the superradiant mode's frequency decreases and consequently the superradiant condition is fulfilled for smaller values of the spin. This can be clearly noticed by examining the left panel of Fig.~\ref{fig:Dima}, where the real part of the mode frequency is plotted vs the BH spin for different values of $\alpha$. By making the coupling ${\alpha}$ decrease we get smaller values for the real part, eventually satisfying the superradiant condition. Therefore, while in GR a small increase of the coronal mass is sufficient to quench plasma-driven superradiant instabilities\cite{Dima_2020}, in scalar-tensor theories we can avoid this problem by decreasing ${\alpha}$ in order to recover an efficient superradiant regime, as we also discuss in more detail below. 

Nevertheless, if we exaggerate in decreasing ${\alpha}$ the potential barrier becomes too low for being able to confine the modes. For the case of Model I, we numerically found that if we go below ${\alpha}=0.15$ the eigen-functions start losing their confinement by showing non-negligible amplitudes even after the potential barrier.

\begin{figure}[t]
\centering
\includegraphics[width=0.9\textwidth]{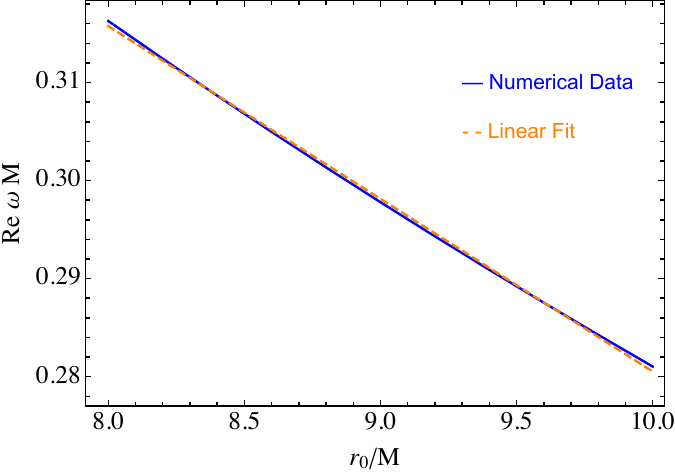}
\caption{Real part of the \textbf{mode frequencies} in Model~I as a function of $r_0$ for $\alpha \rho_H M^2=4$, $\alpha \rho_C M^2=0$, and $a=0$. The real part decreases linearly with $1/r_0$, as can be observed by comparing the numerical result with a linear fit.}
\label{fig:BHBomb}
\end{figure}

Assuming highly spinning BH, we can consequently quench the superradiant instability in the following ways:
\begin{itemize}
\item{by increasing the coronal density up to the point the system gets stabilized, which, for the chosen parameters, happens in Model I if $\sqrt{\alpha \rho_C}M>0.42$;}
\item{by decreasing the confinement, which starts happening when $\sqrt{\alpha \rho_H}M<0.76$;}
\item{by decreasing the width of the cavity, down to the point it cannot support QBSs inside it}
\end{itemize}
Indeed, if the effective mass within the cavity is negligble, i.e. $\sqrt{\alpha \rho_C}M\ll0.1$, this system resembles the original BH bomb, in which case the mode frequencies scale as the inverse of the width of the cavity, $\omega_R\sim 1/r_0$\cite{Cardoso_superradiant_instability}. In Fig. \ref{fig:BHBomb} we show that our system can recover the same scaling.

In the opposite case, instead, i.e. if the barrier and the cavity are high and wide enough, we get efficient confinement for the modes. Moreover, for a sufficiently tenuous coronal density, providing modes in the cavity having a not too large effective mass, an efficient superradiant instability can develop around an accreting spinning BH. We will examine this aspect in detail in the next paragraph, where we will show the resulting constraints to scalar-tensor theories arising from superradiance.

We stress how the main difference with respect to the analysis made by \textit{Dima et al.} in\cite{Dima_2020} is in the presence of the scalar-tensor free parameter ${\alpha}$. \textit{Dima et al.} in fact have show how an extremely tenuous plasma inside this cavity, of the order of $n_e\sim 10^{-2} {\rm cm}^{-3}$ for $M=10M_\odot$, is sufficient to suppress the instability, despite the disk can create a cavity where superradiant modes could develop. Given that realistic coronal densities are orders of magnitude higher, the instability is strongly suppressed. On the other hand, as we will discuss in detail in the next paragraph, in scalar-tensor theories the coupling $\alpha$ has large unconstrained ranges in which the effective mass due to the corona is negligible while at the same time the disk barrier is yet sufficiently high.

\subsection{Model II and III: truncation of the corona and smoothness of the profiles}
\begin{figure}[h]
\centering
\includegraphics[width=0.65\textwidth]{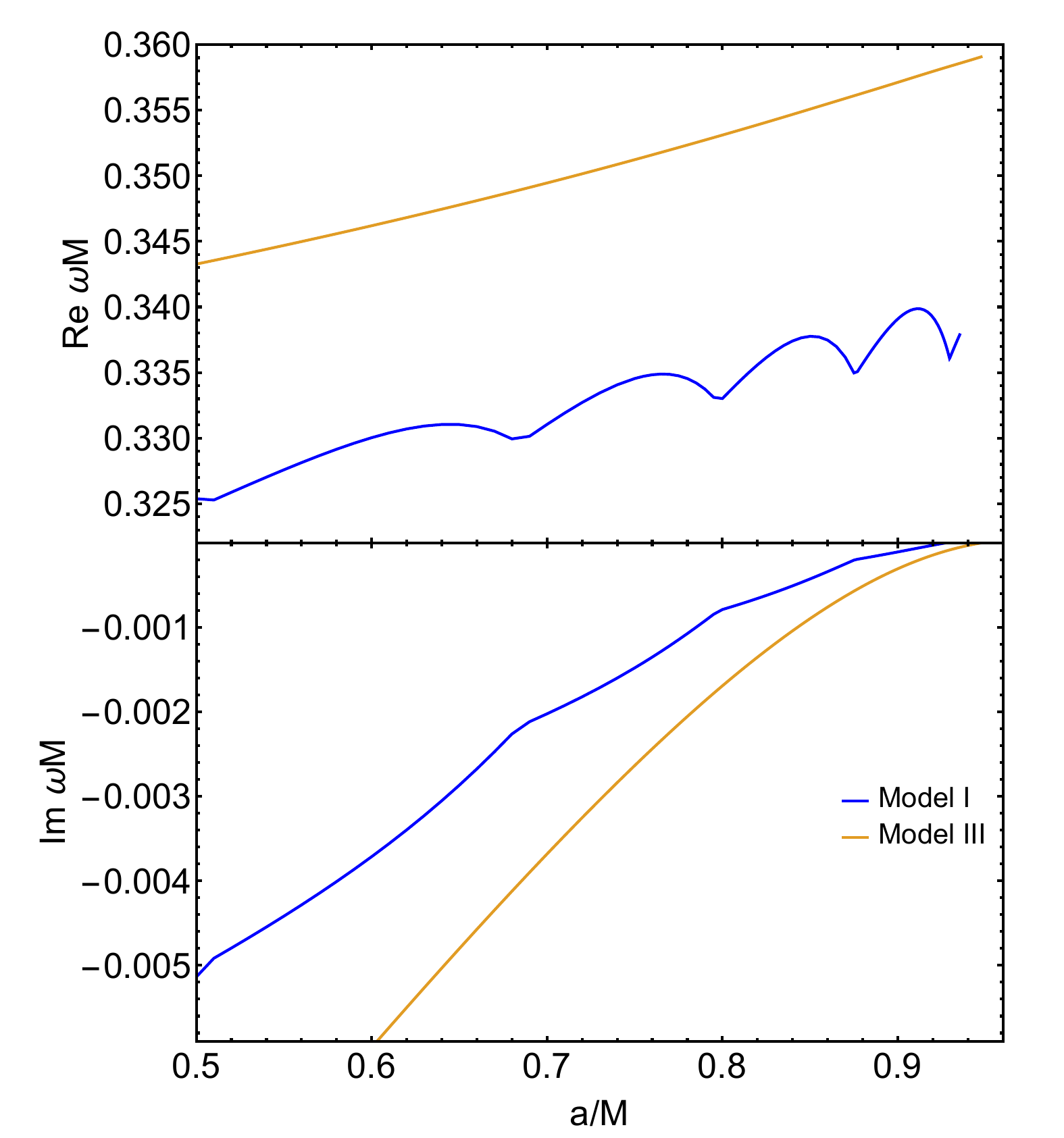}
\caption{Comparison between Model I (blue) and Model III (orange) for the real and imaginary parts of the frequency modes. The replacement of the step function with a sigmoid makes the profile become more regular and the corners disappear.}
\label{fig:ComparisonModes}
\end{figure}
We used Model I in order to quantitatively verify that only inside the cavity the coronal density is relevant for providing an additional effective mass. This is why we truncate the corona at $r=r_0$, i.e. where the disk begins. The resulting numerical solutions we found almost coincide with those of Model I, thus confirming that only the density inside the cavity is really relevant to increase the effective mass and, hence, to possibly quench the instability. 

Finally, in Model III we use a sigmoid-like function for replacing the step function of the inner edge, in order to show that the Heaviside function used in modelling the density profile is what produces the corners in both the real and imaginary parts shown in Fig. \ref{fig:Dima}. In Fig. \ref{fig:ComparisonModes} we show that when we described the barrier with a smooth sigmoid the corners disappear, and the resulting modes are also smooth functions of the model parameters.

\subsection{Model IV and V: role of the corona density}

In these models, by parametrizing the coronal density as $\rho_C=\gamma \rho_H$, we varied the parameter $\gamma$ in the realistic range $10^{-6}-10^{-1}$ \cite{Bisnovatyi-Kogan:1976fbc, DeVilliers:2003gr, Meyer-Hofmeister:2017ott} in order to see the impact of different coronal densities.
\begin{figure}
\begin{subfigure}[f]{1\textwidth}
\centering
\includegraphics[height=0.75\textwidth]{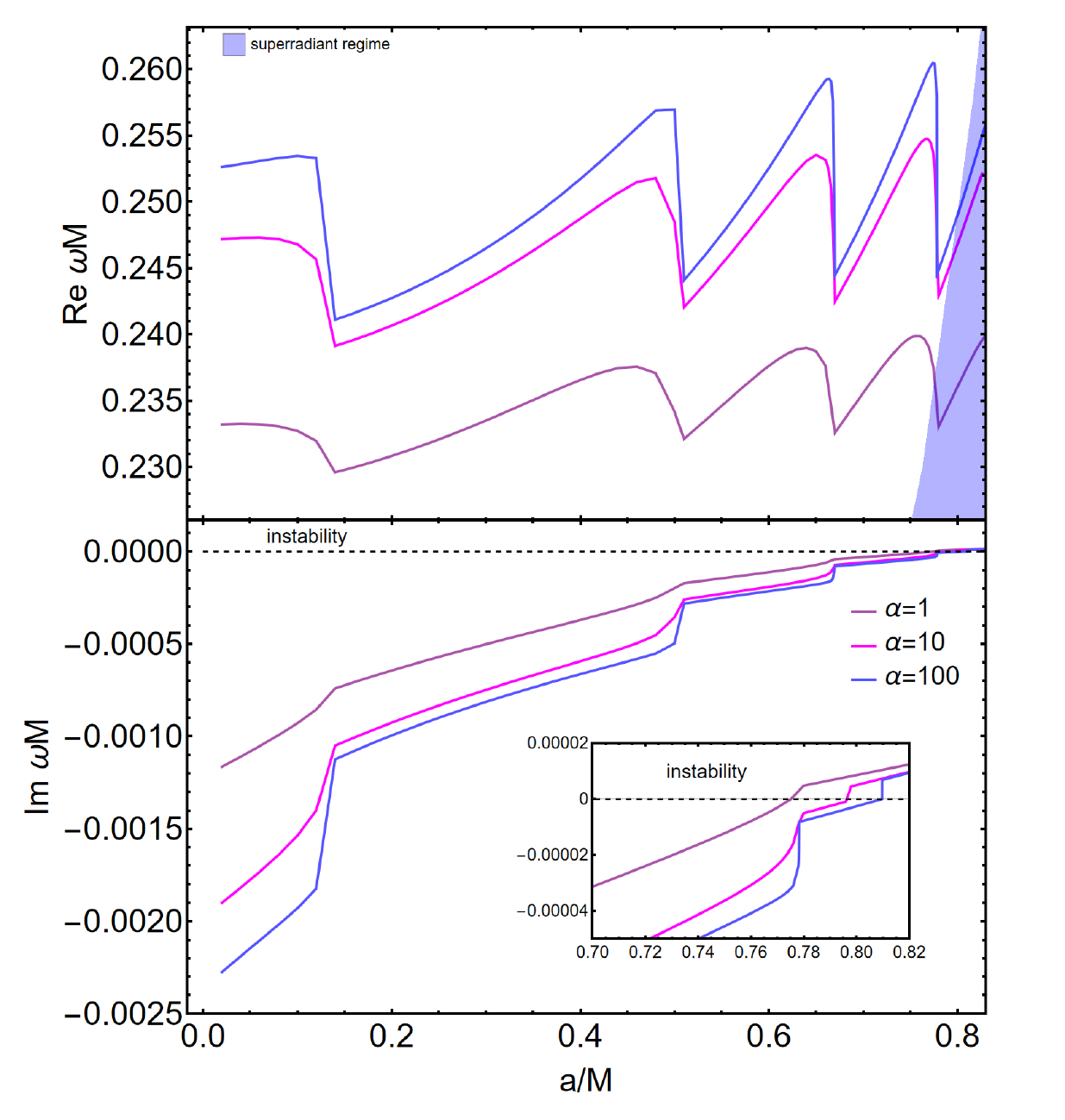}
\end{subfigure}
\begin{subfigure}[f]{1\textwidth}
\centering
\includegraphics[height=0.75\textwidth]{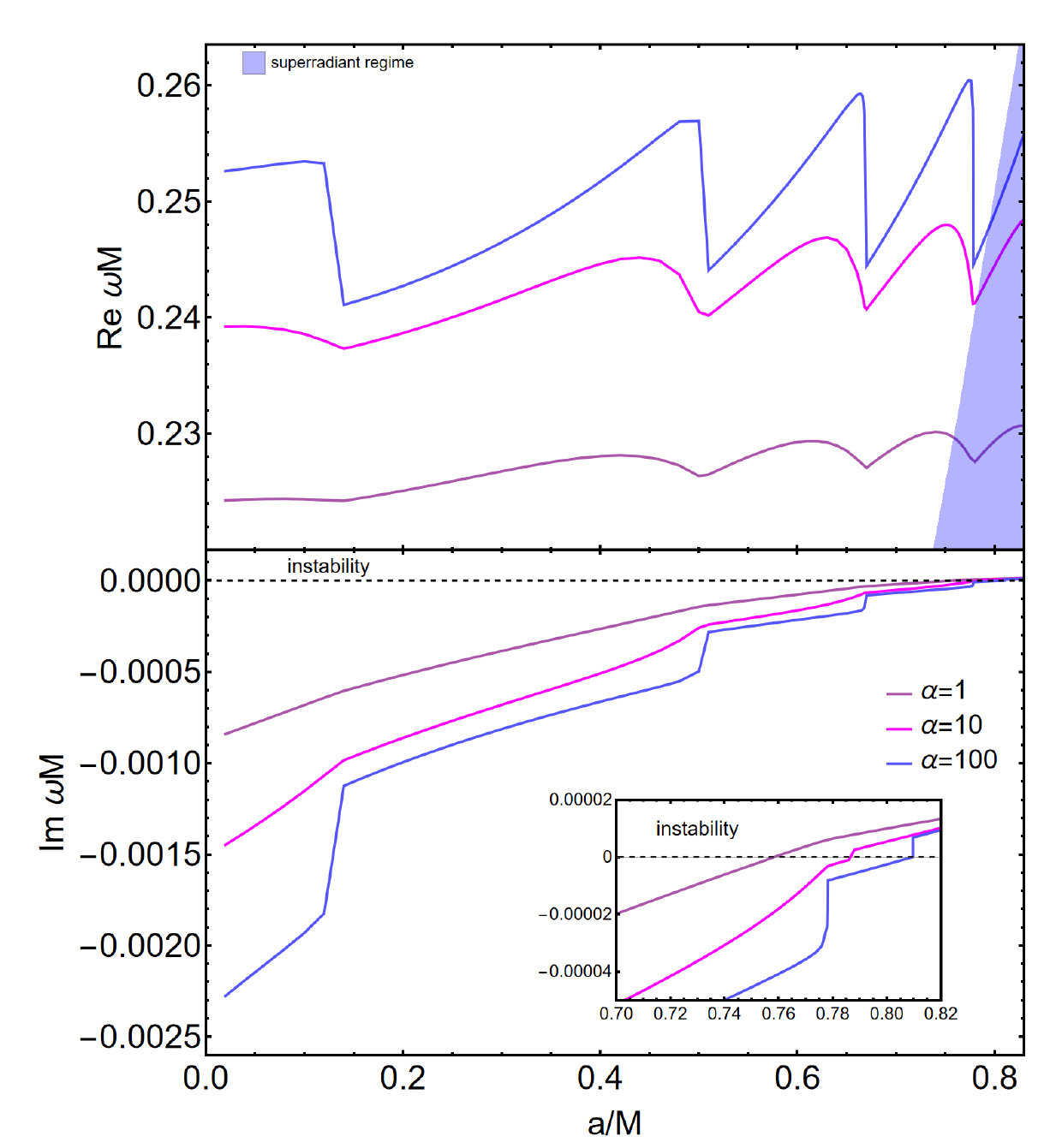}
\end{subfigure}
\caption{Model IV (up) and Model V (down) for $r_0=14M$ and $\gamma=10^{-6}$: the curves are functions of the dimensionless spin parameter, we show the results for different values of ${\alpha}$ and highlight the superradiant regime. Even by varying ${\alpha}$ across two orders of magnitude, the instability is preserved.}
\label{fig:realistic}
\end{figure}
In Figure \ref{fig:realistic} we show the imaginary part of the solutions for $\gamma=10^{-6}$, $r_0=14M$, $\rho_H=4/M^2$, that we got by varying the parameter ${\alpha}$ in Model IV and Model V. Despite the variation of ${\alpha}$ across two orders of magnitude, the instability is preserved with qualitatively similar features: the reason is that the coronal density is so low that it remains negligible, while the disk density is high enough to confine the modes in this range of $\alpha$. Hence, when the coronal density is strongly suppressed with respect to the disk one, we can have an instability in a wide range of values of $\alpha$. We also noted that assuming a larger truncation radius gives a smaller spin threshold for the instability. This happens in analogy to the original BH bomb phenomenon, where the real part of the frequency decreases with the truncation radius $\omega_R \sim 1/r_0$~\cite{Cardoso_superradiant_instability} (see Fig. \ref{fig:BHBomb}). 
\begin{figure}[h]
\centering
\includegraphics[width=1\textwidth]{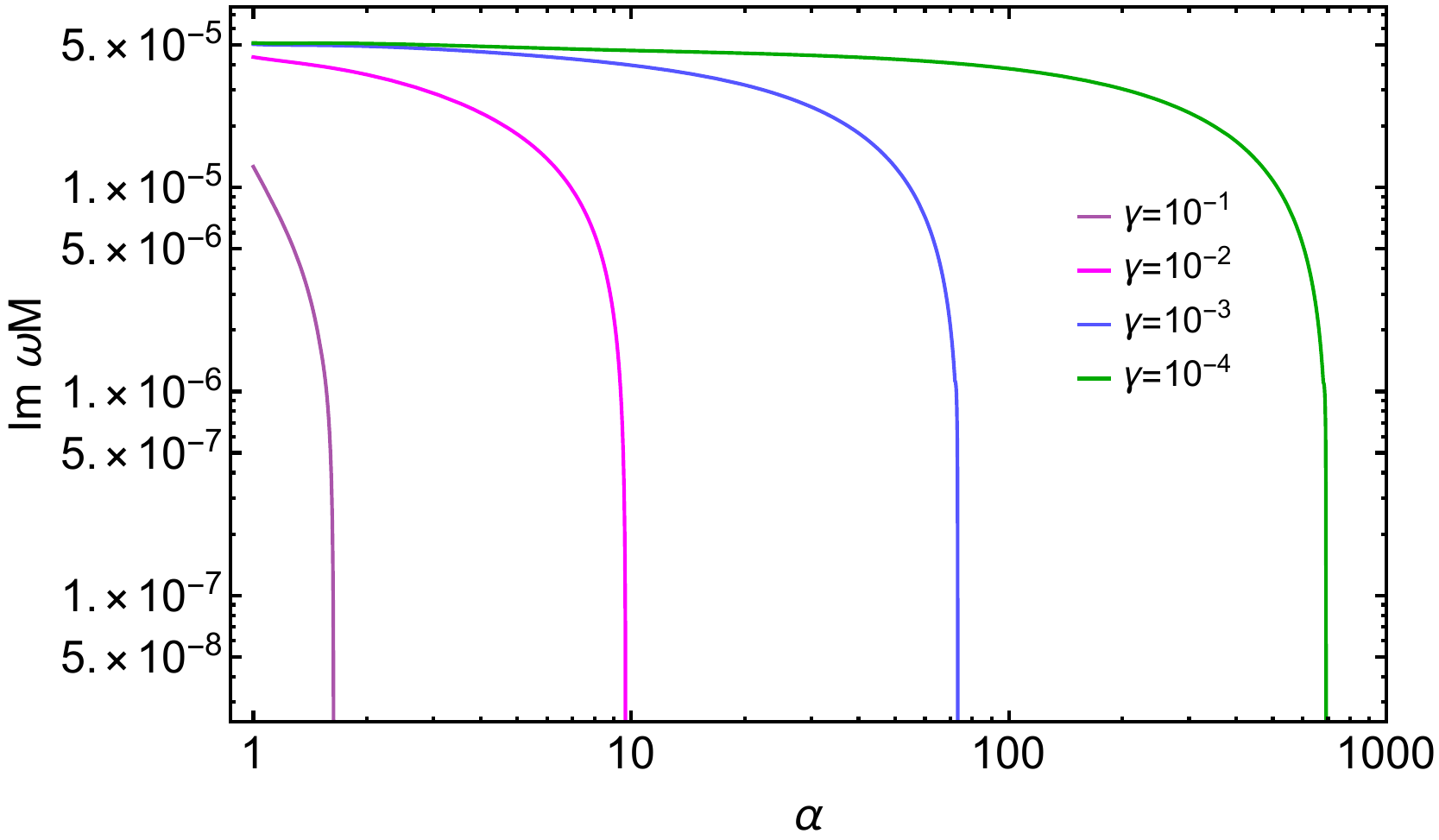}
\caption{Imaginary part of the mode frequency vs $\alpha$ in Model V (with $\rho_H=4/M^2$ and $r_0=8M$) for different values of the corona-disk density ratio $\gamma$ for a spinning BH with $a=0.97M$. The instability is suppressed if $\alpha\gamma\gtrsim O(10^{-1})$. Hence, the lower $\gamma$ it is, the more efficient the instability iwill be across several orders of magnitudes in $\alpha$.}
\label{fig:realistic_alpha1}
\end{figure}
Finally, in Fig. \ref{fig:realistic_alpha1} we show the imaginary part of the modes vs $\alpha$ for different density ratios $\gamma$ in Model V. Note that, for certain values of $\alpha$ (e.g. $\alpha\approx 1$ for the parameters chosen in Fig.~\ref{fig:realistic_alpha1}) the modes are independent of $\gamma$ in the $\gamma\ll1$ limit. This is because the coronal density in this regime is subdominant and does not affect the mode. On the other hand, as the $\alpha$ parameter grows, the coronal effective mass eventually becomes relevant and quenches the instability. In particular, for the chosen parameters the instability is suppressed when $\alpha\gamma\gtrsim O(10^{-1})$.

\section{Constraints on scalar-tensor theories from spinning BH superradiance}
\label{sec:contraints}

In the previous paragraph we explored the parameter space of our models and identified the salient features of the plasma-triggered superradiant instability in scalar-tensor theories. Now, thus, we are able to draw a general picture and apply it to the identification of the parameter space of scalar-tensor theories in which the instability is effective.

The first key requirement is a sufficiently dense disk extending down to some truncation radius $r_0>{\cal O}({\rm few})M$, as predicted in various models. The disk can efficiently confine scalar modes if the following condition is satisfied:
\begin{equation}
    \sqrt{\alpha \rho_H}M\gtrsim 1\,.\label{cond1}
\end{equation}
The typical outer density for a standard thin disk  is\cite{Shakura-Sunyaev, Barausse_environmental_effects}: 
\begin{equation}
    \rho \approx 169 \frac{f_{\rm Edd}^\frac{11}{20}}{(r/M)^\frac{15}{8}}\Bigg(1-\sqrt{\frac{r_0}{r}}\Bigg)^{\frac{11}{20}}\Bigg(\frac{0.1}{\beta}\Bigg)^{\frac{7}{10}}M_6^{-\frac{7}{10}}{\rm kg/m}^3 \, ,
\end{equation}
where $r_0$ is the truncation radius, $\beta$ is the viscosity parameter, $f_{\rm Edd}=\Dot{M}/\Dot{M}_{\rm Edd}$ is the mass accretion Eddington ratio. By using the above expression, Eq. \eqref{cond1} yields the following lower bound on the scalar coupling,
\begin{equation}
   \alpha \gtrsim \alpha_c=\frac{1}{\rho_H M^2}\approx 3\times 10^6 \left(\frac{M}{10^6 M_\odot}\right)^{-13/10}\,,
\end{equation}
i.e. supermassive BHs would yield a smaller lower bound.

The above condition we found is necessary but not sufficient. As a matter of fact, if we also have a corona with characteristic density $\rho_C=\gamma \rho_H$, we should also impose to have a not too large resulting effective mass inside the cavity, thus:
\begin{equation}
    \sqrt{\alpha \rho_C}M\lesssim1\,. \label{cond2}
\end{equation}
This last condition can be rewritten as an upper bound on the scalar coupling:
\begin{equation}
    \alpha\lesssim \frac{\alpha_c}{\gamma}\approx \frac{3}{\gamma}\times 10^6\left(\frac{M}{10^6 M_\odot}\right)^{-13/10}\,.
\end{equation}
Given that the corona density is way smaller than the disk's one, $\gamma\ll1$ and condition \ref{cond1} always partially overlaps with condition \ref{cond2}. In particular, provided a disk truncation not too close to the BH horizon, the superradiant instability can be triggered when
\begin{equation}
  3\times 10^6 \lesssim \alpha \left(\frac{M}{10^6 M_\odot}\right)^{13/10}\lesssim 3 \left(\frac{10^{-4}}{\gamma}\right) 10^{10}\,, \label{condF}
\end{equation}
where we have normalized the typical coronal density such that $\gamma=\rho_C/\rho_H=10^{-4}$.

Remarkably, from the theoretical bounds we found, we deduce that different ranges of $\alpha$ can potentially be constrained by different classes of BHs, extending roughly from $\alpha\sim {\cal O}(100)$ for $M\sim 10^9 M_\odot$ up to $\alpha\sim {\cal O}(10^{17})$ for $M\sim 5 M_\odot$. 
Moreover, as we have shown in the previous section, the instability time scale $\tau=1/\omega_I$ is typically very short if we compare it with astrophysical time scales. Thus, the instability can effectively change the dynamics of the system, given what is known in literature\cite{Brito_evolution_SR_instabilities,Brito_SR} about the phenomenology of BH superradiant instabilities in various systems.

This means that, provided we can accurately model the accretion flow, we can rule out scalar-tensor theories with positive couplings in a very wide range by putting constraints from the observation of highly-spinning accreting BHs. Interestingly, while scalar-tensor theories have stringent constraints for $\alpha<0$ coming from spontaneous scalarization and the absence of observed dipolar radiation in binary pulsars\cite{Cardoso:2013opa, Cardoso:2013fwa,Kramer:2021jcw}, the positive values $\alpha>0$ are essentially unconstrained while being also relevant for cosmology. 

In particular, models having $\alpha\gg1$ are of great interest. For example, in the symmetron model\cite{Hinterbichler:2010es} the conformal factor reads\footnote{We can consider this cosmological model because its bare-mass term and scalar self-interactions are negligible for astrophysical BHs \cite{Davis:2011pj,Davis:2014tea}, so the approximations we assumed are valid also in this case.} $A(\phi)=1+{\alpha} \phi^2/2$ and requiring screening for the Milky Way imposes ${\alpha} \gtrsim 10^6-10^8$\cite{Hinterbichler:2011ca, Davis:2014tea, deAguiar:2021bzg}, which perfectly lies in the range that we can potentially constrain thanks to accretion-driven BH superradiance.

\section{On the role of nonlinearities}\label{sec:nonlinear}

In the previous sections we have discovered how, in scalar-tensor theories, there is a wide range of parameter space prone to trigger matter-driven BH superradiant instabilities. Given that in the instability phase the scalar amplitude grows exponentially in a short timescale, linear theory eventually breaks down. Thus, it is important to understand the modifications introduced by non-linearities in the system.  We can do it by analysing the backreaction of the superradiantly growing scalar modes onto the plasma. From the conservation of its stress-energy tensor we deduce that, in the Jordan frame, matter particles will follow geodesics:
\begin{equation}
    \nabla_\nu T^{\mu\nu}=0 ~\rightarrow ~\frac{D u^\mu}{D\tau}=u^\nu \nabla_\nu u^\mu=0\,,
\end{equation}
where $u^\mu$ is the four-velocity of plasma in the Jordan frame.
By switching to the Einstein frame, we can rewrite this equation in the following form\cite{Fujii:2003pa}:
\begin{equation}
    \frac{D u_E^\mu}{D\tau_E}=f_\nu u_E^\nu u_E^\mu-f_E^\mu (u_{E\,\mu} {u_E}^\mu)\,,
\end{equation}
where $u_E^\mu=d x^\mu /d\tau_E$ and $\tau_E$ are the four-velocity and proper time in the Einstein frame, respectively, while $f_\nu=-\partial_\nu \ln A(\Phi)$ and $f_E^\mu=g_E^{\mu\nu}f_{\nu}$. By expanding the conformal factor around the background as we did before, $\Phi_E\sim \Phi_E^{(0)}$, we can rewrite this equation to the leading order as
\begin{equation}
    \frac{D u_E^\mu}{D\tau_E}= - \alpha\Big(\varphi \partial_\nu \varphi u_E^\mu u_E^\mu - g_E^{\mu\alpha}\varphi \partial_\alpha \varphi ({u_E}^\nu u_{E\,\nu})\Big)\,.
\end{equation}
We can observe from this equation that, in the Einstein frame, the acceleration of the plasma particles depends on non-linear terms in the scalar field $\varphi$, with coupling constant $\alpha$. By solving it, we can then relate the backreaction on the four-velocity with the one on the density by exploiting the continuity equation of the fluid. Therefore, non-linear effects can modify the density of the fluid, which evolves dynamically. The details on the non-linear evolution depend on the specific models due to the higher-order scalar interactions.

Nevertheless, and most importantly, this system is not affected by another non-linear effect, the relativistic transparency, which kills plasma-driven superradiant instabilities in GR \cite{cardoso_plasma_tests}. Due to this specific non-linear correction, we get a modification to the effective photon mass in plasma \cite{relativistic_nonlinear_propagation_plasma,strong_EM_waves_in_plasma,cardoso_plasma_tests}:
\begin{equation}
    \omega_p^2=\frac{4\pi e^2 n}{m_e\sqrt{1+\frac{e^2 E^2}{m_e^2 \omega^2}}}.
\end{equation}
In this case, in the presence of large-amplitude electric fields the effective mass becomes negligible, thus dramatically suppressing plasma-driven GR instabilities before any significant amount of energy can be extracted from the compact object\cite{cardoso_plasma_tests}. We can interpret this effect as an increase of the relativistic electron mass-energy, and, thus, this effect differs totally from the field backreaction on the density distribution. We show how in scalar-tensor theories the effective mass is not affected by this type of suppression. As a matter of fact, in this system the effective mass is the trace of the stress-energy tensor $T^{\mu\nu}=\rho u^\mu u^\nu$.
The crucial point is that the trace of this tensor is always the rest-mass density, no matter what the fluid four-velocity is, given that $u_\mu u^\mu=1$ is a relativistic invariant. Thus, the trace of a tensor is a scalar quantity and therefore must be invariant under Lorentz boosts. Hence, no Lorentz boost factor enters in the expression of the effective scalar mass in the relativistic non-linear regime, at variance with the standard case of plasma-photon interactions in GR.
Consequently, even when the plasma is accelerated to relativistic velocities, the effective mass does not change its expression, although the density becomes a dynamical quantity. 

\chapter*{Conclusions and future prospects}
\addcontentsline{toc}{chapter}{Conclusions and future prospects}

\section*{The numerical technique and its application in General Relativity}
\addcontentsline{toc}{section}{The numerical technique and its application in General Relativity}

In this work we have studied in detail and extended to the tensor case the numerical technique developed by \textit{Baumann et al.} in \cite{Spectra_grav_atom} for the computation of QBSs of massive bosonic perturbations in spinning BH spacetimes in GR. 

We succeeded in developing a fully general-relativistic formalism able to deal with the multipolar decomposition of massive tensor perturbations propagating in spinning BH spacetimes (Chapter \ref{chapter:spherical_harmonic}), extending \textit{Baumann et al.}'s formalism for vector radiation. The technique has the great advantage of not relying on any separation ansätze, therefore it can potentially be modified for the application to any problem involving the computation of QBSs of linear fields in curved spacetime. Massive spin-2 fields in Kerr spacetimes feature field equations whose separability is currently unknown, thus we designed the modifications so that the technique can be applied also to this problem.

The numerical method is able to give precise results with few spherical harmonic modes, but, given that it is based on spherical harmonic decompositions, it requires the computation of a huge number of overlap integrals giving the couplings among angular momentum states. This issue is the reason why we had to look for some technique parametrizing the spherical harmonic couplings (Section \ref{section:overlap_parametrization}), in order not to have to compute them every time we change the values of the parameters characterizing the solution to be found.

We have been able to reproduce known solutions for massive scalar and vector QBSs (Chapter \ref{chapter:numerical_SR_GR}), while the results for the tensor case have not been completed yet. In a preliminary phase of the computation, we have been able to reproduce some monopolar non-spinning states of the massive tensor radiation, thus giving some consistency check to the technique, but failed in getting accurate spinning BH states. We noticed that the issue was related to not having substituted the trace-less constraint $h_{\mu\nu}{\bar g}^{\mu\nu}=0$ into the other spin-2 field equations: hence, we changed the numerical method accordingly, but this operation has greatly increased the expected computation time due to the consequent involved expressions. In this context, the parametrization we found for the computation of spherical harmonic couplings is pivotal for getting results also for massive tensor states, thus we expect we will have them soon.

In Section \ref{tensor_numerical_results} we have shown the results for massive spin-2 states got in \cite{dias2023black} through an alternative technique developed by our collaborators, the first ever to confirm the results found in \cite{Brito_spin-2_SR_slow_rot} regarding the fundamental polar dipolar mode, which is expected to be the most unstable state of massive spin-2 fields (giving the highest instability among all bosonic ultralight fields in general). We expect to confirm this result soon with our technique, and probe the possibility of computing also the spinning monopolar states, now missing from the results in \cite{dias2023black}. This result gives the chance of getting tighter observational constraints for ultralight spin-2 bosons through BH and GWs observations, therefore opening the possibility of discarding or confirming the existence of massive spin-2 fields.

The numerical technique used by our collaborators in \cite{dias2023black} differs from the one we developed, given that it discretizes both $r$ and $\theta$ coordinates through pseudospectral methods. The advantage they get with respect to what we do is not having to deal with any overlap integral, thus requiring less resources and time for the computations. Although, in their case there is the disadvantage of being able to design a proper 2-dimensional ansatz every time the technique is applied to a different system, while in our case finding a consistent ansatz is way more easier\footnote{We work just with the radial coordinate.}.

The numerical technique we have been using requires large resources: given the experience gathered by applying it to the standard GR massive scalar, vector and tensor fields, we do not think it can be applied to scenarios more involved than the spin-2 case, which we think it has to be considered as the upper limit of what is possible through this path. Thus, more involved scenarios might need to apply methods not involving spherical harmonics, such as the pseudospectral ones in $r$ and $\theta$, but there can still be possibilities in exploring also environmental effects\cite{Barausse_environmental_effects, cardoso_plasma_tests}\footnote{Especially if we consider non-minimally coupled fields in beyond-GR theories.} if the perturbations are nothing more than (few) scalar and vector fields.

We expect to give more precise insights on the advantages and disavantages of this technique once the computation of spin-2 modes will be completed.

\section*{Application to "dirty" beyond-General-Relativity scenarios}
\label{sec:discussion_plasma}
\addcontentsline{toc}{section}{Application to "dirty" beyond-General-Relativity scenarios}

In Chapter \ref{chapter:plasma_SR} we applied \textit{Baumann et al.}'s numerical technique to a beyond-GR scenario which involves matter non-minimally coupled to an extra scalar field, arising from the interaction between scalar-tensor theories and plasma in a spinning BH spacetime.

Thus, we have studied in detail the phenomenon of matter-driven BH superradiant instabilities in this context, by considering arbitrarily spinning BHs and realistic models of truncated thin and thick accretion disks. In general the linearized scalar equation is non-separable, therefore it is exactly the type of problem that requires techniques such as the one developed by \textit{Baumann et al.}
We found two interesting results:
\begin{itemize}
    \item{ although the qualitative features of the instability are akin to the case of plasma-driven electromagnetic superradiant instabilities within GR, the obstacles preventing the latter (namely suppression due to the corona~\cite{Dima_2020} and non-linearities~\cite{cardoso_plasma_tests}) can be circumvented in scalar-tensor theories;}
    \item{remarkably, there exists a very wide range of (positive and large) scalar couplings where BH superradiant instabilities can be triggered in realistic scenarios.}
\end{itemize} 
The range triggering superradiance is unconstrained by observations and it actually includes the regime where certain scalar-tensor alternatives to the dark energy, e.g. symmetron models with screening, can evade solar system constraints while remaining cosmologically viable.

Our results suggest that such theories could be ruled out as dark-energy alternatives by the observation of highly spinning BHs, using the same technique adopted to constrain ultralight bosons from BH mass-spin observations~\cite{string_axiverse,string_axiverse_BH,Brito_SR}.

However, at variance with the ultralight boson case, here an accurate modelling of the accretion flow around the BH is needed in order to quantitatively characterize the instability.

Furthermore, the possibility of circumventing nonlinear damping effects suggests that the models proposed for ordinary plasma-driven instabilities (e.g. as a possible explanation for fast radio bursts~\cite{Conlon:2017hhi} or for constraints on primordial BHs~\cite{Pani:2013hpa}) could actually work in the context of scalar-tensor theories.

Although the quantitative features of the instability depend on the geometry of the accretion flow near a BH, the key ingredients are naturally predicted in various models:
\begin{itemize}
    \item{a sufficiently dense disk with a sharp transition from a low-density to a high-density region in the vicinity of the ISCO;}
    \item{a sufficiently tenuous corona in the low-density region, such that its density is much smaller than the one of the disk;}
    \item{a BH spinning sufficiently fast to make the quasi-bound modes unstable against the superradiant instability.}
\end{itemize}

The numerical method implemented to compute the unstable modes in the absence of separable equations is general and robust, and could find applications in other contexts, such as non-linear plasma interactions in the presence of dark photons (which could provide another testing ground for the parametrizations of spherical harmonic overlaps).

Another interesting finding is the fact that the unstable modes of this system resemble a quasi-bound state in the vicinity of the BH but are in fact propagating waves far from it. Therefore, one could imagine situations in which (perhaps during the superradiant growth) the quasi-bound states are not efficiently trapped and could propagate to infinity, possibly after several reflections within the cavity.
The scalar modes in the Einstein frame correspond to a (breathing) scalar polarization of the gravitational waves in the Jordan frame. Therefore, the phenomenology of this effect would be similar to the gravitational-wave echoes predicted for matter fields~\cite{Barausse_environmental_effects}, near-horizon structures~\cite{Cardoso:2016rao}, and exotic compact objects~\cite{Cardoso_ECOs}. A more detailed study of this interesting phenomenon, that we leave to the future, will probably require a time-domain analysis. 

Finally, an important follow-up of this work from this side is to study backreaction effects on the plasma and the full dynamics of the system at the nonlinear level. 

\begin{appendix}
\renewcommand{\theequation}{\thechapter.\arabic{section}.\arabic{equation}}
\chapter{Expressions for the spherical harmonic decomposition}
\section{Couplings in spin-1 and spin-2 wave equations}\label{couplings_wave_eqs}
Here we report the explicit expressions of all the couplings appearing in Equations \ref{radial_wave_eqs}. We will use the spinor components of the spin-eigenstates, i.e. 
\begin{subequations}
 \begin{align}
    &\tau^J=\tau^\mu \left(\tilde\theta_\mu^J\right)^*~,~\tilde\tau_J=\tilde\tau_\mu \left(\theta^\mu_J\right)^*\,,\\    &\tau^J_{s,s_z}=\tau^{\mu\nu}_{s,s_z}\left(\tilde\Theta_{\mu\nu}^J\right)^*~,~\tilde\tau_J^{s,s_z}=\tilde\tau_{\mu\nu}^{s,s_z}\left(\Theta^{\mu\nu}_J\right)^*\,,\\
    &\chi^J_{s,s_z}=\delta_{s,1}\tilde\chi^\mu_{s,s_z} \left(\tilde\theta_\mu^J\right)^*+(1-\delta_{s,1})\tilde\chi^{\mu\nu}_{s,s_z} \left(\tilde\Theta_{\mu\nu}^J\right)^*\,,\\
    &\tilde\chi_J^{s,s_z}=\delta_{s,1}\tilde\chi_\mu^{s,s_z} \left(\theta^\mu_J\right)^*+(1-\delta_{s,1})\chi_{\mu\nu}^{s,s_z} \left(\Theta^{\mu\nu}_J\right)^*\,.
\end{align}
\end{subequations}
For the vector case ($S=1$), the expressions are the following:
\begin{subequations}
{\allowdisplaybreaks[4]
\vspace{-0.5cm}\begin{align}
    &\nonumber\begin{aligned}
    \end{aligned}\\
    &\label{gamma_mixing_s1}
    \begin{aligned}
    \Gamma_{l,s,j}^{l',s',j'}(r)=&-\frac{1}{\zeta'^2(r)}\int d\Omega \left(Y^{(1)I}_{l,s,j,j_z}\right)^* \left[S_I^{~J}+ Q_I^{~J}\mathbfcal{L}_{\boldsymbol z} +R_I^{~J}\frac{F'(r)}{F(r)}\right]\tilde{Y}_{(1)J}^{l',s',j',j_z}=\\
    =&-\frac{1}{\zeta'^2(r)}\sum\limits_{m,s_z, m',s_z'}\left\langle l,m,s,s_z \right| \left. j , j_z \right\rangle\left\langle l',m',s',s_z' \right| \left. j' , j_z \right\rangle   \left[\delta_{0,s} \tau^I\right.\\
    &+\left.\left(1-\delta_{0,s}\right) \chi^I_{s,s_z} \right] \left[\delta_{0,s'} \tilde{\tau}_J+\left(1-\delta_{0,s'}\right) \tilde{\chi}_J^{s',s_z'} \right]\\
    &\times\left[\left\langle l,m \right|\left(S_I^{~J}+R_I^{~J}\frac{F'(r)}{F(r)}\right)\left| l',m'\right\rangle+ m \delta^{l'}_l \delta^{m'}_m Q_I^{~J}\right]\,,
    \end{aligned}\\
    &\label{lambda_mixing_s1}
    \begin{aligned}
    \Lambda_{l,s,j}^{l',s',j'}(r)=&-\frac{1}{\zeta'^2(r)}\int d\Omega \left(Y^{(1)J}_{l,s,j,j_z}\right)^* \frac{a^2 \cos^2\theta(\mu_V^2-\omega^2)}{\Delta(r)}\tilde{Y}_{(1)J}^{l',s',j',j_z}=\\
    =&-\frac{\delta^{s'}_s a^2(\mu_V^2-\omega^2)}{\zeta'^2(r)\Delta(r)}\sum\limits_{m,s_z}\left\langle  l,m,s,s_z\right. \left|j, j_z \right\rangle\left\langle l',m,s,s_z\right.\left|j',j_z\right\rangle c^m_{l l'}\,,
    \end{aligned}\\
    &\label{D_mixing_s1}
    \begin{aligned}
    D_{l,s,j}^{l',s',j'}(r)=&-\frac{1}{\zeta'^2(r)}\int d\Omega \left(Y^{(1)I}_{l,s,j,j_z}\right)^* \left[P_I^{~J}\boldsymbol{D_+}+Z_I^{~J}\boldsymbol{D_0}+M_I^{~J}\boldsymbol{D_-}\right]\tilde{Y}_{(1)J}^{l',s',j',j_z}\\
    =&-\frac{1}{\zeta'^2(r)}\sum\limits_{m,s_z, m',s_z'}\left\langle l,m,s,s_z \right| \left. j , j_z \right\rangle\left\langle l',m',s',s_z' \right| \left. j' , j_z \right\rangle\\
    &\times\left[\sqrt{\frac{(l'+1)^2(l'- m')(l'- m'-1)}{2(2 l'+1)(2 l'-1)}}\left\langle l,m \right|P_I^{~J}\left| l'-1,m'+1\right\rangle\right.\\
    &+\sqrt{\frac{l'^2(l'+ m'+1)(l'+ m'+2)}{2(2 l'+3)(2 l'+1)}} \left\langle l,m \right|P_I^{~J} \left| l'+1,m'+1\right\rangle\\
    &+\sqrt{\frac{l'^2(l'+ m'+1)(l'-m'+1)}{(2 l'+1)(2 l'+3)}}\left\langle l,m \right|Z_I^{~J}\left| l'+1,m'\right\rangle\\
    &-\sqrt{\frac{(l'+1)^2(l'+ m')(l'- m')}{(2 l'+1)(2 l'-1)}} \left\langle l,m \right|Z_I^{~J} \left| l'-1,m'\right\rangle\\
    &+\sqrt{\frac{(l'+1)^2(l'+m')(l'+ m'-1)}{2(2 l'+1)(2 l'-1)}}\left\langle l,m \right|M_I^{~J}\left| l'-1,m'-1\right\rangle\\
    &\left.+\sqrt{\frac{l'^2(l'- m'+1)(l'- m'+2)}{2(2 l'+3)(2 l'+1)}} \left\langle l,m \right|M_I^{~J} \left| l'+1,m'-1\right\rangle \right]\\
    &\times \left[\delta_{0,s'} \tilde{\tau}_J+\left(1-\delta_{0,s'}\right) \tilde{\chi}_J^{s',s_z'} \right]\left[\delta_{0,s} \tau^I+\left(1-\delta_{0,s}\right) \chi^I_{s,s_z} \right] \,,
    \end{aligned}\\
    &\label{c1_mixing_s1}
    \begin{aligned}
    C_{1,l,s,j}^{l',s',j'}(r)=&\delta^{l'}_l\delta^{s'}_s\delta^{j'}_j\left[\left(\frac{1}{r-r_+}+\frac{1}{r-r_-} \right)\frac{1}{\zeta'(r)}+\frac{1}{\zeta'(r)}\frac{2 F'(r)}{F(r)}\right.\\
    &+\left.\frac{\zeta''(r)}{\zeta'^2(r)}\right] -\frac{1}{\zeta'(r)}\int d\Omega \left(Y^{(1)I}_{l,s,j,j_z}\right)^* R_I^{~J}\tilde{Y}_{(1)J}^{l',s',j',j_z}=\\
    =&\delta^{l'}_l\delta^{s'}_s\delta^{j'}_j\left[\left(\frac{1}{r-r_+}+\frac{1}{r-r_-} \right)\frac{1}{\zeta'(r)}+\frac{1}{\zeta'(r)}\frac{2 F'(r)}{F(r)}+\frac{\zeta''(r)}{\zeta'^2(r)}\right]\\
    &-\frac{1}{\zeta'(r)}\sum\limits_{m,s_z, m',s_z'}\left\langle l,m,s,s_z \right| \left. j , j_z \right\rangle\left\langle l',m',s',s_z' \right| \left. j' , j_z \right\rangle   \left[\delta_{0,s} \tau^I\right.\\
    &\left.+\left(1-\delta_{0,s}\right) \chi^I_{s,s_z} \right]\left[\delta_{0,s'} \tilde{\tau}_J+\left(1-\delta_{0,s'}\right) \tilde{\chi}_J^{s',s_z'} \right]\left\langle l,m \right|R_I^{~J}\left| l',m'\right\rangle\,,
    \end{aligned}\\
    &\label{c2_mixing_s1}
    \begin{aligned}
    C_{2,l,s,j}^{l',s',j'}(r)=&\frac{\delta^{l'}_l\delta^{s'}_s\delta^{j'}_j}{\zeta'^2(r)}\left[\frac{ F''(r)}{F(r)}+\left(\frac{1}{r-r_+}+\frac{1}{r-r_-} \right)\frac{F'(r)}{F(r)}+\frac{P_+^2}{(r-r_+)^2}\right.\\
    &\left.+\frac{P_-^2}{(r-r_-)^2}-\frac{A_+}{(r_+-r_-)(r-r_+)}+\frac{A_-}{(r_+-r_-)(r-r_-)}\right.\\
    &\left.-(\mu_V^2-\omega^2)-\frac{l(l+1)}{\Delta(r)}\right]\,.
    \end{aligned}
\end{align}}
\end{subequations}
The couplings for the tensor case ($S=2$), instead, read:
\begin{subequations}
{\allowdisplaybreaks[4]
\vspace{-0.5cm}\begin{align}
    &\nonumber\begin{aligned}
    \end{aligned}\\
    &\label{gamma_mixing_s2}
    \begin{aligned}
    \Gamma_{l,s,j}^{l',s',j'}(r)=&-\frac{1}{\zeta'^2(r)}\int d\Omega \left(Y^{(2)I}_{l,s,j,j_z}\right)^* \left[\tilde S_I^{~J}+\tilde Q_I^{~J}\mathbfcal{L}_{\boldsymbol z} +\tilde R_I^{~J}\frac{F'(r)}{F(r)}\right]\tilde{Y}_{(2)J}^{l',s',j',j_z}\\
    =&-\frac{1}{\zeta'^2(r)}\sum\limits_{m,s_z, m',s_z'}\left\langle l,m,s,s_z \right| \left. j , j_z \right\rangle\left\langle l',m',s',s_z' \right| \left. j' , j_z \right\rangle   \left[\delta_{s,1} \tau^I_{s,s_z}\right.\\
    &+\left.\left(1-\delta_{s,1}\right) \chi^I_{s,s_z} \right] \left[\delta_{s',1} \tilde{\tau}_J^{s',s_z'}+\left(1-\delta_{s',1}\right) \tilde{\chi}_J^{s',s_z'} \right]\\
    &\times\left[\left\langle l,m \right|\left(\tilde S_I^{~J}+\tilde R_I^{~J}\frac{F'(r)}{F(r)}\right)\left| l',m'\right\rangle+ m \delta^{l'}_l \delta^{m'}_m \tilde Q_I^{~J}\right]\,,
    \end{aligned}\\
    &\label{lambda_mixing_s2}
    \begin{aligned}
    \Lambda_{l,s,j}^{l',s',j'}(r)=&-\frac{1}{\zeta'^2(r)}\int d\Omega \left(Y^{(2)J}_{l,s,j,j_z}\right)^* \frac{a^2 \cos^2\theta(\mu_T^2-\omega^2)}{\Delta(r)}\tilde{Y}_{(2)J}^{l',s',j',j_z}=\\
    =&-\frac{\delta^{s'}_s a^2(\mu_T^2-\omega^2)}{\zeta'^2(r)\Delta(r)}\sum\limits_{m,s_z}\left\langle  l,m,s,s_z\right. \left|j, j_z \right\rangle\left\langle l',m,s,s_z\right.\left|j',j_z\right\rangle c^m_{l l'}\,,
    \end{aligned}\\
    &\label{D_mixing_s2}
    \begin{aligned}
    D_{l,s,j}^{l',s',j'}(r)=&-\frac{1}{\zeta'^2(r)}\int d\Omega \left(Y^{(2)I}_{l,s,j,j_z}\right)^* \left[\tilde P_I^{~J}\boldsymbol{D_+}+\tilde Z_I^{~J}\boldsymbol{D_0}+\tilde M_I^{~J}\boldsymbol{D_-}\right]\tilde{Y}_{(2)J}^{l',s',j',j_z}\\
    =&-\frac{1}{\zeta'^2(r)}\sum\limits_{m,s_z, m',s_z'}\left\langle l,m,s,s_z \right| \left. j , j_z \right\rangle\left\langle l',m',s',s_z' \right| \left. j' , j_z \right\rangle\\
    &\times\left[\sqrt{\frac{(l'+1)^2(l'- m')(l'- m'-1)}{2(2 l'+1)(2 l'-1)}}\left\langle l,m \right|\tilde P_I^{~J}\left| l'-1,m'+1\right\rangle\right.\\
    &+\sqrt{\frac{l'^2(l'+ m'+1)(l'+ m'+2)}{2(2 l'+3)(2 l'+1)}} \left\langle l,m \right|\tilde P_I^{~J} \left| l'+1,m'+1\right\rangle\\
    &+\sqrt{\frac{l'^2(l'+ m'+1)(l'-m'+1)}{(2 l'+1)(2 l'+3)}}\left\langle l,m \right|\tilde Z_I^{~J}\left| l'+1,m'\right\rangle\\
    &-\sqrt{\frac{(l'+1)^2(l'+ m')(l'- m')}{(2 l'+1)(2 l'-1)}} \left\langle l,m \right|\tilde Z_I^{~J} \left| l'-1,m'\right\rangle\\
    &+\sqrt{\frac{(l'+1)^2(l'+m')(l'+ m'-1)}{2(2 l'+1)(2 l'-1)}}\left\langle l,m \right|\tilde M_I^{~J}\left| l'-1,m'-1\right\rangle\\
    &\left.+\sqrt{\frac{l'^2(l'- m'+1)(l'- m'+2)}{2(2 l'+3)(2 l'+1)}} \left\langle l,m \right|\tilde M_I^{~J} \left| l'+1,m'-1\right\rangle \right]\\
    &\times \left[ \tilde{\tau}_J^{s',s_z'}+(1-\delta_{s',1}) \tilde{\chi}_J^{s',s_z'} \right]\left[\delta_{s,1} \tau^I_{s,s_z}+(1-\delta_{s,1}) \chi^I_{s,s_z} \right]\,,
    \end{aligned}\\
    &\label{c1_mixing_s2}
    \begin{aligned}
    C_{1,l,s,j}^{l',s',j'}(r)=&\delta^{l'}_l\delta^{s'}_s\delta^{j'}_j\left[\left(\frac{1}{r-r_+}+\frac{1}{r-r_-} \right)\frac{1}{\zeta'(r)}+\frac{1}{\zeta'(r)}\frac{2 F'(r)}{F(r)}\right.\\
    &+\left.\frac{\zeta''(r)}{\zeta'^2(r)}\right] -\frac{1}{\zeta'(r)}\int d\Omega \left(Y^{(2)I}_{l,s,j,j_z}\right)^* \tilde R_I^{~J}\tilde{Y}_{(2)J}^{l',s',j',j_z}=\\
    =&\delta^{l'}_l\delta^{s'}_s\delta^{j'}_j\left[\left(\frac{1}{r-r_+}+\frac{1}{r-r_-} \right)\frac{1}{\zeta'(r)}+\frac{1}{\zeta'(r)}\frac{2 F'(r)}{F(r)}+\frac{\zeta''(r)}{\zeta'^2(r)}\right]\\
    &-\frac{1}{\zeta'(r)}\sum\limits_{m,s_z, m',s_z'}\left\langle l,m,s,s_z \right| \left. j , j_z \right\rangle\left\langle l',m',s',s_z' \right| \left. j' , j_z \right\rangle   \left[\delta_{s,1} \tau^I_{s,s_z}\right.\\
    &\left.+(1-\delta_{s,1}) \chi^I_{s,s_z} \right]\left[\delta_{s',1} \tilde{\tau}_J^{s',s_z'}+(1-\delta_{s,1}) \tilde{\chi}_J^{s',s_z'} \right]\left\langle l,m \right|\tilde R_I^{~J}\left| l',m'\right\rangle\,,
    \end{aligned}\\
    &\label{c2_mixing_s2}
    \begin{aligned}
    C_{2,l,s,j}^{l',s',j'}(r)=&\frac{\delta^{l'}_l\delta^{s'}_s\delta^{j'}_j}{\zeta'^2(r)}\left[\frac{ F''(r)}{F(r)}+\left(\frac{1}{r-r_+}+\frac{1}{r-r_-} \right)\frac{F'(r)}{F(r)}+\frac{P_+^2}{(r-r_+)^2}\right.\\
    &\left.+\frac{P_-^2}{(r-r_-)^2}-\frac{A_+}{(r_+-r_-)(r-r_+)}+\frac{A_-}{(r_+-r_-)(r-r_-)}\right.\\
    &\left.-(\mu_T^2-\omega^2)-\frac{l(l+1)}{\Delta(r)}\right]\,.
    \end{aligned}
\end{align}}
\end{subequations}

\section{Couplings in spin-1 and spin-2 Lorenz constraints}\label{couplings_lorenz}
Here we show the full expressions of the couplings appearing in \ref{radial_lorenz_s1} and \ref{radial_lorenz_s2}:
\begin{subequations}
{\allowdisplaybreaks[4]
\begin{align}
    &\label{c3_constr_s1}
    \begin{aligned}
    C_{3,j}^{l',s',j'}(r)=&\int d\Omega \left(Y_{j,j_z}\right)^* \rho^{J}\tilde{Y}_{(1)J}^{l',s',j',j_z}=\\
    =&\sum\limits_{m',s_z'}\left\langle l',m',s',s_z' \right| \left. j' , j_z \right\rangle\\
    &\times\left[\delta_{0,s'} \tilde{\tau}_J+\left(1-\delta_{0,s'} \right) \tilde{\chi}_J^{s',s_z'} \right]\left\langle j,j_z \right|\rho^{J}\left| l',m'\right\rangle\,,
    \end{aligned}\\
    &\label{c3_constr_s2}
    \begin{aligned}
    C_{3,l,s,j}^{l',s',j'}(r)=&\int d\Omega \left(Y^{(1)I}_{l,s,j,j_z}\right)^* \tilde\rho_I^{~J}\tilde{Y}_{(2)J}^{l',s',j',j_z}=\\
    =&\sum\limits_{m,s_z, m',s_z'}\left\langle l,m,s,s_z \right| \left. j , j_z \right\rangle\left\langle l',m',s',s_z' \right| \left. j' , j_z \right\rangle\left\langle l,m \right|\tilde\rho_I^{~J}\left| l',m'\right\rangle\\
    &\times\left[\delta_{s,0} {\tau}^I+(1-\delta_{s,0}) {\chi}^I_{s,s_z} \right]\left[\delta_{s',1} \tilde{\tau}_J^{s',s_z'}+(1-\delta_{s',1})\tilde{\chi}_J^{s',s_z'} \right]\,,
    \end{aligned}\\
    &\label{c4_constr_s1}
    \begin{aligned}
    C_{4,j}^{l',s',j'}(r)=&\int d\Omega \left(Y_{j,j_z}\right)^* \frac{1}{\zeta'(r)}\left[\rho^{J}\frac{F'(r)}{F(r)}+\sigma^{J}\right]\tilde{Y}_{(1)J}^{l',s',j',j_z}=\\
    =&\sum\limits_{ m',s_z'}\left\langle l',m',s',s_z' \right| \left. j' , j_z \right\rangle\left[\delta_{0,s'} \tilde{\tau}_J+\left(1-\delta_{0,s'}\right) \tilde{\chi}_J^{s',s_z'} \right]\\
    &\times \frac{1}{\zeta'(r)}\left\langle j,j_z \right|\left[\rho^{J}\frac{F'(r)}{F(r)}+\sigma^{J}\right]\left| l',m'\right\rangle\,,
    \end{aligned}\\
    &\label{c4_constr_s2}
    \begin{aligned}
    C_{4,l,s,j}^{l',s',j'}(r)=&\int d\Omega \left(Y^{(1)I}_{l,s,j,j_z}\right)^* \frac{1}{\zeta'(r)}\left[\tilde\rho_I^{~J}\frac{F'(r)}{F(r)}+\tilde\sigma_I^{~J}\right]\tilde{Y}_{(2)J}^{l',s',j',j_z}=\\
    =&\sum\limits_{m,s_z, m',s_z'}\left\langle l,m,s,s_z \right| \left. j , j_z \right\rangle\left\langle l',m',s',s_z' \right| \left. j' , j_z \right\rangle\\
    &\times \left[\delta_{s,0} {\tau}^I+(1-\delta_{s',0}) {\chi}^J_{s,s_z} \right]\left[\delta_{s',1} \tilde{\tau}_J^{s',s_z'}+(1-\delta_{s',1}) \tilde{\chi}_J^{s',s_z'} \right]\\
    &\times \frac{1}{\zeta'(r)}\left\langle l,m \right|\left[\tilde\rho_I^{~J}\frac{F'(r)}{F(r)}+\tilde\sigma_I^{~J}\right]\left| l',m'\right\rangle\,,
    \end{aligned}\\
    &\label{tildeD_constr_s1}
    \begin{aligned}
    \tilde{D}_{j}^{l',s',j'}(r)=&\int d\Omega \left(Y_{j,j_z}\right)^* \frac{1}{\zeta'(r)}\left[\pi^{J}\boldsymbol{D_+}+\zeta^{J}\boldsymbol{D_0}+\mu^{J}\boldsymbol{D_-}\right]\tilde{Y}_{(1)J}^{l',s',j',j_z}=\\
    =&\sum\limits_{m',s_z'}\frac{1}{\zeta'(r)}\left\langle l',m',s',s_z' \right| \left. j' , j_z \right\rangle \\
    &\times \left[ \sqrt{\frac{(l'+1)^2(l'- m')(l'- m'-1)}{2(2 l'+1)(2 l'-1)}}\left\langle j,j_z \right|\pi^{J}\left| l'-1,m'+1\right\rangle+\right.\\
    &+\sqrt{\frac{l'^2(l'+ m'+1)(l'+ m'+2)}{2(2 l'+3)(2 l'+1)}} \left\langle j,j_z \right|\pi^{J} \left| l'+1,m'+1\right\rangle\\
    &+\sqrt{\frac{l'^2(l'+ m'+1)(l'-m'+1)}{(2 l'+1)(2 l'+3)}}\left\langle j,j_z \right|\zeta^{J}\left| l'+1,m'\right\rangle\\
    &-\sqrt{\frac{(l'+1)^2(l'+ m')(l'- m')}{(2 l'+1)(2 l'-1)}} \left\langle j,j_z \right|\zeta^{J} \left| l'-1,m'\right\rangle\\
    &+\sqrt{\frac{(l'+1)^2(l'+m')(l'+ m'-1)}{2(2 l'+1)(2 l'-1)}}\left\langle j,j_z \right|\mu^{J}\left| l'-1,m'-1\right\rangle\\
    &\left.+\sqrt{\frac{l'^2(l'- m'+1)(l'- m'+2)}{2(2 l'+3)(2 l'+1)}} \left\langle j,j_z \right|\mu^{J} \left| l'+1,m'-1\right\rangle \right]\\
    &\times \left[\delta_{0,s'} \tilde{\tau}_J+\left(1-\delta_{0,s'}\right) \tilde{\chi}_J^{s',s_z'} \right]\,,
    \end{aligned}\\
    &\label{tildeD_constr_s2}
    \begin{aligned}
    \tilde{D}_{l,s,j}^{l',s',j'}(r)&=\int d\Omega \left(Y^{(1)I}_{l,s,j,j_z}\right)^* \frac{1}{\zeta'(r)}\left[\tilde\pi_I^{~J}\boldsymbol{D_+}+\tilde\zeta_I^{~J}\boldsymbol{D_0}+\tilde\mu_I^{~J}\boldsymbol{D_-}\right]\tilde{Y}_{(2)J}^{l',s',j',j_z}=\\
    &=\sum\limits_{m,s_z, m',s_z'}\frac{1}{\zeta'(r)}\left\langle l,m,s,s_z \right| \left. j , j_z \right\rangle\left\langle l',m',s',s_z' \right| \left. j' , j_z \right\rangle \\
    &\times \left[ \sqrt{\frac{(l'+1)^2(l'- m')(l'- m'-1)}{2(2 l'+1)(2 l'-1)}}\left\langle l,m \right|\tilde\pi_I^{~J}\left| l'-1,m'+1\right\rangle+\right.\\
    &+\sqrt{\frac{l'^2(l'+ m'+1)(l'+ m'+2)}{2(2 l'+3)(2 l'+1)}} \left\langle l,m \right|\tilde\pi_I^{~J} \left| l'+1,m'+1\right\rangle\\
    &+\sqrt{\frac{l'^2(l'+ m'+1)(l'-m'+1)}{(2 l'+1)(2 l'+3)}}\left\langle l,m \right|\tilde\zeta_I^{~J}\left| l'+1,m'\right\rangle\\
    &-\sqrt{\frac{(l'+1)^2(l'+ m')(l'- m')}{(2 l'+1)(2 l'-1)}} \left\langle l,m \right|\tilde\zeta_I^{~J} \left| l'-1,m'\right\rangle\\
    &+\sqrt{\frac{(l'+1)^2(l'+m')(l'+ m'-1)}{2(2 l'+1)(2 l'-1)}}\left\langle l,m \right|\tilde\mu_I^{~J}\left| l'-1,m'-1\right\rangle\\
    &\left.+\sqrt{\frac{l'^2(l'- m'+1)(l'- m'+2)}{2(2 l'+3)(2 l'+1)}} \left\langle l,m \right|\tilde\mu_I^{~J} \left| l'+1,m'-1\right\rangle \right]\\
    &\times \left[\delta_{s',1} \tilde{\tau}_J^{s',s_z'}+(1-\delta_{s',1}) \tilde{\chi}_J^{s',s_z'} \right]\left[\delta_{s,0} {\tau}^I+(1-\delta_{s,0}) {\chi}^I_{s,s_z} \right]\,,
    \end{aligned}
\end{align}}
\end{subequations}

\section{Couplings of the spin-2 second-order constraint}\label{couplings_second-order}
In this section we show the full formulae of the radial couplings appearing in equations \ref{decomposed_second-order_constraint}:
\begin{subequations}
{\allowdisplaybreaks[4]
\begin{align}
    &\label{c5_constr}
    \begin{aligned}
    C_{5,j}^{l',s',j'}(r)=&\int d\Omega \left(Y_{j,j_z}\right)^* R^{J}\tilde{Y}_{(2)J}^{l',s',j',j_z}=\\
    =&\sum\limits_{m',s_z'}\left\langle l',m',s',s_z' \right| \left. j' , j_z \right\rangle\\
    &\times\left[\delta_{1,s'} \tilde{\tau}^{s',s_z}_J+\left(1-\delta_{1,s'} \right) \tilde{\chi}_J^{s',s_z'} \right]\left\langle j,j_z \right|R^{J}\left| l',m'\right\rangle\,,
    \end{aligned}\\
    &\label{c6_constr}
    \begin{aligned}
    C_{6,j}^{l',s',j'}(r)=&\int d\Omega \left(Y_{j,j_z}\right)^* \left[R^{J}\frac{2 F'(r)}{F(r)\zeta'(r)}+\frac{\tilde R^{J}}{\zeta'(r)}\right]\tilde{Y}_{(2)J}^{l',s',j',j_z}=\\
    =&\sum\limits_{ m',s_z'}\left\langle l',m',s',s_z' \right| \left. j' , j_z \right\rangle\left[\delta_{1,s'} \tilde{\tau}^{s',s_z'}_J+\left(1-\delta_{1,s'}\right) \tilde{\chi}_J^{s',s_z'} \right]\\
    &\times \left\langle j,j_z \right|\left[R^{J}\frac{2 F'(r)}{F(r)\zeta'(r)}+\frac{\tilde R^{J}}{\zeta'(r)}\right]\left| l',m'\right\rangle\,,
    \end{aligned}\\
    &\label{c7_constr}
    \begin{aligned}
    C_{7,j}^{l',s',j'}(r)=&\frac{1}{\zeta'^2(r)}\int d\Omega \left(Y_{j,j_z}\right)^* \left[A^J l'(l'+1)+\tilde S^{J}+\tilde Q^{J}\mathbfcal{L}_{\boldsymbol z} +\tilde R^{J}\frac{F'(r)}{F(r)}\right.\\
    &\left.+R^{J}\frac{F''(r)}{F(r)}\right]\tilde{Y}_{(2)J}^{l',s',j',j_z}=\\
    =&\frac{1}{\zeta'^2(r)}\sum\limits_{m',s_z'}\left\langle l',m',s',s_z' \right| \left. j' , j_z \right\rangle\left[\delta_{s',1} \tilde{\tau}_J^{s',s_z'}+\left(1-\delta_{s',1}\right) \tilde{\chi}_J^{s',s_z'} \right]\times\\
    &\left\langle j,j_z\right|\left[l'(l'+1) A^J +\tilde S^{J}+m'\tilde Q^{J} +\tilde R^{J}\frac{F'(r)}{F(r)}+R^{J}\frac{F''(r)}{F(r)}\right]\left| l',m'\right\rangle\,,
    \end{aligned}\\
    &\label{L_constr}
    \begin{aligned}
    L_{j}^{l',s',j'}(r)=&\frac{1}{\zeta'^2(r)}\int d\Omega \left(Y_{j,j_z}\right)^* \left[\tilde P_0^{~J}\boldsymbol{D_+}+\tilde Z_0^{~J}\boldsymbol{D_0}+\tilde M_0^{~J}\boldsymbol{D_-}\right.\\
    &\left.+\tilde P^{J}\mathbfcal{L}_++\tilde M^{J}\mathbfcal{L}_-\right]\tilde{Y}_{(2)J}^{l',s',j',j_z}=\\
    =&\sum\limits_{m',s_z'}\left[ \sqrt{\frac{(l'+1)^2(l'- m')(l'- m'-1)}{2(2 l'+1)(2 l'-1)}}\left\langle j,j_z \right|\tilde P_0^{~J}\left| l'-1,m'+1\right\rangle+\right.\\
    &+\sqrt{\frac{l'^2(l'+ m'+1)(l'+ m'+2)}{2(2 l'+3)(2 l'+1)}} \left\langle j,j_z \right|\tilde P_0^{~J} \left| l'+1,m'+1\right\rangle\\
    &+\sqrt{\frac{l'^2(l'+ m'+1)(l'-m'+1)}{(2 l'+1)(2 l'+3)}}\left\langle j,j_z \right|\tilde Z_0^{~J}\left| l'+1,m'\right\rangle\\
    &-\sqrt{\frac{(l'+1)^2(l'+ m')(l'- m')}{(2 l'+1)(2 l'-1)}} \left\langle j,j_z \right|\tilde Z_0^{~J} \left| l'-1,m'\right\rangle\\
    &+\sqrt{\frac{(l'+1)^2(l'+m')(l'+ m'-1)}{2(2 l'+1)(2 l'-1)}}\left\langle j,j_z \right|\tilde M_0^{~J}\left| l'-1,m'-1\right\rangle\\
    &+\sqrt{\frac{l'^2(l'- m'+1)(l'- m'+2)}{2(2 l'+3)(2 l'+1)}} \left\langle j,j_z \right|\tilde M_0^{~J} \left| l'+1,m'-1\right\rangle\\
    &+\sqrt{(l'-m')(l'+m'+1)}\left\langle j,j_z \right|\tilde P^{J}\left| l',m'+1\right\rangle\\
    &+\sqrt{(l'+m')(l'-m'+1)}\left\langle j,j_z \right|\tilde M^{J}\left| l',m'-1\right\rangle \Bigg]\\
    &\times \left[\delta_{1,s'} \tilde{\tau}^{s',s_z'}_J+\left(1-\delta_{1,s'}\right) \tilde{\chi}_J^{s',s_z'} \right]\frac{1}{\zeta'^2(r)}\left\langle l',m',s',s_z' \right| \left. j' , j_z \right\rangle\,,
    \end{aligned}
\end{align}}
\end{subequations}

\chapter{Complements for the bra-ket overlap integrals}
\section{Alternative derivation of the trigonometric spherical harmonic bra-kets}\label{referee_alternative}

An anonymous referee, to whom we express our gratitude, suggested the following alternative derivation for integrals (\ref{bra_ket1}) and (\ref{bra_ket2}). Legendre polynomials can be expanded as linear combinations of Čebyšëv polynomials via a Fourier transform, whose coefficients can be found in Example~15.1.2 of \cite{whittaker_watson_1996}:
\begin{equation}\label{Lagrange_expand}
    P_l(\cos\theta)=\sum\limits_{j=0}^{\floor{l/2}} a_{l,j} T_{l-2j}(\cos\theta),\quad 
    a_{l,j}=\frac{2 (2l-2j-1)!!(2j-1)!!}{(1+\delta_{l-2j,0})(2l-2j)!!(2j)!!}\,.
\end{equation}
Thus, by exploiting the properties of the product of Čebyšëv polynomials\cite{special_polynomials} and using the fact that
\begin{equation}
\begin{aligned}
    \int_{-1}^1 dx T_n(x)&=\int_0^\pi d\theta\sin\theta \cos(n\theta)=\\
    &=\int_0^\pi d\theta\frac{\sin((n+1)\theta)-\sin((n-1)\theta)}{2}=\\
    &=\frac{(-1)^n+1}{1-n^2}(1-\delta_{n,1}) \,,
\end{aligned}
\end{equation}
we can solve the following integral:
\begin{equation}\label{proj_int}
\begin{aligned}
    I_{n,l}&=\int_0^\pi d\theta \sin\theta \cos(n\theta)P_l(\cos\theta)=\\
    &=\sum\limits_{j=0}^{\floor{l/2}} a_{l,j}\int_{-1}^1 dx T_n(x) T_{l-2j}(x)=\\
    &=\sum\limits_{j=0}^{\floor{l/2}} \frac{a_{l,j}}{2}\int_{-1}^1 dx \left[T_{n+l-2j}(x) +T_{|n-l+2j|}(x)\right]=\\
    &=\sum\limits_{j=0}^{\floor{l/2}} a_{l,j}\left[\frac{1}{1-(n+l-2j)^2}+\frac{1}{1-(n-l+2j)^2}\right]\delta_{((n+l)\bmod 2),0}.
\end{aligned}
\end{equation}
Because of the orthogonality of Legendre polynomials\cite{special_polynomials},
\begin{equation}
    \int_{-1}^1 dx P_l(x) P_{l'}(x)=\frac{2}{2l+1}\delta_{l,l'}\,,
\end{equation}
the coefficients $b^n_l$ and $a^n_l$ appearing in Eq.~(\ref{cos-sin-legendre0}) can be written in terms of the integrals (\ref{proj_int}) as follows
\begin{equation}
    b^n_l=\left(l+\frac{1}{2}\right)I_{n,l}~,~\qquad a^n_l=-\frac{1}{n}\left(l+\frac{1}{2}\right)I_{n,l}\,.
\end{equation}
Therefore, we get the following alternative expressions for the bra-ket integrals:
\begin{subequations}
\begin{align}
    &\label{cos_braket_var}\begin{aligned}
        &\left\langle l_1, m_1 \right| e^{i k \phi} \cos(n \theta) \left| l_2 , m_2\right\rangle=\\
        &=\sum\limits_{l=|l_1-l_2|}^{l_1+l_2} \sum\limits_{j=0}^{\floor{n/2}}\frac{(-1)^{m_1}}{2}\left(n-2j+\frac{1}{2}\right)I_{n,n-2j}\sqrt{\frac{(2 l_1+1)(2 l_2+1)(l+k)!}{(l-k)!}}\\
        &\times  \left\langle l_1 , -m_1 , l_2 , m_2 \right| \left. l , -k\right\rangle \left\langle l_1 , 0 , l_2 , 0 \right| \left. l , 0\right\rangle I(n-2 j, 0, l,-k)\,,
    \end{aligned}\\
    &\label{sin_braket_var}\begin{aligned}
        &\left\langle l_1, m_1 \right| e^{i k \phi} \sin(n \theta) \left| l_2 , m_2\right\rangle=\\
        &=\sum\limits_{l=|l_1-l_2|}^{l_1+l_2} \sum\limits_{j=0}^{\floor{n/2}}\frac{(-1)^{m_1+1}}{2 n}\left(n-2j+\frac{1}{2}\right)I_{n,n-2j}\sqrt{\frac{(2 l_1+1)(2 l_2+1)(l+k)!}{(l-k)!}}\\
        &\times \left\langle l_1 , -m_1 , l_2 , m_2 \right| \left. l , -k\right\rangle \left\langle l_1 , 0 , l_2 , 0 \right| \left. l , 0\right\rangle I(n-2 j, 1, l,-k)\,.
    \end{aligned}
\end{align}
\end{subequations}
This alternative derivation is more straightforward than the one we found, but it gives less compact expressions involving double factorials and an additional sum instead of the gamma functions appearing in our solutions. With these alternative formulas we observe a slightly higher computation time with respect to the expressions we found, which can become relevant when there are many integrals of the type (\ref{bra_ket1}) and (\ref{bra_ket2}) to be computed.

\section{Simplified solutions for the axisymmetric trigonometric spherical harmonic bra-kets}\label{simplified_axisym_braket}

The formulas appearing in the alternative derivation described in \ref{referee_alternative} can be used also for finding simplified expressions in the axisymmetric case ($k=0$). In this simplified case, the key integrals appearing in Eq.~(\ref{cos_braket0}) involve just Legendre polynomials instead of Legendre associated functions, thus coinciding with integrals \ref{proj_int}. Consequently, Eq.~(\ref{cos_braket0}) for $k=0$ can be reduced to
\begin{equation}\label{cos_braket0k0}
\begin{aligned}
    \left\langle l_1, m_1 \right| \cos(n \theta) \left| l_2 , m_2\right\rangle&=\sum\limits_{l=|l_1-l_2|}^{l_1+l_2}\sqrt{\left( l_1+\frac{1}{2}\right)\left( l_2+\frac{1}{2}\right)}(-1)^{m_1}\\
    &\times\left\langle l_1 , 0 , l_2 , 0 \right| \left. l , 0\right\rangle\left\langle l_1 , -m_1 , l_2 , m_2 \right| \left. l , 0\right\rangle I_{n,l}\,.
\end{aligned}
\end{equation}
We can find a similar procedure also for the sine case, shown in Eq.~(\ref{sin_braket0}). From the trigonometric product-to-sum formulas we have
\begin{equation}\label{product-to-sum}
\begin{aligned}
    \sin\theta\sin(n\theta)&=\frac{1}{2}\left[\cos((n-1)\theta)-\cos((n+1)\theta) \right]=\\
    &=\frac{T_{n-1}(\cos\theta)-T_{n+1}(\cos\theta)}{2}
\end{aligned}    
\end{equation}
and, therefore, the key integrals appearing in (\ref{sin_braket0}) become
\begin{small}
\begin{equation}
\begin{aligned}
    \int_0^\pi d\theta \sin\theta \sin(n\theta) P_l(\cos\theta)&=\sum\limits_{j=0}^{\floor{l/2}} \frac{a_{l,j}}{2}\int_0^\pi d\theta \left[T_{n-1}(\cos\theta)-T_{n+1}(\cos\theta)\right] T_{l-2j}(x)=\\
    &=\sum\limits_{j=0}^{\floor{l/2}}\frac{a_{l,j}}{4}\int_0^\pi d\theta\Bigl[T_{n-1+l-2j}(\cos\theta) -T_{|n-1-l+2j|}(\cos\theta)\\
    &-T_{n+1+l-2j}(\cos\theta)+T_{|n+1-l+2j|}(\cos\theta)\Bigr]=0\,.
\end{aligned} 
\end{equation}
\end{small}
Hence, equation (\ref{sin_braket0}) in this case reduces to
\begin{equation}
    \left\langle l_1, m_1 \right| \sin(n \theta) \left| l_2 , m_2\right\rangle=0\,,
\end{equation}
which can also be deduced by considering the parity of the functions involved.

\section{Formulas for the spherical harmonic decomposition of the Regge-Wheeler tensor basis}\label{appendix_Regge-Wheeler}

From \ref{L+L-}, \ref{L+L-_spherical_coordinates}, \ref{angular_operators}, \ref{D+-} and \ref{D0} we get the following expressions for the angular functions appearing in \ref{Regge-Wheeler}:
\begin{subequations}{\allowdisplaybreaks[4]
\begin{align}&\begin{aligned}
    X_{j, j_z}=& i\left[ (j_z + 1) \sqrt{(j - j_z) (j + j_z + 1)} e^{-i \phi}Y_{j,j_z+1} \right.\\
    &\left.- (j_z - 1) \sqrt{(j+j_z) (j-j_z + 1)}  e^{i \phi}Y_{j,j_z-1}\right]
\end{aligned}\\
&\begin{aligned}
W_{j,j_z}=&\frac{1}{2} \left\lbrace\sqrt{(j - j_z) [j^2 - (j_z + 1)^2] (j + j_z + 2)}e^{-2 i \phi}Y_{j, j_z + 2} +\right.\\
&\left.+ 2 j_z^2 Y_{j, j_z} +\sqrt{(j + j_z) [j^2 - (j_z - 1)^2] (j - j_z + 2)} e^{2 i\phi}Y_{j, j_z - 2}\right\rbrace
\end{aligned}\\
&\begin{aligned}
    \csc\theta \partial_\phi Y_{j,j_z}=& -\frac{i}{2} \left\lbrace e^{-i \phi} \left[\sqrt{\frac{(j+ 1)^2 (j - j_z) (j - j_z - 1)}{(2 j + 1) (2 j - 1)}}Y_{j - 1, j_z + 1}\right.\right.\\
    &\left.+\sqrt{\frac{j^2 (j + j_z + 1) (j + j_z + 2)}{(2 j + 3) (2 j + 1)}}Y_{j + 1, j_z + 1}\right]\\
    & +e^{i \phi}\left[\sqrt{\frac{(j + 1)^2 (j + j_z) (j + j_z - 1)}{(2 j + 1) (2 j-1)}}\right.\\
    &\left.\left.\times Y_{j - 1, j_z - 1} + \sqrt{\frac{j^2 (j - j_z + 1) (j - j_z + 2)}{(2j + 3) (2 j + 1)}}Y_{j + 1, j_z - 1}\right]\right\rbrace
\end{aligned}\\
&\begin{aligned}
    \sin\theta\partial_\theta Y_{j,j_z}=&\sqrt{\frac{j^2 (j + j_z + 1) (j - j_z + 1)}{(2 j + 1) (2 j + 3)}}Y_{j + 1, j_z} \\&- \sqrt{\frac{(j + 1)^2 (j + j_z) (j - j_z)}{(2 j + 1) (2 j - 1)}}Y_{j - 1, j_z}
\end{aligned}\\
&\begin{aligned}
    \partial_\theta Y_{j,j_z}= \frac{1}{2}\left[\sqrt{(j - j_z) (j + j_z + 1)}e^{-i \phi}Y_{j,j_z+1} -\sqrt{(j + j_z) (j - j_z + 1)}e^{i \phi}Y_{j, j_z - 1}\right]
\end{aligned}
\\
&\begin{aligned}
    \csc\theta X_{j,j_z}=&\frac{i}{2} \left\lbrace-\sqrt{(j - j_z) (j + j_z + 1)} \left[e^{-2 i \phi}\left(\sqrt{\frac{(j + 1)^2 (j - j_z - 1) (j - j_z - 2)}{(2 j + 1) (2 j - 1)}}\right.\right.\right.\\
    &\times\left. Y_{j - 1, j_z + 2} + \sqrt{\frac{j^2 (j + j_z+ 2) (j + j_z + 3)}{(2 j + 3) (2 j + 1)}}Y_{j + 1, j_z + 2}\right)\\ & + \sqrt{\frac{j^2 (j - j_z) (j - j_z + 1)}{(2 j + 3) (2 j + 1)}} \left. Y_{j + 1, j_z}+\sqrt{\frac{(j + 1)^2 (j + j_z + 1) (j + j_z)}{(2j + 1) (2j - 1)}}Y_{j - 1, j_z}\right]\\
    &+\sqrt{(j + j_z) (j - j_z + 1)} \Bigg[ e^{2 i \phi}\left(\sqrt{\frac{(j + 1)^2 (j + j_z - 1) (j + j_z -2)}{(2 j + 1) (2 j - 1)}}Y_{j - 1, j_z - 2}\right.\\
    &+\sqrt{\frac{j^2 (j - j_z + 2) (j - j_z + 3)}{(2 j + 3) (2 j + 1)}} Y_{j + 1, j_z - 2}\Bigg)+\sqrt{\frac{(j + 1)^2 (j - j_z + 1) (j - j_z)}{(2 j + 1) (2 j - 1)}} \\
    &\times Y_{j - 1, j_z}+\sqrt{\frac{j^2 (j+j_z) (j+j_z + 1)}{(2 j + 1) (2 j + 3)}} Y_{j + 1, j_z}\Bigg]\Bigg\rbrace
\end{aligned}
\end{align}}
\end{subequations}

\end{appendix}

\backmatter
\cleardoublepage
\phantomsection

\bibliographystyle{utphys}
\bibliography{mybibfile}

\end{document}